\documentclass[eqsecnum, aps, amsmath, amssymb, singlecolumn, superscriptaddress, 11pt ]{revtex4-1}

\usepackage{graphicx}
\usepackage{dcolumn}
\usepackage[colorlinks=true,
            linkcolor=blue,
            urlcolor=blue,
            citecolor=blue]{hyperref}
\usepackage{bm}
\usepackage{enumerate}
\usepackage{lipsum}
\usepackage{threeparttable}
\usepackage{epstopdf}
\usepackage{float}
\usepackage{xcolor,colortbl}
\usepackage[caption = false]{subfig}
\usepackage{tabularx}
\usepackage{chngpage}

\begin{document}


\title{Impact of the transverse direction on the many-body tunneling dynamics in a two-dimensional bosonic Josephson junction}
\author{Anal Bhowmik}
\email{abhowmik@campus.haifa.ac.il}
\affiliation{Department of Mathematics, University of Haifa, Haifa 3498838, Israel}
\affiliation{Haifa Research Center for Theoretical Physics and Astrophysics, University of Haifa, Haifa 3498838, Israel}
\author{Sudip Kumar Halder}
\affiliation{Department of Mathematics, University of Haifa, Haifa 3498838, Israel}
\affiliation{Haifa Research Center for Theoretical Physics and Astrophysics, University of Haifa, Haifa 3498838, Israel}
\affiliation{Department of Physics, SRM University Delhi-NCR, Plot No. 39 Rajiv Gandhi education city, Sonipat 131029, India}
\author{Ofir E. Alon}
\affiliation{Department of Mathematics, University of Haifa, Haifa 3498838, Israel}
\affiliation{Haifa Research Center for Theoretical Physics and Astrophysics, University of Haifa, Haifa 3498838, Israel}




\date{\today}

\begin{abstract}
Tunneling in a many-body system appears as one of the novel implications of quantum physics, in which particles move in space under an otherwise classically-forbidden potential barrier. Here, we theoretically describe the quantum dynamics of the tunneling phenomenon of a few intricate bosonic clouds in a closed system of a two-dimensional symmetric double-well potential. We examine how the inclusion of the transverse direction, orthogonal to the junction of the double-well, can intervene in the tunneling dynamics of bosonic clouds. We use a well-known many-body numerical method, called the multiconfigurational time-dependent Hartree for bosons (MCTDHB) method. MCTDHB allows one to obtain accurately the time-dependent many-particle wavefunction of the bosons which in principle entails all the information of interest about the system under investigation. We analyze the tunneling dynamics by preparing the initial state of the bosonic clouds in the left well of the double-well either as the ground, longitudinally or transversely excited, or a vortex state. We unravel the detailed mechanism of the tunneling process by analyzing the evolution in time of the survival probability, depletion and fragmentation, and the many-particle position, momentum, and angular-momentum expectation values and their variances. As a general rule, all objects lose coherence while tunneling through the barrier and the states which include transverse excitations do so faster.  Implications are briefly discussed.
\end{abstract}

\maketitle
\section{INTRODUCTION}
After the experimental observations of Bose-Einstein condensation (BEC) \cite{Streltsov1995, Bradley1995, Davis1995},
ultra-cold quantum gases have emerged as one of the most advanced platforms to mimic a wide variety of typical
models of condensed-matter physics, optics, high-energy physics,
and even of quantum biology and chemistry \cite{Lewenstein2007, Lewenstein2012, Lee2006, Hild2014, Fukuhara2013, Buchler2005, Jaksch2002, Ferlaino2011}.
One of such well known paradigms of quantum physics is the existence of Josephson effect which is a clear manifestation of the macroscopic quantum coherence, originally predicted for superconductors \cite{Josephson1962}
and later observed in superfluid $^3$He \cite{Davis2002} and gaseous BECs \cite{Smerzi1997, Albiezet2005}. When ultra-cold bosons are tunneling in a double-well potential, the system is usually referred to as bosonic Josephson junction (BJJ) \cite{Gati2007}.

An extensive theoretical study of trapped BECs in one-dimensional double-well potentials
is available using 
a variety of theoretical approaches
\cite{Dobrzyniecki2016, Menotti2001, Salgueiro2007, Zollner2008, Carr2010, He2012, Liu2015,  Dobrzyniecki2018, Dobrzyniecki2018a, Ferrini2008, Jia2008, Burchinati2017, Pawlowski2011, Griffin2020}.
Some of the interesting features, such as Josephson oscillations \cite{Smerzi1997, Gillet2014, Levy2007, Burchinati2017} and self trapping \cite{Smerzi1997, Levy2007} have been reported using a two-mode theory.
A full many-body Schr\"{o}dinger dynamics 
starting from the ground state of the BEC in one of the wells
(hereafter for brevity, ground state of the BEC)
has been studied in one-dimensional double-well potentials 
and shows the development of fragmentation and loss of coherence
in the BEC \cite{Sakmann2009, Sakmann2010, Sakmann2014, Halder2018, Halder2019}.
The uncertainty product of the many-particle position and momentum operators \cite{Klaiman2016}
as well as the evolution in time of the position and momentum variances \cite{Halder2019} have been studied
by solving the full many-body Schr\"{o}dinger equation.
The development of fragmentation of the ground state of the BEC validates the necessity of a many-body treatment 
in order to obtain the accurate dynamical behavior of bosons in the BJJ.

Tunneling dynamics of the ground state of a trapped BEC 
in higher dimensions has also been explored using two-mode or improved two-mode models \cite{Ananikian2006, Spagnolli2017}. Ananikian and Bergeman showed that when the extent of the wave function in each well vary appreciably with time,
the nonlinear interaction term creates a temporal change in the tunneling energy or rate \cite{Ananikian2006}.
Spagnolli \textit{et al.} reported a detailed study of the transition from Rabi to plasma oscillations
by crossing over from the attractive to repulsive inter-atomic interaction in terms of the evolution of atomic imbalance \cite{Spagnolli2017}. Moreover, in two dimensions (2D), the tunneling dynamics of trapped vortices were studied using
Gross-Pitaevskii  mean-field model in 2D superfluids \cite{Arovas2008},
in an harmonic potential with a Gaussian potential barrier \cite{Martin2007},
and between two pinning potentials \cite{Fialko2012}. 
Salgueiro \textit{et al.} proposed a method of generating replicas of a vortex state in
 a double-well potential formed by conjoining two Gaussian potentials
using the mean-field approach \cite{Salgueiro2009}. 
Garcia-March and Carr showed a comparative study of the tunneling of axisymmetric and transverse vortex structures \cite{March2015}. The most of the literature in relation with ground and vortex states of a BEC
in a two-dimensional double-well potential are devoted to the density oscillations between the wells.
There is a recent study of tunneling dynamics of the vortex state using an in-principle numerically-exact many-body theory 
in a 2D radial double-well trap \cite{Beinke2015}.
Beinke \textit{et al.} showed that the development of the fragmentation of the vortex state
is accompanied by damping of the amplitude of the survival probability,
thereby indicating the importance of the accurate many-body theoretical treatment \cite{Beinke2015}. Moreover, on a different note,   the hidden vortices in a rotating double-well potential \cite{Wen2010}, excitation of non-zero angular-momentum modes in tunnel-coupled two-dimensional Bose gas \cite{Montgomery2010}, and creation of vortices in a BEC by external laser beam with orbital angular-momentum \cite{Schmiegelow2016, Bhowmik2016, Bhowmik2018} are   studied in the literature.
Although there is some literature discussing the dynamics of the ground and vortex states in a double-well potential, there is no detailed investigation of the inter-connection of the density oscillations with the time evolution of quantum mechanical observables and their variances, let alone beyond one spatial dimension. Furthermore,  to the best of our knowledge, 
there is no available literature which discusses 
the tunneling dynamics of complicated bosonic objects  in a double-well potential by solving the many-particle problem at the many-body level of theory.

The main focus of this work is to explain the physics behind the tunneling dynamics of a few intricate bosonic clouds
in a 2D double-well potential by analyzing the time evolution of various physical quantities,
focusing on tunneling scenarios and research questions 
which require at least a 2D geometry to investigate.
In order to explore the tunneling dynamics in a 2D symmetric double-well with the junction along the $x$ direction,
we consider four basic structures of bosonic clouds in the harmonic potential,  namely, ground, $x$-excited, $y$-excited, and vortex states.
Per definition, the $y$-excited and vortex states have no one-dimensional analogs.
Although, there are one-dimensional analogs to the ground and $x$-excited states, 
we ask how the inclusion of the transverse direction can affect the overall dynamics of all four initial states. A general question we ask is if there is any  difference between the many-body and mean-field dynamics in the 2D BJJ.
We ask whether and how quantum correlations develop in the process of tunneling
for the initial states considered here.
Will there be any qualitative and quantitative differences in the correlations due to the different
initial structures of the bosonic clouds?
Therefore, to investigate the tunneling dynamics in detail in a 2D double-well,
we need to solve the many-body Schr\"{o}dinger equation numerically accurately.
A particularly suitable approach to solve the full Schr\"{o}dinger equation is called the multiconfigurational time-dependent
Hartree for bosons (MCTDHB) method \cite{Streltsov2007, Alon2008,  Lode2020}.

{
In this paper, we show that the ground and excited states can tunnel through the barrier without destroying their initial structures. But the vortex state creates two vortex dipoles in  the tunneling process  and the dipoles rotate around the minima of the respective well. We find that the creation of the dipoles from the vortex states  relies on the tunneling of the excited states considered here. We observe a difference between the mean-field and many-body density oscillations in the long-time dynamics for all objects due to the growing degree of quantum correlations in the later. We show that the fragmentation develops faster when there is transverse excitation in the system. Moreover, the mechanism  of the development of fragmentation exhibits significant differences when there are transverse excitations.    All in all, we have studied the time evolution of a purely many-body quantity, fragmentation, and discussed its  impact on the survival probability, expectation values and  variances.  We  find an interconnection between the density oscillations and some quantum mechanical quantities by accurately calculating the time evolution of the survival probability and  the many-particle position, momentum, and angular-momentum  expectation values and their variances, both at the mean-field and  many-body levels. As the variance is a sensitive probe of correlations \cite{Klaiman2015}, even when the bosons are fully condensed,  comparisons of the mean-field and many-body variances show that the correlations have different impact on the different physical quantities  depending on the initial structure of the bosonic cloud and the presence of transverse excitations in the system.

\section{System and methodology}
According to the time-dependent many-body Schr\"{o}dinger equation,
the dynamics of $N$ interacting structureless bosons are governed by 
\begin{equation}\label{1}
\hat{H}\Psi = i\dfrac{\partial \Psi}{\partial t}, \qquad
\hat{H}(\textbf{r}_1, \textbf{r}_2,\ldots, \textbf{r}_N)=
\sum_{j=1}^{N} \hat{h}(\textbf{r}_j)+\sum_{j<k} \hat{W}(\textbf{r}_j-\textbf{r}_k).
\end{equation}
Here, $\hat{h}(\textbf{r})=\hat{T}(\textbf{r})+\hat{V}(\textbf{r})$ is the one-particle Hamiltonian where $\hat{T}(\textbf{r})$ and $\hat{V}(\textbf{r})$ represent the kinetic energy and trap potential, respectively.
$\hat{W}(\textbf{r}_j-\textbf{r}_k)$ is a short-range repulsive inter-particle interaction
modeled by a Gaussian function, $W(\textbf{r}_1-\textbf{r}_2)=\lambda_0\dfrac{e^{-(\textbf{r}_1-\textbf{r}_2)^2/2\sigma^2}}{2\pi\sigma^2}$ with $\sigma=0.25$,  to avoid the regularization problems of the zero-ranged contact potential in two spatial dimensions \cite{Doganov2013, Christensson2009, Klaiman2014, Beinke2015, Beinke2018}.  The particular shape of the inter-particle interaction model potential does not impact the physics of the bosons to be described below.}
To quantify the interaction strength, 
the mean-field interaction parameter $\Lambda=\lambda_0(N-1)$ is standardly introduced.
Throughout this work, $\textbf{r}=(x,y)$ is the position vector in two spatial dimensions 
and the natural units $\hbar=m=1$ are employed. 

We solve the time-dependent many-boson Schr\"odinger equation presented in Eq.~\ref{1}
using the MCTDHB method \cite{Streltsov2007, Alon2008,  Sakmann2009, Grond2009, Grond2011, Streltsov2013, Streltsova2014, Klaiman2014, Fischer2015, Tsatsos2015, Weiner2017, Beinke2015, Schurer2015, Lode2016, Lode2017, Weiner2017, Lode2018, Klaiman2018, Halder2018, Alon2018, Chatterjee2019, Halder2019, Alon2019a, Alon2019b, Bera2019, Lin2020}. The method is well documented and applied in the literature \citep{Lode2020}. Detailed derivation of the MCTDHB equation of motions is described in \cite{Alon2008}.  For our numerical computations, we use the numerical implementation in \cite{Package_1, Package_2}.  MCTDHB uses the ansatz 
\begin{equation}\label{2}
|\Psi(t)\rangle =\sum_{{\{n\}}}C_\textbf{n}(t)|\textbf{n};t\rangle
\end{equation}
where $|\textbf{n};t\rangle= |n_1,n_2,\ldots,n_M;t\rangle$ are the time-dependent permanents
obtained by distributing $N$ bosons in $M$ time-adaptive single-particle orbitals.
In the limit $M\rightarrow \infty$, the permanents $|\textbf{n};t\rangle$ span
the complete $N-$particle Hilbert space and the expansion in Eq.~\ref{2} becomes formally exact. The usage of time-adaptive permanents allows one to solve the time-dependent Schr\"odinger equation numerically accurately with finite, often quite small number of orbitals $M$ \cite{Lode2012}.  At the opposite end, for $M=1$, Eq.~\ref{2} becomes the Gross-Pitaevskii ansatz and solves the time-dependent Gross-Pitaevskii equation. 

The main theme of this work is to explore the  dynamical behavior of the ground, longitudinally and transversely excited, and vortex states in a symmetric 2D double-well in terms of different physical quantities such as the survival probability, depletion and fragmentation, and the many-particle position, momentum, and angular-momentum variances. These would help us to extract relevant information embedded in the $N$-boson time-dependent wavefunction and shed light on the physics of tunneling in the junction. We begin our analysis by preparing the initial state either as the ground ($\Psi_G$),  $x$-excited ($\Psi_X$) or $y$-excited ($\Psi_Y$),
or a linear combination of $\Psi_X$ and $\Psi_Y$, i.e. a vortex state ($\Psi_V$),
of non-interacting many bosons at the left well of a 2D symmetric double-well potential.
The double well potential is formed by fusing together two harmonic potentials,
$V_L(x,y)=\dfrac{1}{2}(x+2)^2+\dfrac{1}{2}y^2$ and
$V_R(x,y)=\dfrac{1}{2}(x-2)^2+\dfrac{1}{2}y^2$,
where $V_L(x,y)$ and $V_R(x,y)$ represent the left and right wells of the trap potential, respectively,
with a quadratic polynomial $\dfrac{3}{2}(1-x^2)+\dfrac{1}{2}y^2$ in the region $|x|\leq \dfrac{1}{2}$ and $-\infty< y<\infty$,
and is given by 
\begin{equation}\label{3}
V_T(x,y)=
 \begin{cases}
 \dfrac{1}{2}(x+2)^2+\dfrac{1}{2}y^2, \hspace{0.5cm}  x<-\dfrac{1}{2}, \hspace{0.2cm} \hspace{0.5cm} -\infty< y<\infty,  \\
      \dfrac{3}{2}(1-x^2)+\dfrac{1}{2}y^2, \hspace{0.6cm} |x|\leq \dfrac{1}{2},  \hspace{0.2cm} \hspace{0.5cm} -\infty< y<\infty,  \\
      \dfrac{1}{2}(x-2)^2+\dfrac{1}{2}y^2, \hspace{0.5cm}  x>+\dfrac{1}{2}, \hspace{0.2cm}   \hspace{0.5cm} -\infty< y<\infty.
 \end{cases}
\end{equation}
 $V_T(x,y)$ is a natural 2D generalization of the one-dimensional potential used, e.g., in \cite{Klaiman2016}.
The mathematical forms of the initial conditions, $\Psi_G$, $\Psi_X$, $\Psi_Y$, and $\Psi_V$, are taken as $\Psi_G=\dfrac{1}{\sqrt{\pi}}F(x,y)$, $\Psi_X=\sqrt{\dfrac{2}{\pi}}(x+2)F(x,y)$, $\Psi_Y=\sqrt{\dfrac{2}{\pi}}y F(x,y)$, and $\Psi_V=\frac{1}{\sqrt{2}}(\Psi_X+i\Psi_Y)$, where $F(x,y)=exp[-\{(x+2)^2+y^2\}/2]$.
In order to investigate the time evolution of the prepared initial states,
we suddenly quench the inter-particle interaction
at $t=0$ from $\Lambda=0$ to $\Lambda=0.01$ accompanied by the change of trapping potential 
from the initial single-well, $V_L(x,y)$, to the final double-well, $V_T(x,y)$, potential.
The consistency of the initial-state preparation is discussed in the supplemental material.
Now, we will investigate the tunneling dynamics of the considered bosonic clouds in the symmetric double-well potential $V_T(x,y)$.

\section{The tunneling dynamics and its analysis}

In this section, we explore in detail the time evolution of various physical quantities
for a collection of bosons trapped in a 2D symmetric double-well.
In particular, we are interested to show the time variation of the survival probability in the left well, 
the degree of fragmentation of the bosons, and the expectation values and variances of the position, momentum, and angular-momentum 
many-particle operators
for the initial states, $\Psi_G$, $\Psi_X$, $\Psi_Y$, and $\Psi_V$. These quantities draw increasingly more involved information from the time-dependent many-boson wave function,
namely, from the density, reduced one-particle density matrix,
and reduced two-particle density matrix, respectively. As the double-well potential is symmetric,  preparation of the initial state either in left or right well does not affect the quantities discussed here. 

Our research approach is a combined
investigation of the dynamics at the mean-field and many-body levels of theory. The MCTDHB theory incorporates the correlations among the bosons,
therefore to highlight the many-body effects, we compare the many-body survival probability, expectation  values,  and variances computed using the MCTDHB method with the corresponding mean-field ($M=1$ time-adaptive orbitals) results. 
In our work, we have performed all the many-body computations for $\Psi_{G}$ and $\Psi_{X}$ 
using $M=6$ time-adaptive orbitals, 
while for $\Psi_{Y}$ and $\Psi_{V}$ using $M=10$ time-adaptive orbitals. We shall see later on that the inclusion of the transverse excitations generally requires
more time-adaptive orbitals to faithfully represent the many-body dynamics. In order to check the convergence with respect to the orbital numbers,
we have repeated our computations with $M=10$ orbitals for $\Psi_{G}$ and $\Psi_{X}$ 
and with $M=12$ orbitals for $\Psi_{Y}$ and $\Psi_{V}$; all the results are found to be well converged, see  the supplemental material for more details. 
For the numerical solution, we use a grid of $64^2$ points in a box of size $[-10, 10) \times [-10, 10)$ with periodic boundary conditions. Convergence of the results with respect to the number of grid points has been verified using a grid of $128^2$ points,
see the supplemental material.
All many-body computations are carried out for a finite number of bosons, $N=10$, with the inter-boson interaction $\Lambda=0.01$. The mean-field computations are done for the same interaction parameter $\Lambda=0.01$.
Therefore, one can relate the tunneling dynamics of the bosonic clouds between the many-body and mean-field levels.
Furthermore, it is instructive to mention that all the systems mentioned here are weakly interacting,
which allows us to mimic the so-called infinite-particle limit of the  interacting bosons,
at least for very short times. We set the time-scale for the dynamics equal to 
the period of the Rabi oscillations $(t_{\text{Rabi}})$ in the double well trap presented in Eq.~\ref{3}.
Here $t_{\text{Rabi}}=\dfrac{2\pi}{\Delta E}=132.498$, where $\Delta E$ is the energy difference between the ground state and first excited state, calculated by diagonalizing the single-particle Hamiltonian using discrete variable representation method.  The ground and excited states in the 2D double-well are the even and odd functions along the $x$-direction. 
We shall use the same time scale for all tunneling processes discussed below,
to facilitate a direct comparison between them.

\subsection{Density and survival probability}
\begin{figure*}[!h]
{\includegraphics[trim = 0.1cm 0.5cm 0.1cm 0.2cm, scale=.60]{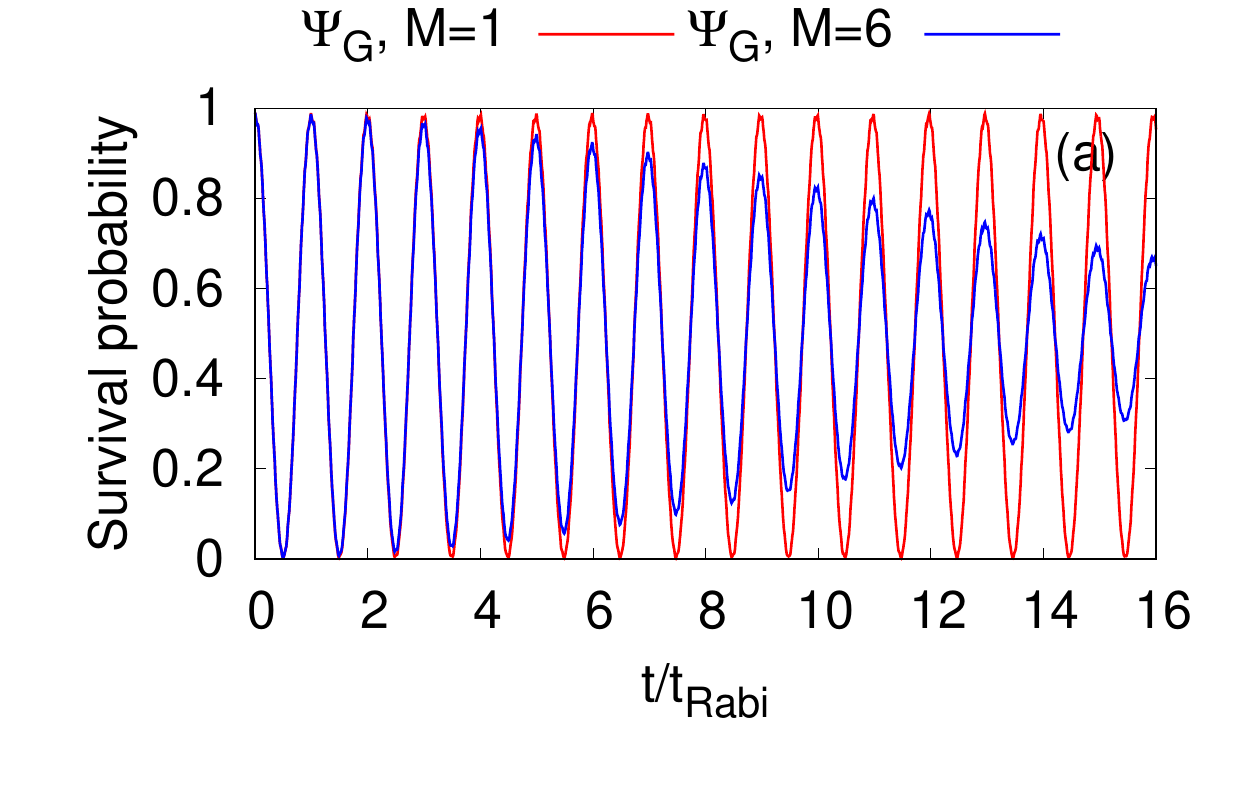}}
{\includegraphics[trim =  0.1cm 0.5cm 0.1cm 0.2cm, scale=.60]{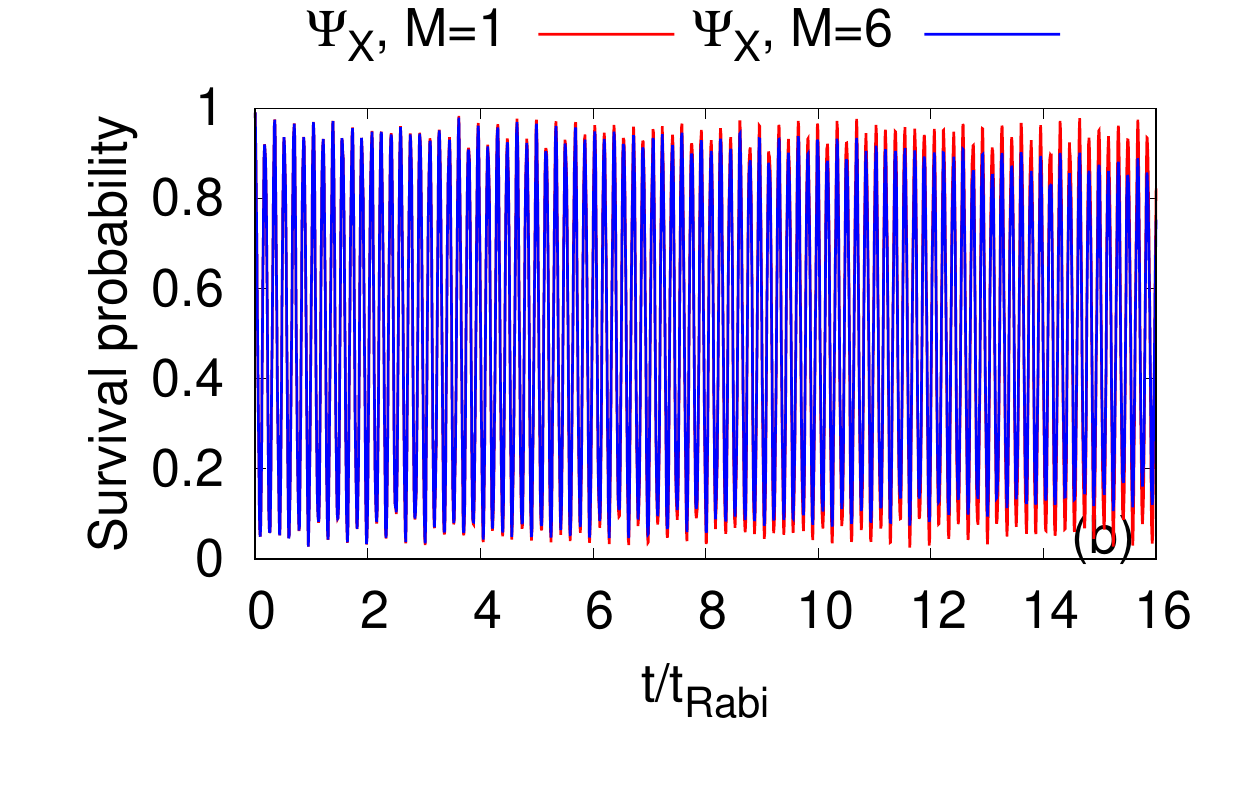}}\\
\vglue 0.25 truecm 
{\includegraphics[trim =  0.1cm 0.5cm 0.1cm 0.2cm, scale=.60]{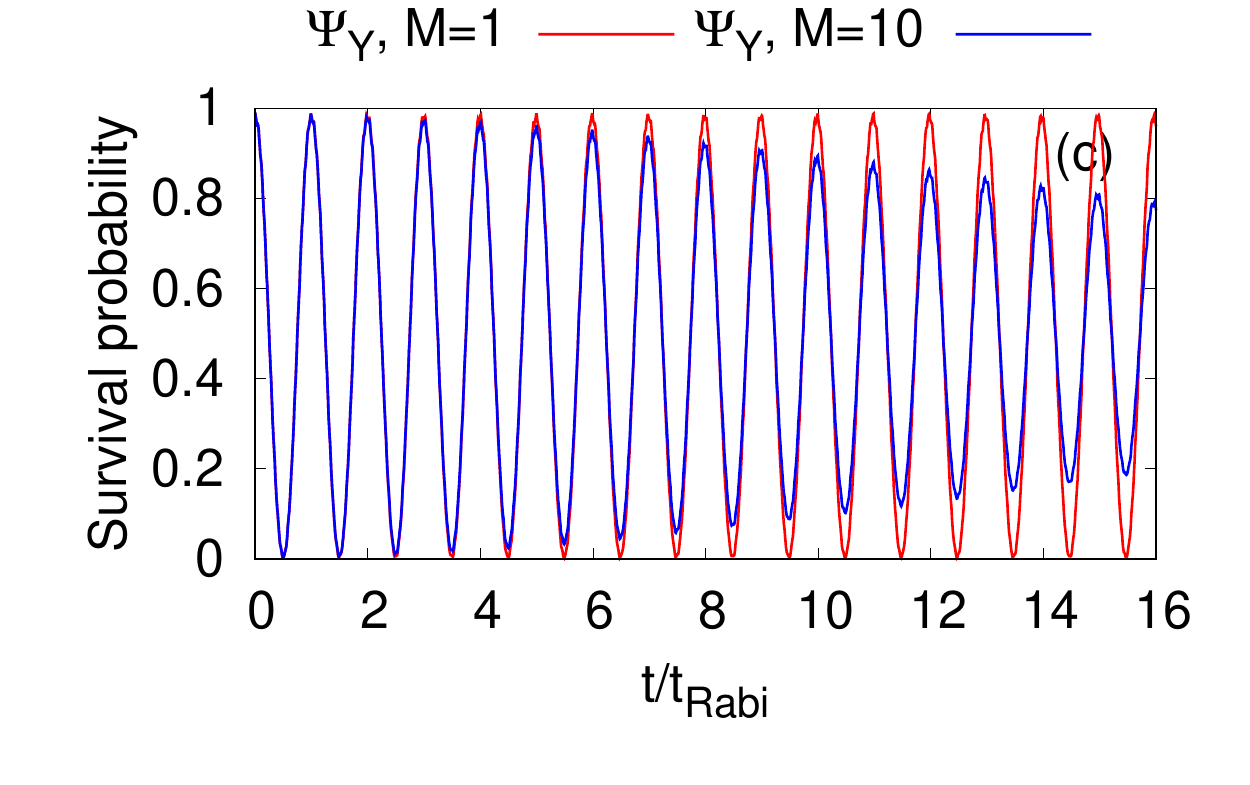}}
{\includegraphics[trim =  0.1cm 0.5cm 0.1cm 0.2cm, scale=.60]{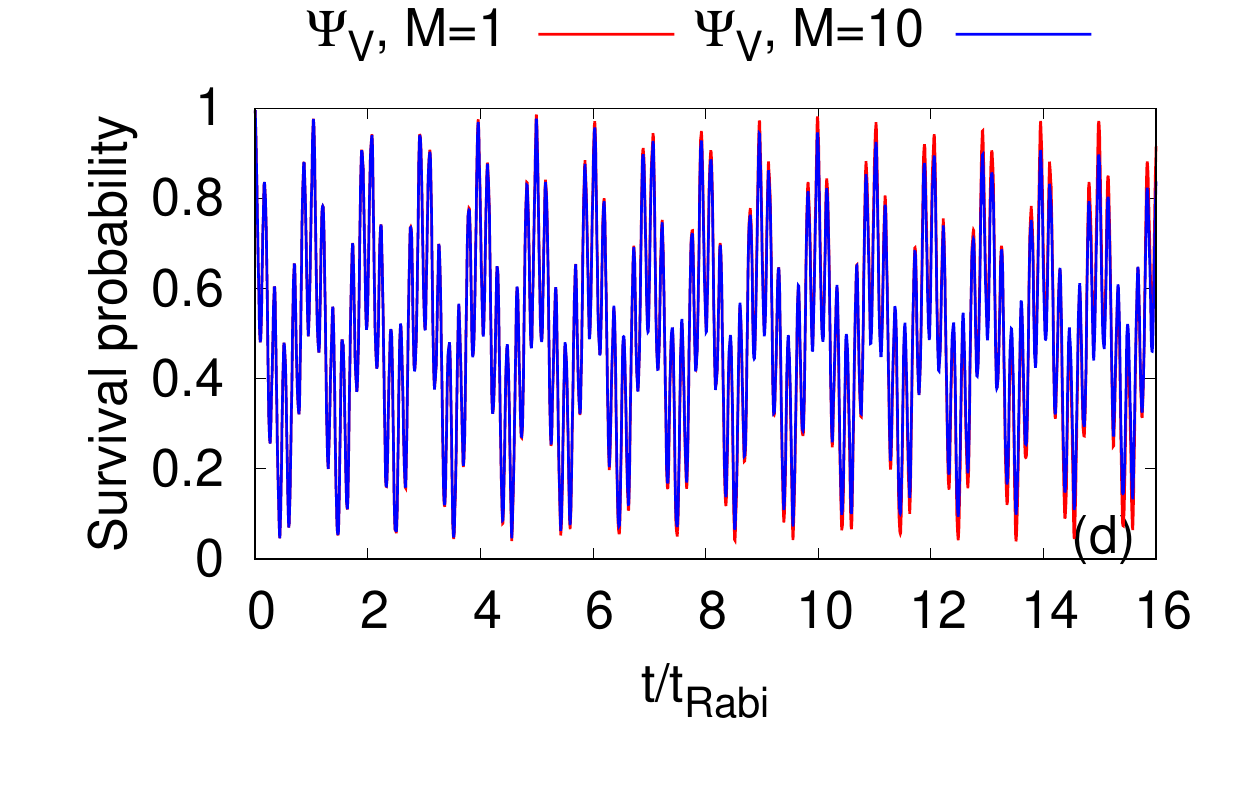}}\\
\caption{Time evolution of the survival probability in the left well, $P_L(t)$, of a symmetric 2D double-well potential for the initial states (a)  $\Psi_G$, (b) $\Psi_X$, (c) $\Psi_Y$, and (d) $\Psi_V$.  Mean-field results are  in red solid line and corresponding many-body results are  in blue solid line. The interaction parameter is $\Lambda=0.01$ and the number of bosons is $N=10$. The many-body time evolutions are computed  using the MCTDHB method
with $M=6$ time-adaptive orbitals for $\Psi_{G}$ and $\Psi_{X}$ and 
$M=10$ time-adaptive orbitals for $\Psi_{Y}$ and $\Psi_{V}$.
See the text for more details.
The quantities shown are dimensionless.}
\label{Fig1}
\end{figure*}

\begin{figure*}[!h]
{\includegraphics[trim = 4.9cm 0.5cm 3.1cm 0.2cm,scale=.65]{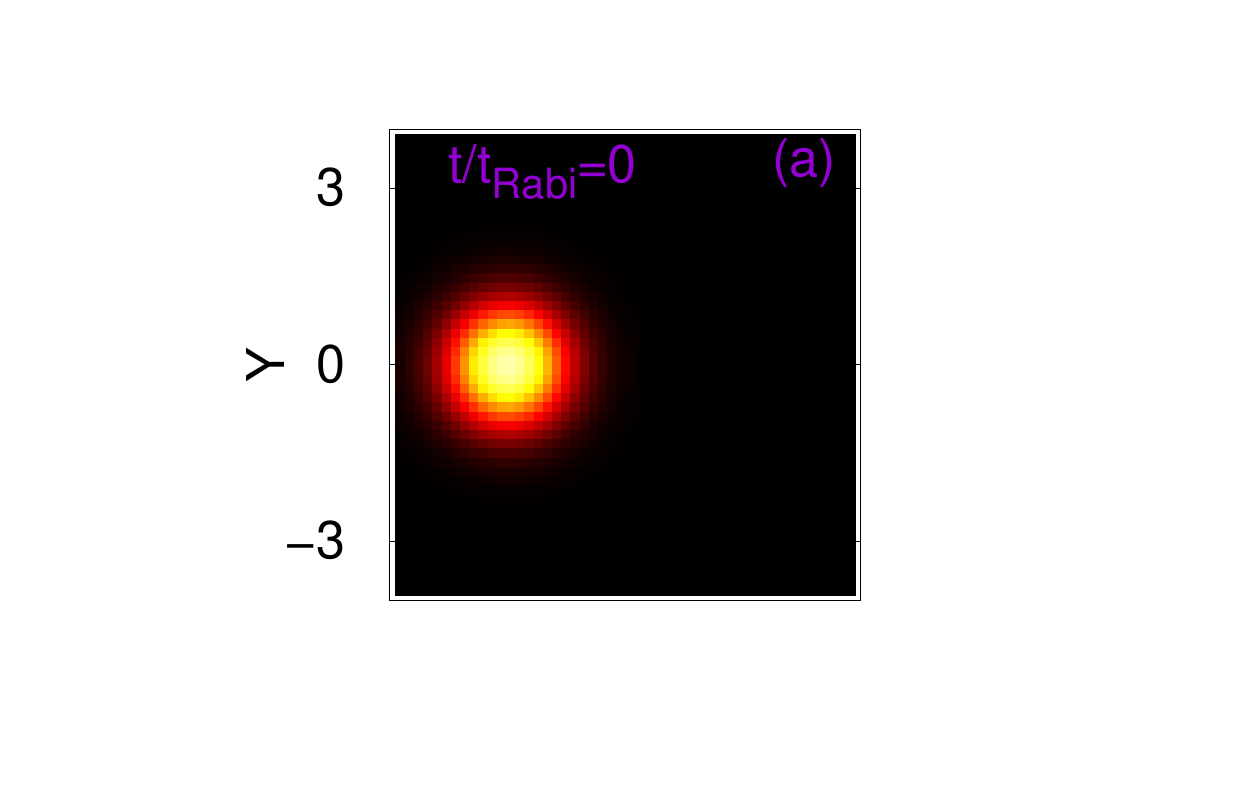}}
{\includegraphics[trim =  4.9cm 0.5cm 3.1cm 0.2cm, scale=.65]{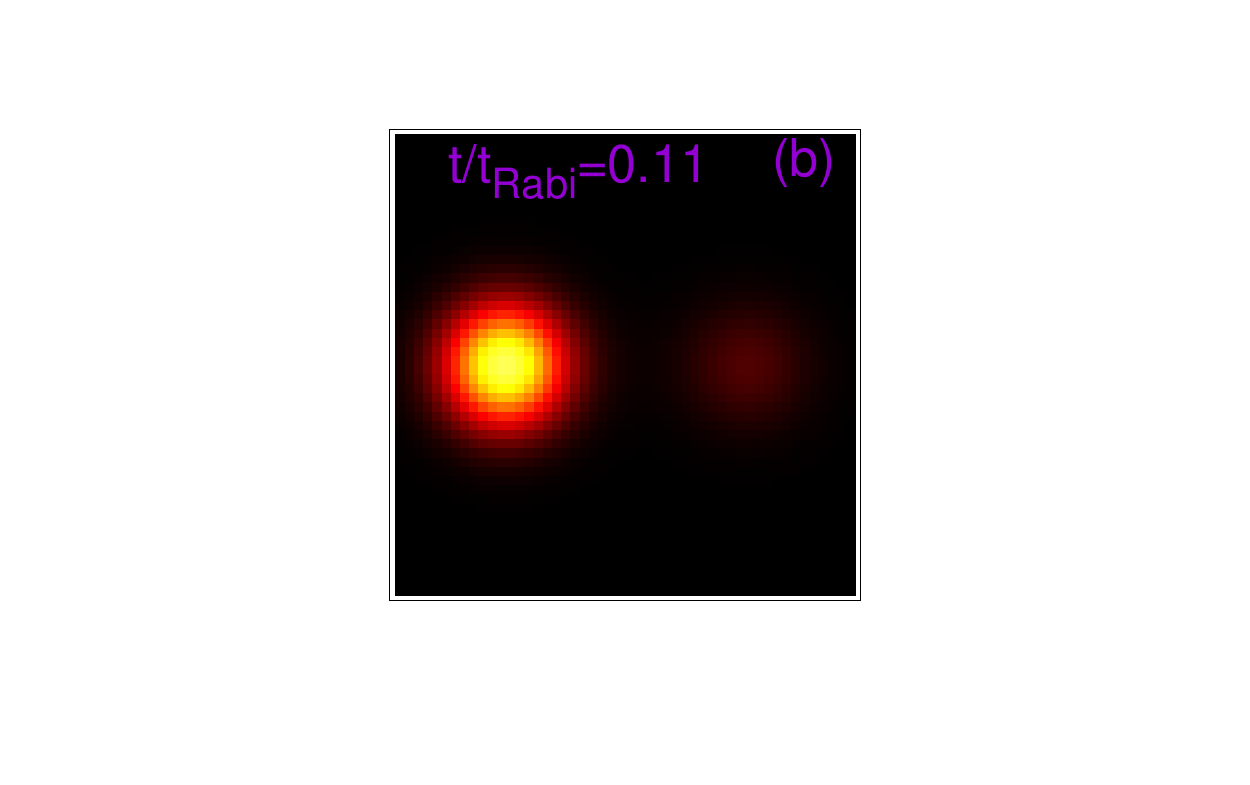}}
{\includegraphics[trim =  4.9cm 0.5cm 3.1cm 0.2cm, scale=.65]{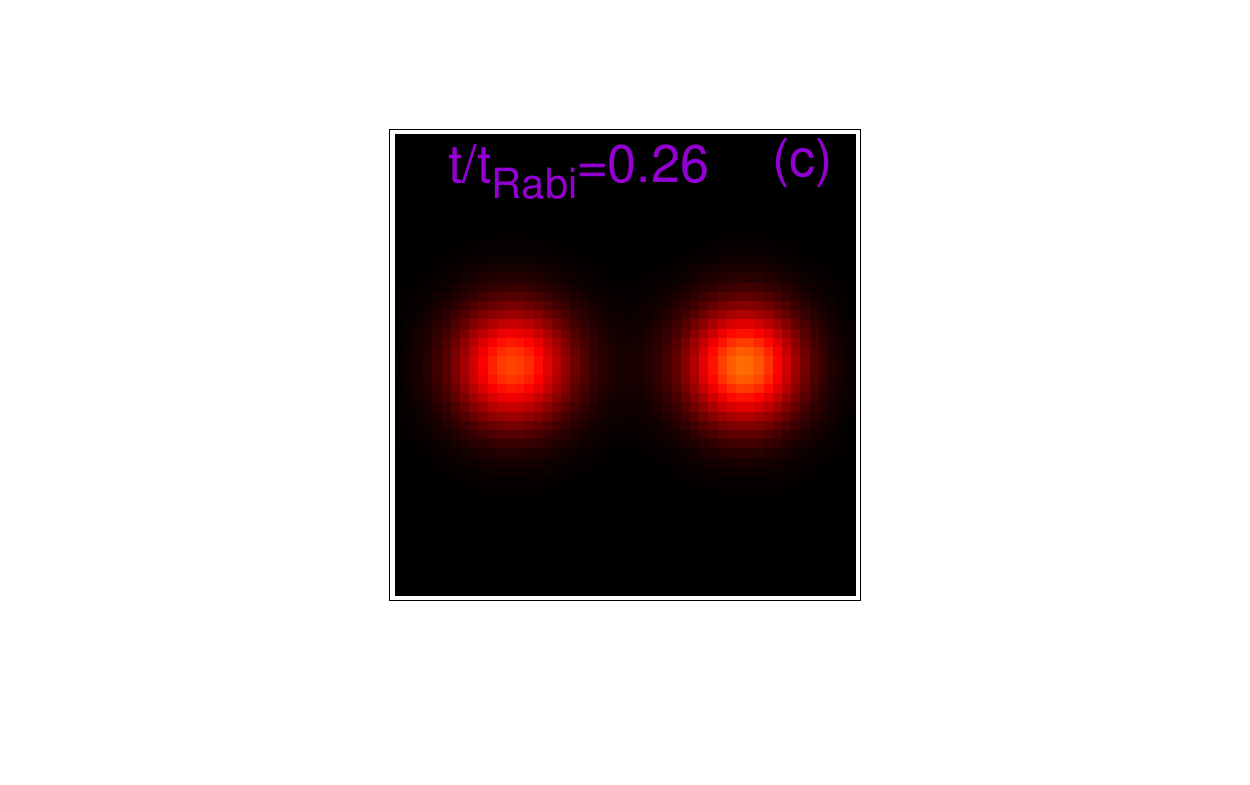}}
{\includegraphics[trim =  4.9cm 0.5cm 3.1cm 0.2cm, scale=.65]{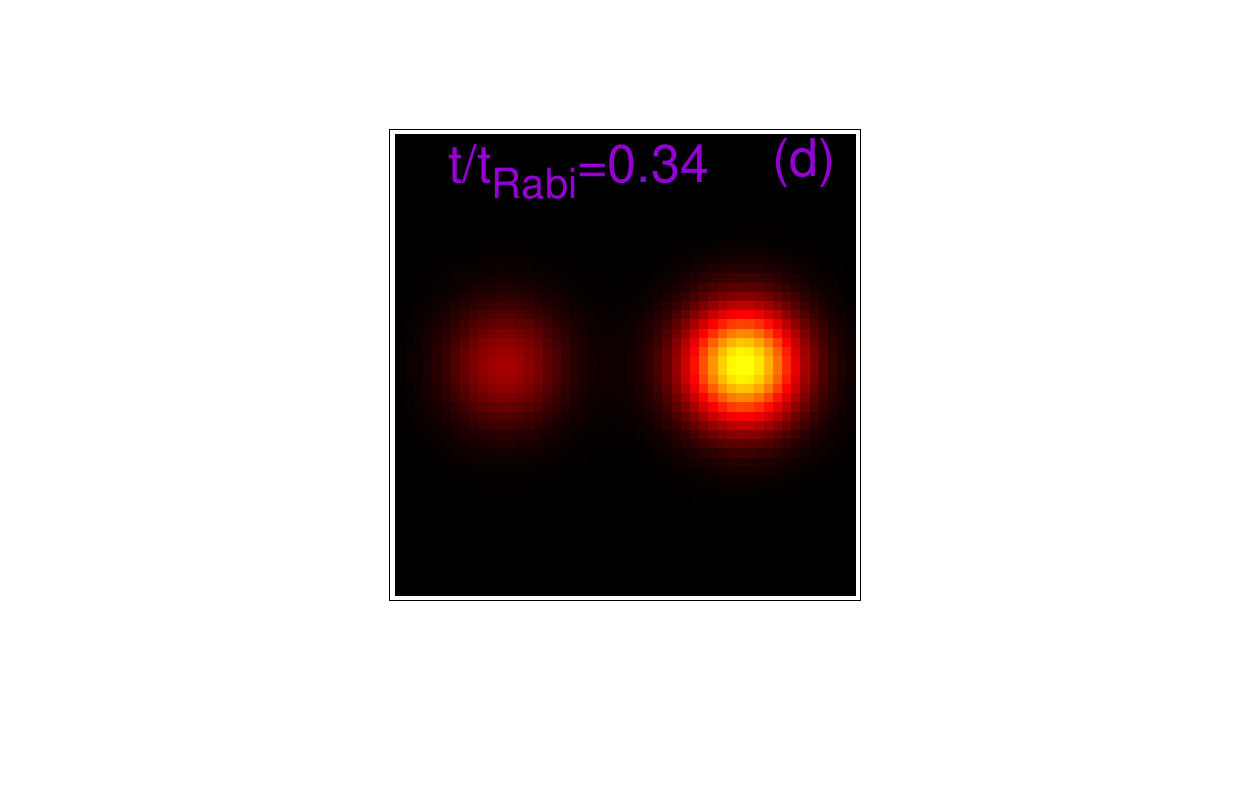}}
{\includegraphics[trim =  4.9cm 0.5cm 3.1cm 0.2cm, scale=.65]{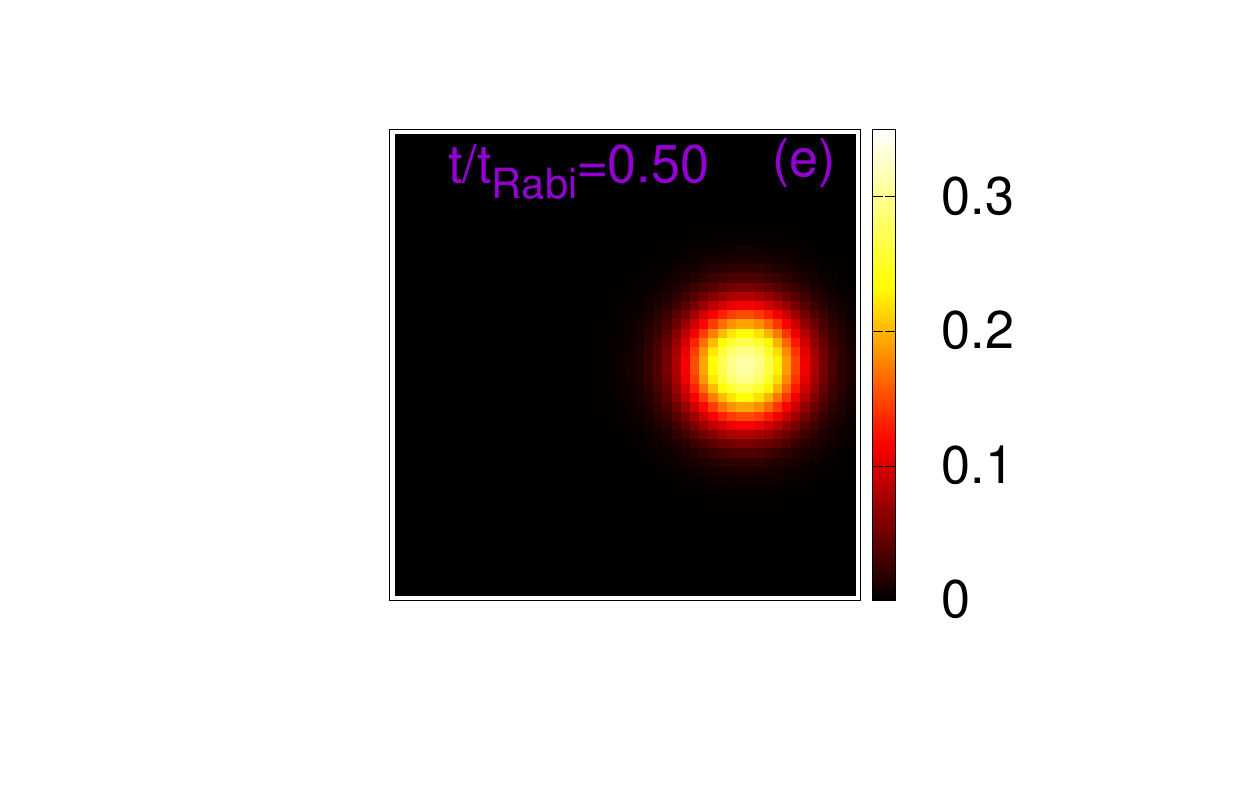}}\\
{\includegraphics[trim = 4.9cm 0.5cm 3.1cm 2.5cm,scale=.65]{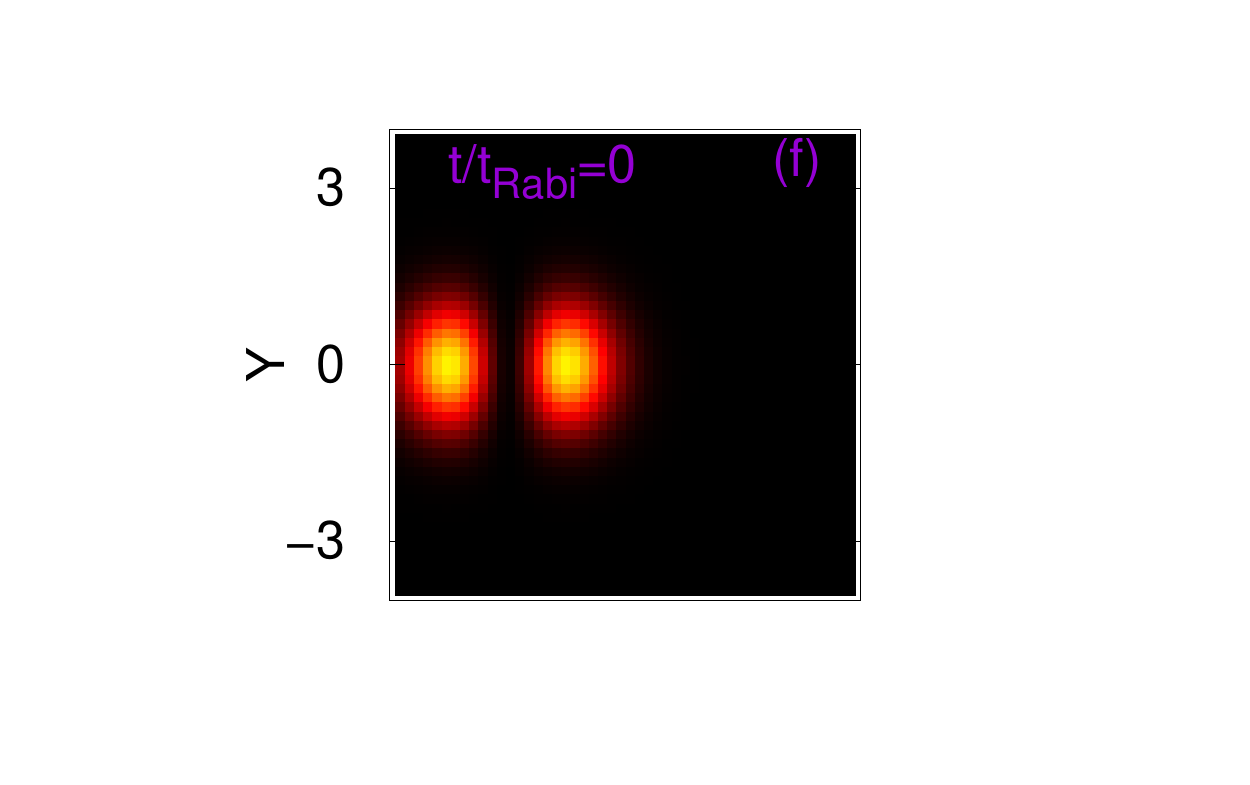}}
{\includegraphics[trim =  4.9cm 0.5cm 3.1cm 2.5cm, scale=.65]{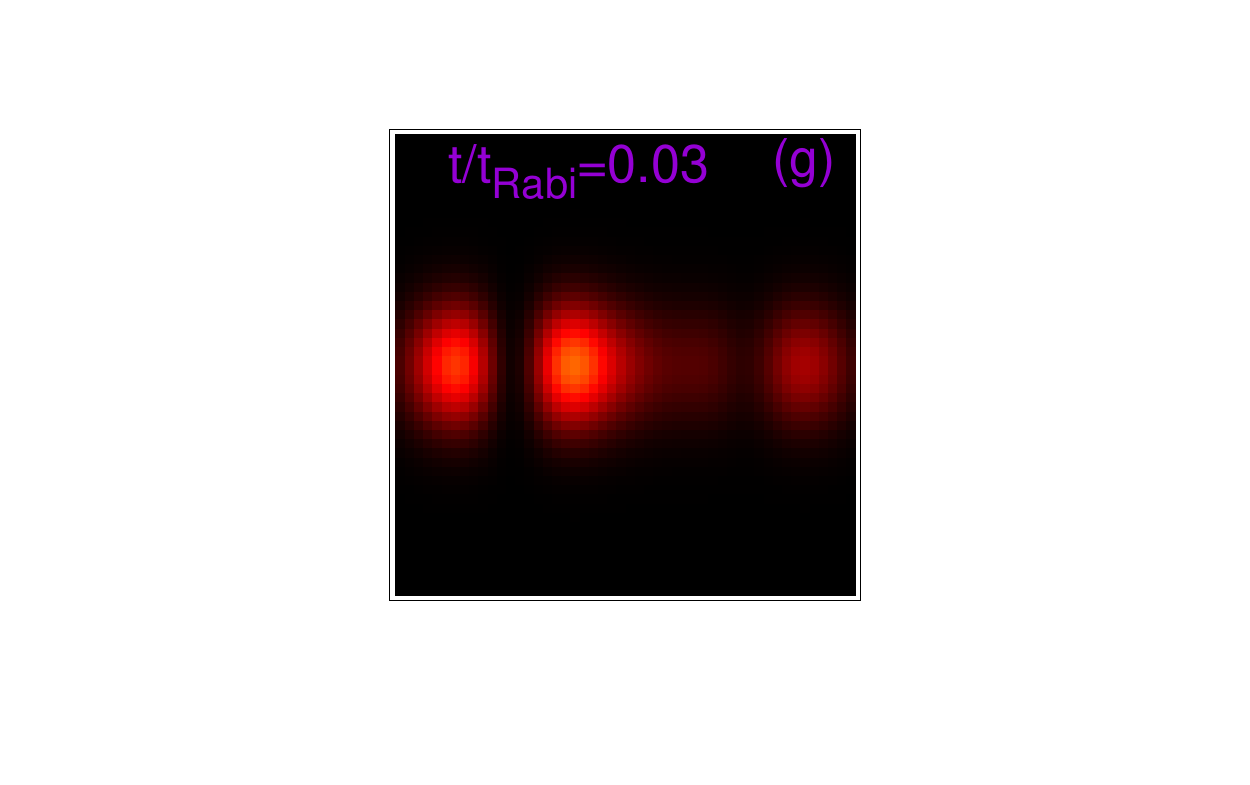}}
{\includegraphics[trim =  4.9cm 0.5cm 3.1cm 2.5cm, scale=.65]{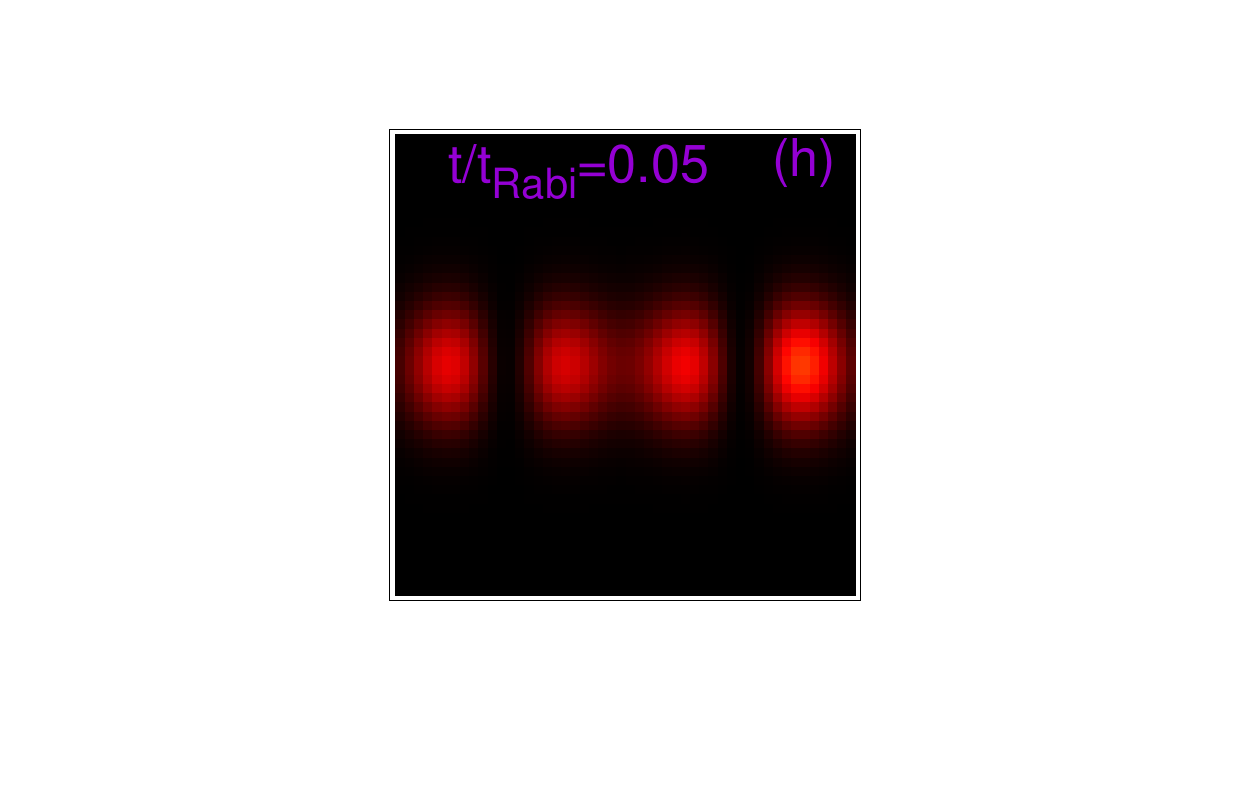}}
{\includegraphics[trim =  4.9cm 0.5cm 3.1cm 2.5cm, scale=.65]{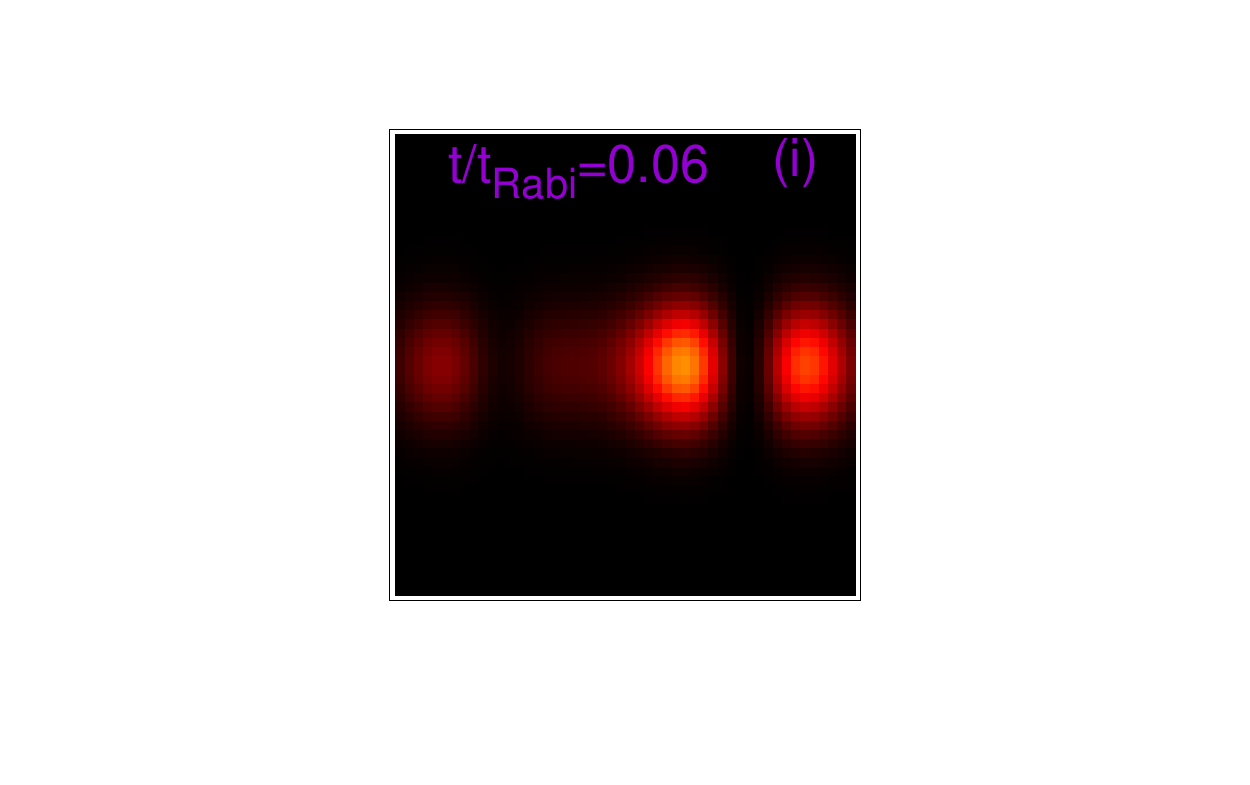}}
{\includegraphics[trim =  4.9cm 0.5cm 3.1cm 2.5cm, scale=.65]{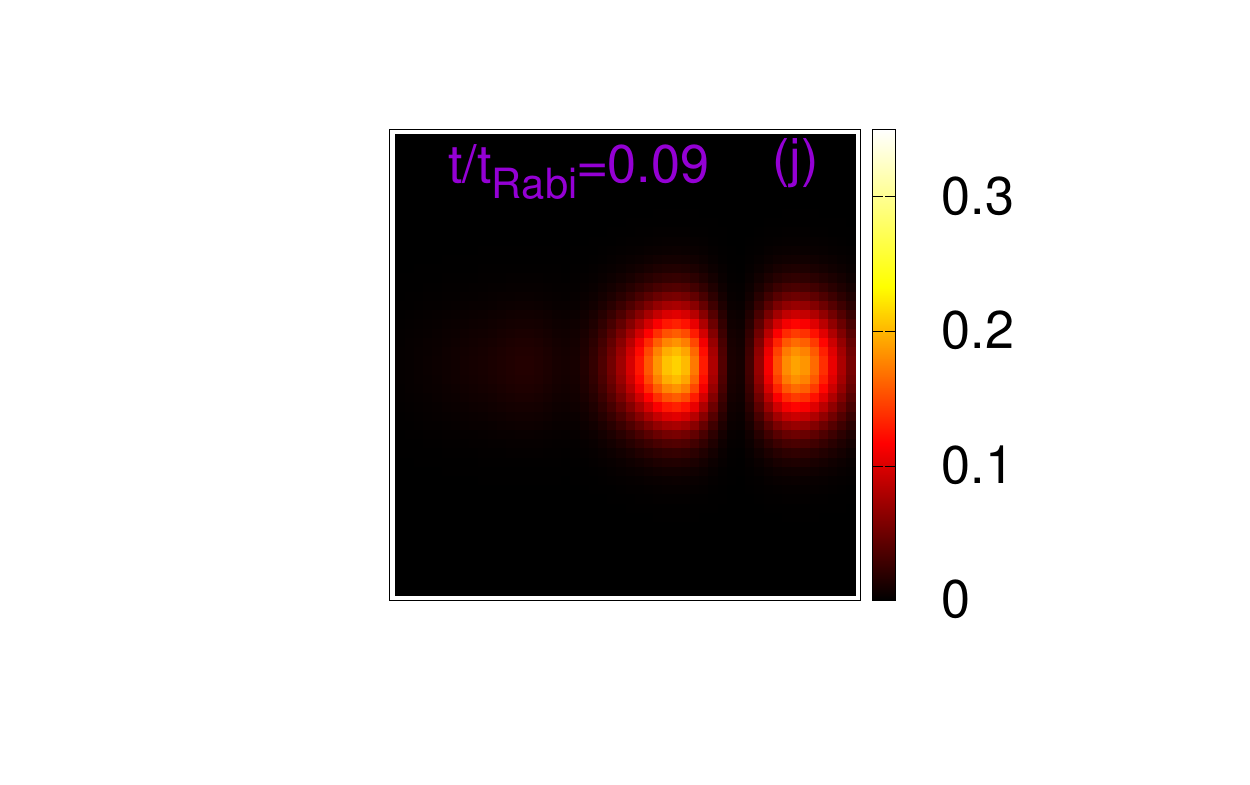}}\\
{\includegraphics[trim = 4.9cm 0.5cm 3.1cm 2.5cm,scale=.65]{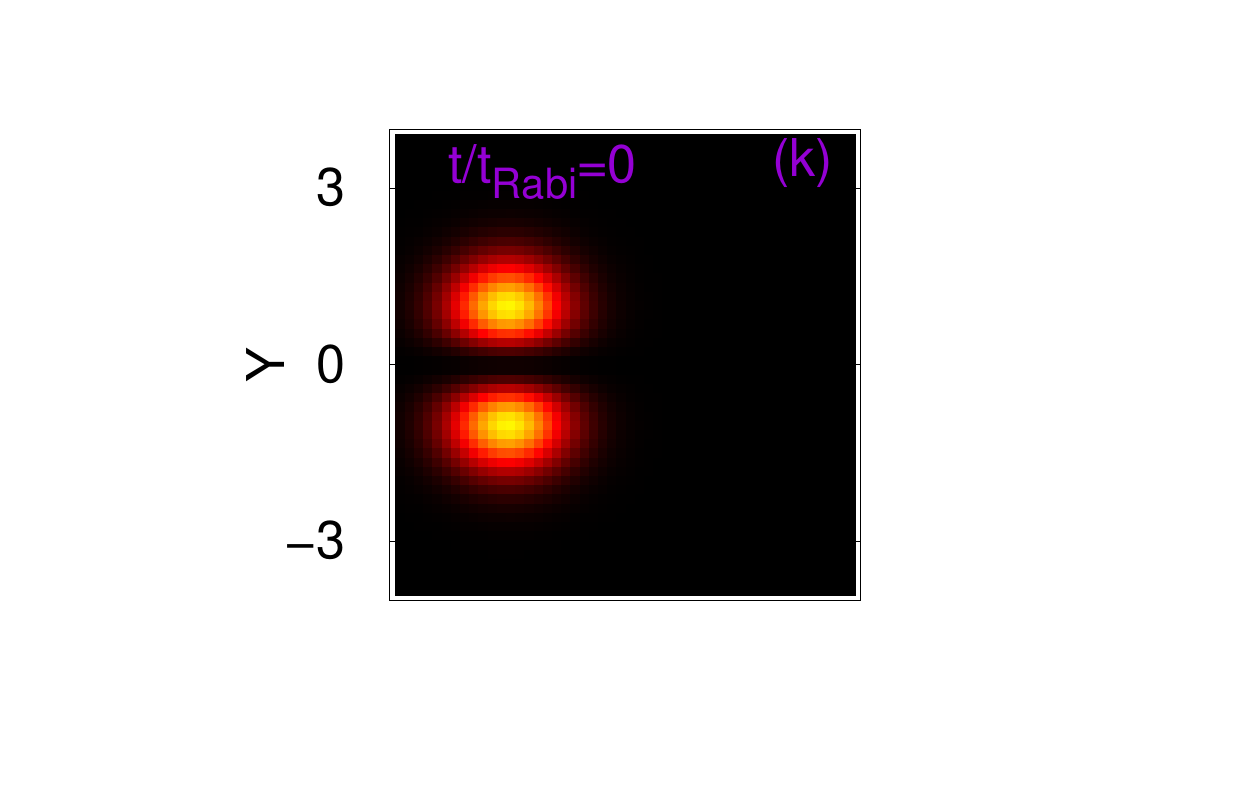}}
{\includegraphics[trim =  4.9cm 0.5cm 3.1cm 2.5cm, scale=.65]{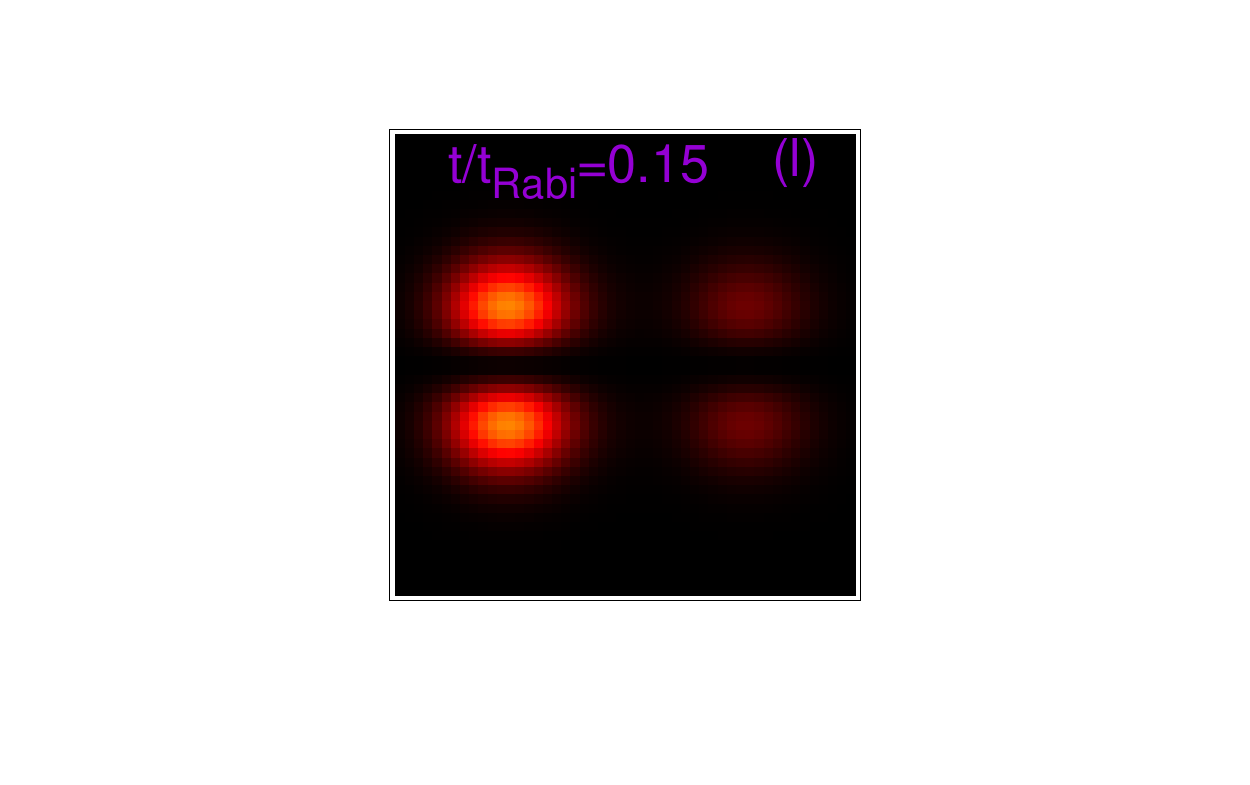}}
{\includegraphics[trim =  4.9cm 0.5cm 3.1cm 2.5cm, scale=.65]{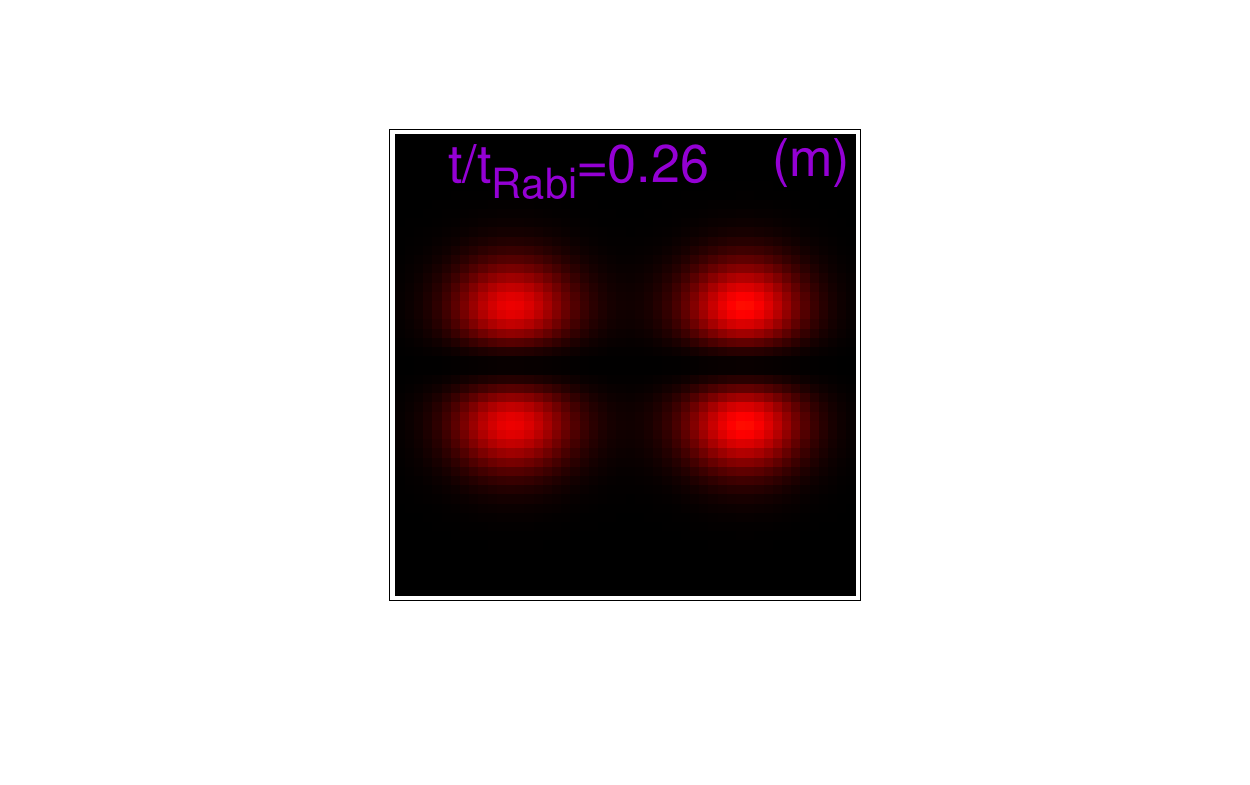}}
{\includegraphics[trim =  4.9cm 0.5cm 3.1cm 2.5cm, scale=.65]{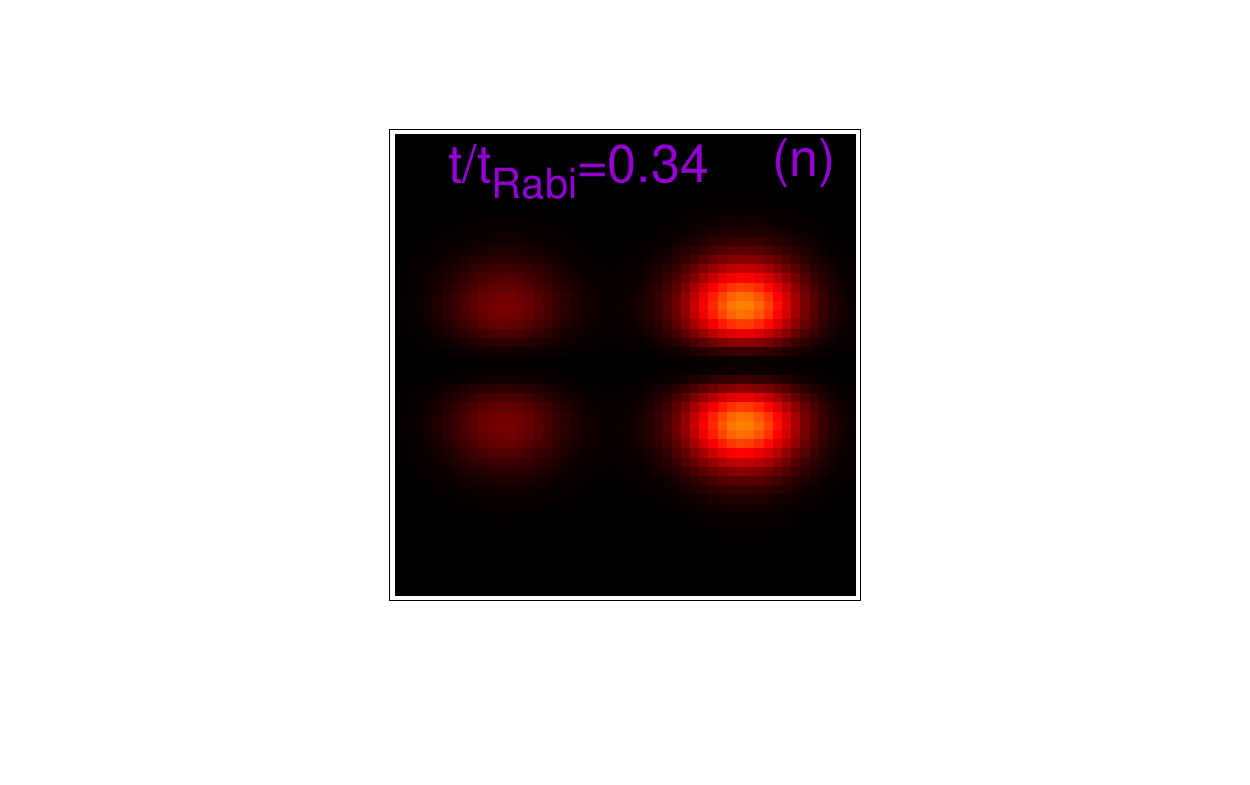}}
{\includegraphics[trim =  4.9cm 0.5cm 3.1cm 2.5cm, scale=.65]{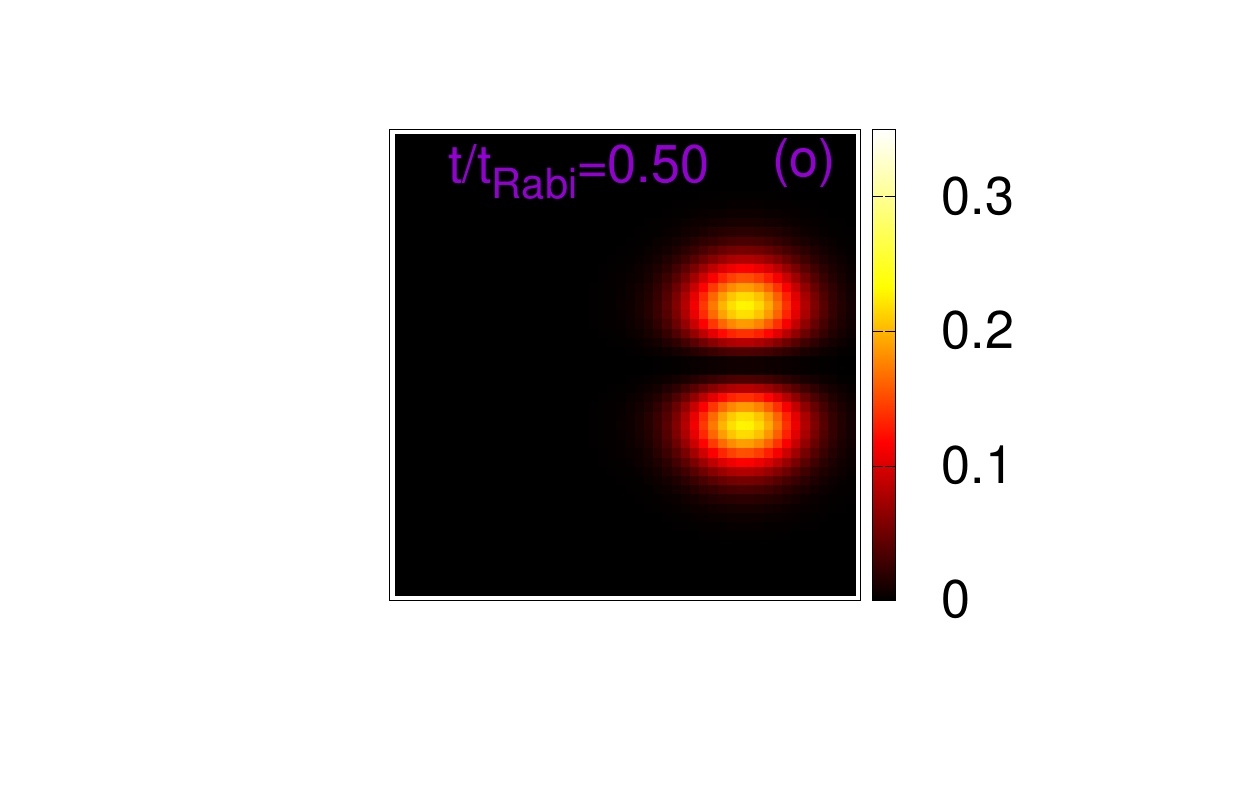}}\\
{\includegraphics[trim = 4.9cm 0.5cm 3.1cm 2.5cm,scale=.65]{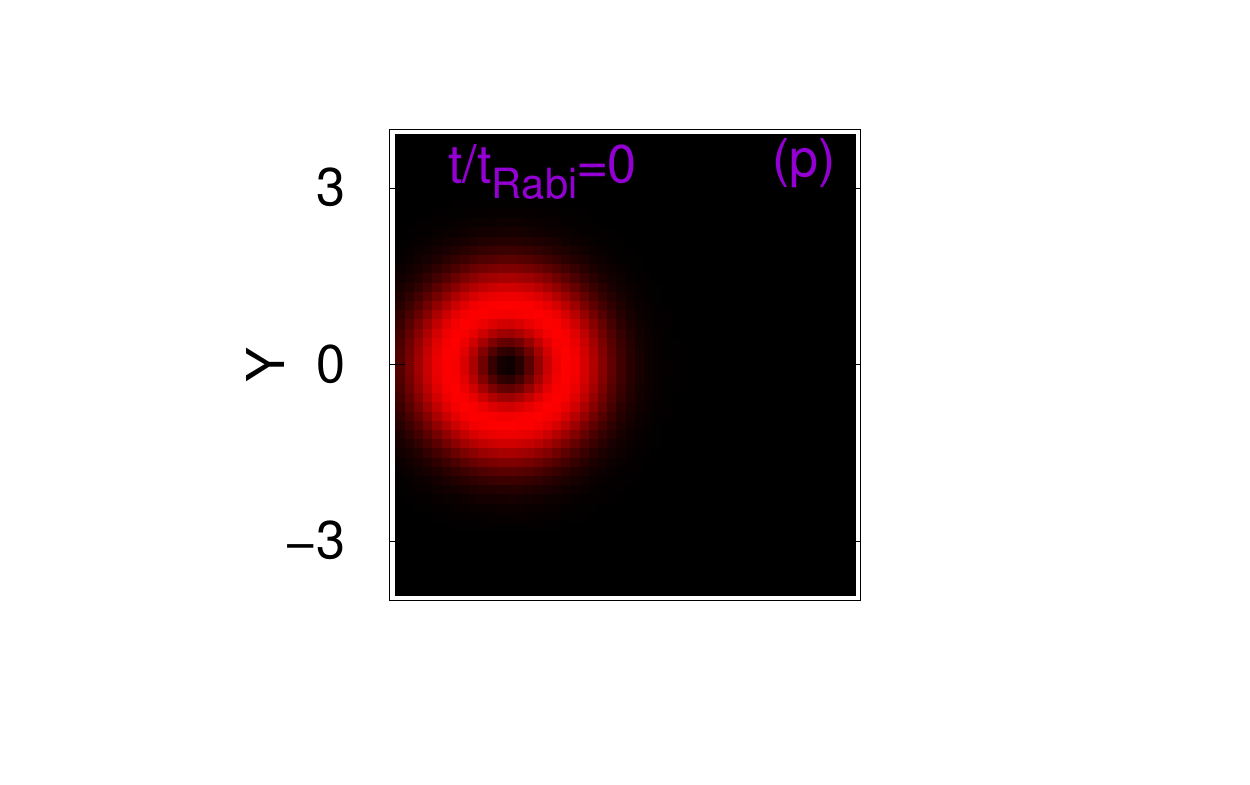}}
{\includegraphics[trim =  4.9cm 0.5cm 3.1cm 2.5cm, scale=.65]{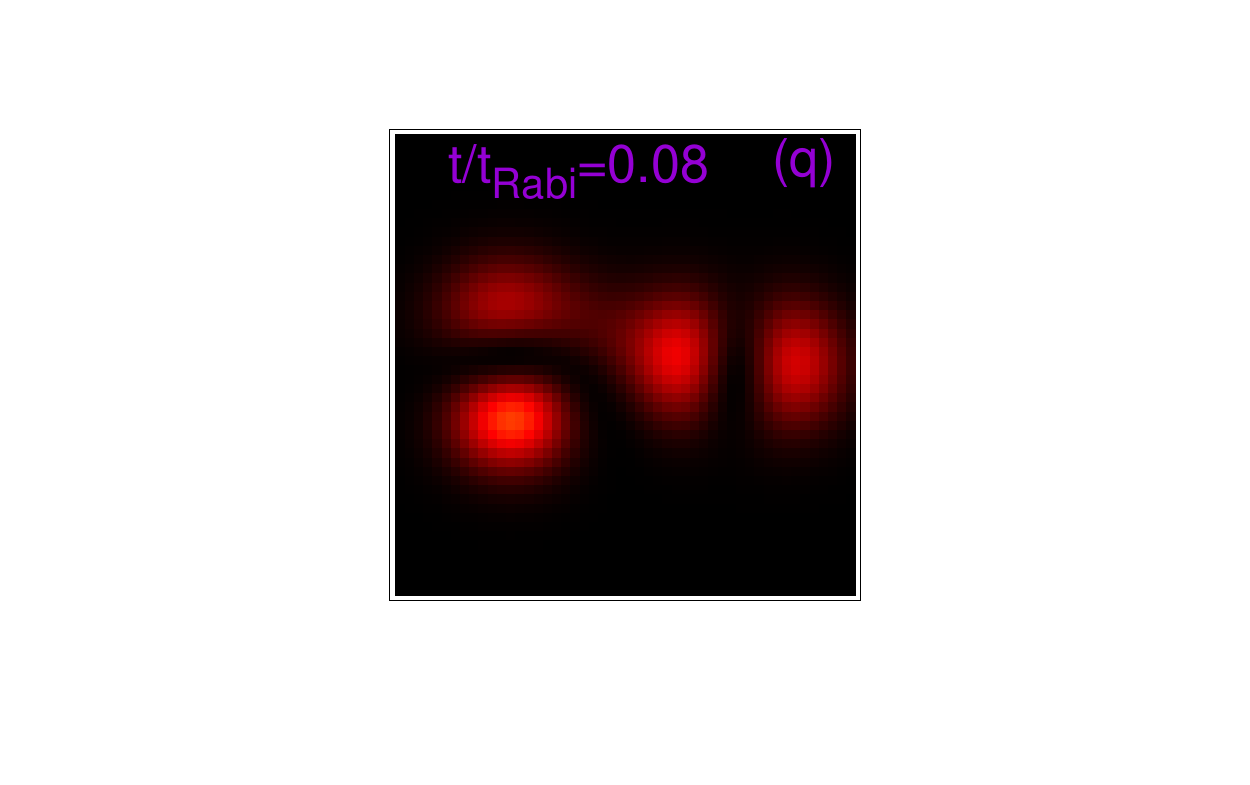}}
{\includegraphics[trim =  4.9cm 0.5cm 3.1cm 2.5cm, scale=.65]{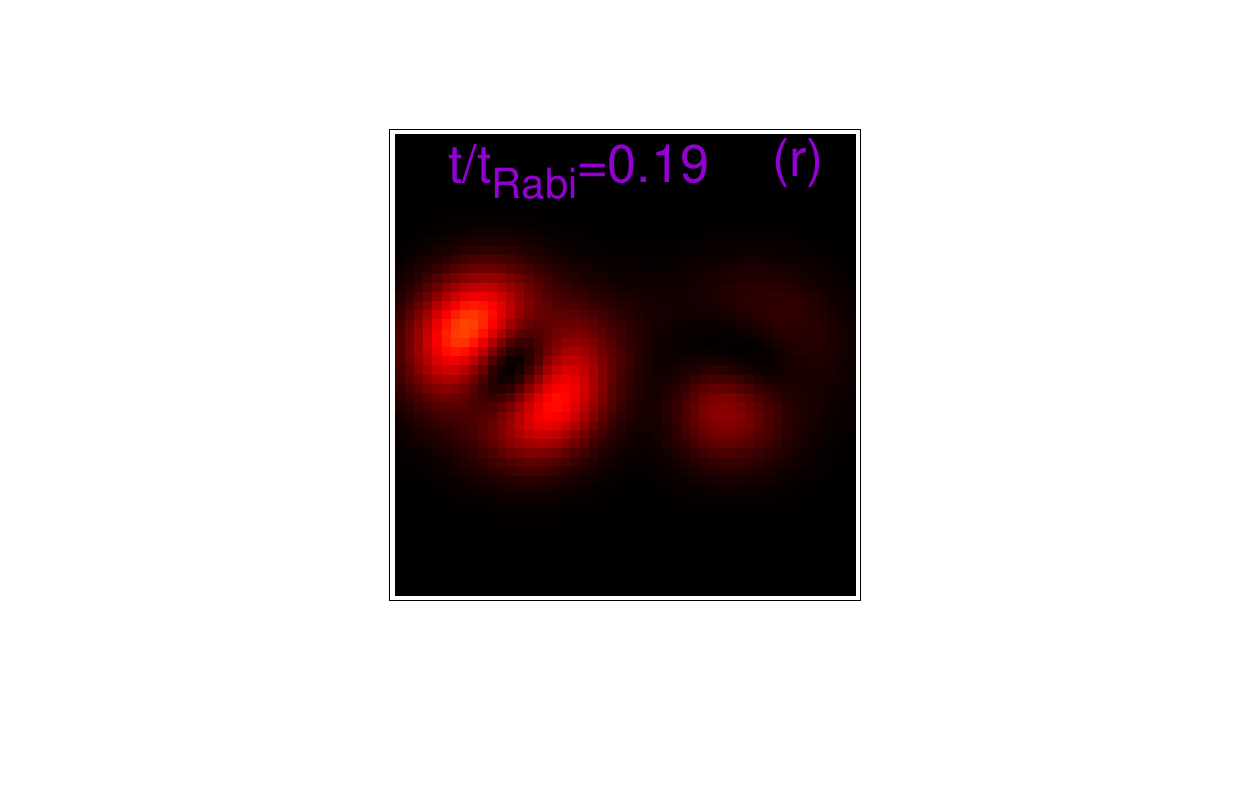}}
{\includegraphics[trim =  4.9cm 0.5cm 3.1cm 2.5cm, scale=.65]{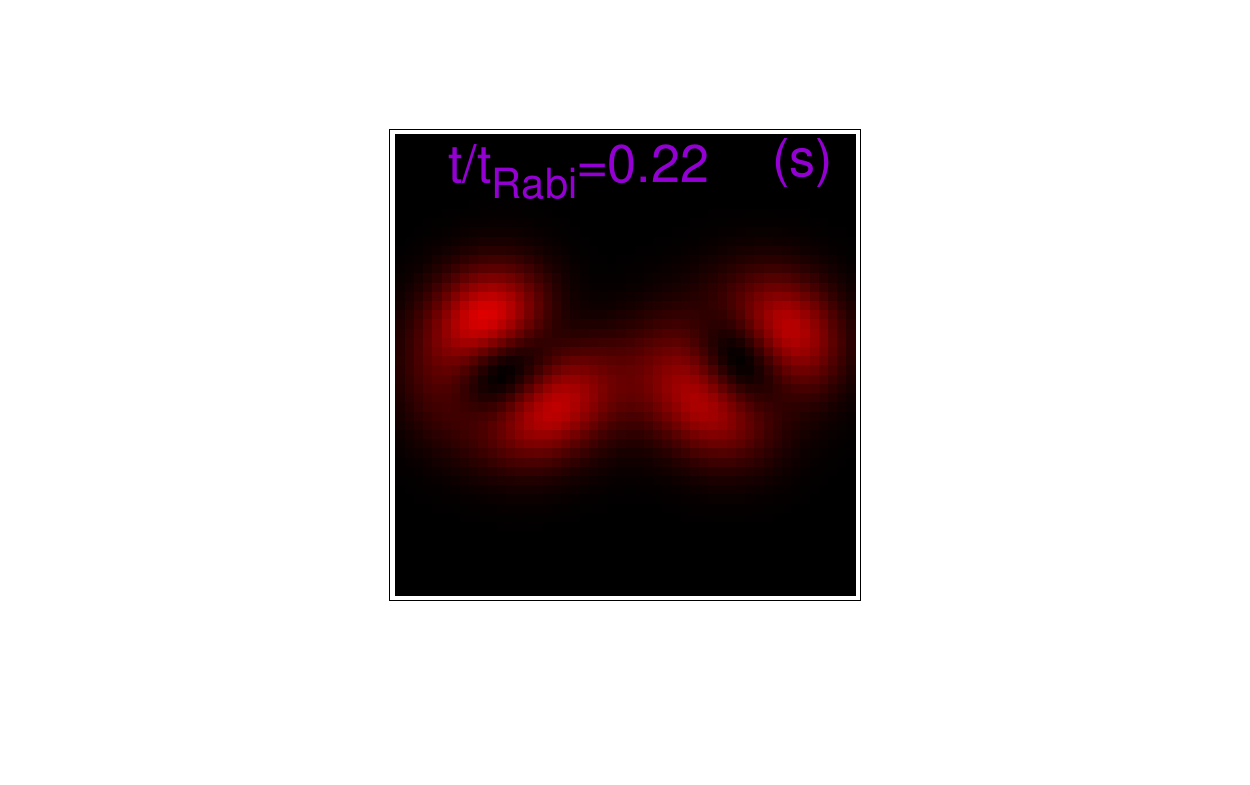}}
{\includegraphics[trim =  4.9cm 0.5cm 3.1cm 2.5cm, scale=.65]{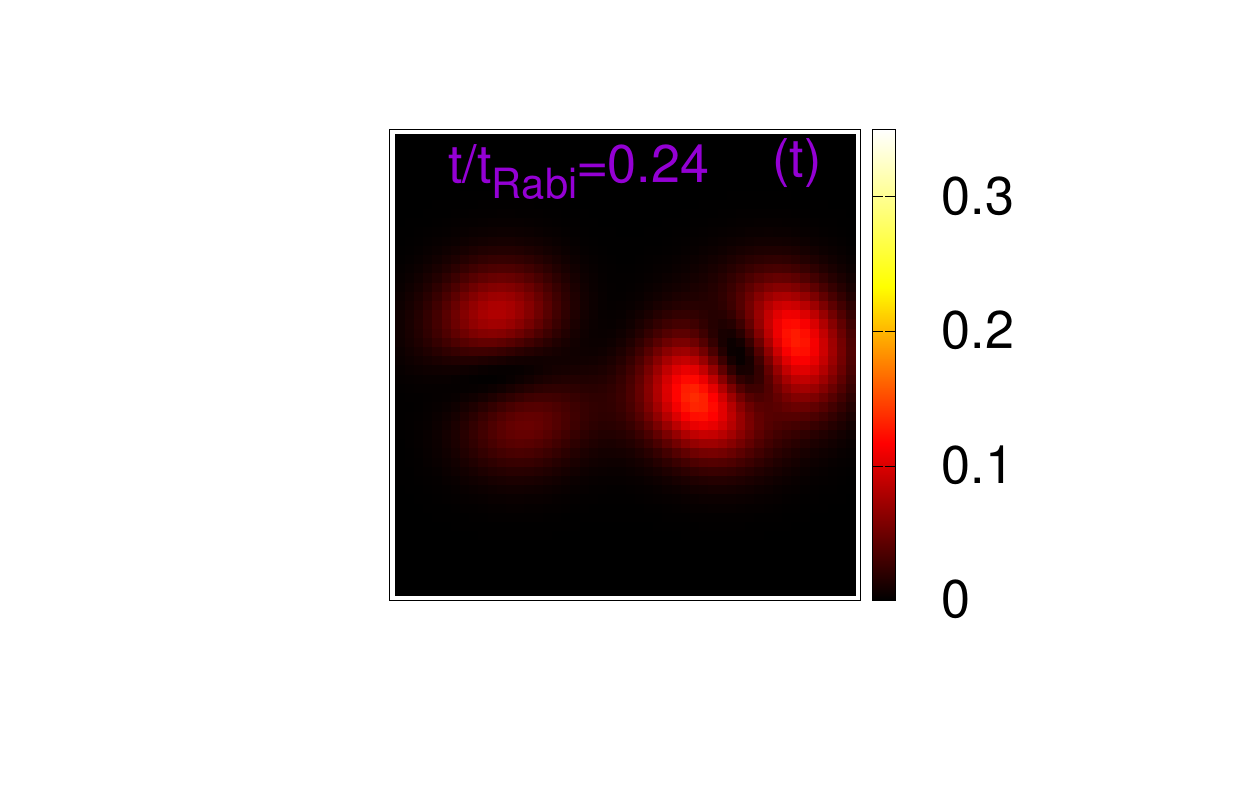}}\\
{\includegraphics[trim = 4.9cm 0.5cm 3.1cm 2.5cm,scale=.65]{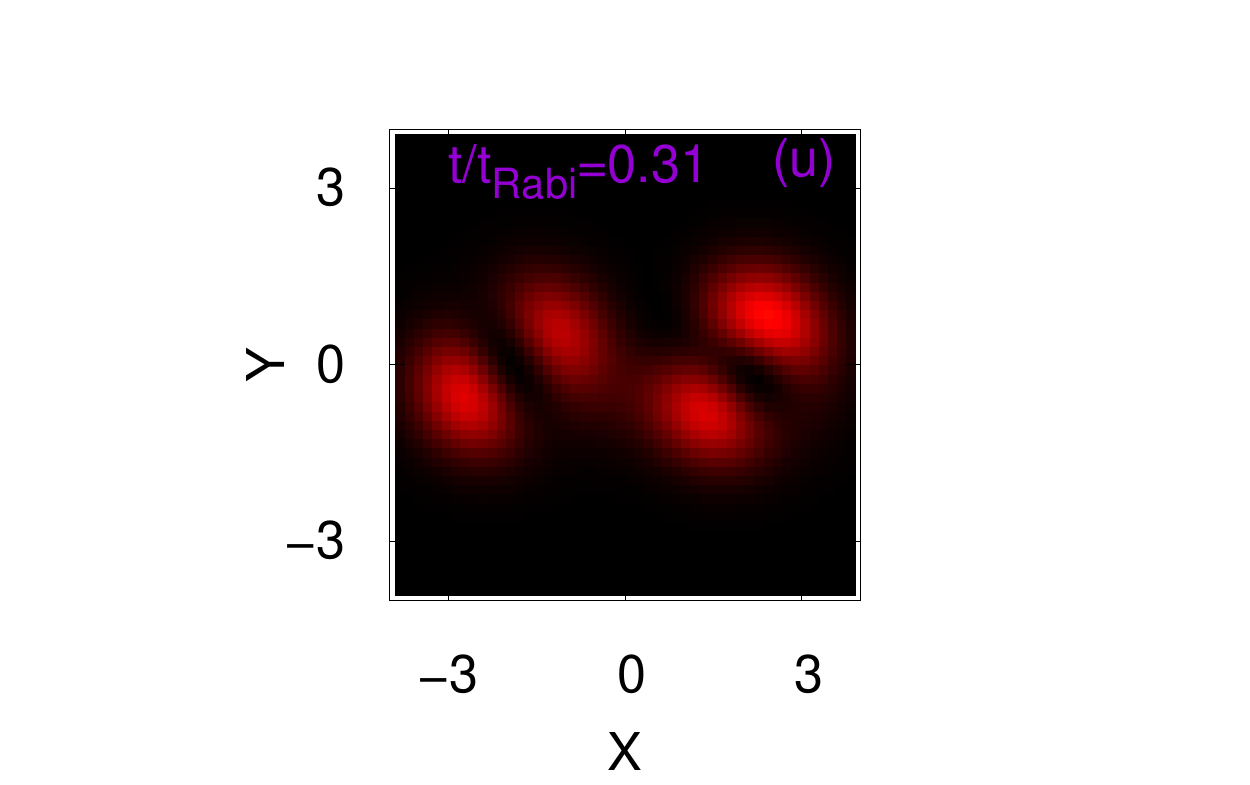}}
{\includegraphics[trim =  4.9cm 0.5cm 3.1cm 2.5cm, scale=.65]{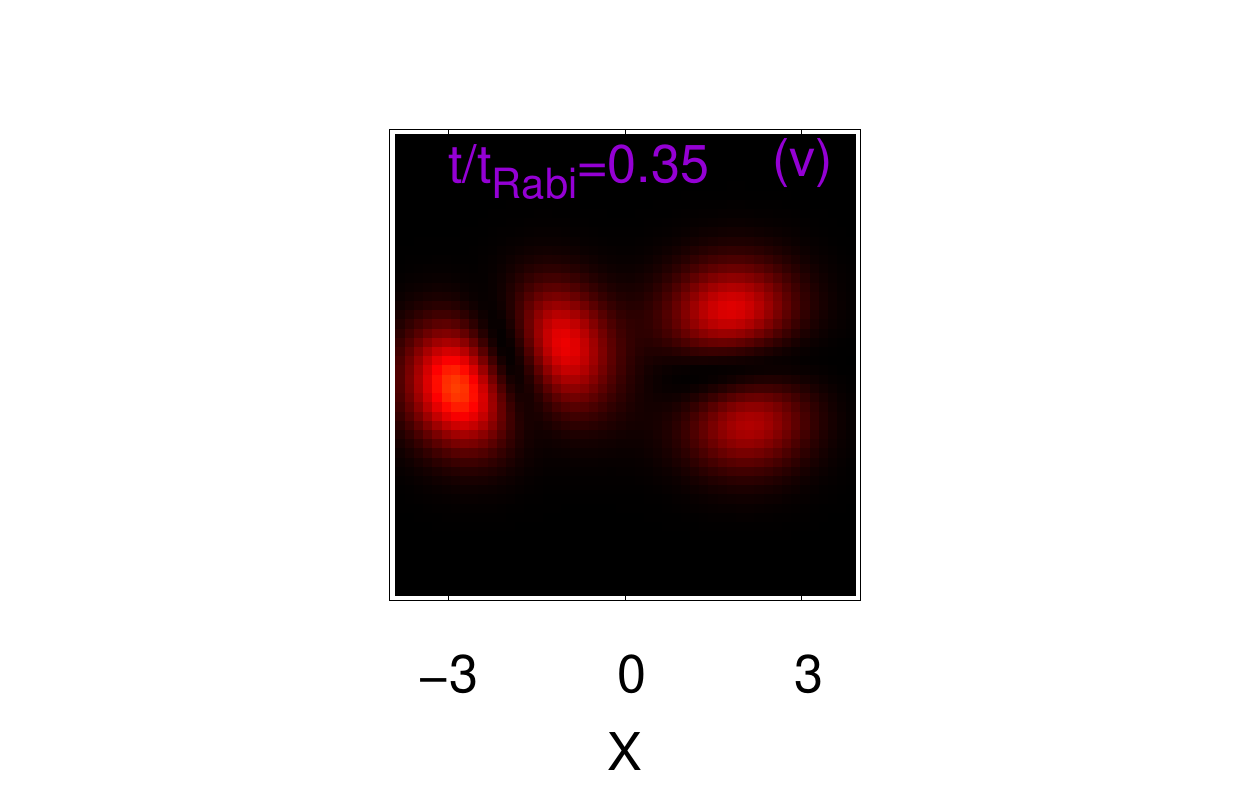}}
{\includegraphics[trim =  4.9cm 0.5cm 3.1cm 2.5cm, scale=.65]{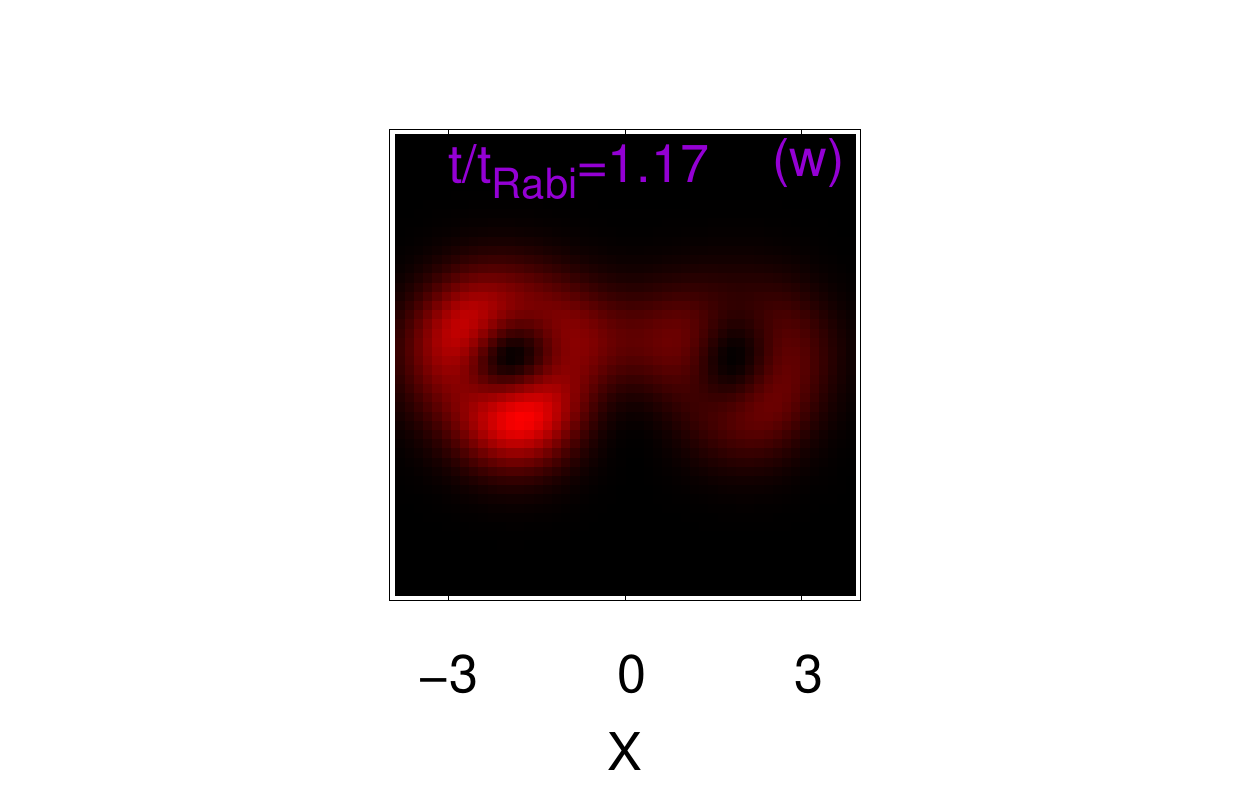}}
{\includegraphics[trim =  4.9cm 0.5cm 3.1cm 2.5cm, scale=.65]{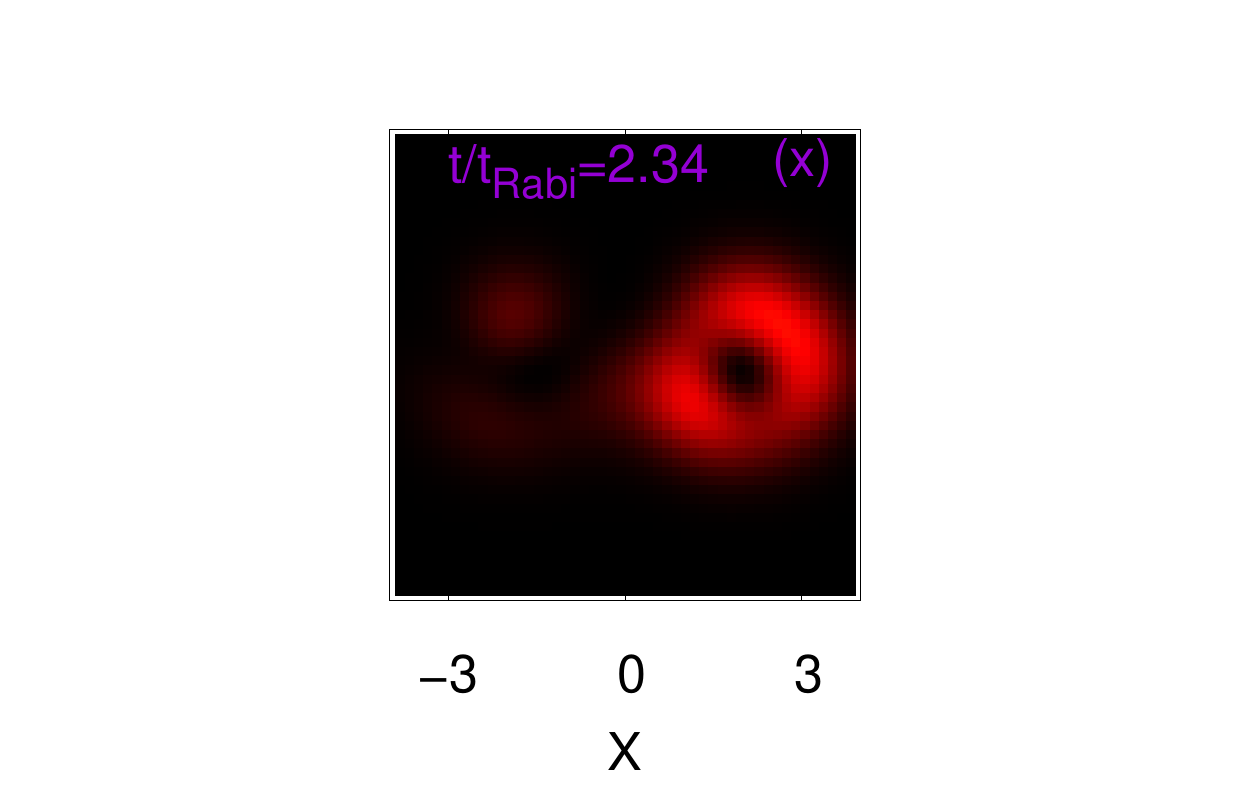}}
{\includegraphics[trim =  4.9cm 0.5cm 3.1cm 2.5cm, scale=.65]{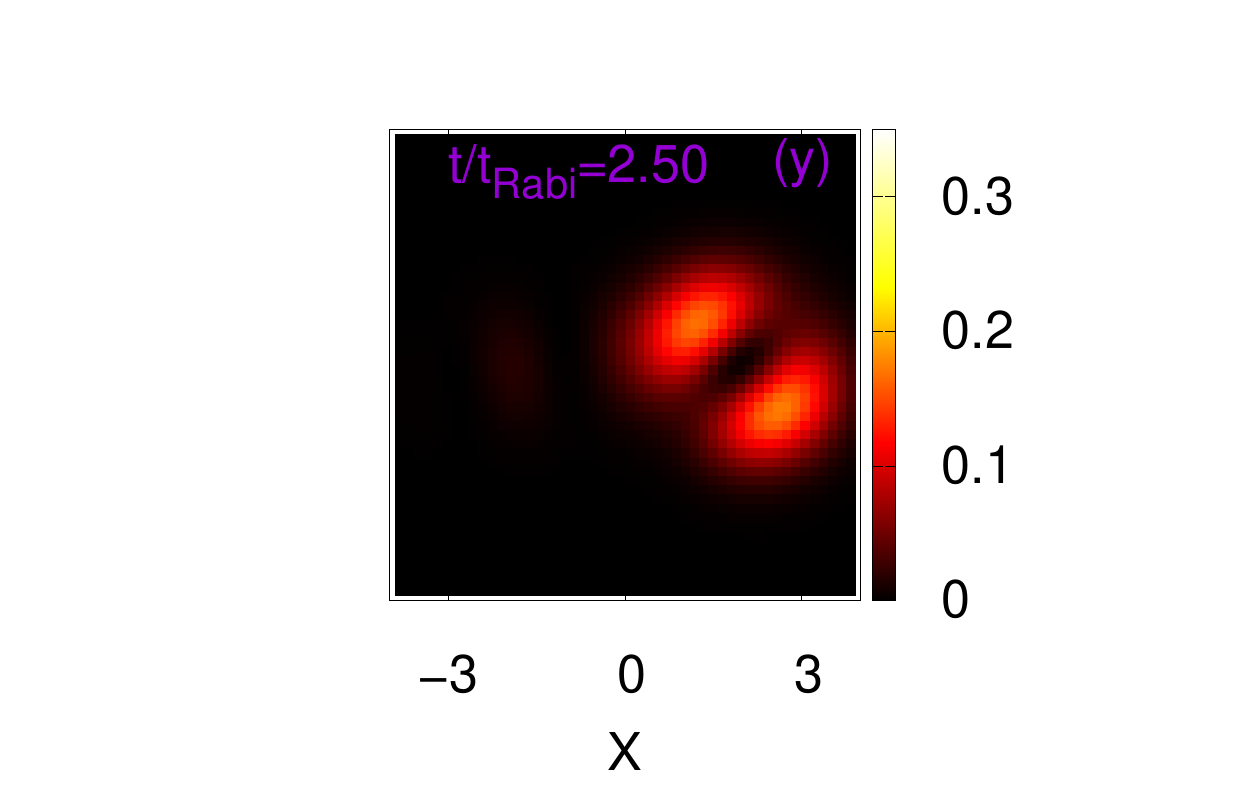}}\\
\caption{Time evolution of the many-body density oscillations in a symmetric 2D double-well.
The interaction parameter is $\Lambda=0.01$ and the number of bosons is $N=10$. Shown are the densities per particle for the initial states 
$\Psi_G$ (first row, from left to right, at $\frac{t}{t_{Rabi}}=0, 0.11, 0.26, 0.34$, and $0.50$); 
$\Psi_X$ (second row, at $\frac{t}{t_{Rabi}}=0, 0.03, 0.05, 0.06$, and $0.09$); 
$\Psi_Y$ (third row, at $\frac{t}{t_{Rabi}}=0, 0.15, 0.26, 0.34$, and $0.50$); and 
$\Psi_V$ (fourth and fifth rows, at $\frac{t}{t_{Rabi}}=0, 0.08, 0.19, 0.22, 0.24$ and $0.31, 0.35, 1.17, 2.34, 2.50$).
The many-body time evolutions are computed using the MCTDHB method
with $M=6$ time-adaptive orbitals for $\Psi_{G}$ and $\Psi_{X}$ and 
$M=10$ time-adaptive orbitals for $\Psi_{Y}$ and $\Psi_{V}$.
See the text for more details.
The quantities shown are dimensionless.}
\label{Fig2}
\end{figure*}

We begin our investigation with the time evolution of the most basic quantity,
the density $\rho(x,y;t)$ of the bosonic clouds $\Psi_G$, $\Psi_X$, $\Psi_Y$, and $\Psi_V$. 
To this end, we present the interconnection between the four densities 
with respect to the survival probability in the left well,
$P_L(t)=\int\limits_{x=-\infty}^{0}\int\limits_{y=-\infty}^{+\infty}dx dy \dfrac{\rho(x,y;t)}{N}$. 
In Fig.~\ref{Fig1},
we compare the many-body dynamics of $P_L(t)$ with the corresponding
mean-field  results.
In order to have a more detailed description of the tunneling dynamics of the considered initial states,
we depict surface plots of the density oscillations at the many-body level in Fig.~\ref{Fig2}.
We observe in Fig.~\ref{Fig1}  the tunneling of the density back and forth between 
the left and right wells for all bosonic clouds,
but the frequency of the tunneling oscillations are distinct for the different initial states,
apart from $\Psi_G$ and $\Psi_Y$ which have essentially the same frequency of oscillations.  The tunneling dynamics are consistent with the density oscillations shown in Fig.~\ref{Fig2}.
For a particular initial state the frequency of the tunneling oscillations is practically identical at the mean-field and many-body levels,
but certainly it  does not remain so for the amplitudes of the tunneling oscillations in the course of time evolution. As $\Psi_G$ and $\Psi_Y$ both lie (in the non-interacting system) in the lowest band along the $x$-direction (the direction along which the barrier is formed), the tunneling oscillations of $\Psi_Y$ are very similar to those of $\Psi_G$ at the mean-field level, both in frequency and in amplitude. On the other hand, $\Psi_X$ lies (in the non-interacting system) in the first excited band along the $x$-direction, it 'feels' a smaller potential barrier, and therefore its tunneling oscillations are faster. Furthermore, the effect of coupling between the lowest energy band and the higher excited states produces high-frequency  breathing oscillations for $\Psi_X$.

We observe the  complete tunneling of bosons from the left well to the right well  without destroying the
structure of the initial states in Fig.~\ref{Fig2} 
for $\Psi_G$, $\Psi_X$, and $\Psi_Y$ at about $t=0.50 t_{\text{Rabi}}$, $t=0.09 t_{\text{Rabi}}$ and $t=0.50 t_{\text{Rabi}}$, respectively. Recall the interaction is weak $(\Lambda=0.01)$ and the tunneling period is very close to that of non-interacting bosons. It is clear from Figs.~\ref{Fig1} and \ref{Fig2} that the nature of tunneling for the vortex state is very different and intricate compared to the other initial bosonic clouds.  As $\Psi_V$ is a linear combination of $\Psi_X$ and $\Psi_Y$ at $t=0$ and the interaction is weak, the tunneling of $\Psi_V$ can be interpreted by combining the resulting dynamics of $\Psi_X$ and $\Psi_Y$. In Fig.~\ref{Fig2}, it is shown that the vortex state initially  destroys its structure in the process of tunneling and creates two dipole states.  It is noted that the essentially full tunneling of $\Psi_V$ happens at about $t=2.50 t_{\text{Rabi}}$ but partial tunneling (around 95\%) of $\Psi_V$ is observed at about $t=0.50 t_{\text{Rabi}}$.  With progress of time, the dipole structures rotate and change their relative phase. The collapse of the vortex structure into dipoles and the rotation of these dipoles take place due to the different tunneling 
frequencies of $\Psi_X$ and $\Psi_Y$. 
At $t=2.50 t_{Rabi}$ practically tunnels to the right well. It is one of that definite moments in time when each of the clouds of $\Psi_X$ and $\Psi_Y$ 
individually and practically  completely tunnels  to the right well after around 14 and 2.5 oscillations, respectively. Similarly, in the mean-field, $\Psi_X$, $\Psi_Y$  and hence $\Psi_V$  completely tunnel back to the left well at about $t=5.00 t_{Rabi}$ (not shown).

The beginning of the time evolution  in all panels of Fig.~\ref{Fig1} shows a  complete overlap between the mean-field and many-body results, 
thereby confirming that the survival probability can be accurately described by the mean-field theory at short time scales 
for the interaction strength considered here.
Correspondingly, the densities per particle computed at the many-body level in Fig.~\ref{Fig2}
 match the mean-field densities per particles of the bosonic clouds. This situation emulates the so-called infinite-particle limit of the time-dependent
many-boson Schr\"odinger equation,
in which the time-dependent density per particle coincides
with the respective density obtained from the time-dependent Gross-Pitaevskii equation,
see in this respect \cite{Erdos2007, Klaiman2016}.
As time progresses,
we observe incomplete tunneling of the densities of all the systems at the many-body level,
a first  signature of the build up of  many-body correlations,
resulting in a gradual decrease in the amplitudes of the oscillations.
The decay in the amplitudes of the tunneling oscillations can not be seen at the level of mean-field theory. The results generalize what is known in the literature for tunneling from the ground state (of the left well) in BJJs \cite{Sakmann2009}.
 Looking at Fig.~\ref{Fig1},
the decay rates of the density oscillations for $\Psi_X$ and $\Psi_V$ are rather similar
and quite smaller from the decay rates of $\Psi_G$ and $\Psi_Y$.
The intuition suggests that since $\Psi_X$ and $\Psi_V$ 'feel' a smaller barrier when tunneling,
many-body effects would develop slower, and hence the above-discussed decay rates are smaller.
Correspondingly, $\Psi_G$ has the highest decay rate and hence reaches the smallest amplitude of
oscillations at the largest time presented here.
We shall analyze 
further measures and signatures of the many-body dynamics and 
return to this intuitive reasoning in the next subsection.
As the transversely-excited and vortex states cannot be created in one spatial dimensions,
it will be particularly interesting to dig deeper into the many-body as well as the mean-field
dynamics of the $\Psi_Y$ and $\Psi_V$ states of BECs in the two-dimensional geometry.
Furthermore, we shall be looking for
signatures of the impact of the transverse direction in the tunneling dynamics of
$\Psi_G$ and $\Psi_X$ which do have one-dimensional analogs.
\clearpage

\subsection{Dynamics of the condensate fraction and fragmentation}
We have already found a difference between the mean-field and many-body time developments of the survival probability, $P_L(t)$. 
This difference implies that there are many-body correlations which gradually appear in the tunneling process.
To study the effect of the quantum correlations on the tunneling dynamics, 
we would like to discuss how the depletion or fragmentation emerges, depending on the shape of the different initial states. To this end, we compute the reduced one-particle density matrix from the time-dependent many-boson wave-function (Eq.~\ref{2})  and diagonalize the former for obtaining the  time-dependent occupation numbers $n_j(t)$ and natural orbitals $\phi_j(x,y;t)$ \cite{Coleman2000, Sakmann2008}. Here we present the time evolution of the condensate fraction, $\dfrac{n_1(t)}{N}$,
and the details of the depletion $\dfrac{n_{j>1}(t)}{N}$ of the initial states in terms
of the occupation numbers of the natural orbitals.
The change in the occupation number of the first natural orbital signifies the loss of coherence in the initial state.
We use the term fragmentation in a broad manner, to indicate a large amount of depletion,
rather than only in its strict meaning of a macroscopic occupation of more than a single natural orbital.

\begin{figure*}[!h]
{\includegraphics[trim = 0.1cm 0.5cm 0.1cm 0.2cm, scale=.80]{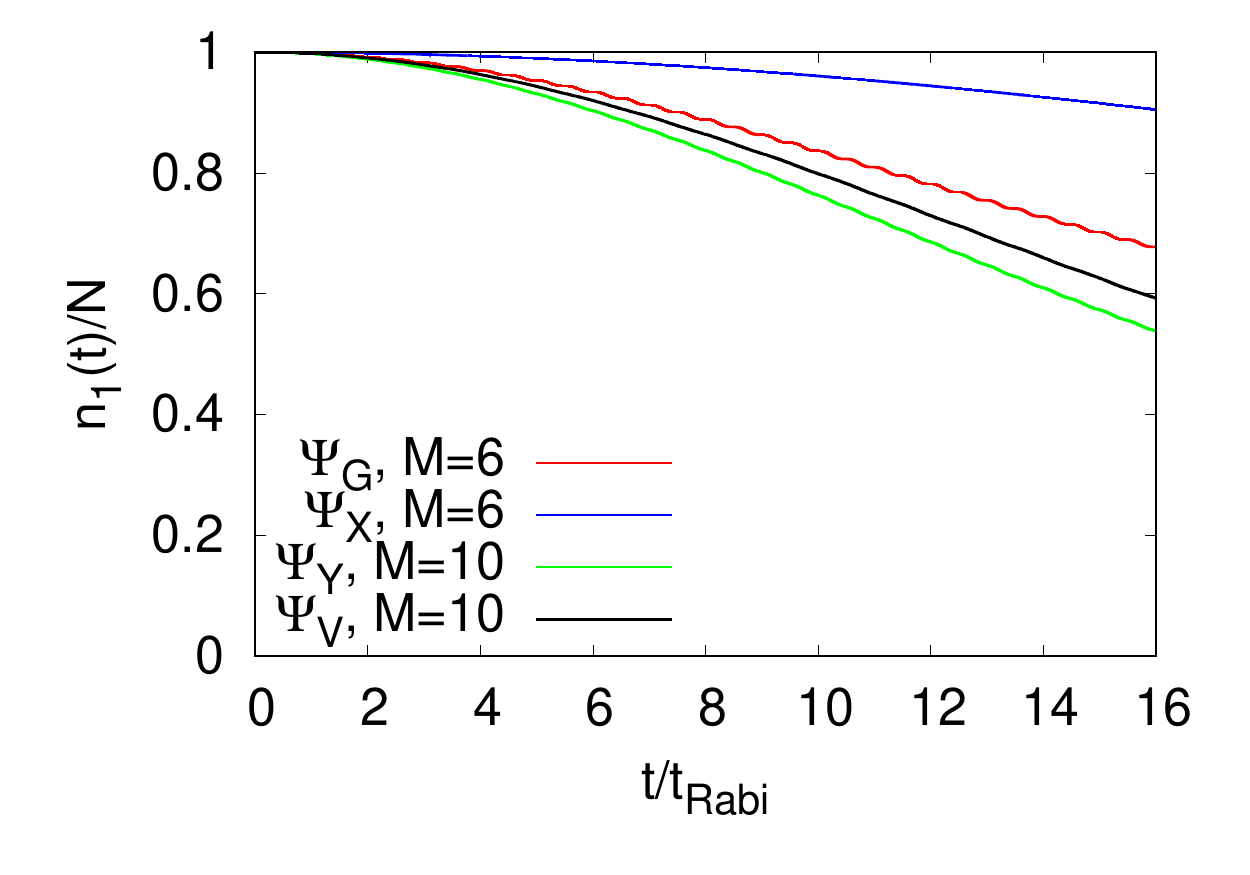}}
\caption{Time-dependent condensate fraction, $n_1(t)/N$, in a symmetric 2D double-well 
for the initial states $\Psi_G$, $\Psi_X$, $\Psi_Y$, and $\Psi_V$.
The number of bosons is $N=10$ and the interaction parameter $\Lambda=0.01$.
The results have been obtained by the MCTDHB method with $M=6$ time-adaptive orbitals
for $\Psi_{G}$ and $\Psi_{X}$ and $M=10$  time-adaptive orbitals for $\Psi_{Y}$ and $\Psi_{V}$.
See the text for more details. Color codes are explained in the panel.
The quantities shown  are dimensionless.}
\label{Fig3}
\end{figure*}

\begin{figure*}[!h]
{\includegraphics[trim = 0.1cm 0.5cm 0.1cm 0.2cm, scale=.65]{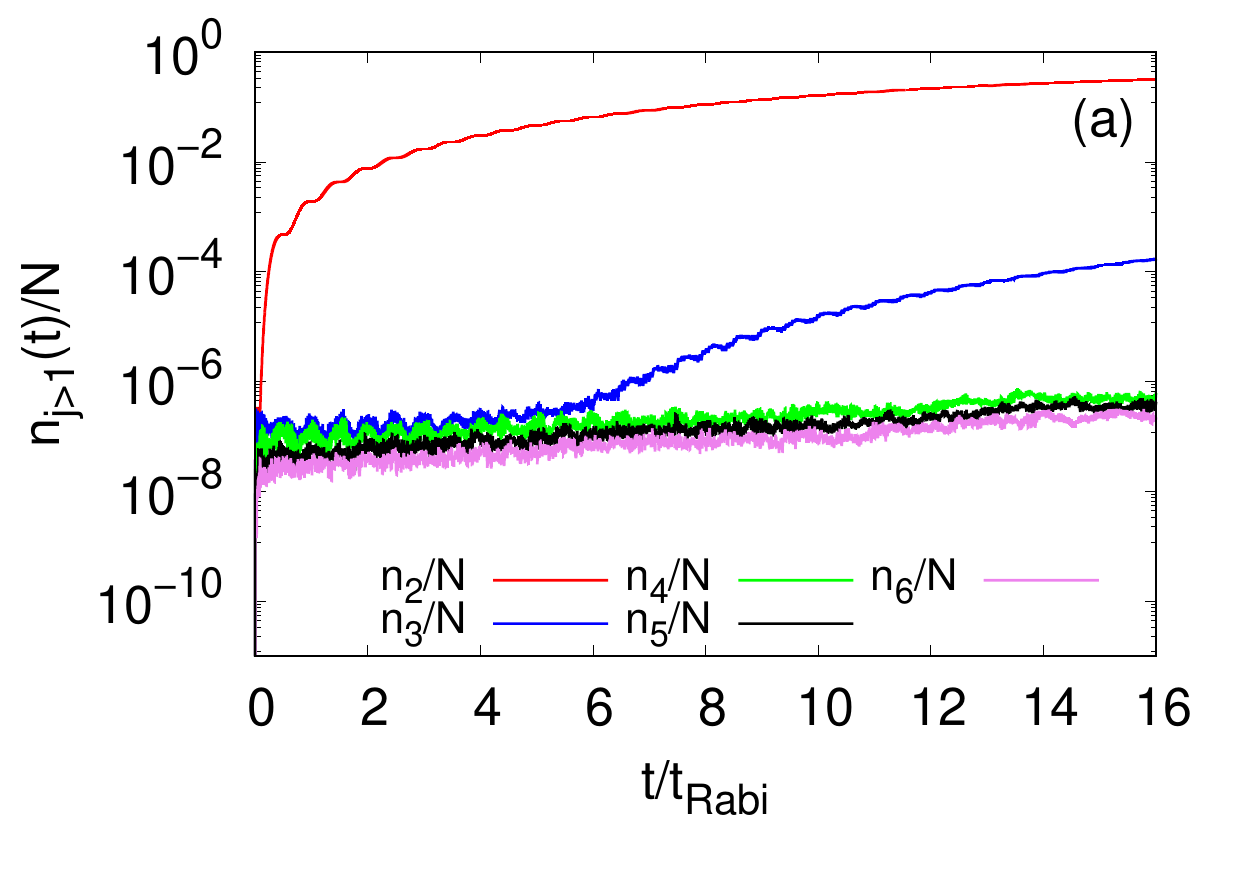}}
{\includegraphics[trim = 0.1cm 0.5cm 0.1cm 0.2cm, scale=.65]{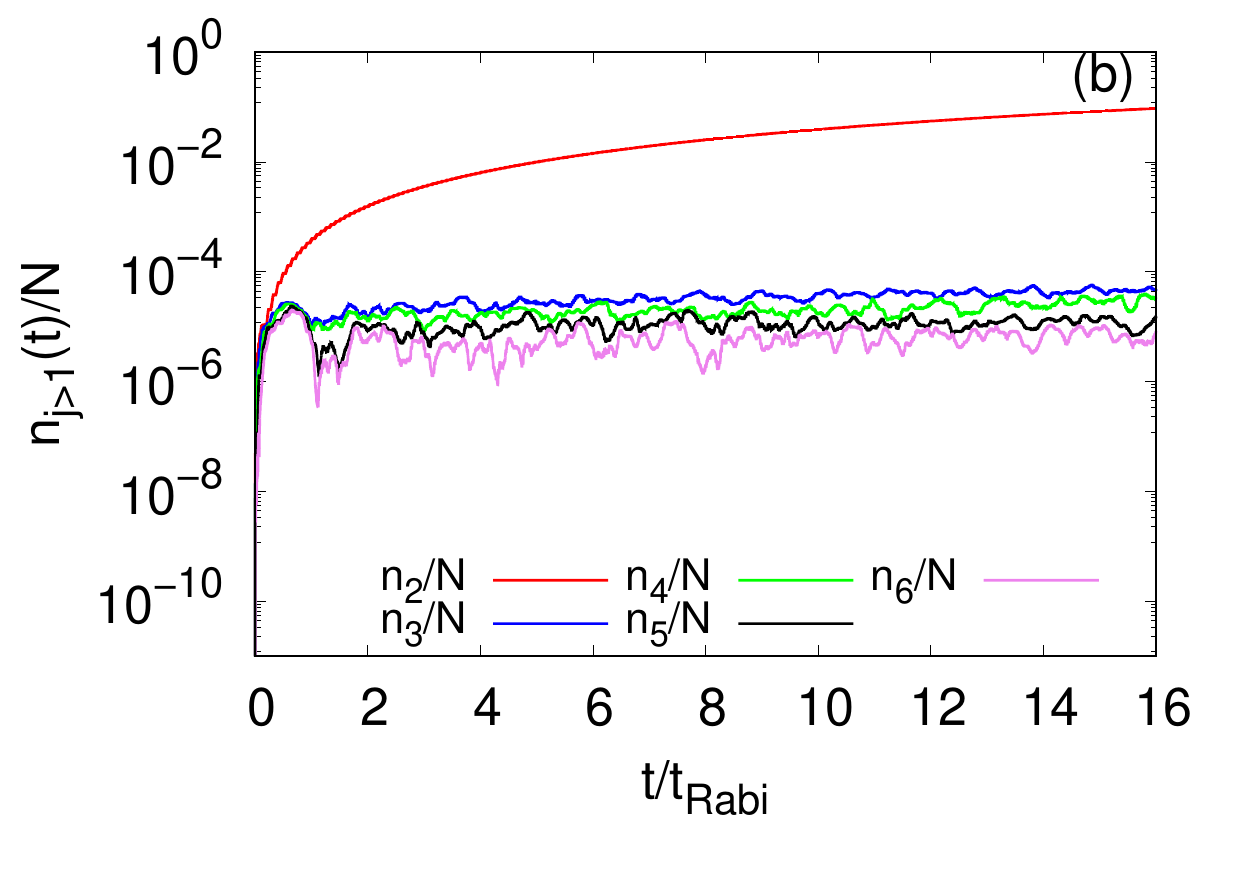}}\\
{\includegraphics[trim = 0.1cm 0.5cm 0.1cm 0.2cm, scale=.65]{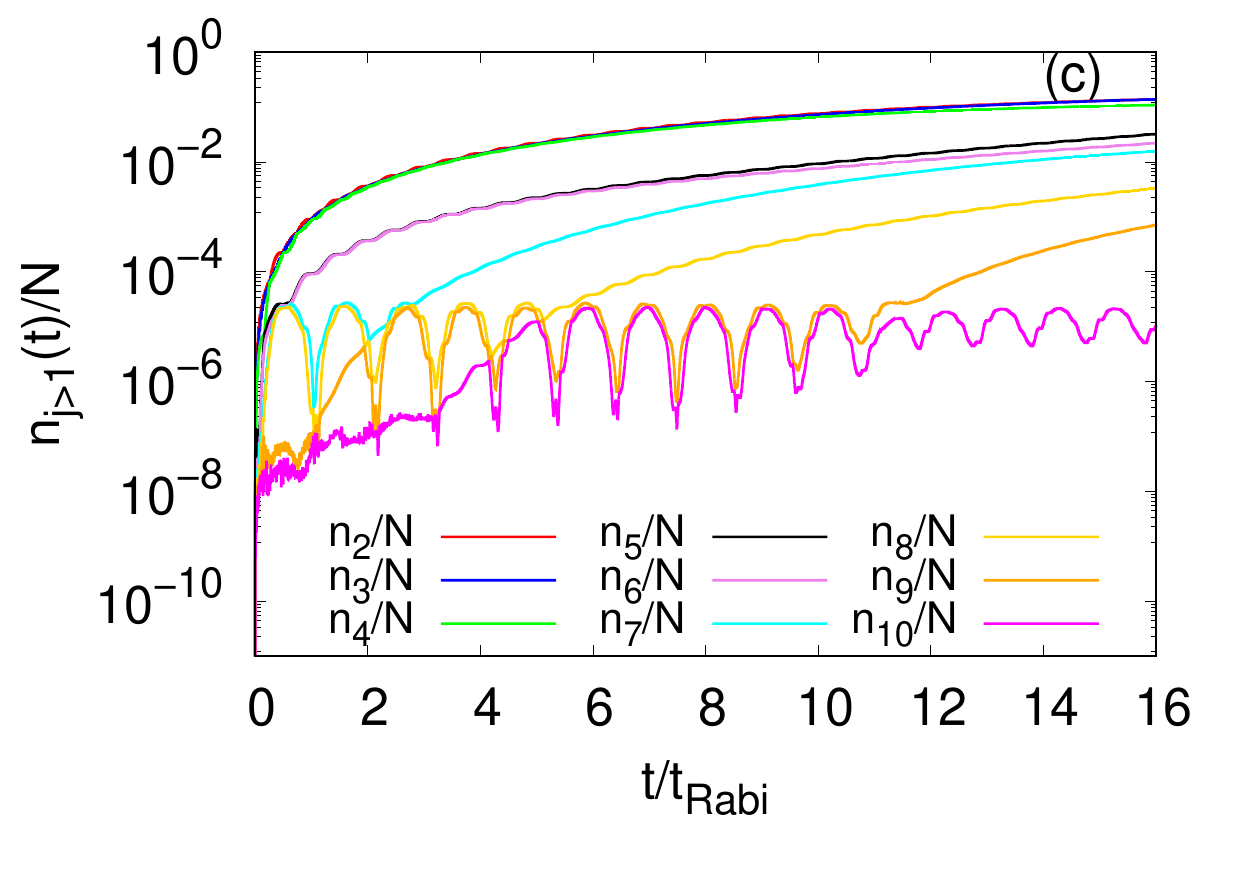}}
{\includegraphics[trim = 0.1cm 0.5cm 0.1cm 0.2cm, scale=.65]{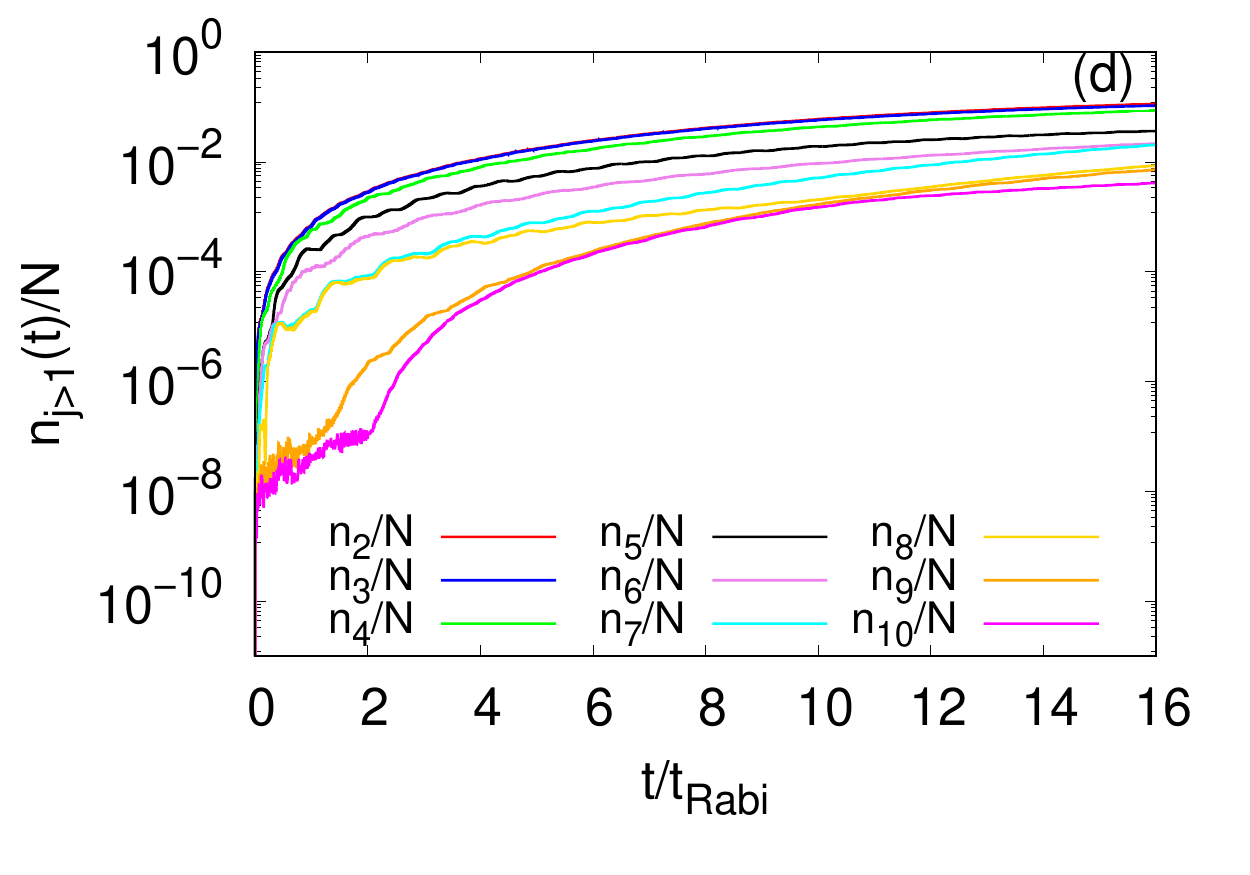}}\\
\caption{Details of the depletion. Time evolution of the occupation numbers per particle of the higher natural orbitals, $n_{j>1}(t)/N$,
in a symmetric 2D double-well for the initial states (a)  $\Psi_G$, (b) $\Psi_X$, (c) $\Psi_Y$, and (d)  $\Psi_V$.
The results have been obtained by the MCTDHB method with $M=6$ time-adaptive orbitals
for $\Psi_{G}$ and $\Psi_{X}$ and $M=10$  time-adaptive orbitals for $\Psi_{Y}$ and $\Psi_{V}$.
See the text for more details. Color codes are explained in the panel. The quantities shown  are dimensionless.}
\label{Fig4}
\end{figure*}

Fig.~\ref{Fig3} presents the time-dependent occupation
of the first natural orbital for the initial states 
$\Psi_G$, $\Psi_X$, $\Psi_Y$, and $\Psi_V$.
The corresponding occupations of the higher natural orbitals are collected in Fig.~\ref{Fig4}. As the MCTDHB computations have been performed with $M=6$ self-consistent orbitals for $\Psi_{G}$ and $\Psi_{X}$ and $M=10$ self-consistent orbitals for $\Psi_{Y}$ and $\Psi_{V}$, we have plotted the occupancies of the higher natural orbitals to have a comparative study among all the initial states and show  how coherence is lost. For the dynamics at longer times and  convergence of the individual occupation numbers see the supplemental material. Overall, here it is observed that as  time increases the occupation of the first natural orbital  decreases  with a weak oscillatory background, and the occupations of all the higher natural orbitals gradually increase, generally and in particular the lower ones in an oscillatory manner. Some of the smaller ones, e.g., for $\Psi_Y$, are oscillatory first  and then increasing. The oscillatory background atop of the global time-evolution of the occupation numbers is in reminiscence of
the tunneling back and forth in the junction.
Furthermore, one can see some high-frequency oscillations in the profiles of the natural occupancy of the higher orbitals.
These types of oscillations are the consequence of the time-dependent density oscillations.
Figs.~\ref{Fig3} and \ref{Fig4} demonstrate
that all the initial states start depleting and eventually become fragmented with time.

Examination of the respective occupation numbers of $\Psi_G$, $\Psi_X$, $\Psi_Y$, and $\Psi_V$ reveal a few trends. The first and perhaps the most prominent one, is that transverse excitation enhances fragmentation. Indeed, as time passes by, $\Psi_Y$ losses coherence faster than $\Psi_G$ and, analogously, $\Psi_V$ losses coherence faster than $\Psi_X$, see Fig.~\ref{Fig3}. On the other hand, longitudinal excitations suppress fragmentation, namely, $\Psi_X$ losses coherence slower than $\Psi_G$ and, similarly, $\Psi_V$ losses coherence slower than $\Psi_Y$. All in all, $\Psi_Y$ is the fastest to fragment and $\Psi_X$ is the slowest.  The second is a comparison of the fragmentation dynamics in Fig.~\ref{Fig3} to the decay of the amplitude of the density oscillations, see Fig.~\ref{Fig1}. Since fragmentation can develop due to the transverse excitations, also see below, there is no one-to-one correlation between the two properties of the junction, as is the case in one spatial dimension \cite{Sakmann2009, Sakmann2014}. For instance, $\Psi_Y$ is more fragmented than $\Psi_G$, but the density oscillations of the former decay slower than the latter.

Finally, we discuss and compare how the higher natural orbitals become occupied in the four initial states, see Fig.~\ref{Fig4}.
We notice that whenever the initial state is transversely excited, $\Psi_Y$, 
or is a linear combination consisting of a transversely excited state, $\Psi_V$,
it requires a larger number of self-consistent orbitals to accurately represent its dynamical behavior.
Furthermore, examining for each state 
the largest higher natural densities sheds light on the microscopic mechanism of fragmentation. In particular, we find  that the second and third  natural densities for $\Psi_G$ and   the second natural density  for $\Psi_X$ have reflection symmetry with no-node in the $y$-direction, they only have excitation in the $x$-direction at $t=10t_{Rabi}$ (not shown). For $\Psi_Y$, we observe that the second, third, and fourth natural densities have  one, zero and two nodes at $t=10t_{Rabi}$ in the $y$-direction, respectively (see Fig. S7 of supplemental material). Unlike the other three initial states,  the  natural densities of the three larger  (second, third, and fourth)  orbitals  for $\Psi_V$ show complex structures having zero, two and one nodes, respectively, in the $x$-$y$ plane at $t=10t_{Rabi}$ (see Fig. S8 of supplemental material). However, the shape of the natural orbitals of  $\Psi_V$ exhibits  the presence of the longitudinal and  transverse excitations in the system.

\subsection{Observables and the dynamics of their expectation values and variances}

So far, we discussed the transport of the bosons in the junction in terms of the survival probability and the loss of coherence and development of fragmentation in the reduced one-particle density matrix. To shed further light on the time-dependent many-particle wavefunction and on possible geometrical and dimensional effects, we resort to further quantities, the position operator along the $x$- and $y$-direction,  the  momentum operator along the $x$- and $y$-direction, and the angular-momentum  operator of $z$ component. Here we demonstrate the dynamics of the expectation values and the variances of  the above-mentioned  operators  and draw a connection with the survival probability and fragmentation.

Let us start with a brief discussion about the expectation values of observables and their dynamics. We find that the expectation value of the $\hat X=\sum_{j=1}^N \hat x_j$ position operator, $\dfrac{1}{N}\langle\Psi|{\hat{X}}|\Psi\rangle(t)$, for all  initial states possesses a similar structure as found for the respective survival probability profile, see Fig.~\ref{Fig1}. Namely, at the mean-field level, $\dfrac{1}{N}\langle\Psi|{\hat{X}}|\Psi\rangle(t)$ are oscillating in between the two minima of the double-well potential starting from the initial value $-2$ at $t=0$, see Table~\ref{I}. While at the many-body level, we find numerically that $\dfrac{1}{N}\langle\Psi|{\hat{X}}|\Psi\rangle(t)$ for all initial states eventually vanish with  time due to the gradual increase of many-body correlations as described in the many-body survival probability (results are not shown).

 For $t=0$, we note that the expectation value of the $\hat P_X=\sum_{j=1}^N \frac{1}{i}\frac{\partial}{\partial x_j}$
momentum operator vanishes, $\dfrac{1}{N}\langle\Psi|{\hat{P}_X}|\Psi\rangle(0)=0$,
due to parity (reflection in $x$ for $\Psi_G$, $\Psi_X$, and $\Psi_Y$; inversion through the origin for $\Psi_V$) and translation.
Similarly, the expectation values of $\hat Y=\sum_{j=1}^N \hat y_j$
and $\hat P_Y=\sum_{j=1}^N \frac{1}{i}\frac{\partial}{\partial y_j}$ along the transverse direction vanish,
$\dfrac{1}{N}\langle\Psi|{\hat{Y}}|\Psi\rangle(0)=\dfrac{1}{N}\langle\Psi|{\hat{P}_Y}|\Psi\rangle(0)=0$,
due to parity (reflection in $y$ for $\Psi_G$, $\Psi_X$, and $\Psi_Y$; inversion for $\Psi_V$).
Table~\ref{I} summarizes the results.
At $t>0$ some of these symmetries are exactly conserved,  and the  expectation values $\dfrac{1}{N}\langle\Psi|{\hat{Y}}|\Psi\rangle$ and $\dfrac{1}{N}\langle\Psi|{\hat{P}_Y}|\Psi\rangle$ vanish. But the expectation value $\dfrac{1}{N}\langle\Psi|{\hat{P}_X}|\Psi\rangle$ shows oscillatory behavior for all  states at $t>0$ for the mean-field as well as  the many-body dynamics. For $\Psi_G$ and $\Psi_Y$, $\dfrac{1}{N}\langle\Psi|{\hat{P}_X}|\Psi\rangle$ keeps on oscillating between $+0.15$ and $-0.15$, while for $\Psi_X$, the values are $+0.70$ and $-0.70$, and   for $\Psi_V$, it oscillates between $+0.45$ and $-0.45$ at the mean-field level.  The dynamics of $\dfrac{1}{N}\langle\Psi|{\hat{P}_X}|\Psi\rangle$ at the many-body level  overlaps only initially  with the respective expectation values at the mean-field level and eventually shows a decay in amplitude  at  long-time as  many-body correlations develop contrary to the mean-field level.

We now move to the angular-momentum which is a fundamental property in 2D
taking the operator form 
$\hat{L}_Z=\sum_{j=1}^N \dfrac{1}{i}\left(x_j\dfrac{\partial}{\partial y_j}-y_j\dfrac{\partial}{\partial x_j}\right)$.
In connection with the above-discussed quantities, angular-momentum is a combination of position and momentum operators. As  is expected, at $t=0$ the expectation value of the angular-momentum operator,
$\dfrac{1}{N}\langle\Psi|{{\hat{L}_Z}}|\Psi\rangle(0)$,
is one for the vortex state while it has a null value for all other states. For $t>0$,  $\dfrac{1}{N}\langle\Psi|{{\hat{L}_Z}}|\Psi\rangle$ for $\Psi_G$, $\Psi_X$, and $\Psi_Y$ vanishes at the mean-field as well as  the many-body level due to the reflection symmetry in $y$. But for $\Psi_V$,  $\dfrac{1}{N}\langle\Psi|{{\hat{L}_Z}}|\Psi\rangle (t>0)$ shows an interesting oscillatory motion.  Recall that angular-momentum is not a conserved quantity in the junction.  At the mean-field level this oscillatory behavior is with values  between  $+1$ and $-1$, where as at the many-body level, there is a decay of the amplitude of oscillations 
due to the loss of coherence of the vortex state (see Fig~\ref{Fig5}) and decay of density oscillations (see Fig. S1(d) of supplemental material). Connecting the expectation value of the angular momentum with the density profile of $\Psi_V$, it is found that Fig~\ref{Fig2}(p)-(y) have $\dfrac{1}{N}\langle\Psi|{{\hat{L}_Z}}|\Psi\rangle$ values $+1$, $0$, $-0.5$, $0$, $0.5$, $0.5$, $0$, $-1$, $+1$, and $-0.7$, respectively, at the many-body level. It is noted that the  $\dfrac{1}{N}\langle\Psi|{{\hat{L}_Z}}|\Psi\rangle$ values at the many-body level initially (until about $t=4t_{Rabi}$) overlap with the corresponding mean-field results (see Fig~\ref{Fig5}), therefore we observe the same values of  $\dfrac{1}{N}\langle\Psi|{{\hat{L}_Z}}|\Psi\rangle$ at the mean-field  and many-body levels at times when the snapshots of Fig~\ref{Fig2}(p)-(y) are taken.  Fig~\ref{Fig2}(x)  finds  the value $+1$ for $\dfrac{1}{N}\langle\Psi|{{\hat{L}_Z}}|\Psi\rangle$ even though the density profile does not produce a pure vortex state. This phase difference in the vortex structure occurs due the different tunneling frequency of $\Psi_X$ and $\Psi_Y$. Also, it is noticed that the many-body $\dfrac{1}{N}\langle\Psi|{{\hat{L}_Z}}|\Psi\rangle$ is  $-0.7$ when the vortex state practically tunnels to the right well [see Fig~\ref{Fig2}(y)].

\begin{table}[htb]
\caption{Expectation values of the observables and their variances at $t=0$.
See the text for more details.
The quantities shown are dimensionless.}
\centering
\vspace{0.5cm}
\begin{tabular}{cccccc|ccccc}
\hline
\hline
\vspace{0.5cm}
State  &$\dfrac{1}{N}\langle{\hat{X}}\rangle$&$\dfrac{1}{N}\langle{\hat{Y}}\rangle$ &$\dfrac{1}{N}\langle{\hat{P}_X}\rangle$& $\dfrac{1}{N}\langle{\hat{P}_Y}\rangle$&$\dfrac{1}{N}\langle{\hat{L}_Z}\rangle$& $\dfrac{1}{N}\Delta_{{\hat{X}}}^2$&  $\dfrac{1}{N}\Delta_{{\hat{Y}}}^2$&$\dfrac{1}{N}\Delta_{{\hat{P}_X}}^2$ & $\dfrac{1}{N}\Delta_{{\hat{P}_Y}}^2$ &$\dfrac{1}{N}\Delta_{{\hat{L}_Z}}^2$ 
         \\ [0.2ex]
\hline 
  \vspace{0.5cm}
$\Psi_G$	&	-2	&	0	&	0	&	0	&	0	&	$\dfrac{1}{2}$&	$\dfrac{1}{2}$	&	$\dfrac{1}{2}$&	$\dfrac{1}{2}$	&	2\\
	\vspace{0.5cm}
$\Psi_X$	&	-2	&	0	&	0	&	0	&	0	&	$\dfrac{3}{2}$	&	$\dfrac{1}{2}$	&	$\dfrac{3}{2}$	&	$\dfrac{1}{2}$	& 3\\
\vspace{0.5cm}
$\Psi_Y$	&	-2	&	0	&	0	&	0	&	0	&	$\dfrac{1}{2}$	&	$\dfrac{3}{2}$	&	$\dfrac{1}{2}$	&	$\dfrac{3}{2}$	& 7\\
\vspace{0.5cm}
$\Psi_V$	&	-2	&	0	&	0	&	0	&	1	&	1	&	1	&	1	&	1	&4\\
\hline
\hline
\label{I}
\end{tabular}
\end{table}
\begin{figure*}[!h]
{\includegraphics[trim = 0.1cm 0.5cm 0.1cm 0.2cm, scale=.80]{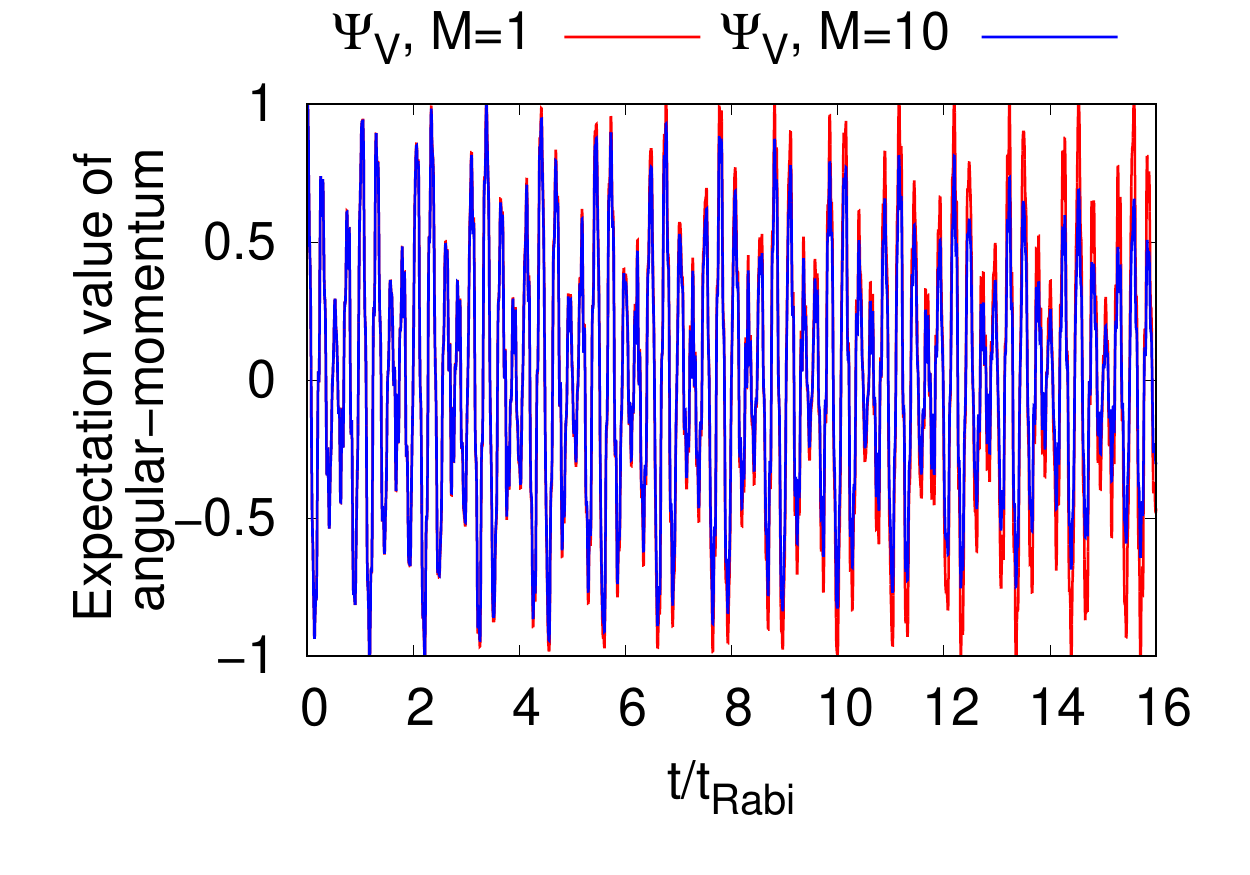}}
\caption{Dynamics of the angular-momentum expectation value   per particle, $\dfrac{1}{N}\langle\Psi_V|{{\hat{L}_Z}}|\Psi_V\rangle$, in a symmetric 2D double-well for the vortex state $(\Psi_V)$ with  $N=10$ bosons. Mean-field ($M=1$ time-adaptive orbitals) result is presented  in  red and corresponding many-body result ($M=10$ time-adaptive orbitals)  in  blue.   See the text for more details. The quantities shown  are dimensionless.}
\label{Fig5}
\end{figure*}

 Now we start the discussion about the many-particle variances of the previously discussed observables at the mean-field and many-body levels. The variances at $t=0$ are analytically calculated and  presented in Table~\ref{I}. At $t=0$, the center-of-mass of the bosonic clouds are at the position $(a,b)=(-2,0)$. To calculate the  variances, we have used the general relation between the variances at $(a,b)$ and  at the origin. It is known that for the position and momentum operators, the variances do not change with the position of the center-of-mass of the clouds, i.e., $\dfrac{1}{N}\Delta_{{\hat{X}}}^2\Big|_{\Psi(a,b)}=\dfrac{1}{N}\Delta_{{\hat{X}}}^2\Big|_{\Psi(0,0)}$,  $\dfrac{1}{N}\Delta_{{\hat{Y}}}^2\Big|_{\Psi(a,b)}=\dfrac{1}{N}\Delta_{{\hat{Y}}}^2\Big|_{\Psi(0,0)}$, $\dfrac{1}{N}\Delta_{{\hat{P}_X}}^2\Big|_{\Psi(a,b)}=\dfrac{1}{N}\Delta_{{\hat{P}_X}}^2\Big|_{\Psi(0,0)}$, and  $\dfrac{1}{N}\Delta_{{\hat{P}_Y}}^2\Big|_{\Psi(a,b)}=\dfrac{1}{N}\Delta_{{\hat{P}_Y}}^2\Big|_{\Psi(0,0)}$.  For our system, the variance of the angular-momentum operator  boils down to (see supplemental material and \cite{Alon2019b})
\begin{equation}\label{4}
\dfrac{1}{N}\Delta_{{\hat{L}_Z}}^2\Big|_{\Psi(a,b)}=\dfrac{1}{N}\Delta_{{\hat{L}_Z}}^2\Big|_{\Psi(0,0)}+a^2\dfrac{1}{N}\Delta_{{\hat{P}_Y}}^2\Big|_{\Psi(0,0)}.
\end{equation}

\begin{figure*}[!h]
{\includegraphics[trim = 0.1cm 0.5cm 0.1cm 0.2cm, scale=.60]{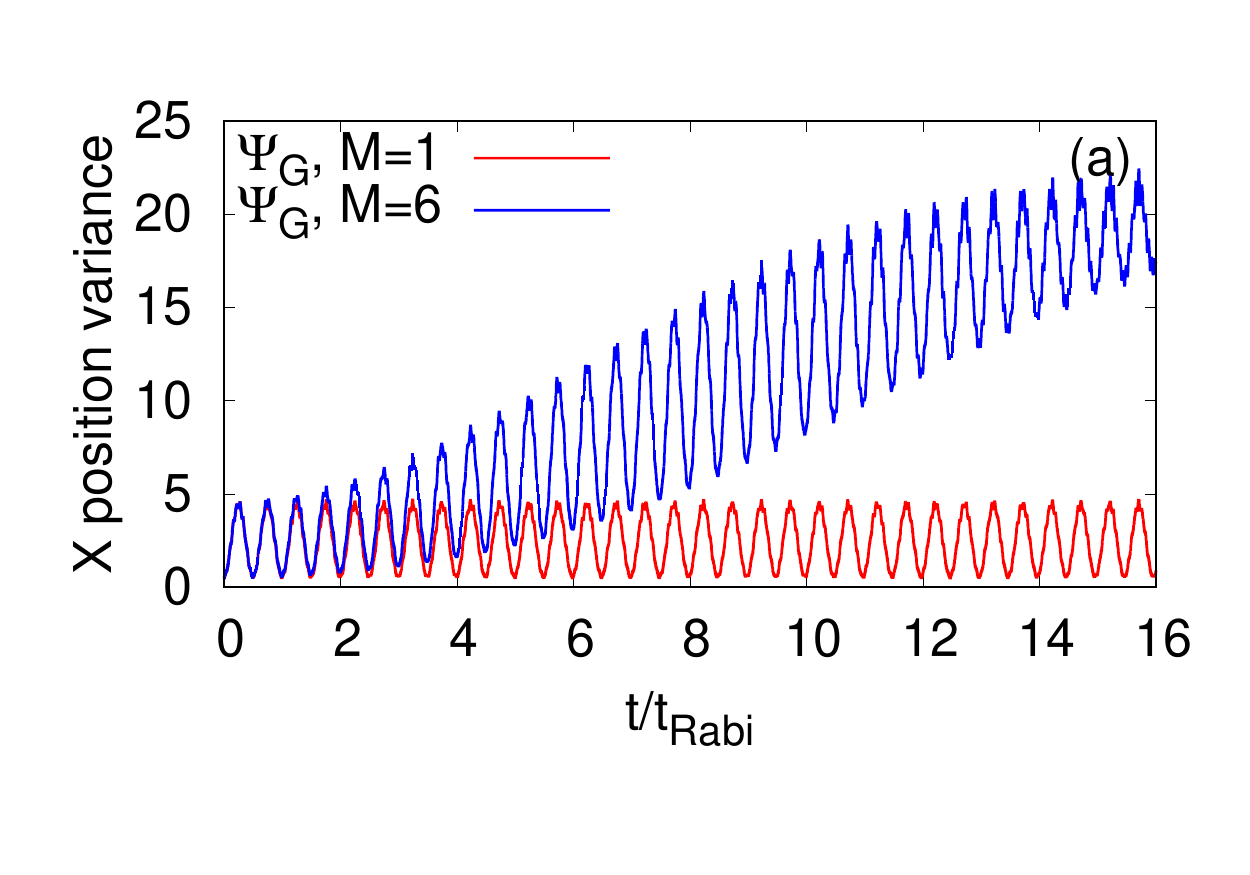}}
{\includegraphics[trim = 0.1cm 0.5cm 0.1cm 0.2cm, scale=.60]{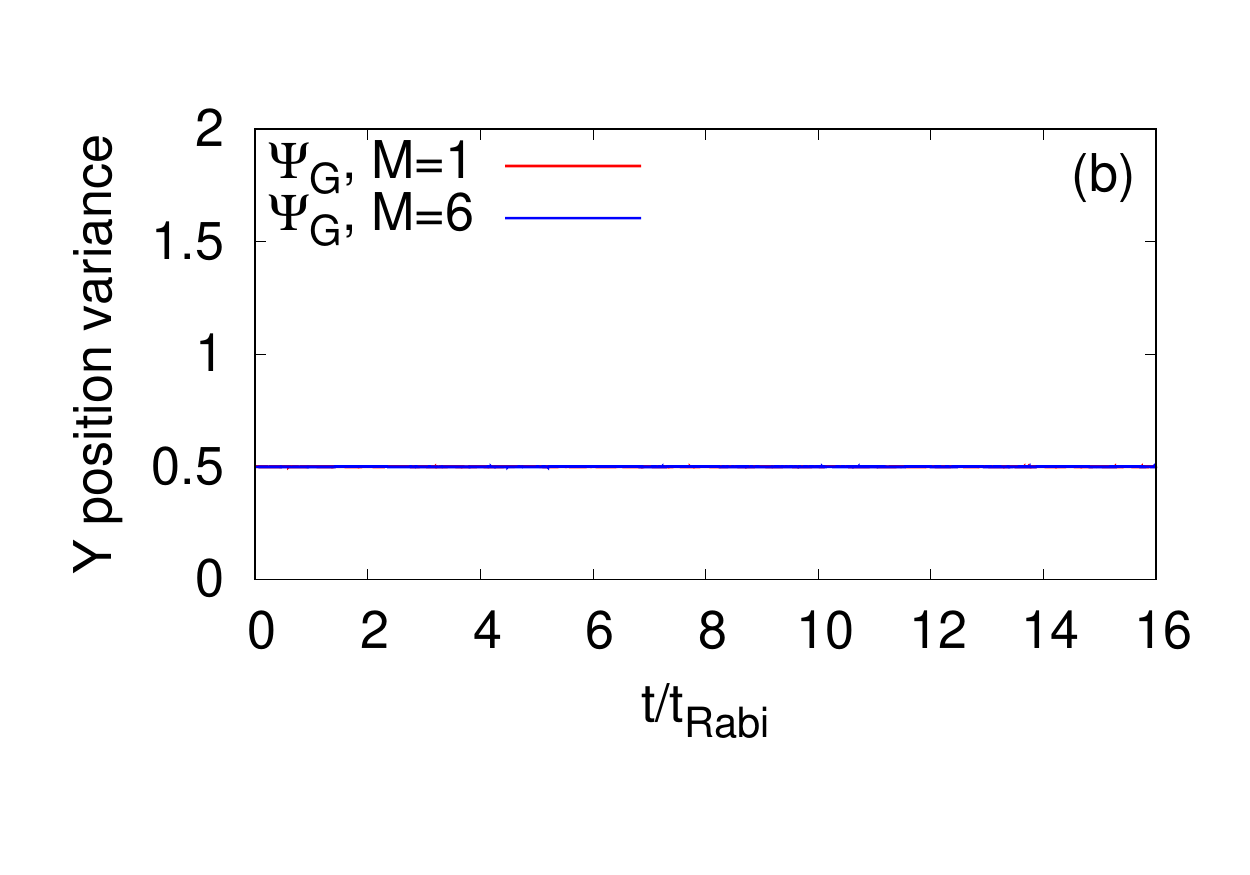}}\\
\vglue -0.75 truecm
{\includegraphics[trim = 0.1cm 0.5cm 0.1cm 0.2cm, scale=.60]{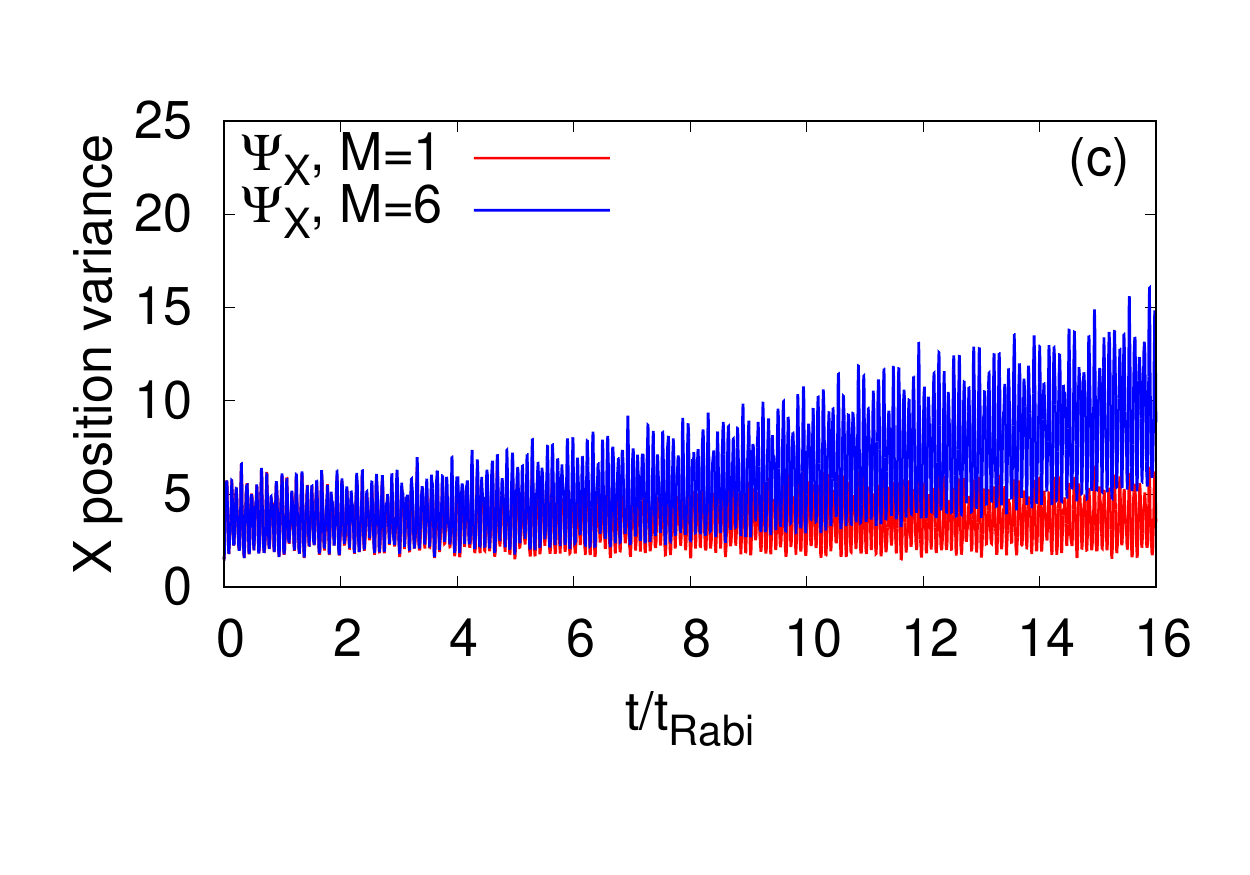}}
{\includegraphics[trim = 0.1cm 0.5cm 0.1cm 0.2cm, scale=.60]{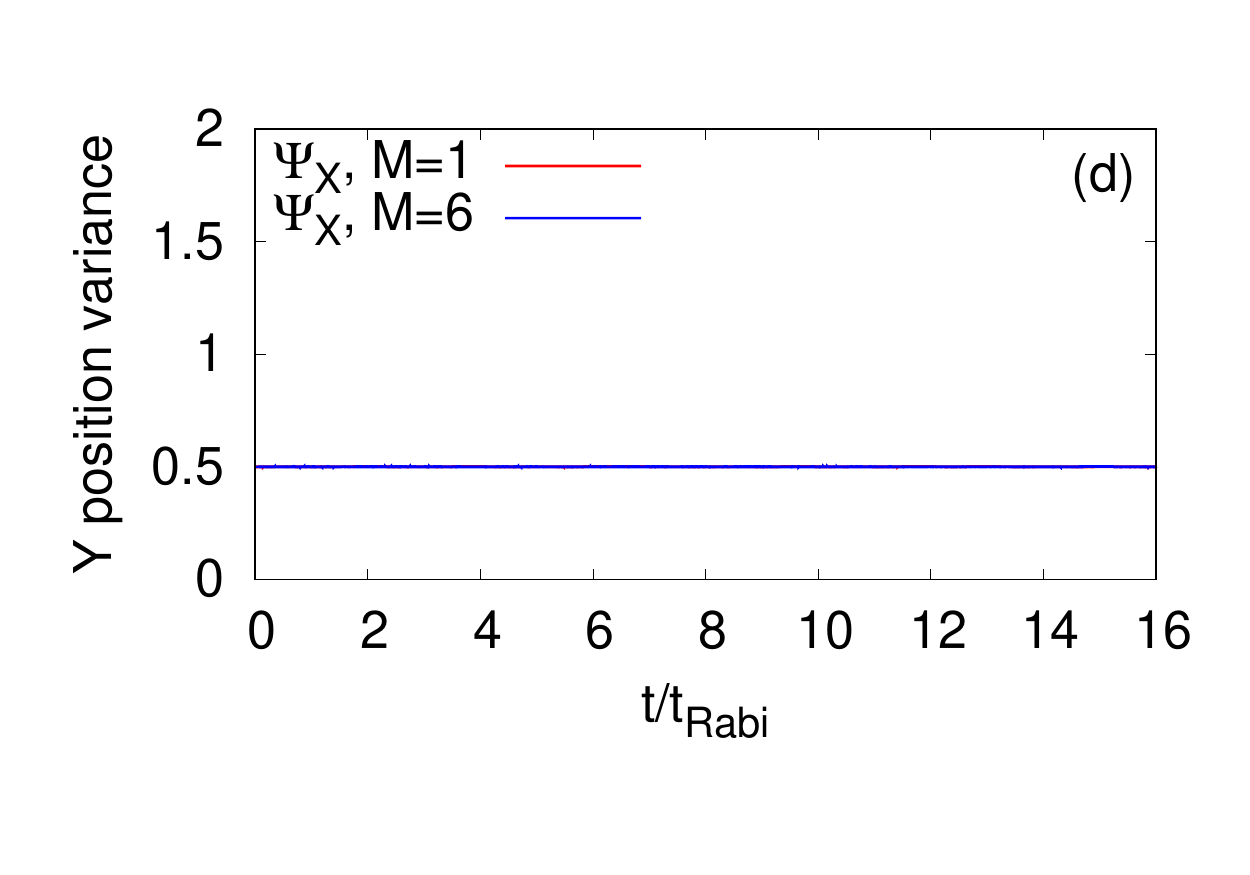}}\\
\vglue -0.75 truecm
{\includegraphics[trim = 0.1cm 0.5cm 0.1cm 0.2cm, scale=.60]{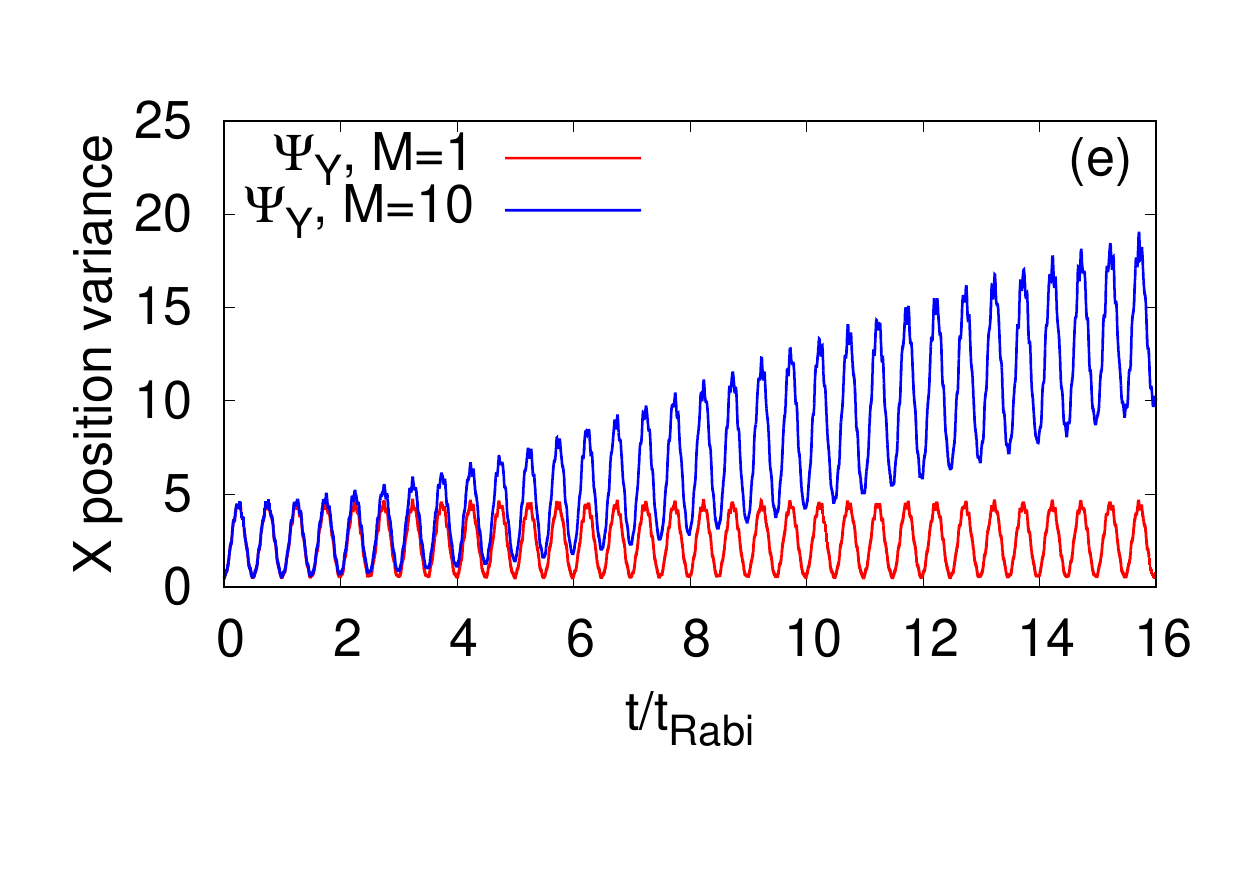}}
{\includegraphics[trim = 0.1cm 0.5cm 0.1cm 0.2cm, scale=.60]{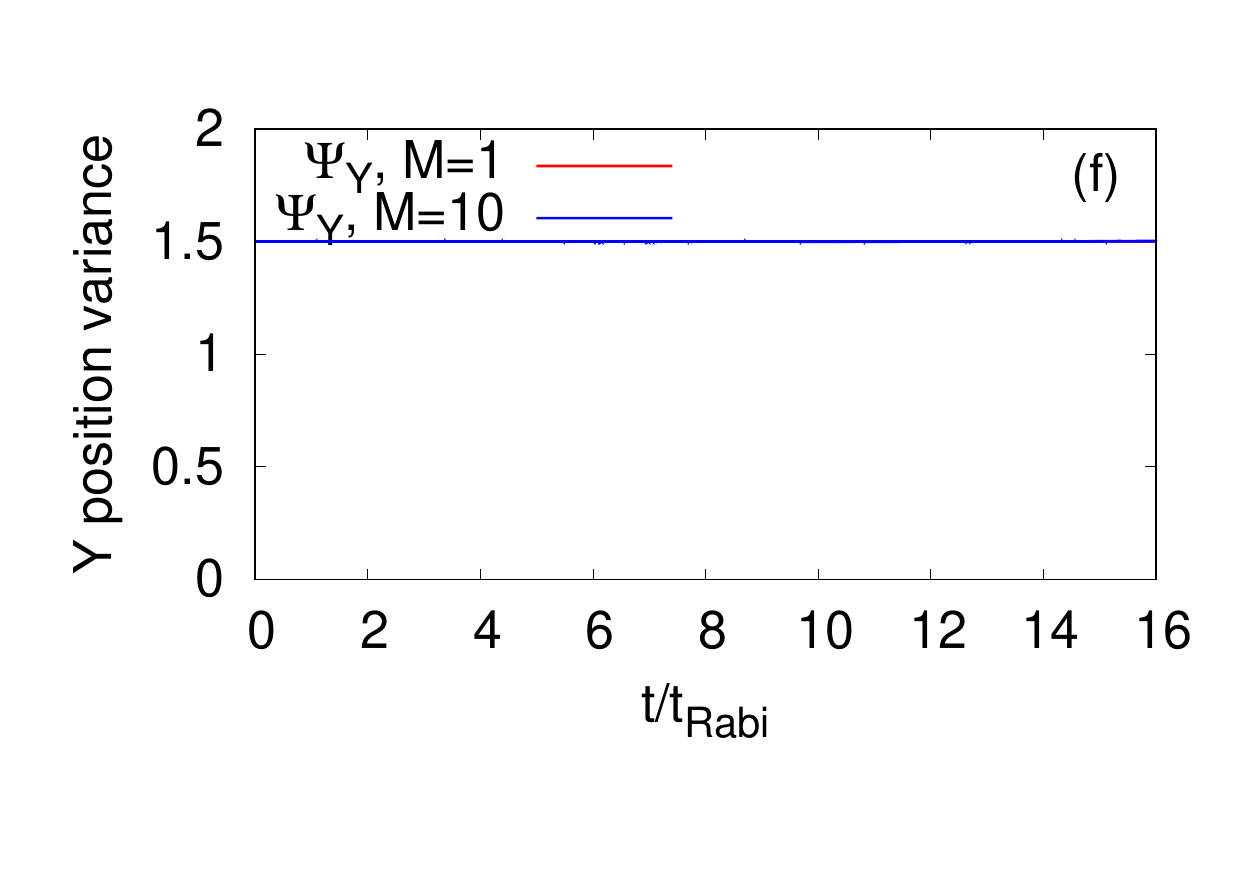}}\\
\vglue -0.75 truecm
{\includegraphics[trim = 0.1cm 0.5cm 0.1cm 0.2cm, scale=.60]{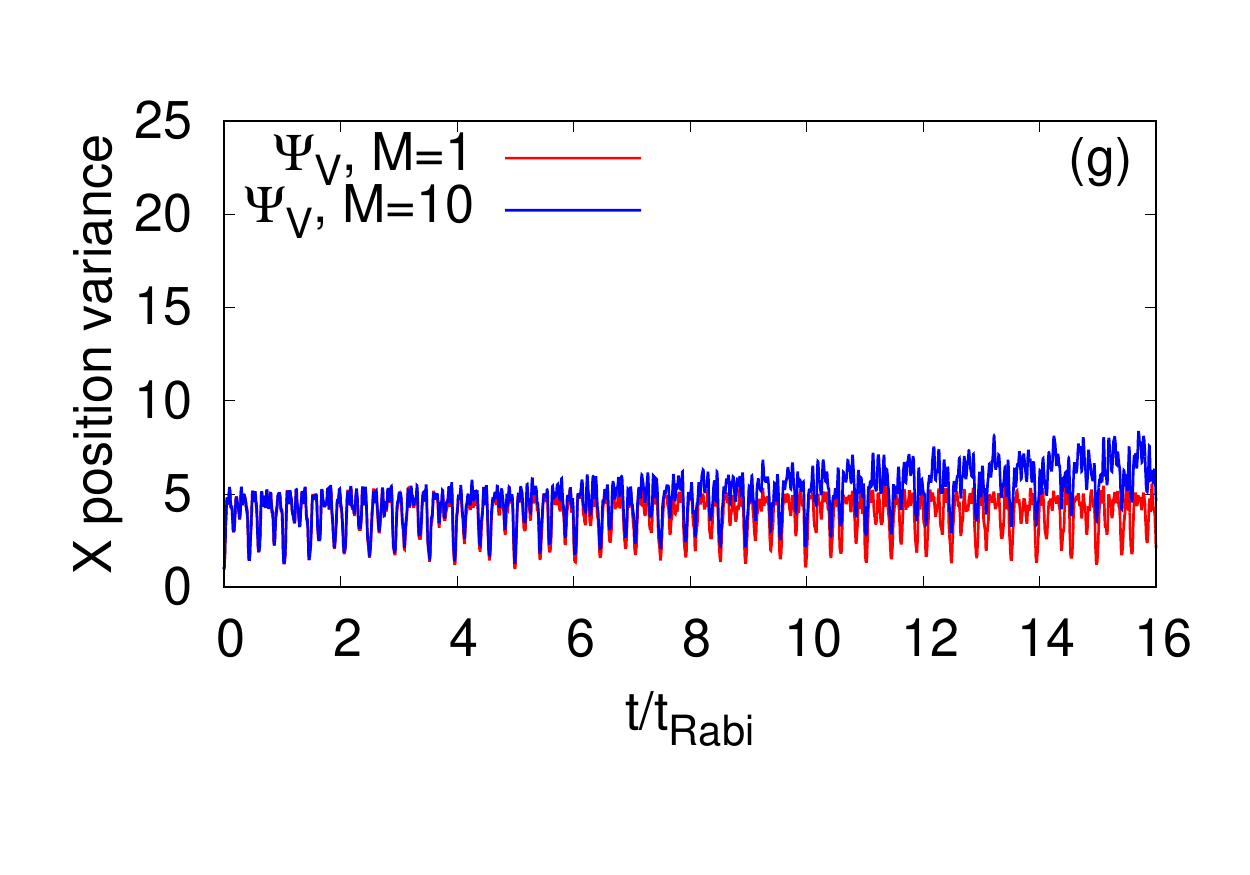}}
{\includegraphics[trim = 0.1cm 0.5cm 0.1cm 0.2cm, scale=.60]{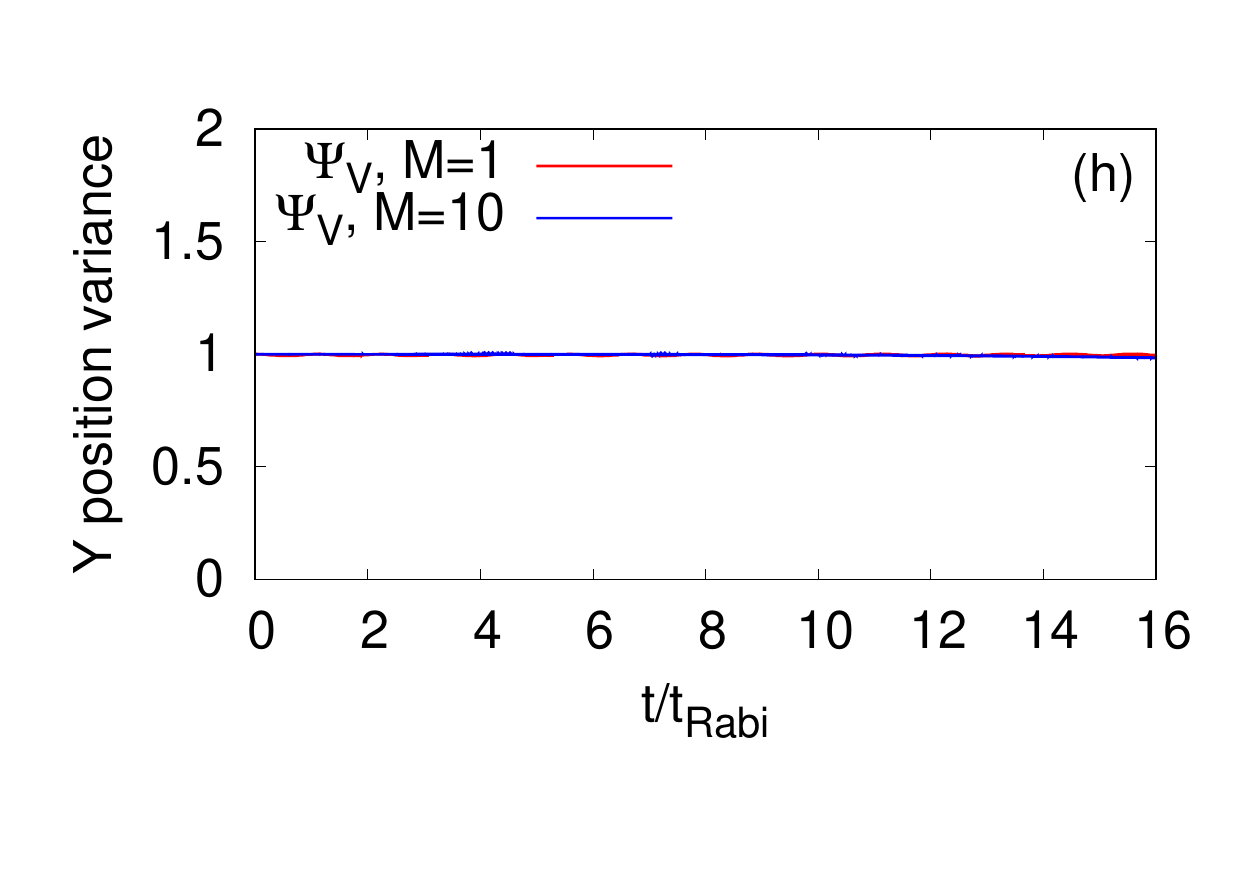}}\\
\caption{The mean-field ($M=1$ time-adaptive orbitals,  in  red) and many-body (in  blue) time-dependent position variances per particle, $\dfrac{1}{N}\Delta_{\hat{X}}^2(t)$  and  $\dfrac{1}{N}\Delta_{\hat{Y}}^2(t)$, are presented in the left and right columns, respectively. The different initial states,  $\Psi_G$, $\Psi_X$, $\Psi_Y$, and $\Psi_V$,  for  $N=10$ bosons are plotted row-wise.  The many-body results are computed using the MCTDHB method  with  $M=6$  orbitals for $\Psi_{G}$ and $\Psi_{X}$ and  $M=10$   orbitals for $\Psi_{Y}$ and $\Psi_{V}$.  See the text for more details. The quantities shown  are dimensionless.}
\label{Fig6}
\end{figure*}

\begin{figure*}[h]
{\includegraphics[trim = 0.1cm 0.5cm 0.1cm 0.2cm, scale=.60]{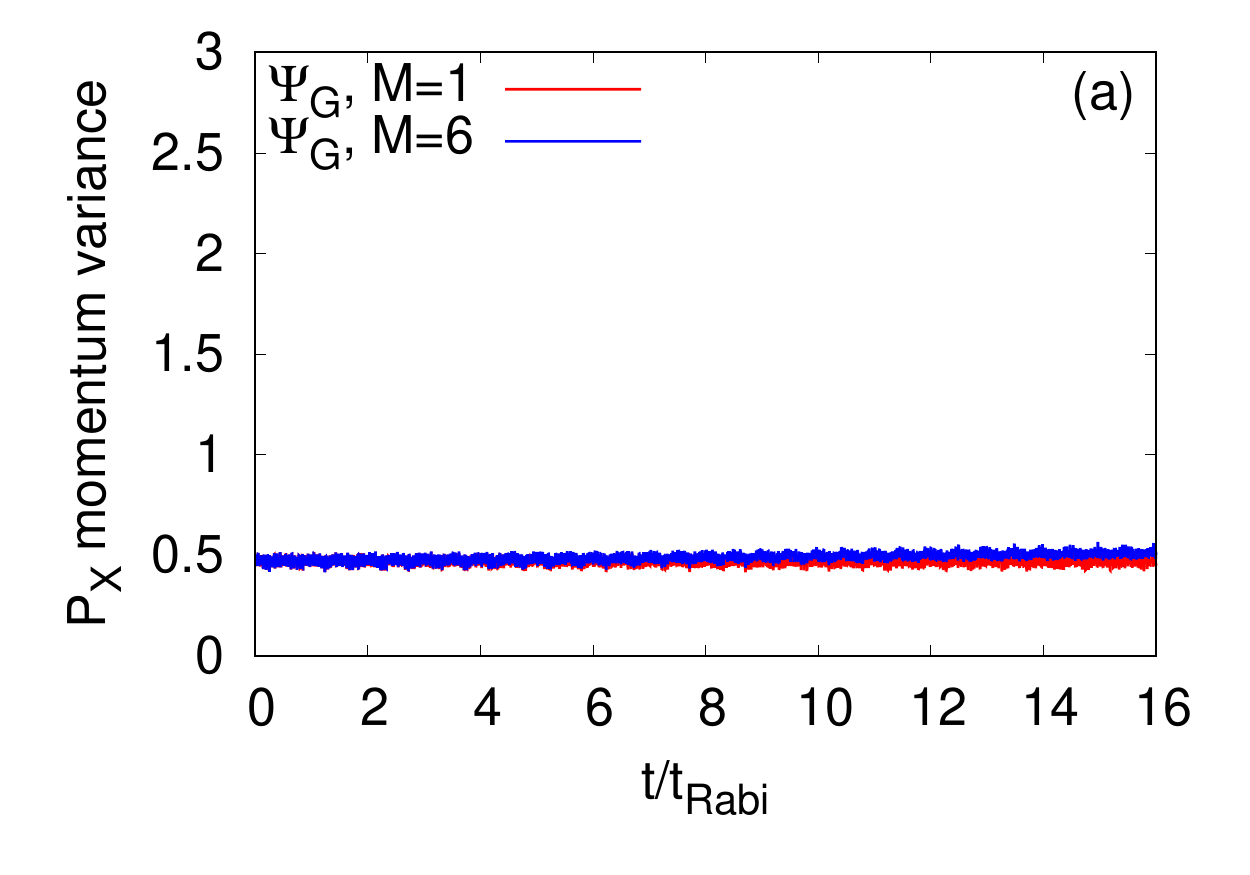}}
{\includegraphics[trim = 0.1cm 0.5cm 0.1cm 0.2cm, scale=.60]{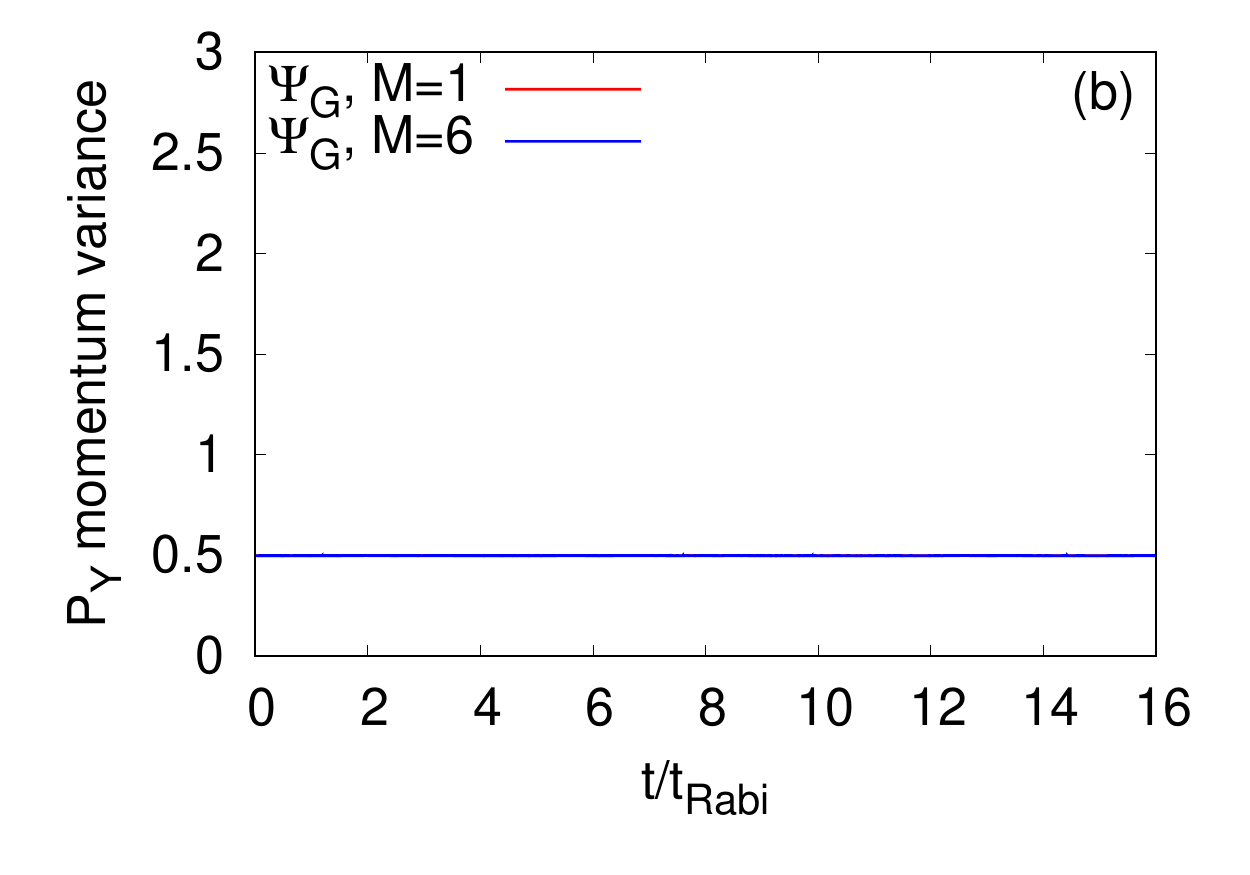}}\\
{\includegraphics[trim = 0.1cm 0.5cm 0.1cm 0.2cm, scale=.60]{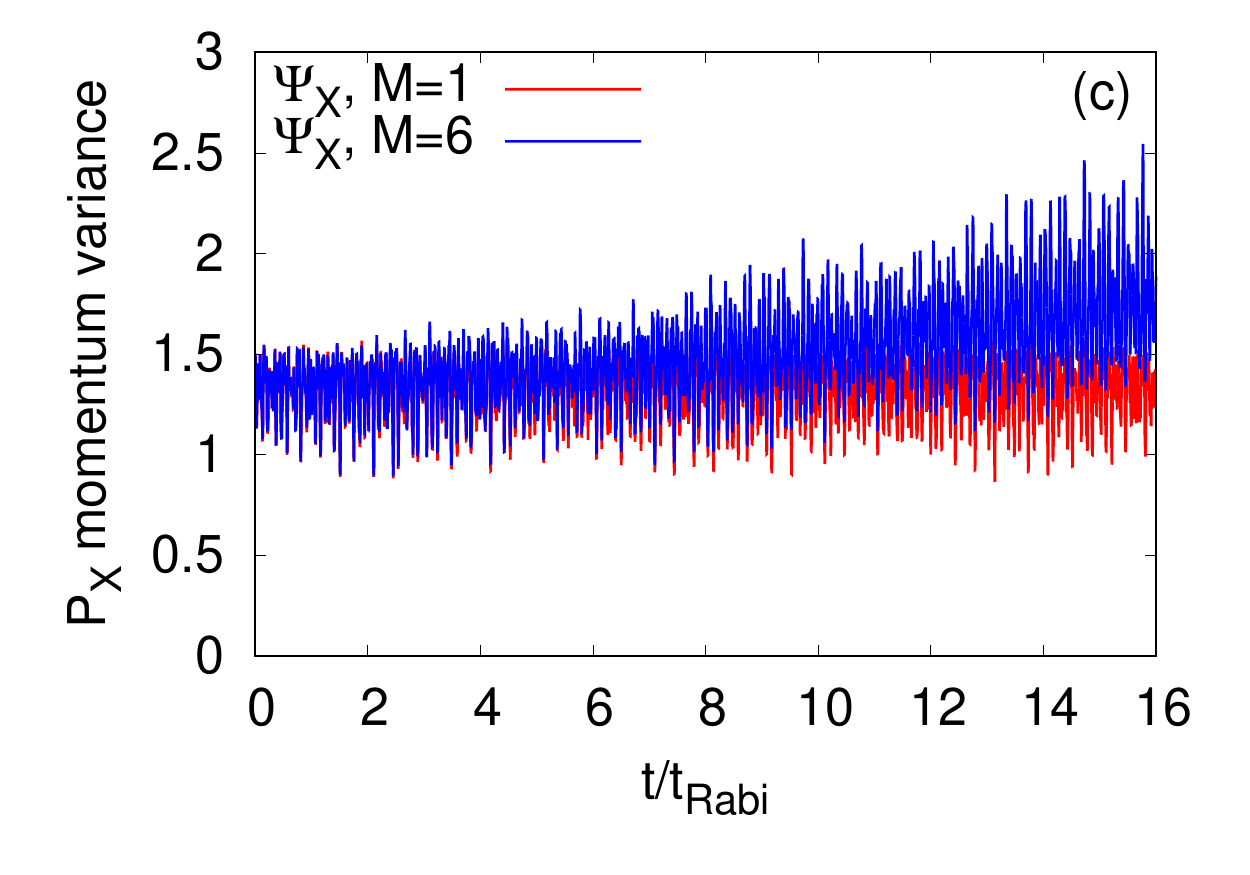}}
{\includegraphics[trim = 0.1cm 0.5cm 0.1cm 0.2cm, scale=.60]{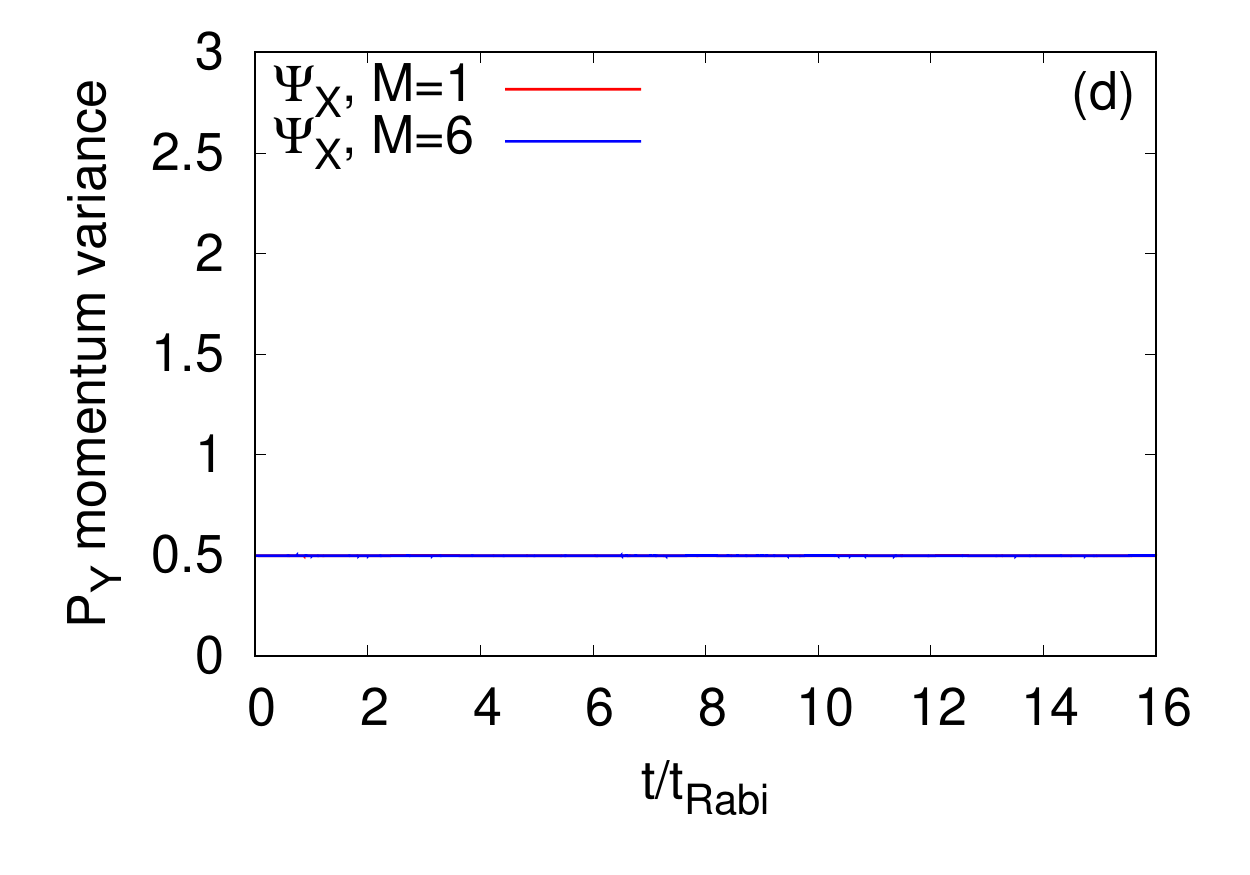}}\\
{\includegraphics[trim = 0.1cm 0.5cm 0.1cm 0.2cm, scale=.60]{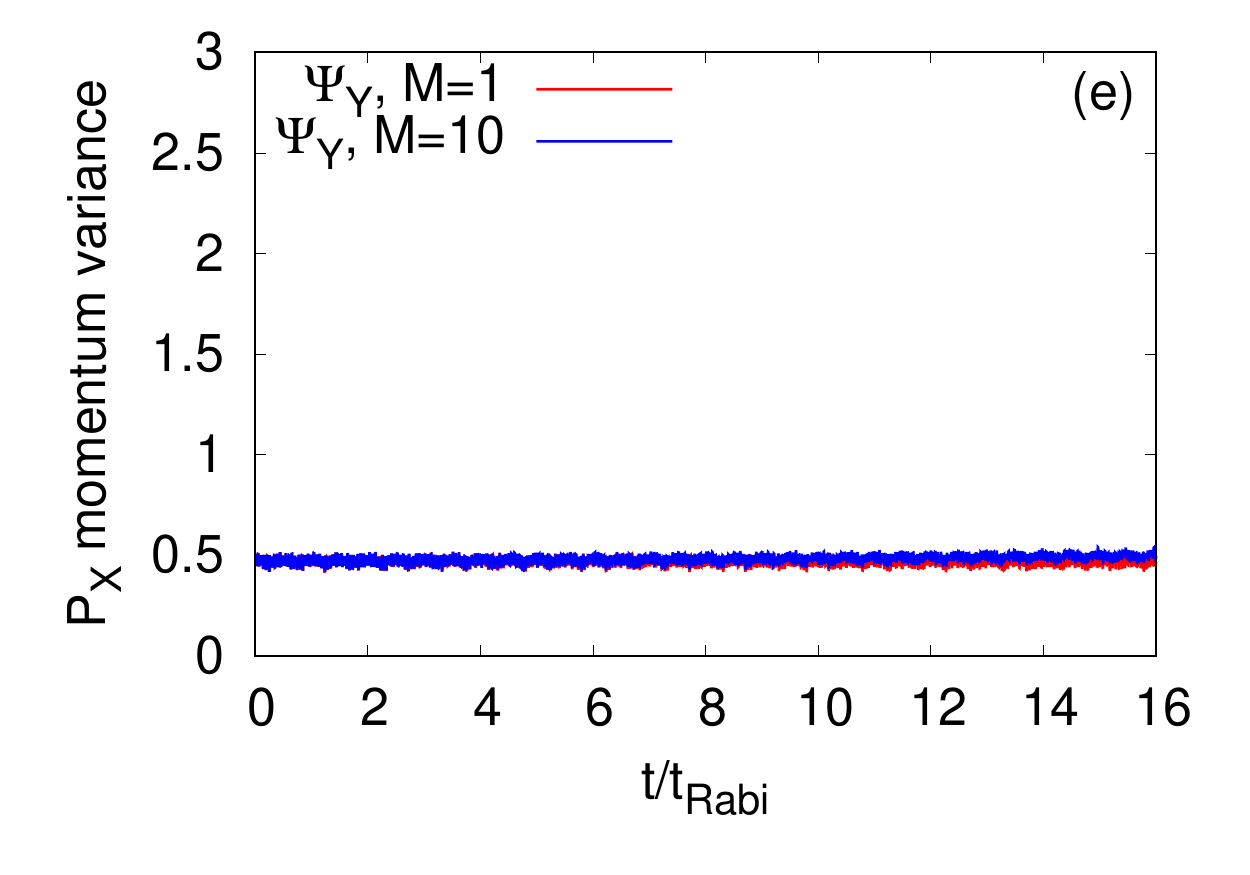}}
{\includegraphics[trim = 0.1cm 0.5cm 0.1cm 0.2cm, scale=.60]{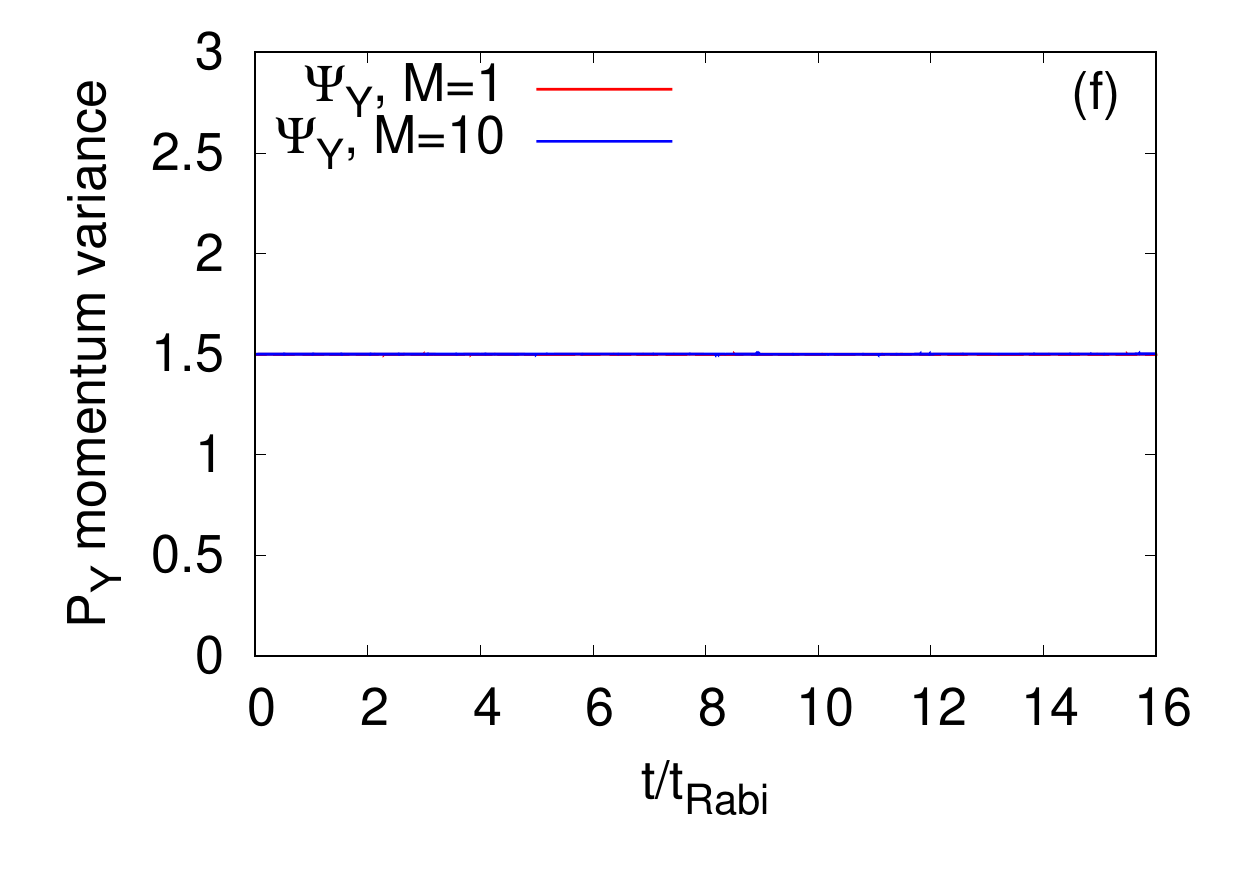}}\\
{\includegraphics[trim = 0.1cm 0.5cm 0.1cm 0.2cm, scale=.60]{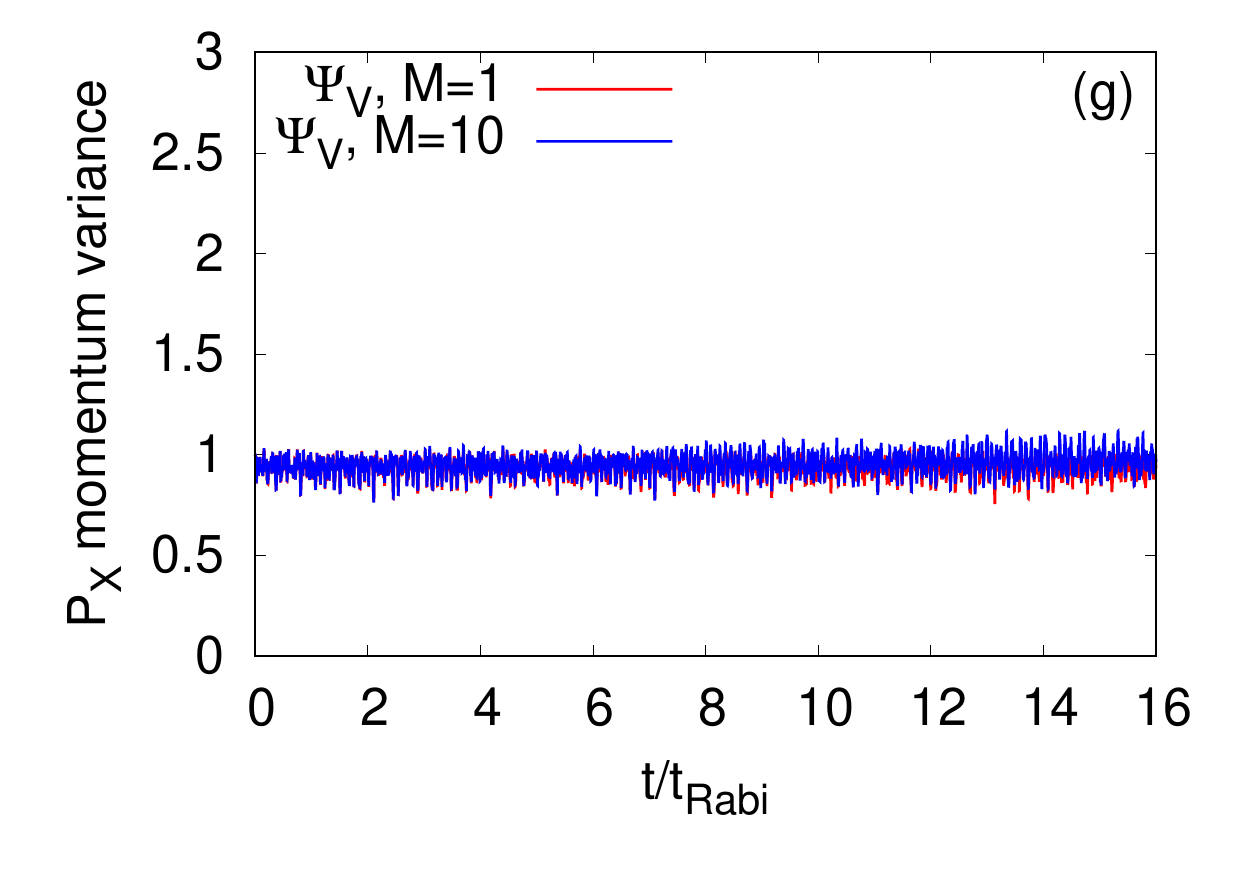}}
{\includegraphics[trim = 0.1cm 0.5cm 0.1cm 0.2cm, scale=.60]{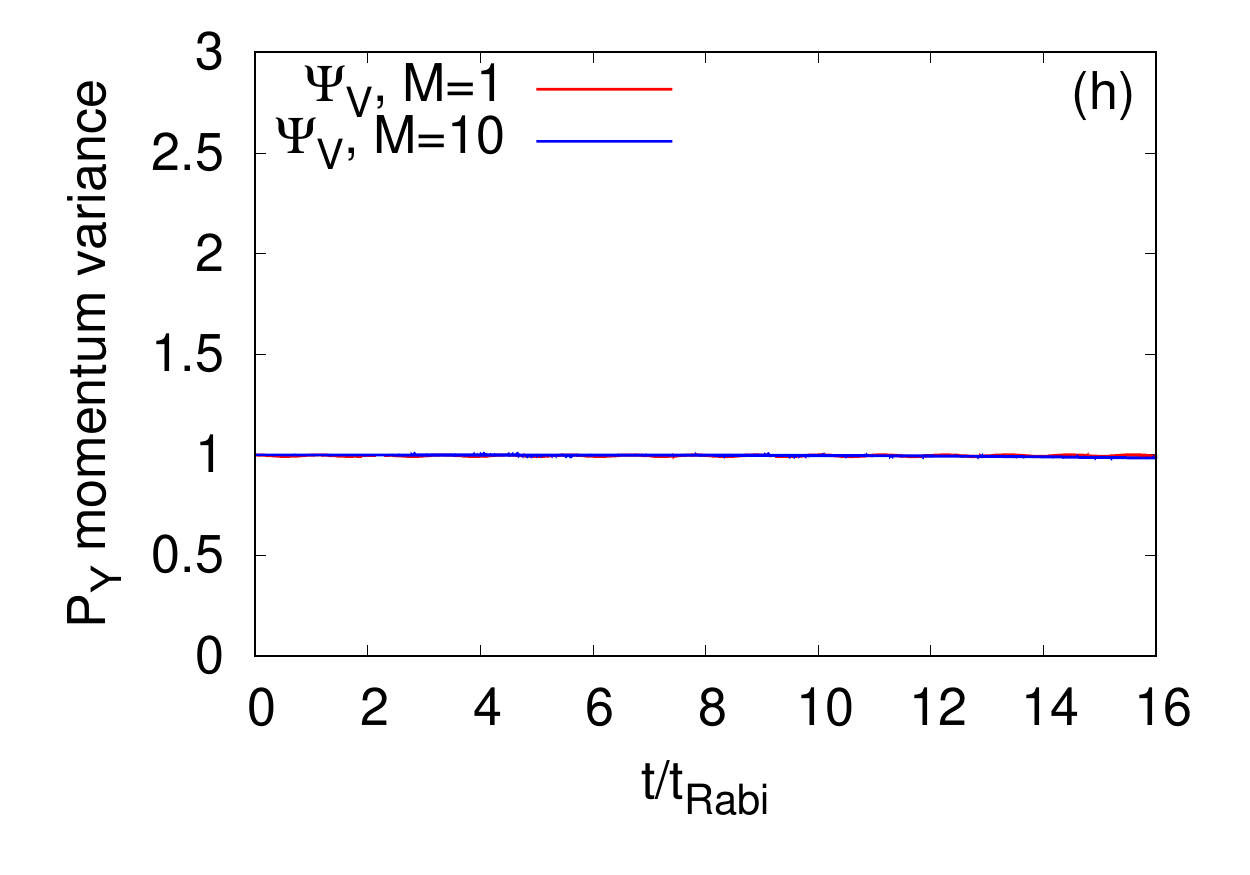}}\\
\caption{ The mean-field ($M=1$ time-adaptive orbitals,  in  red) and many-body (in  blue) time-dependent position variances per particle, $\dfrac{1}{N}\Delta_{{\hat{P}_X}}^2(t)$  and  $\dfrac{1}{N}\Delta_{{\hat{P}_Y}}^2(t)$, are presented in the left and right columns, respectively. The different initial states,  $\Psi_G$, $\Psi_X$, $\Psi_Y$, and $\Psi_V$,  for  $N=10$ bosons are plotted row-wise.  The many-body results are computed using the MCTDHB method  with  $M=6$  orbitals for $\Psi_{G}$ and $\Psi_{X}$ and  $M=10$   orbitals for $\Psi_{Y}$ and $\Psi_{V}$.  See the text for more details. The quantities shown  are dimensionless.}
\label{Fig7}
\end{figure*}
\begin{figure*}[!h]
{\includegraphics[trim = 0.1cm 0.5cm 0.1cm 0.2cm, scale=.60]{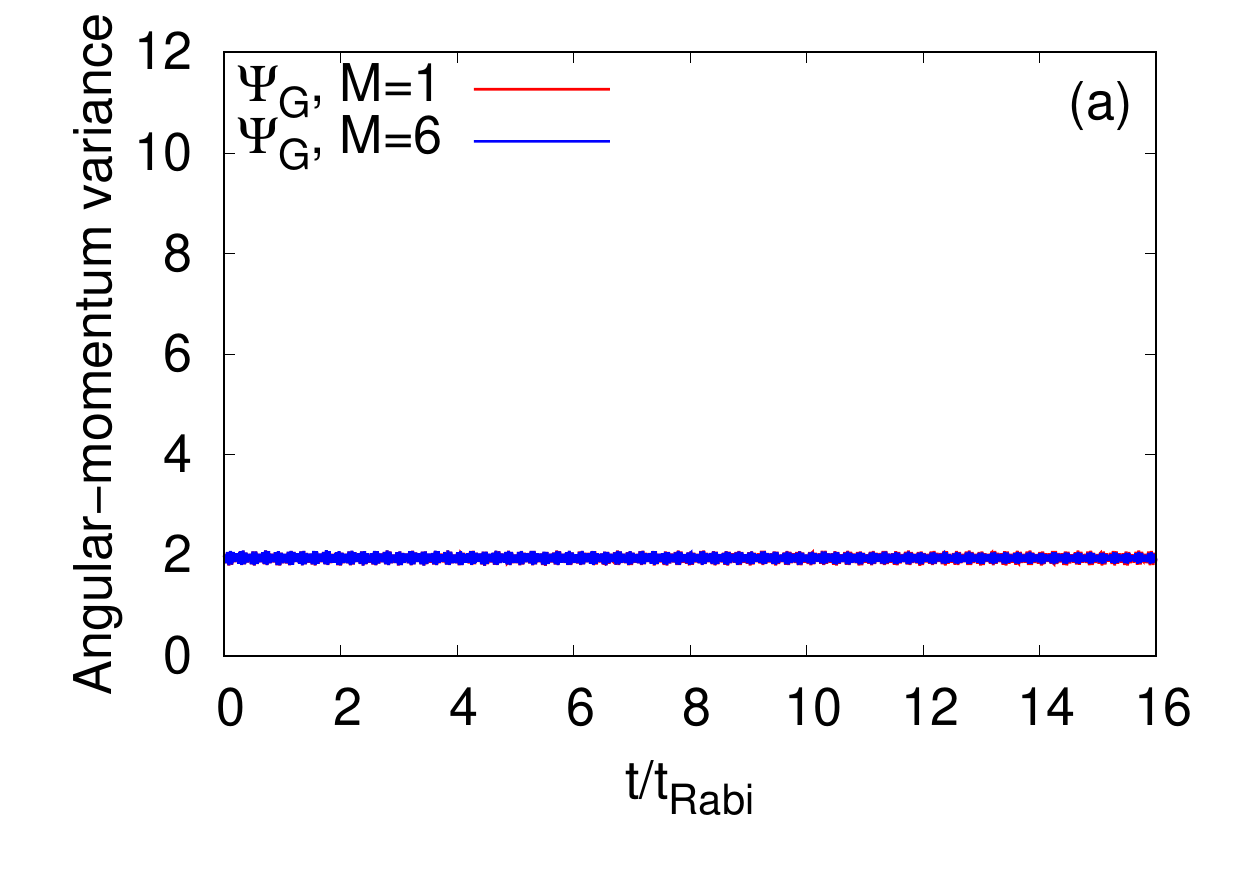}}
{\includegraphics[trim = 0.1cm 0.5cm 0.1cm 0.2cm, scale=.60]{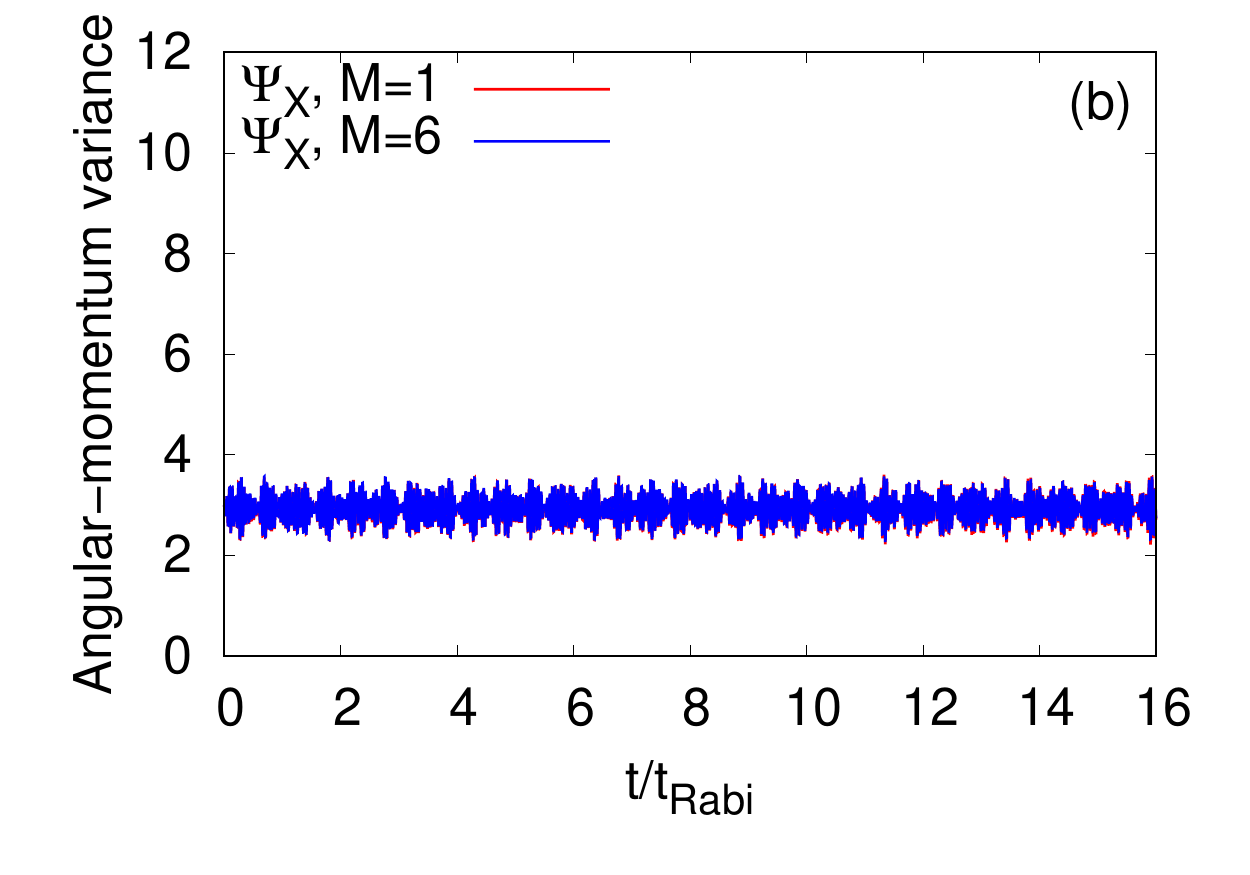}}\\
{\includegraphics[trim = 0.1cm 0.5cm 0.1cm 0.2cm, scale=.60]{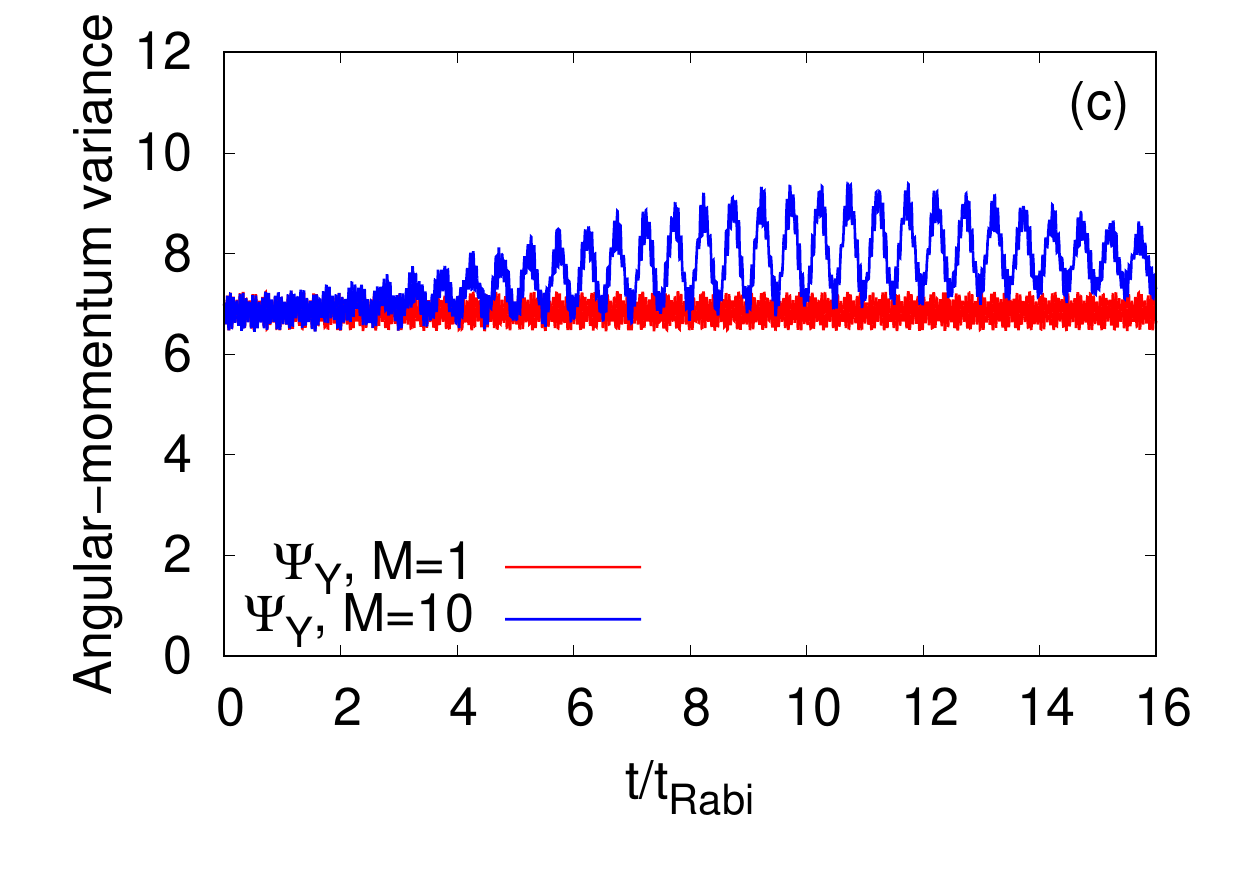}}
{\includegraphics[trim = 0.1cm 0.5cm 0.1cm 0.2cm, scale=.60]{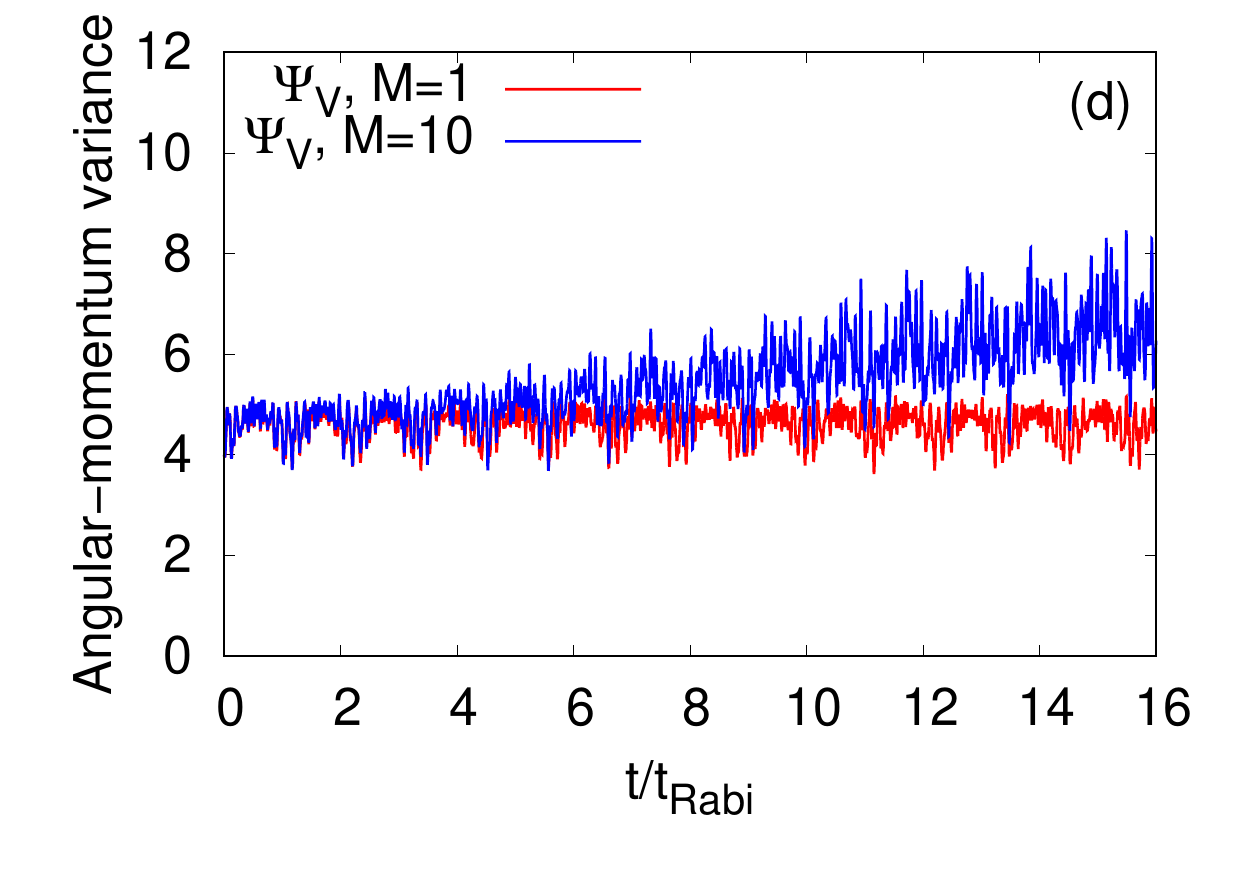}}\\
\caption{Dynamics of the angular-momentum variance  per particle, $\dfrac{1}{N}\Delta_{{\hat{L}_Z}}^2(t)$, in a symmetric 2D double-well for the initial states (a)  $\Psi_G$, (b)  $\Psi_X$, (c) $\Psi_Y$, and (d) $\Psi_V$ with  $N=10$ bosons. The interaction parameter $\Lambda=0.01$. Mean-field ($M=1$ time-adaptive orbitals) results  are presented in  red and corresponding many-body results are shown in  blue. MCTDHB results are computed with  $M=6$ time-adaptive orbitals for $\Psi_{G}$ and $\Psi_{X}$, and  $M=10$  time-adaptive orbitals for $\Psi_{Y}$ and $\Psi_{V}$.  See the text for more details. The quantities shown  are dimensionless.}
\label{Fig8}
\end{figure*}

In  Fig.~\ref{Fig6}, we present the time-dependent many-particle position variances per particle, $\dfrac{1}{N}\Delta_{\hat{X}}^2(t)$ and $\dfrac{1}{N}\Delta_{\hat{Y}}^2(t)$, in a symmetric 2D double-well potential for all the initial states of the bosonic clouds.   We show that the many-body correlations can lead to a deviation in $\dfrac{1}{N}\Delta_{\hat{X}}^2(t)$ which can not be seen at the mean-field level.  Both the many-body and mean-field values of $\dfrac{1}{N}\Delta_{\hat{X}}^2(t)$  vary in time in an oscillatory manner for all the initial states considered here with the highest  frequency of oscillations  for  $\Psi_X$ which is consistent with the respective survival probability.  However, there are  couple of clear  differences that can be seen due to the many-body correlations.   Here the mean-field $\dfrac{1}{N}\Delta_{\hat{X}}^2(t)$ oscillates with a  constant amplitude, whereas the many-body  $\dfrac{1}{N}\Delta_{\hat{X}}^2(t)$ oscillates with a  growing amplitude. Moreover, the pace of growth of  $\dfrac{1}{N}\Delta_{\hat{X}}^2(t)$  is  different for different initial states. One of the interesting features shown in Fig.~\ref{Fig6} is that the minima values of  the many-body  $\dfrac{1}{N}\Delta_{\hat{X}}^2(t)$ increase with time for each of the initial states with a maximal deviation occurs for $\Psi_G$. The increase of minima values of $\dfrac{1}{N}\Delta_{\hat{X}}^2(t)$ due to the growing degree of fragmentation can be found in the literature but  only for the ground state in one-dimensional double-well potentials \cite{Halder2018, Halder2019}.  Also, we notice high-frequency small-amplitude oscillations, specially for the vortex state, on top of the peaks of the large-amplitude oscillations of $\dfrac{1}{N}\Delta_{\hat{X}}^2(t)$. Such high-frequency oscillations  occur due to the breathing-mode oscillations of the system.

Unlike $\dfrac{1}{N}\Delta_{\hat{X}}^2(t)$, the mean-field and many-body values of  $\dfrac{1}{N}\Delta_{\hat{Y}}^2(t)$  have very small fluctuations, of the order of $10^{-3}$, and therefore   their dynamics look more of a constant at the presented scale.  The overlap of the mean-field and many-body values of   $\dfrac{1}{N}\Delta_{\hat{Y}}^2(t)$  tells us that the mean-field results are a good approximation of the many-body results for the position variance in the transverse direction.  $\dfrac{1}{N}\Delta_{\hat{Y}}^2(t)$ suggests that  even though there  is practically no motion along the $y$-direction, the combination of the motion along the $x$-direction with the existence of the almost-frozen transverse degree of freedom could lead to dynamics of the angular-momentum variance in the junction, as will be shown below, which can not be accounted in  the one-dimensional geometry.

To show whether the many-body correlations have any effect on  the variance of  momentum operator,  we compare the many-body $\dfrac{1}{N}\Delta_{{\hat{P}_X}}^2(t)$ and $\dfrac{1}{N}\Delta_{{\hat{P}_Y}}^2(t)$  with the corresponding mean-field results. Here it is worthwhile to mention that the momentum variance is comparatively a more complex quantity than the position variance in the junction as the former one is more sensitive to changes in the shape of the orbitals. In Fig.~\ref{Fig7}, we see that the mean-field   $\dfrac{1}{N}\Delta_{{\hat{P}_X}}^2(t)$ oscillates around a certain value  for each of the initial states considered here. But the many-body $\dfrac{1}{N}\Delta_{{\hat{P}_X}}^2(t)$ shows oscillations with a slowly growing values. It is found that  $\dfrac{1}{N}\Delta_{{\hat{P}_X}}^2(t)$ for $\Psi_X$ are always higher   than for the other states.  We observe that $\dfrac{1}{N}\Delta_{{\hat{P}_X}}^2(t)$ for $\Psi_V$ is  in between the respective results of  $\Psi_X$ and $\Psi_Y$ till the time considered here, which is more evident in the momentum variance along the $y$-direction,  discussed below.  The high frequency oscillations occurring in the momentum variance, which are more prominent for the states $\Psi_X$ and $\Psi_V$, are due to the stronger breathing oscillations of the system along the $x$-direction.

As presented in the discussion of  $\dfrac{1}{N}\Delta_{\hat{Y}}^2(t)$, the momentum variance along the $y$-direction, $\dfrac{1}{N}\Delta_{{\hat{P}_Y}}^2(t)$ also exhibits very small fluctuations, of the order of $10^{-3}$,  for all the initial states of bosonic clouds both at the mean-field as well as the many-body level. The mean-field and many-body values of $\dfrac{1}{N}\Delta_{{\hat{P}_Y}}^2(t)$ practically overlap with each other with almost constant values, being 0.5, 0.5, 1.5, and 1.0 for the states $\Psi_G$, $\Psi_X$, $\Psi_Y$, and $\Psi_V$, respectively. As the vortex state is the combination of $\Psi_X$ and $\Psi_Y$, $\dfrac{1}{N}\Delta_{{\hat{P}_Y}}^2(t)$ for $\Psi_V$  is exactly in between the corresponding results of  $\Psi_X$ and $\Psi_Y$.  Similarly to  $\dfrac{1}{N}\Delta_{\hat{Y}}^2(t)$, the  non-zero values of $\dfrac{1}{N}\Delta_{{\hat{P}_Y}}^2(t)$ with small fluctuations indicate  the existence of transverse motion of the bosonic clouds which will have a consequential effect to the dynamics of the angular-momentum variance in the junction.

Now, we move to the discussion of the angular-momentum variance per particle, $\dfrac{1}{N}\Delta_{{\hat{L}_Z}}^2(t)$,  presented in Fig.~\ref{Fig8}. We observe a marginal difference in the angular-momentum variance calculated at the many-body and mean-field levels for the states $\Psi_G$ and $\Psi_X$, implying that the mean-field theory will be enough to discuss $\dfrac{1}{N}\Delta_{{\hat{L}_Z}}^2(t)$ for these two states. Fig.~\ref{Fig8}(a) finds that $\dfrac{1}{N}\Delta_{{\hat{L}_Z}}^2(t)$ of $\Psi_G$  oscillates  with  amplitude of fluctuations in the order of $10^{-1}$. In comparison with  $\Psi_G$,   $\Psi_X$ has   a larger amplitude of oscillations of  $\dfrac{1}{N}\Delta_{{\hat{L}_Z}}^2(t)$ which varies from the value $2.4$ to $3.6$.  The  fluctuations in the angular-momentum  at the many-body level   are  governed by the  structure of the many-body wave-function, shapes of the time-dependent orbitals, and the mechanism  of fragmentation. Therefore, without excitation  in the $y$-direction, fragmentation occurring due to the barrier in the $x$-direction does hardly impact the fluctuations in the angular-momentum for $\Psi_G$ and   $\Psi_X$.  From Fig.~\ref{Fig8}(a) and (b), one can find that the amount of fluctuations atop  the base-line are practically same at the mean-field and many-body levels. These fluctuations are around $5\%$ and $20\%$ for the states $\Psi_G$ and $\Psi_X$, respectively.

Contrary to  $\Psi_G$ and   $\Psi_X$,  exciting many-body features have been found for the angular-momentum variance of $\Psi_Y$ and $\Psi_V$. Fig.~\ref{Fig8}(c) and (d) show that the many-body  $\dfrac{1}{N}\Delta_{{\hat{L}_Z}}^2(t)$ for $\Psi_Y$ and $\Psi_V$ are oscillatory in nature with a growing amplitude. Also, their minima values  are increasing with time. The maximal fluctuations on top of the baseline of the angular-momentum variance for $\Psi_Y$ and $\Psi_V$ at the mean-field level are found around $7\%$ and $25\%$, respectively, while at the many-body level they are approximately $34\%$ and $112\%$, respectively.  A difference in the onset of the angular-momentum fluctuations for $\Psi_Y$ and $\Psi_V$ at the many-body level can be described by analyzing how fragmentation develops in the system. Unlike $\Psi_G$ and $\Psi_X$, transverse excitation is involved  in the lowest excited natural orbitals, $\phi_2$, $\phi_3$, and $\phi_4$ (see Figs. S7 and S8 in the supplemental material) of $\Psi_Y$ and $\Psi_V$, leading to  large fluctuations in the angular-momentum variance. It is found that after $t\approx 12 t_{\text{Rabi}}$, the amplitude of the oscillations of  $\dfrac{1}{N}\Delta_{{\hat{L}_Z}}^2(t)$ starts decaying for $\Psi_Y$.  $\dfrac{1}{N}\Delta_{{\hat{L}_Z}}^2(t)$ of $\Psi_V$ shows two types of oscillations very prominently, one with a larger amplitude and smaller frequency which arises due to the density oscillations  and the second one with a smaller amplitude but  higher frequency due to the breathing oscillations of the system. Another  feature of $\dfrac{1}{N}\Delta_{{\hat{L}_Z}}^2(t)$ is that  the mean-field values of $\dfrac{1}{N}\Delta_{{\hat{L}_Z}}^2(t)$ for $\Psi_Y$ are always larger than  the respective values  for $\Psi_V$, but the many-body $\dfrac{1}{N}\Delta_{{\hat{L}_Z}}^2(t)$ of $\Psi_V$ eventually becomes    in time larger compared to the corresponding values of $\Psi_Y$. The features of all the quantum mechanical observables, expectation values, and their variances discussed above certainly determine that the  mean-field level of theory is not sufficient to accurately explain the dynamics of a trapped   system in a two-dimensional geometry.

\subsection{Long-time dynamics}

\begin{figure*}[!h]
{\includegraphics[trim = 0.1cm 0.5cm 0.1cm 0.2cm, scale=.65]{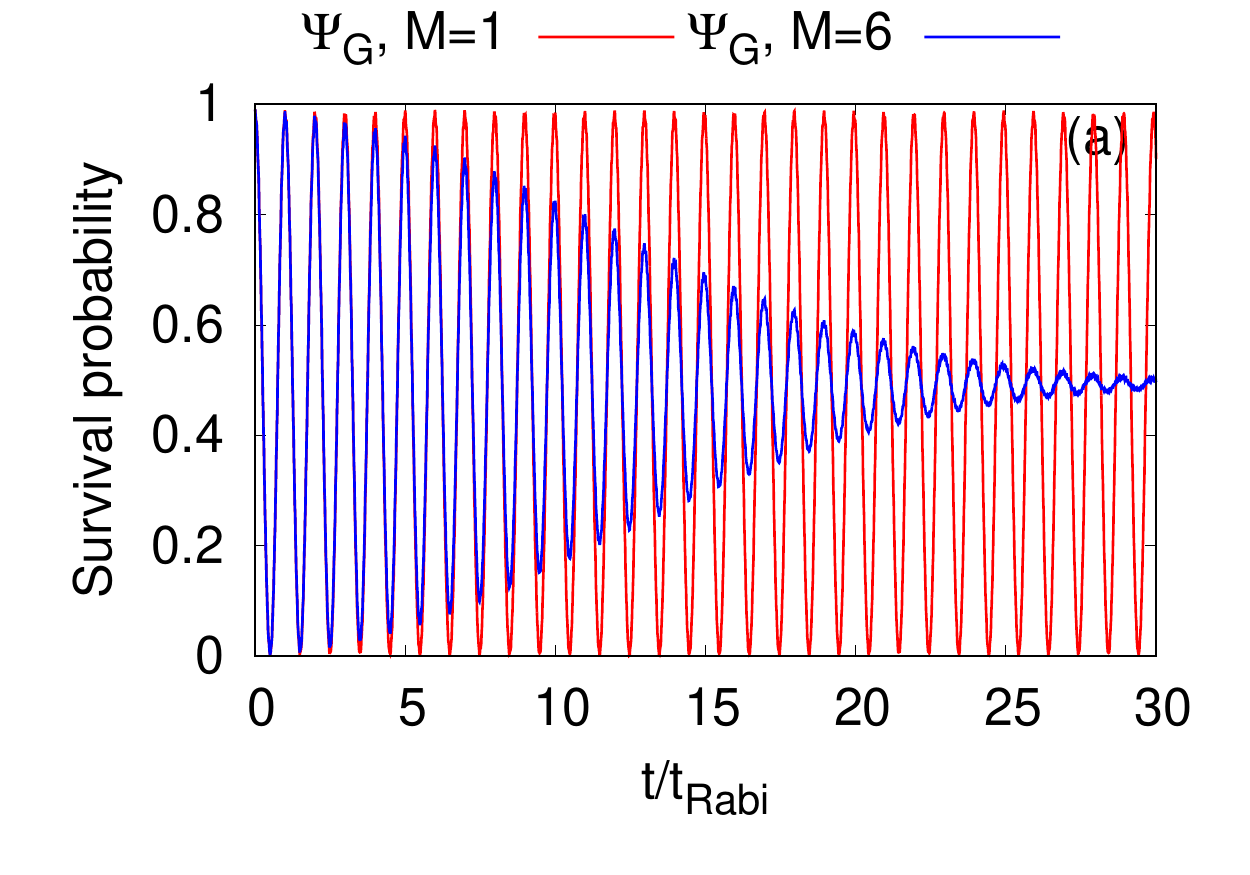}}
{\includegraphics[trim = 0.1cm 0.5cm 0.1cm 0.2cm, scale=.65]{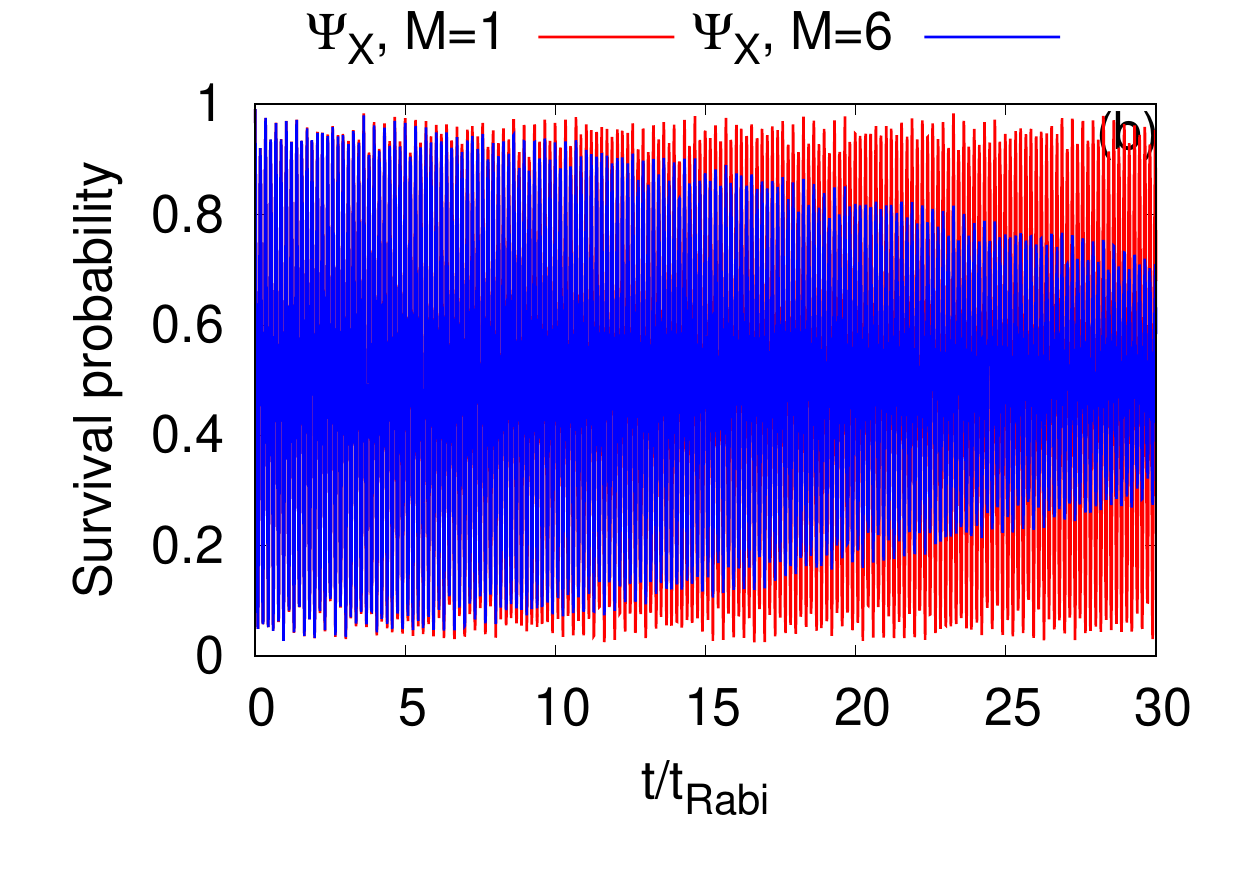}}\\
\vglue 0.25 truecm
{\includegraphics[trim = 0.1cm 0.5cm 0.1cm 0.2cm, scale=.65]{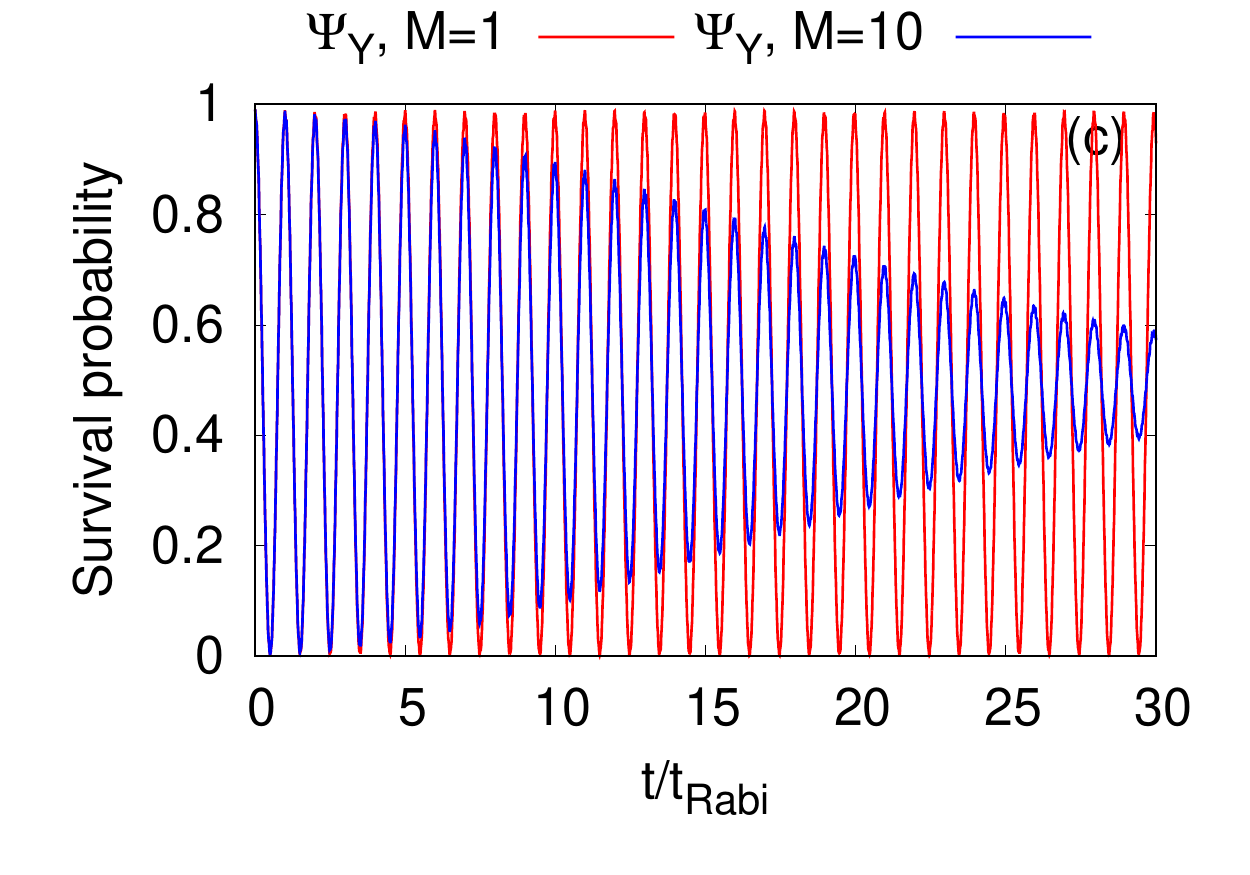}}
{\includegraphics[trim = 0.1cm 0.5cm 0.1cm 0.2cm, scale=.65]{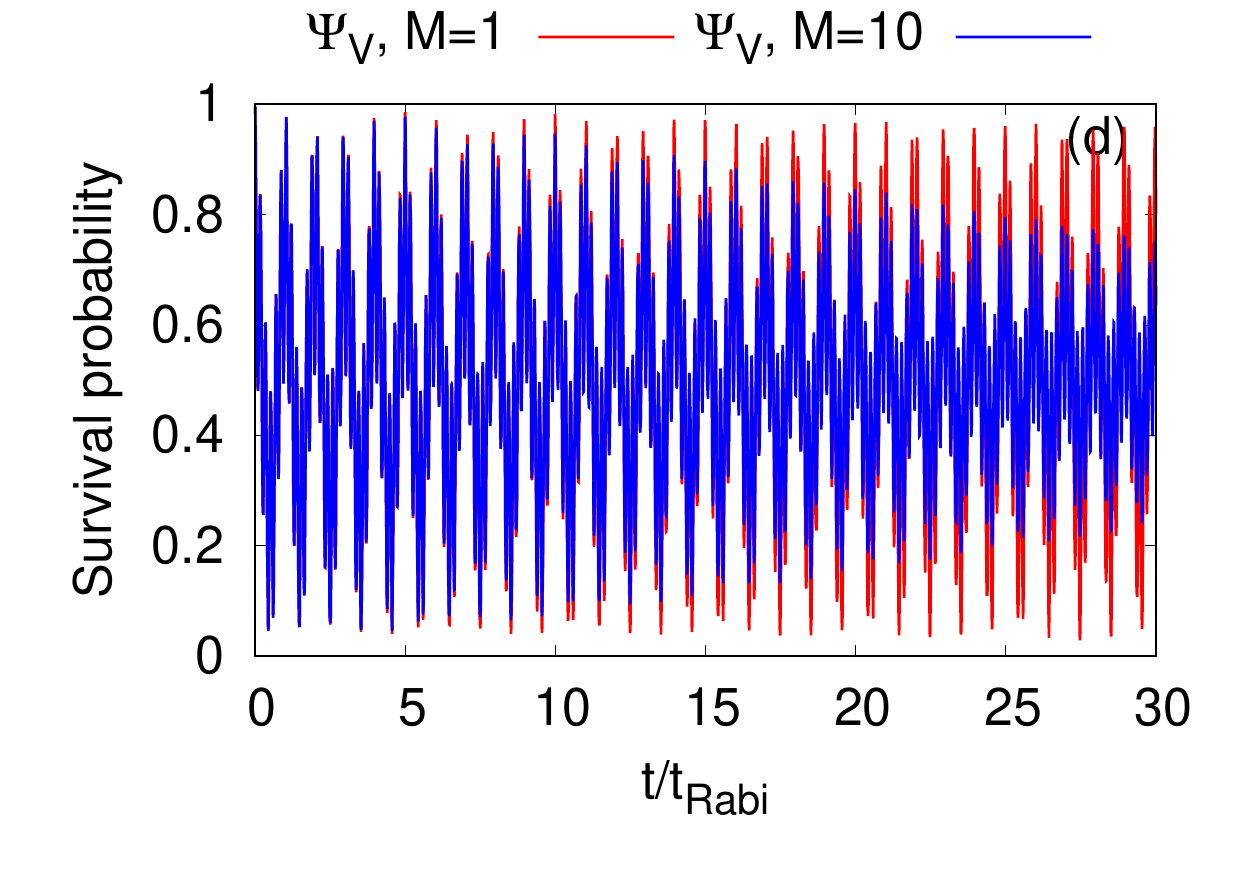}}\\
\caption{Long-time dynamical evolution of the survival probability of $N=10$ bosons in the left well of a symmetric 2D double-well   for the initial states (a) $\Psi_G$, (b) $\Psi_X$, (c) $\Psi_Y$, and (d) $\Psi_V$.    The interaction parameter $\Lambda=0.01$. Mean-field  results   (in  red) and corresponding many-body results (in  blue). MCTDHB results are computed with  $M=6$ time-adaptive orbitals for $\Psi_{G}$ and $\Psi_{X}$, and  $M=10$ time-adaptive orbitals for $\Psi_{Y}$ and $\Psi_{V}$.   The quantities shown  are dimensionless.}
\label{Fig9}
\end{figure*}

So far, we have displayed in a detail study  the dynamical behavior of the density oscillations, loss of coherence,  development of fragmentation,   expectation values, and variances of a few  basic quantum mechanical operators in a symmetric 2D double-well in the short to intermediate time domain  $(t=0$ to $16t_{Rabi})$. The results show that the presence of transverse excitations requires a larger number of time adaptive orbitals to  accurately represent the many-body effects of the quantities discussed here.  Before ending this section, it is worthwhile to include a flavour of the long-time dynamics of the most basic property, $P_L(t)$.  In Fig.~\ref{Fig9}, we have registered the long-time dynamics of $P_L(t)$ for  the four initial states  $\Psi_G$, $\Psi_X$, $\Psi_Y$, and $\Psi_V$. Snapshots of the  density oscillations are shown in Fig.~\ref{Fig10}.     The long-time dynamics of other quantities along with their convergence are discussed in the supplemental material.  The plots show that the densities of the systems tunnel back and forth without changing the amplitude and frequency at  the mean-field level, even for the long-time dynamics. One can clearly observe that the many-body  $P_L(t)$ displays  a collapse in the oscillations, for all the initial bosonic clouds. The collapse of the oscillations  is already shown   only for the ground state, both in theoretically \cite{Sakmann2014} and experimentally \cite{Sakmann2014}. However,  we find that rate of collapse is different for different initial states. Among the four initial states, the collapse of $\Psi_G$ is the  quickest and of  $\Psi_V$ is the slowest for this symmetric 2D double-well. 

The collapse of the initial states are consistent with the density oscillations shown in Fig.~\ref{Fig10}. The snapshots of the mean-field and many-body density oscillations are taken at  $t=10t_{Rabi}$, $20t_{Rabi}$,  and $30t_{Rabi}$.  Unlike the dynamics at the mean-field level, the many-body dynamics of $\Psi_G$, $\Psi_X$, and $\Psi_Y$ show the generation of replicas of the respective initial states in the process of tunneling.  But the time evolution of the vortex state in a  double-well is completely different in comparison with the other three initial states. Both at the mean-field and  many-body levels, $\Psi_V$  generates two dipoles at the two potential minima and they change their orientation  in the process of evolution. It can be seen from the figure that in spite of the generation of vortex dipoles in two dissimilar level of theory,  the evolution of the dipoles are different in terms of shape and orientation due to the development of the many-body fragmentation in the system.

\begin{figure*}[!h]
{\includegraphics[trim = 4.9cm 0.5cm 3.1cm 0.2cm,scale=.50]{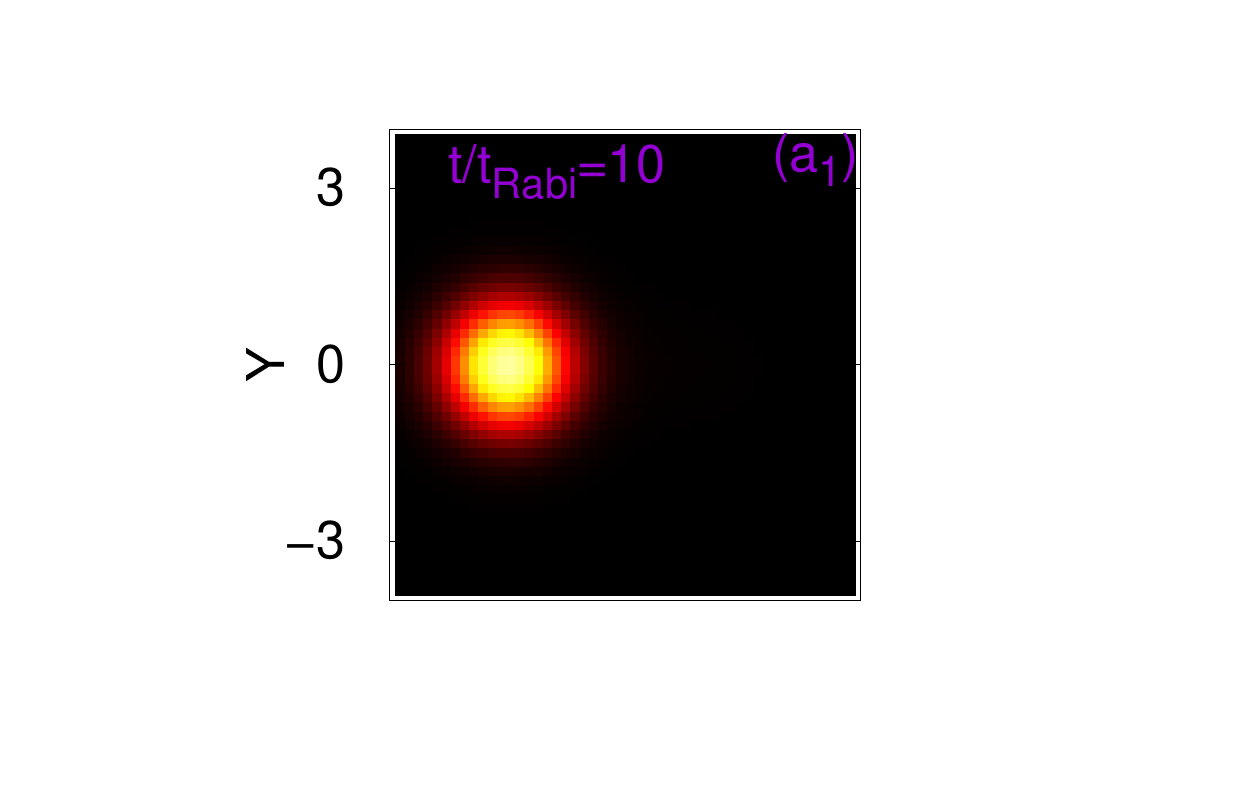}}
{\includegraphics[trim =  4.9cm 0.5cm 3.1cm 0.2cm, scale=.50]{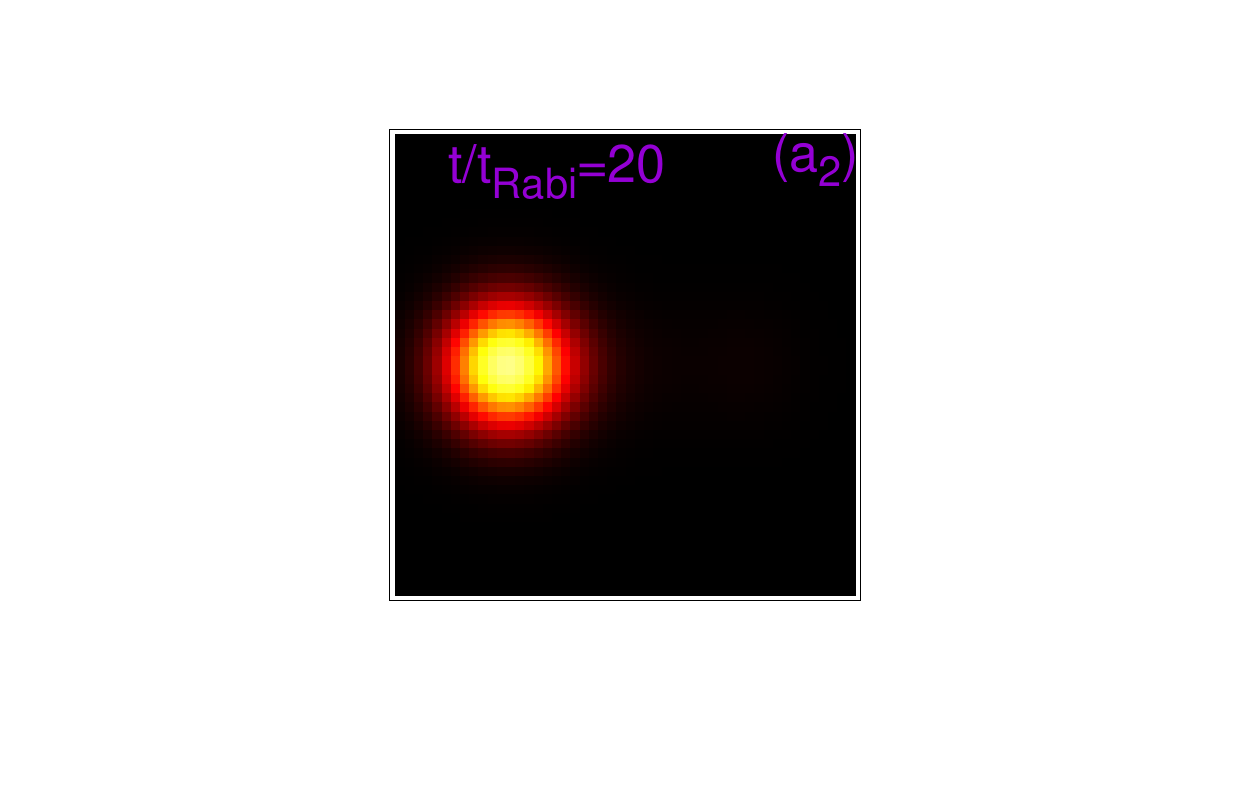}}
{\includegraphics[trim =  4.9cm 0.5cm 3.1cm 0.2cm, scale=.50]{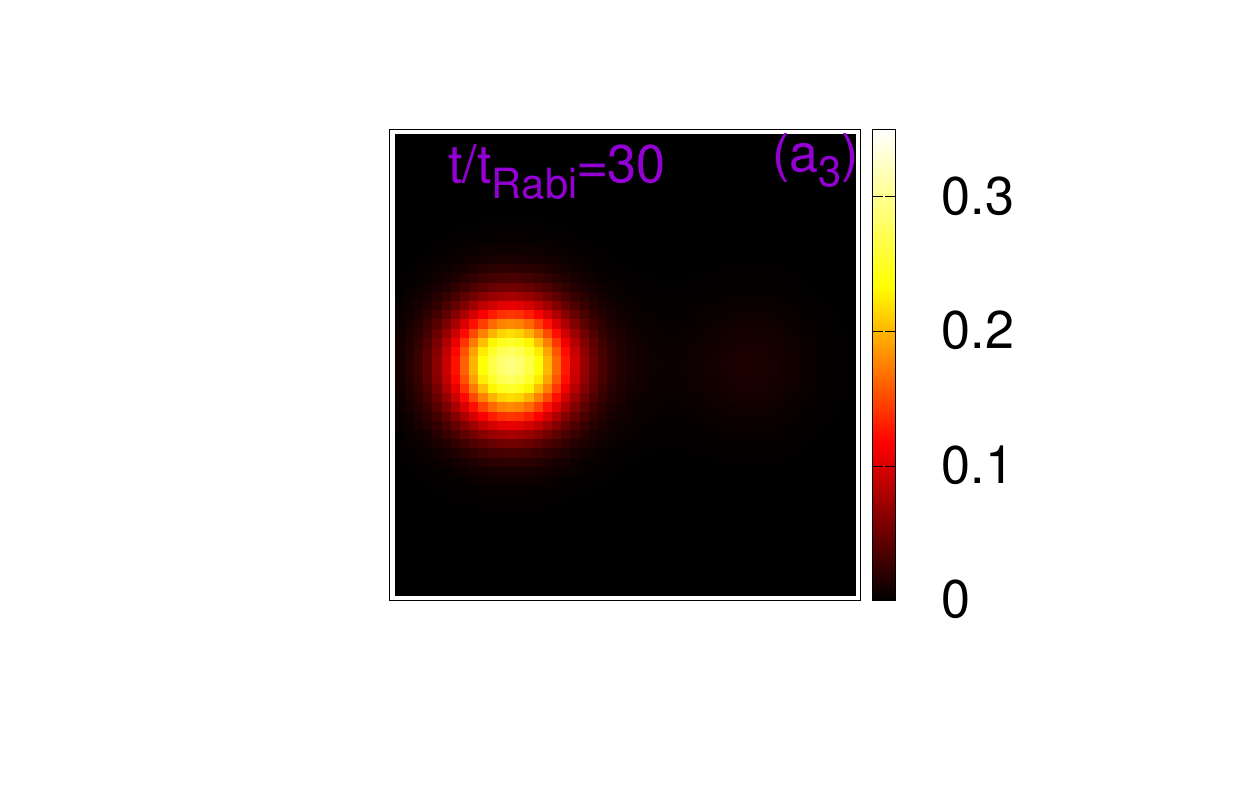}}
{\includegraphics[trim = 2.5cm 0.5cm 3.1cm 2.6cm,scale=.50]{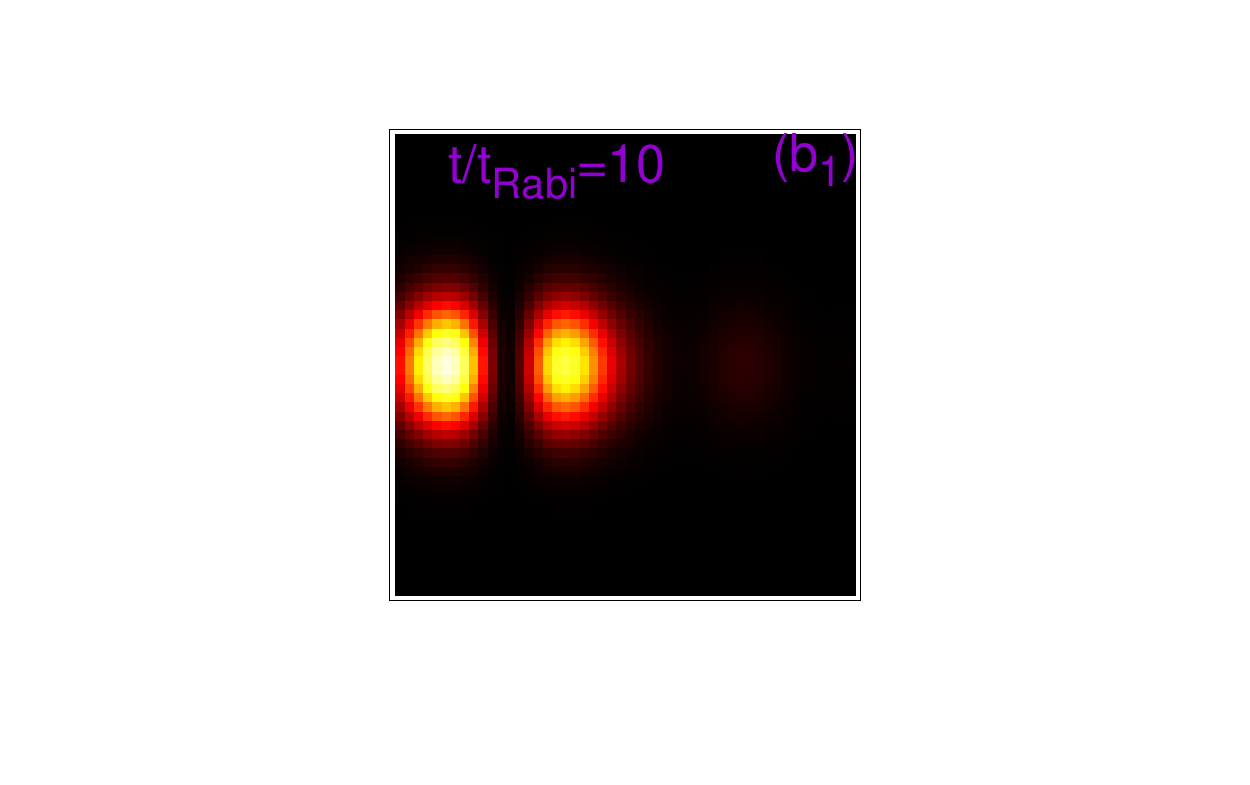}}
{\includegraphics[trim =  4.9cm 0.5cm 3.1cm 2.6cm, scale=.50]{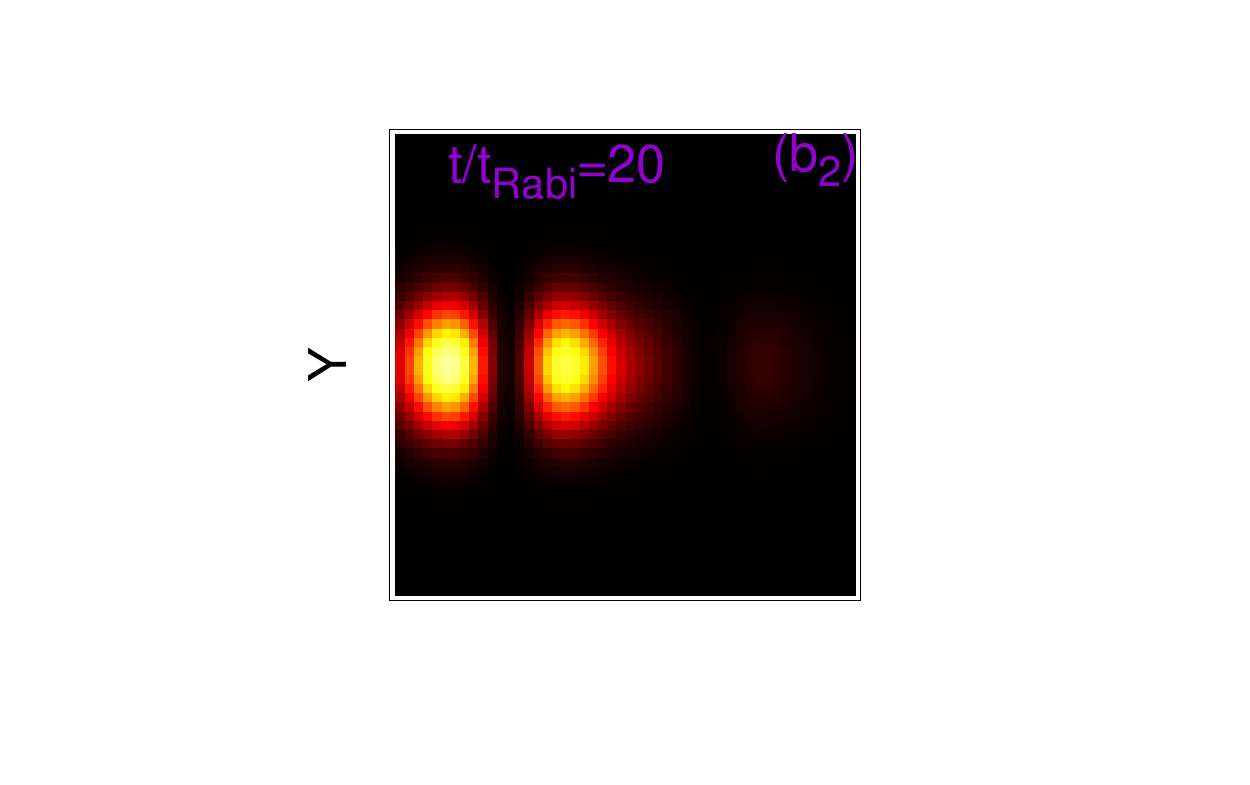}}
{\includegraphics[trim =  4.9cm 0.5cm 3.1cm 2.6cm, scale=.50]{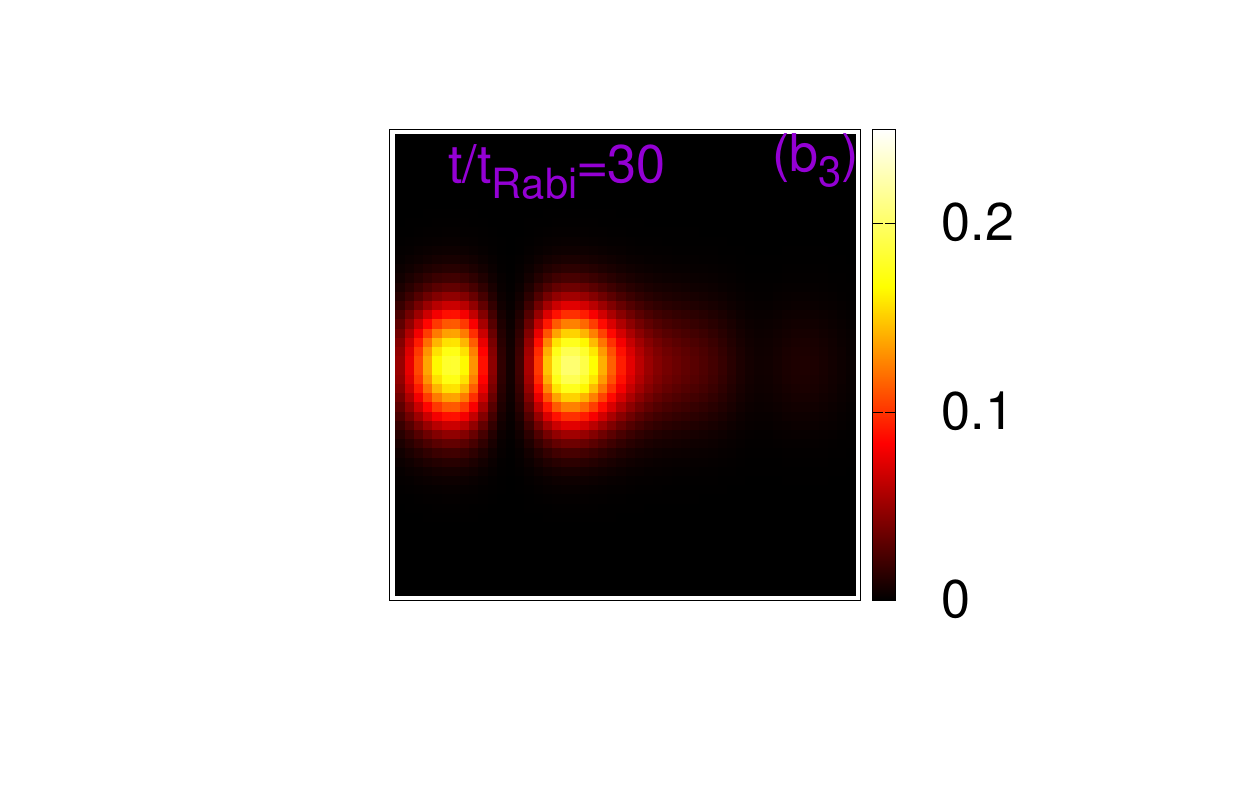}}\\
{\includegraphics[trim =  4.9cm 0.5cm 3.1cm 2.6cm, scale=.50]{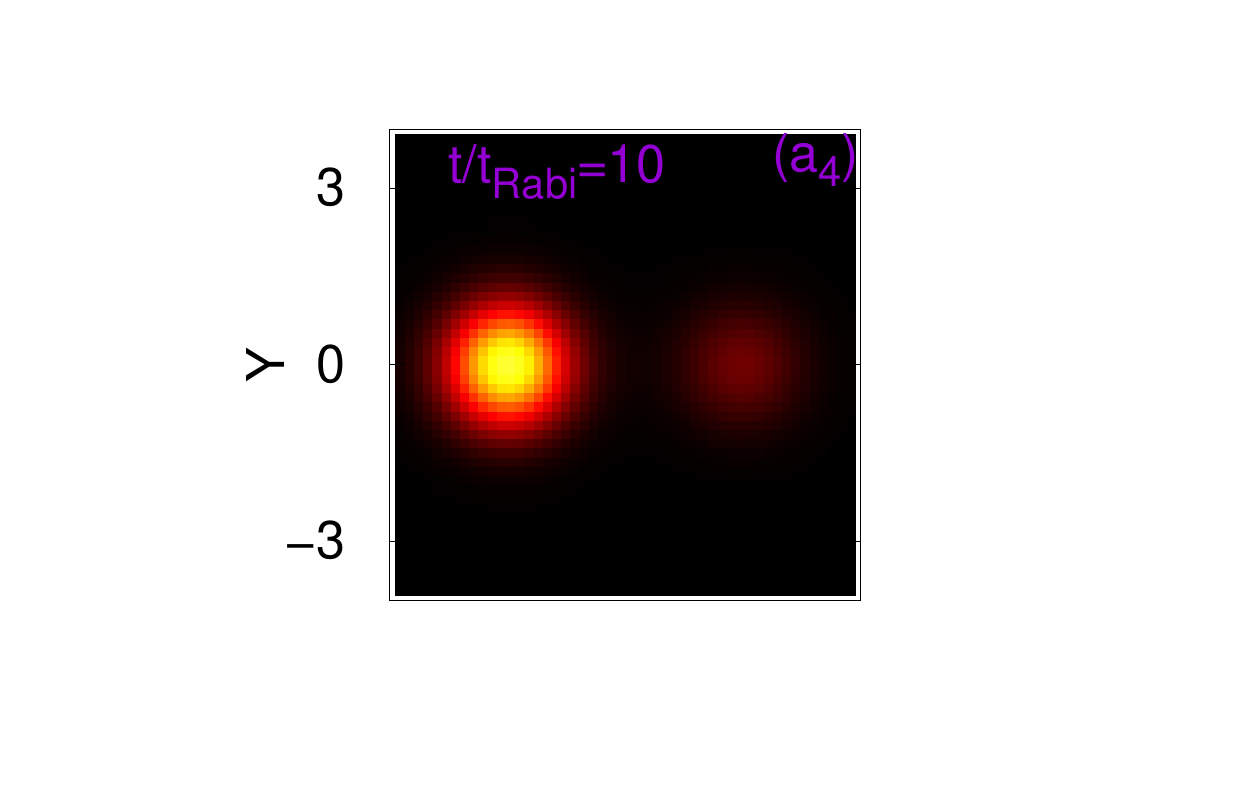}}
{\includegraphics[trim =  4.9cm 0.5cm 3.1cm 2.6cm, scale=.50]{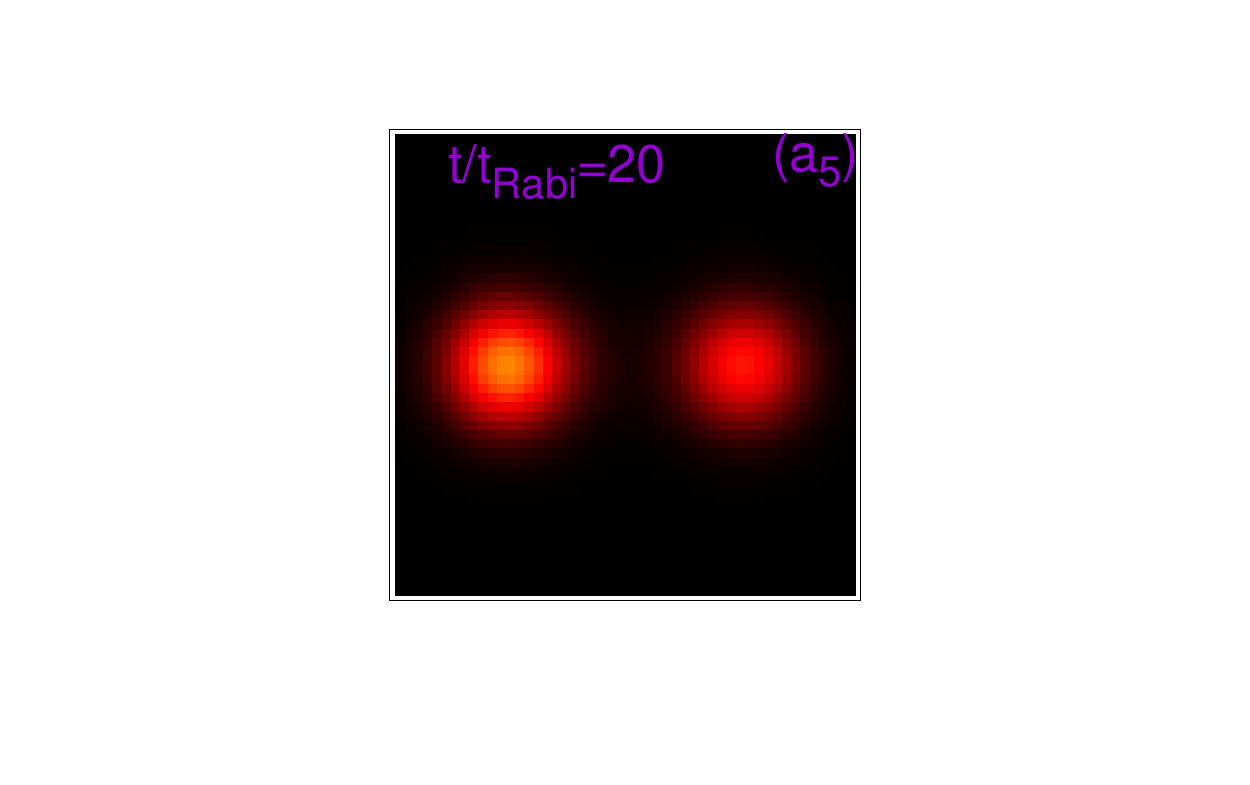}}
{\includegraphics[trim =  4.9cm 0.5cm 3.1cm 2.6cm, scale=.50]{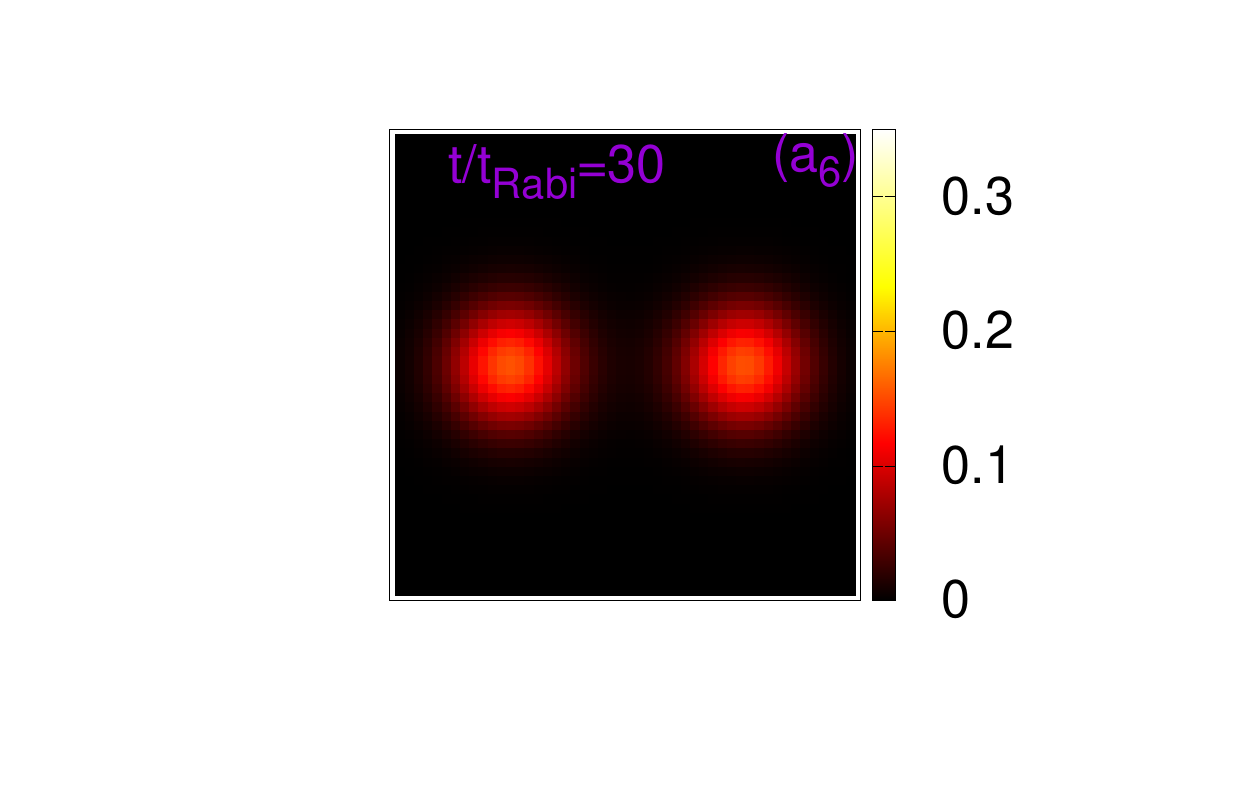}}
{\includegraphics[trim =  2.5cm 0.5cm 3.1cm 2.6cm, scale=.50]{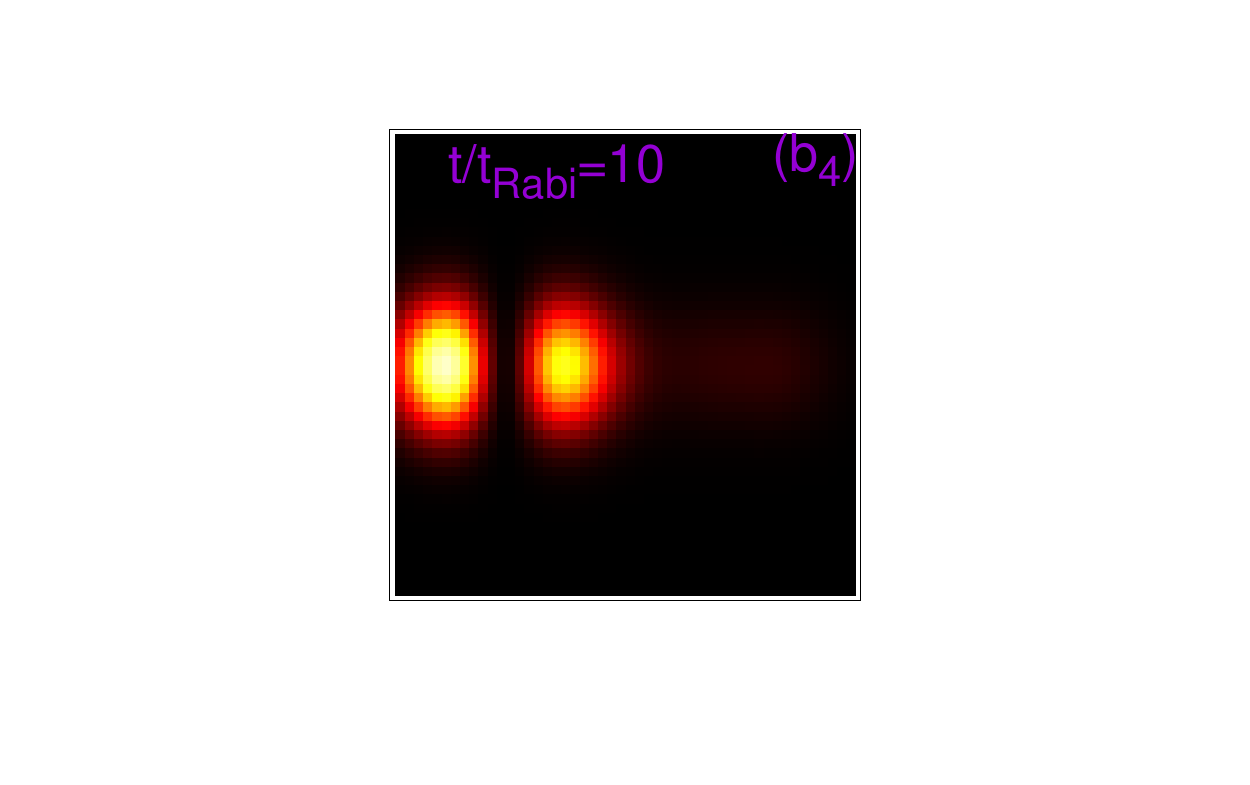}}
{\includegraphics[trim =  4.9cm 0.5cm 3.1cm 2.6cm, scale=.50]{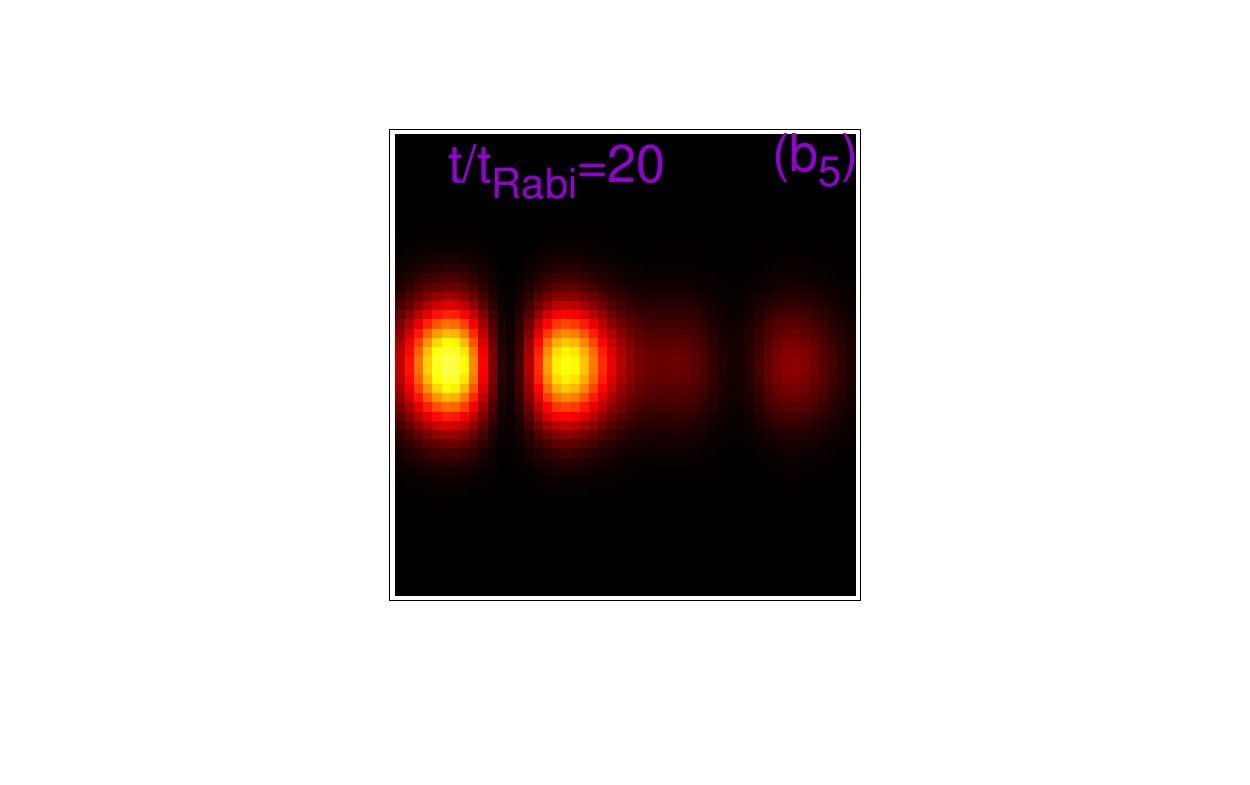}}
{\includegraphics[trim =  4.9cm 0.5cm 3.1cm 2.6cm, scale=.50]{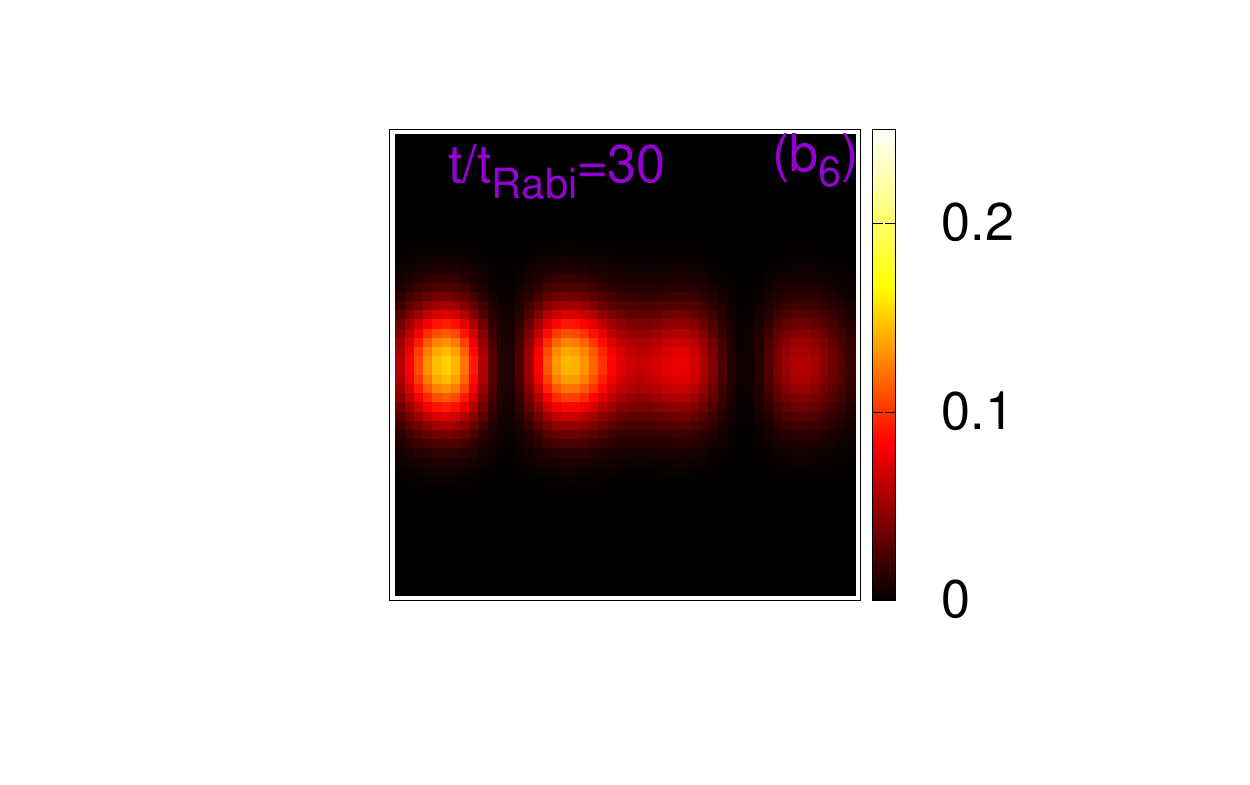}}\\
{\includegraphics[trim = 4.9cm 0.5cm 3.1cm 2.6cm,scale=.50]{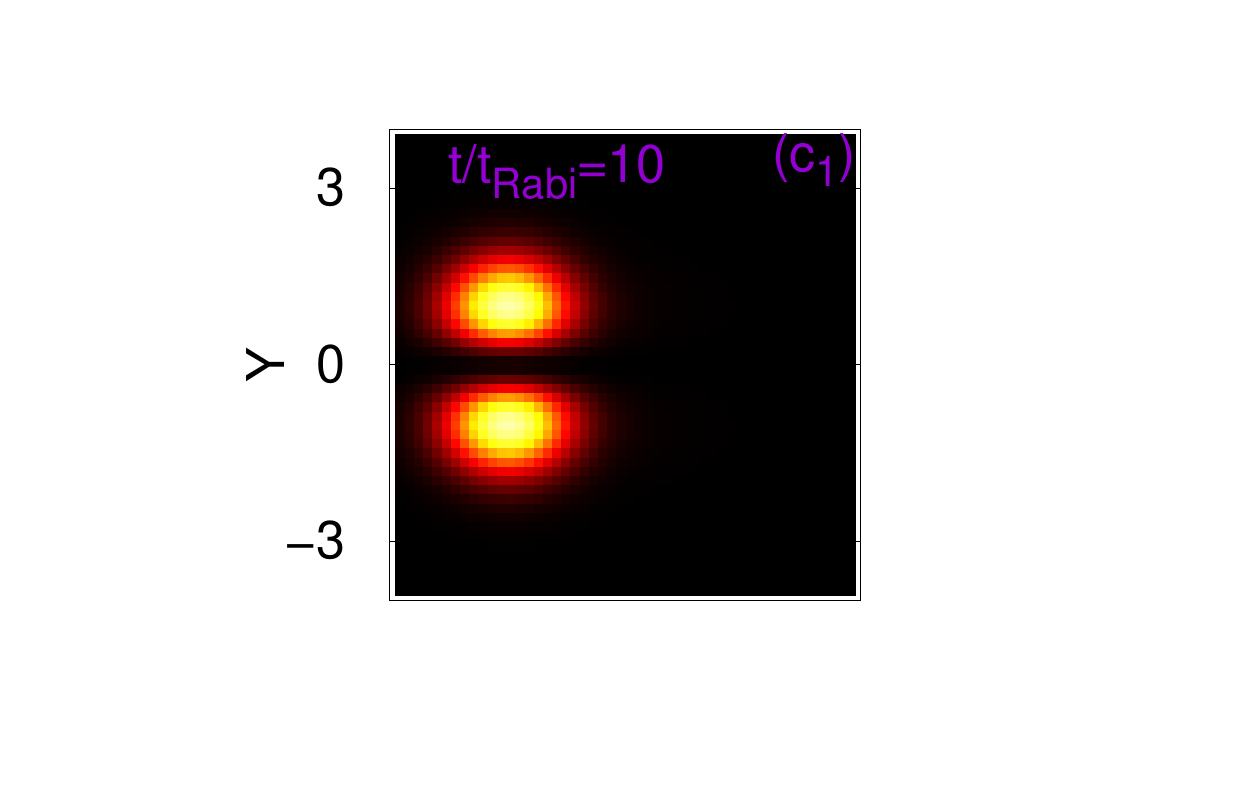}}
{\includegraphics[trim =  4.9cm 0.5cm 3.1cm 2.6cm, scale=.50]{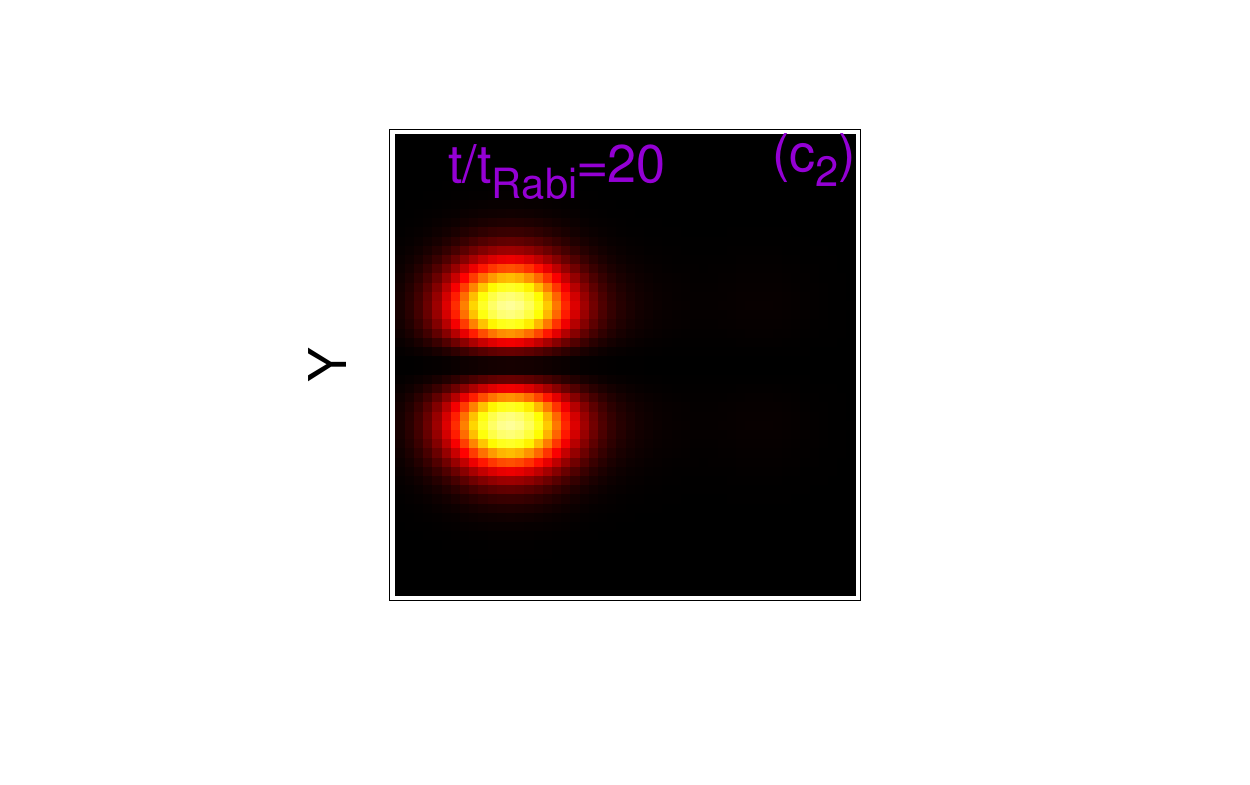}}
{\includegraphics[trim =  4.9cm 0.5cm 3.1cm 2.6cm, scale=.50]{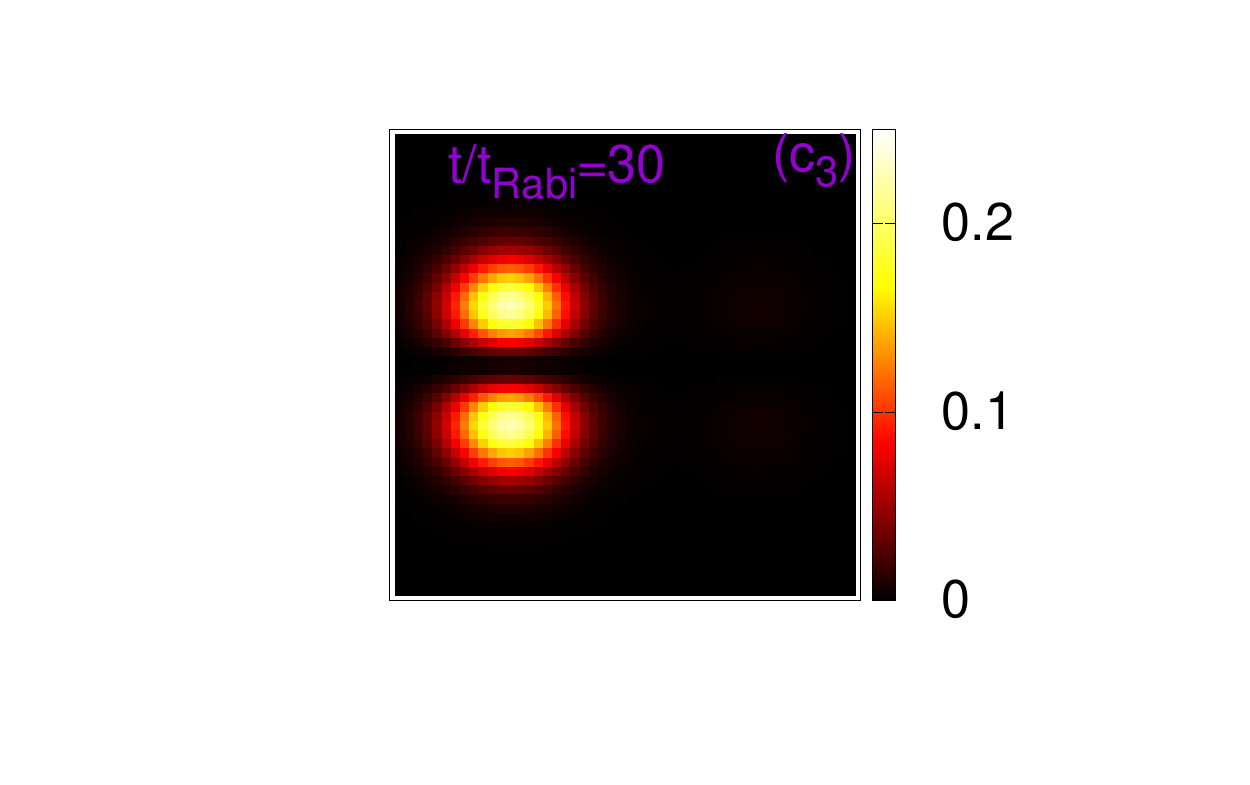}}
{\includegraphics[trim = 2.5cm 0.5cm 3.1cm 2.6cm,scale=.50]{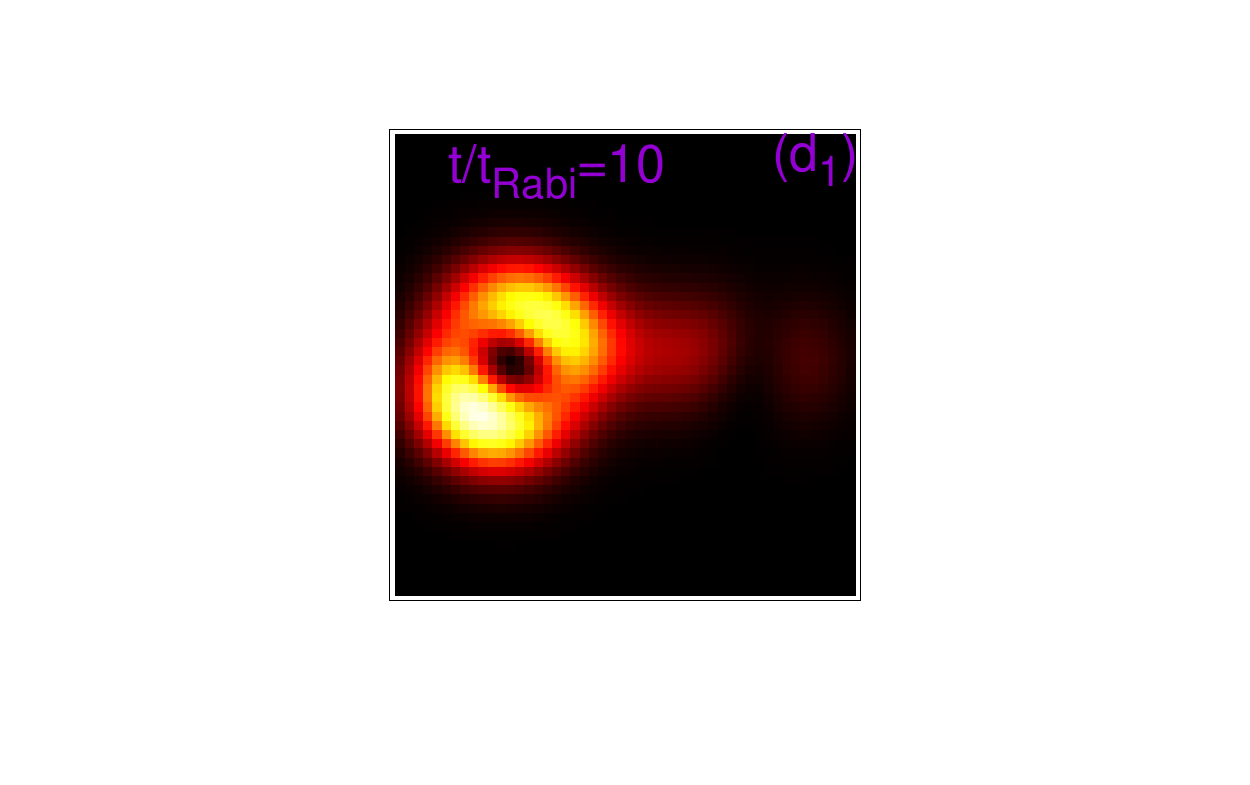}}
{\includegraphics[trim =  4.9cm 0.5cm 3.1cm 2.6cm, scale=.50]{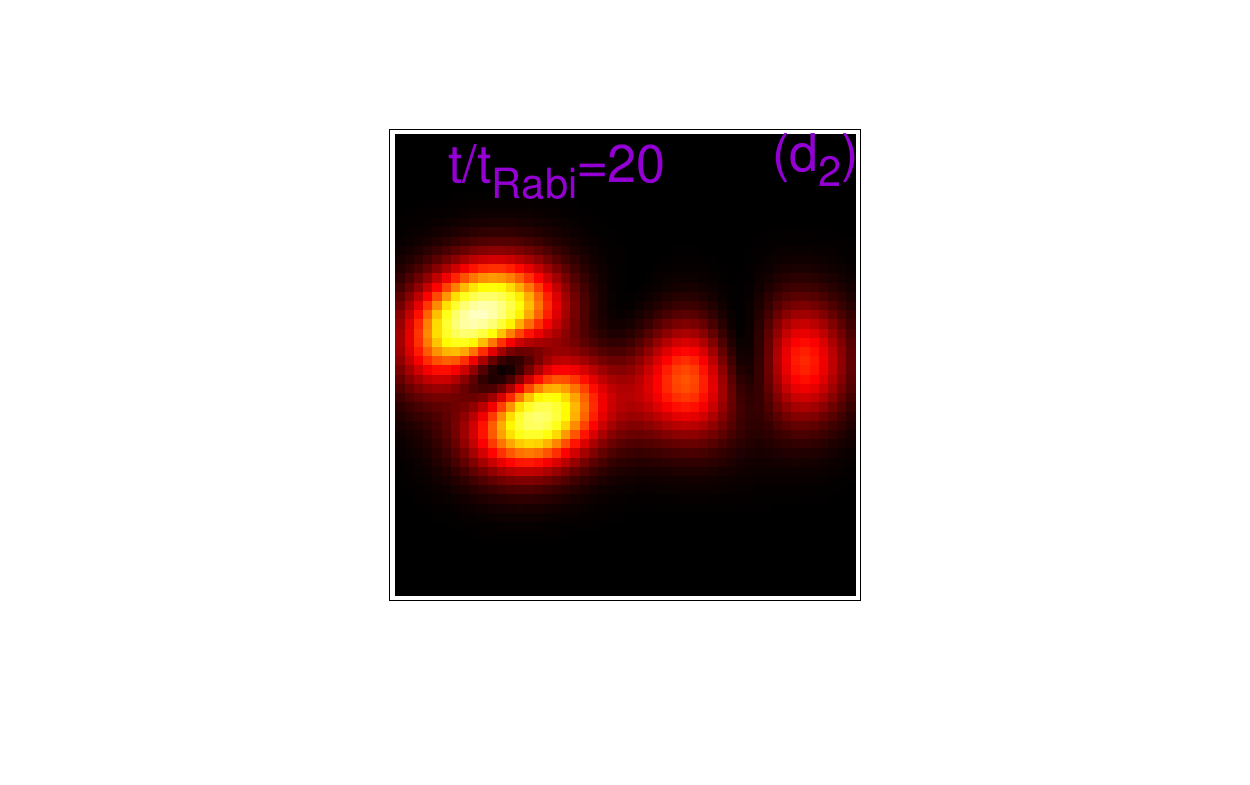}}
{\includegraphics[trim =  4.9cm 0.5cm 3.1cm 2.6cm, scale=.50]{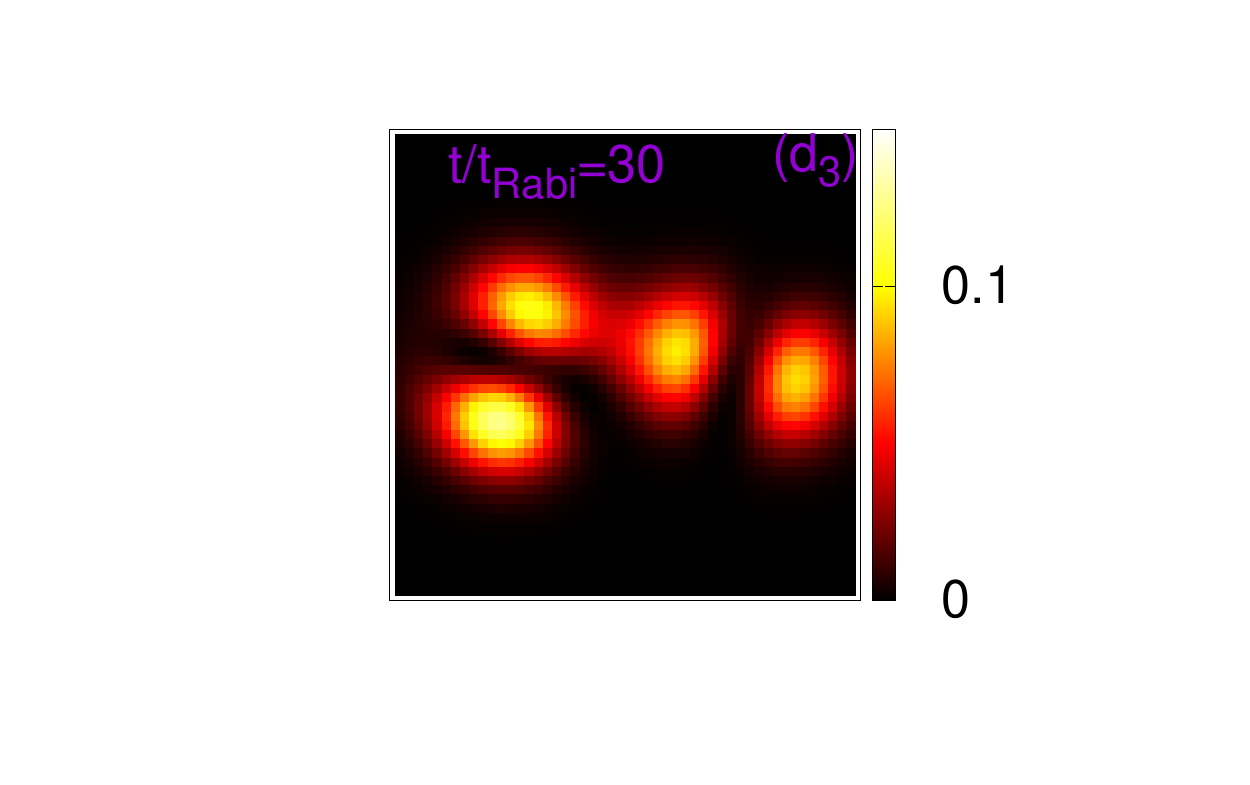}}\\
{\includegraphics[trim =  4.9cm 0.5cm 3.1cm 2.6cm, scale=.50]{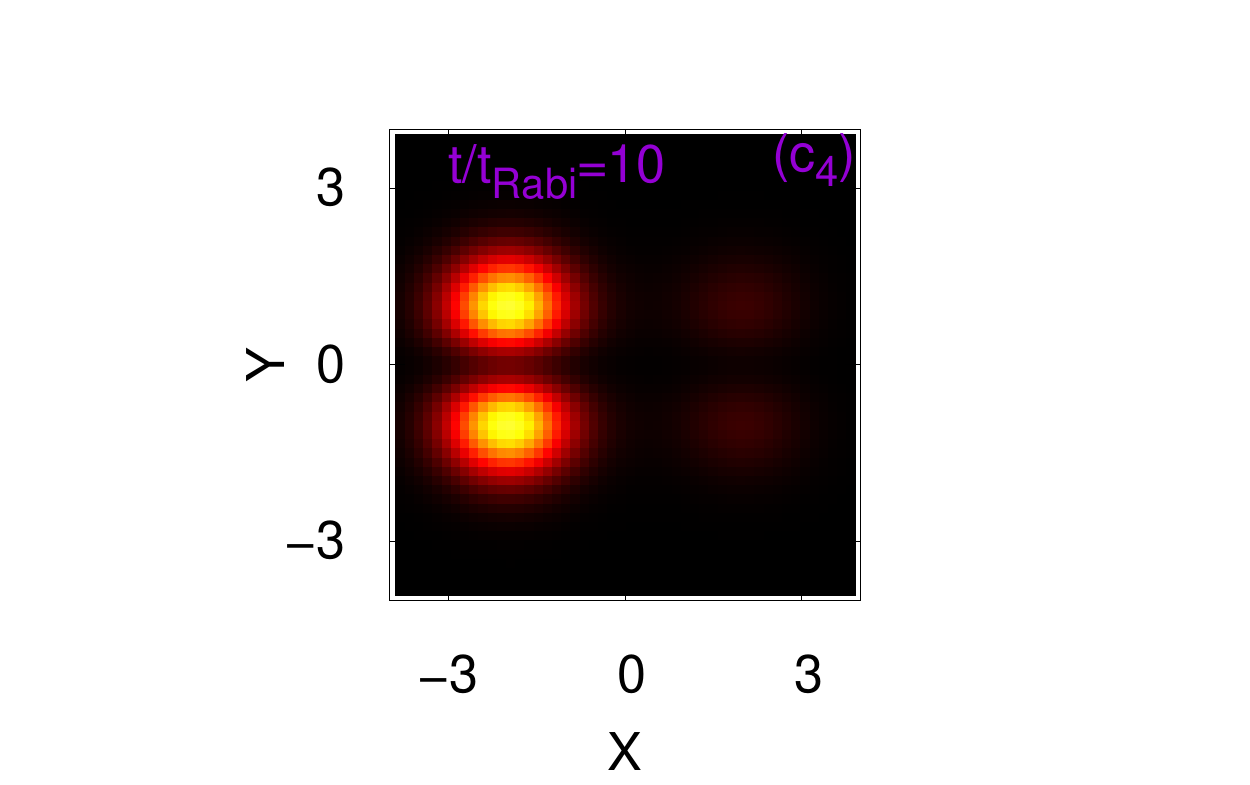}}
{\includegraphics[trim =  4.9cm 0.5cm 3.1cm 2.6cm, scale=.50]{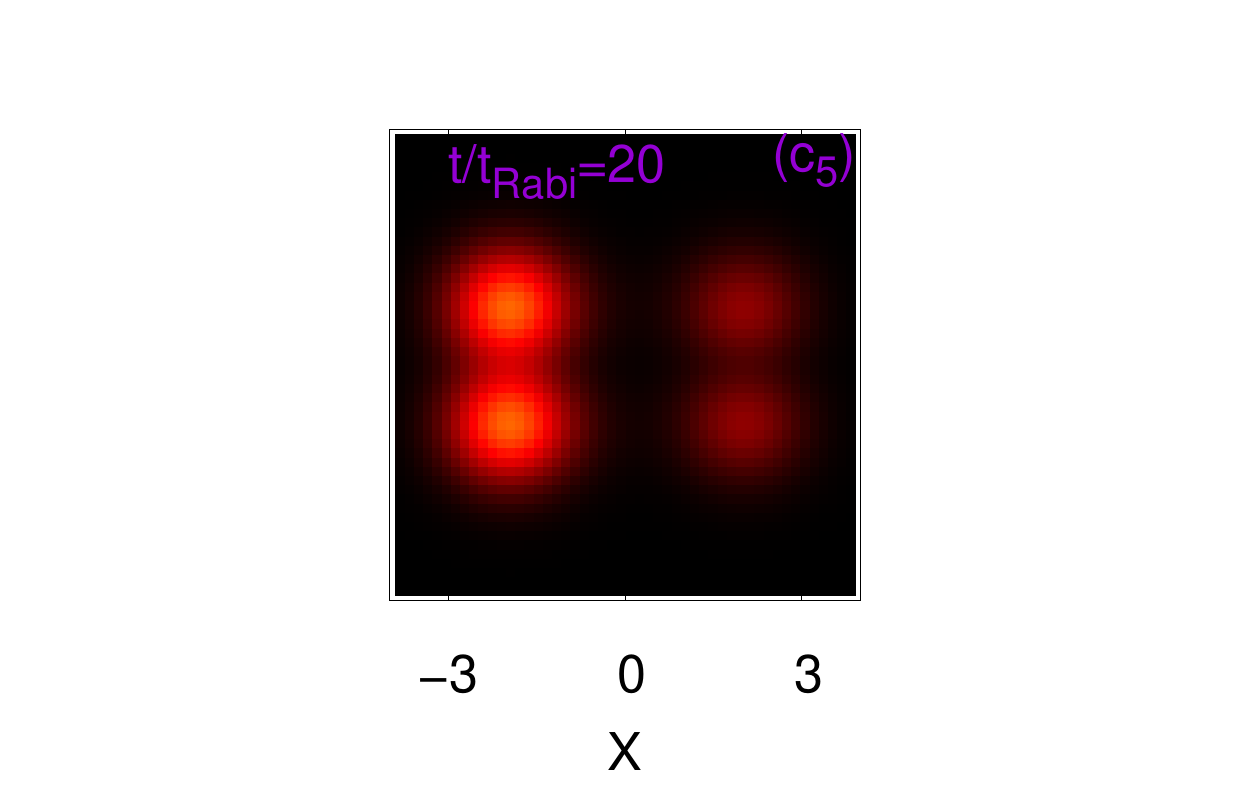}}
{\includegraphics[trim =  4.9cm 0.5cm 3.1cm 2.6cm, scale=.50]{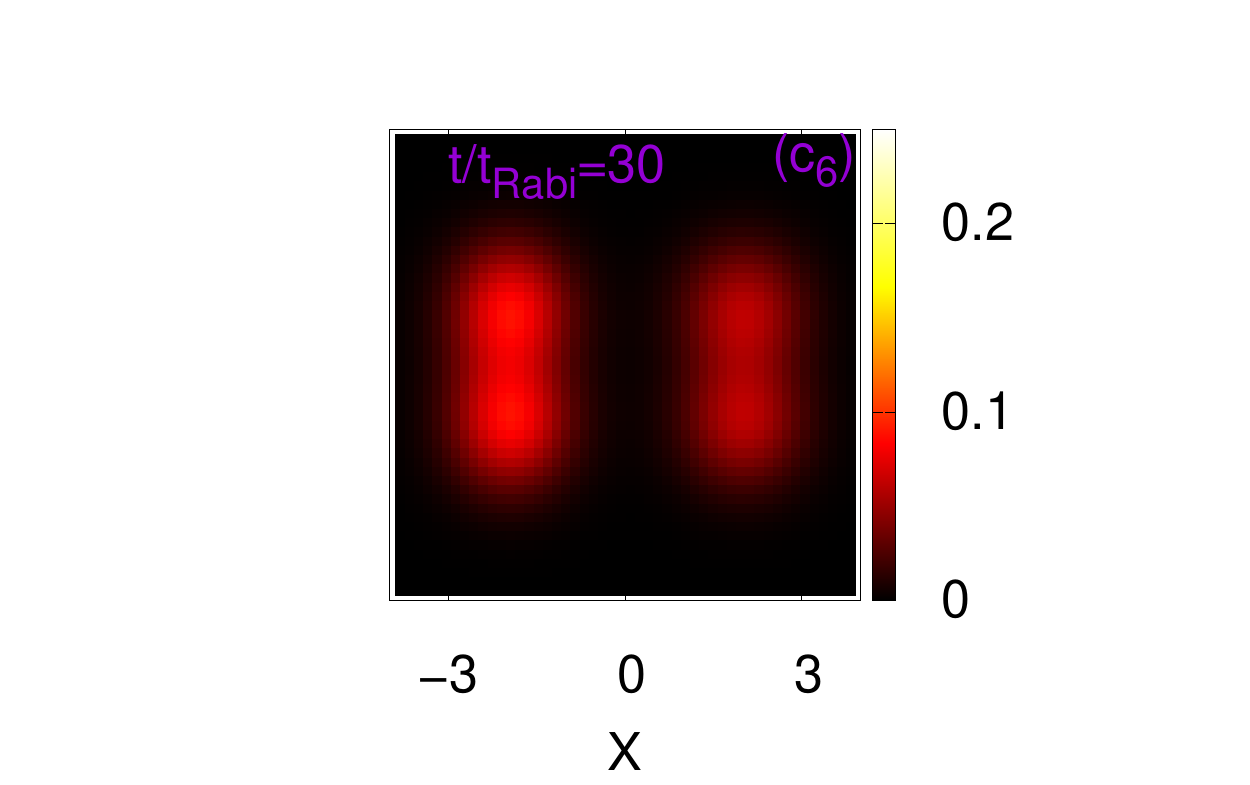}}
{\includegraphics[trim =  2.5cm 0.5cm 3.1cm 2.6cm, scale=.50]{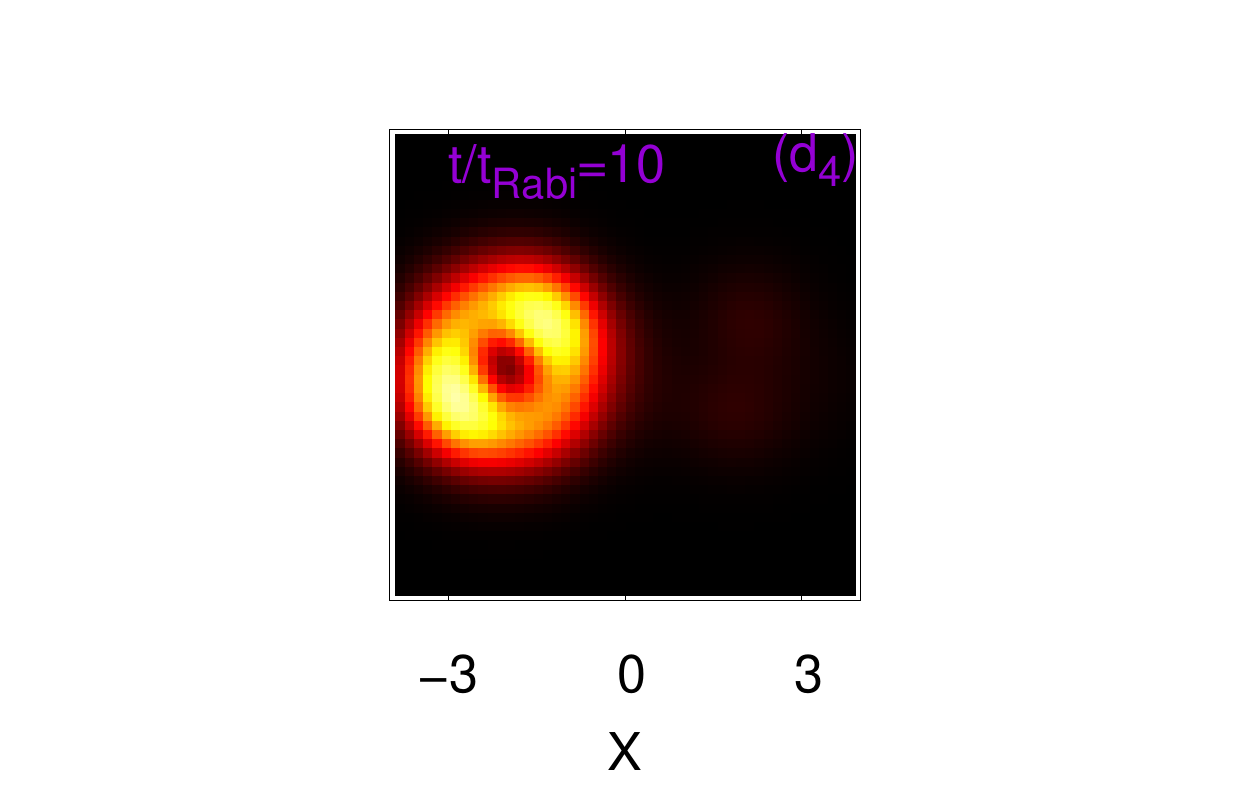}}
{\includegraphics[trim =  4.9cm 0.5cm 3.1cm 2.6cm, scale=.50]{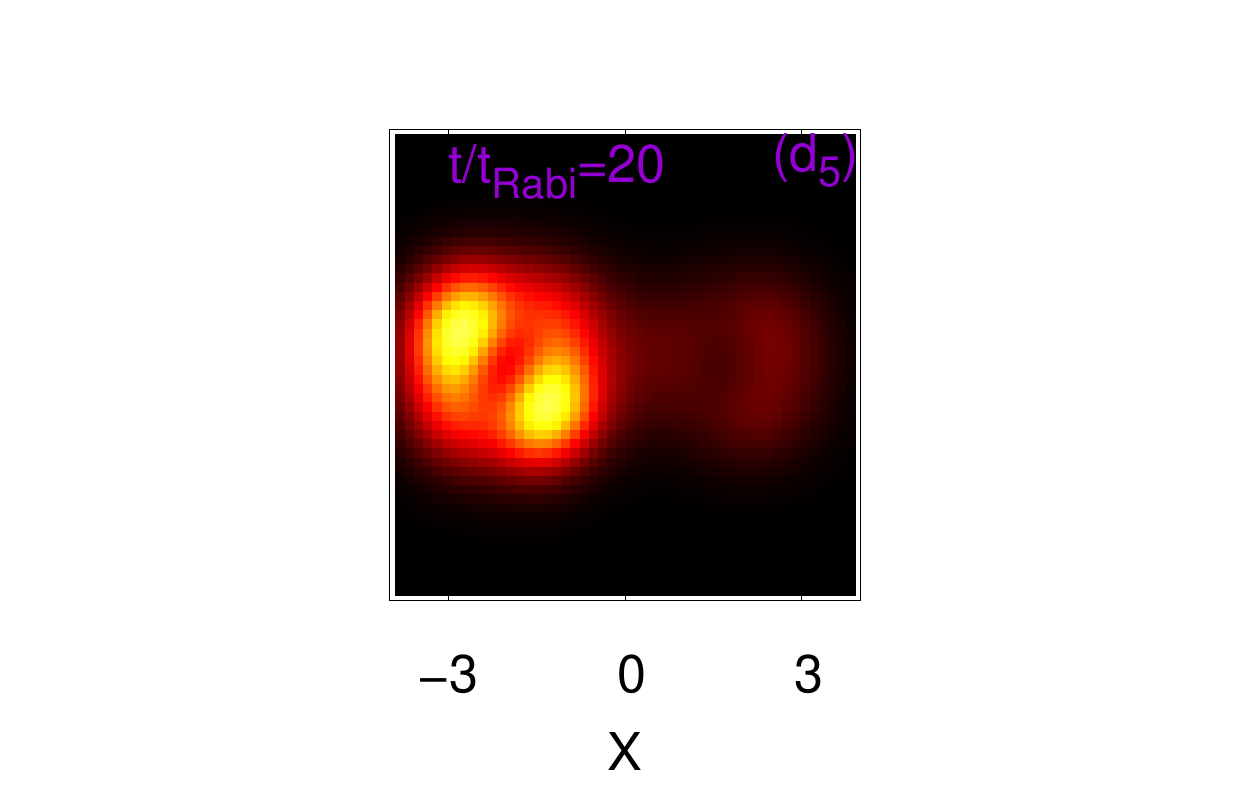}}
{\includegraphics[trim =  4.9cm 0.5cm 3.1cm 2.6cm, scale=.50]{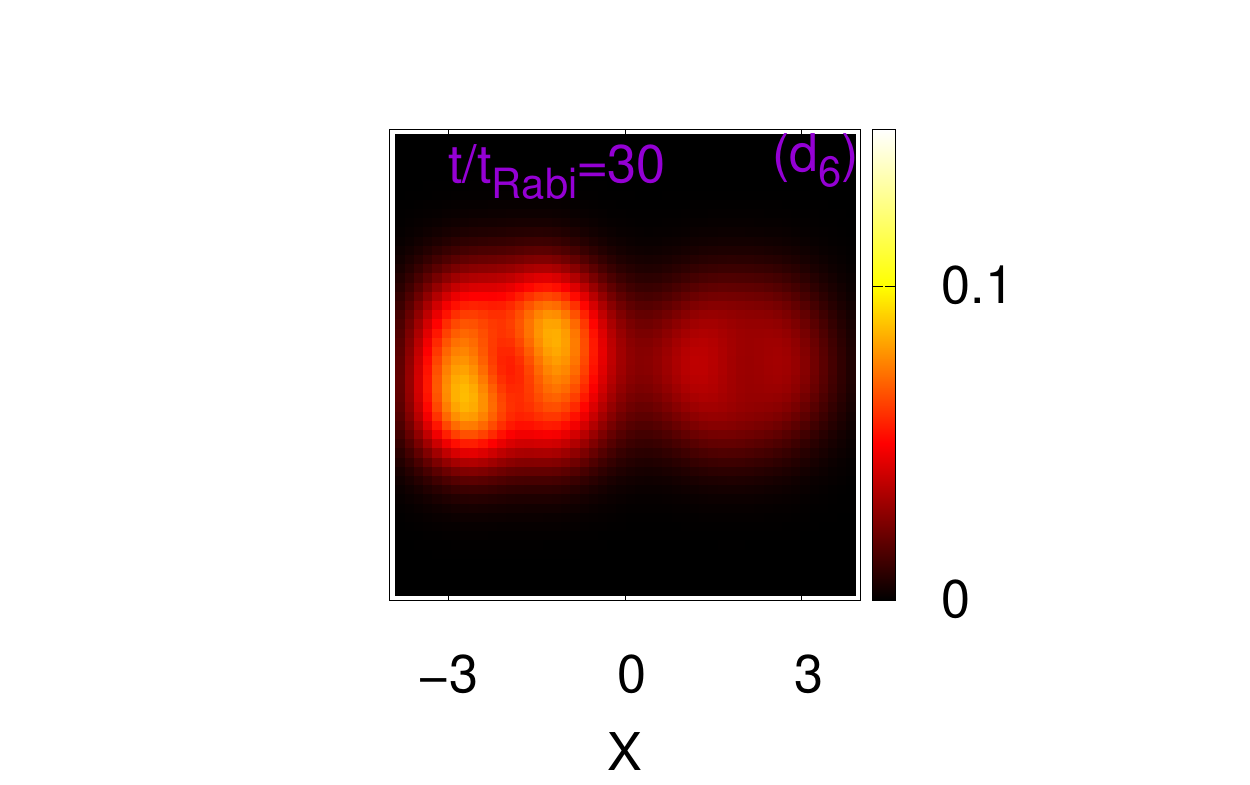}}\\
\caption{Time evolution of the mean-field (first and third rows) and many-body (second and fourth rows) density oscillations in a symmetric 2D double-well. The interaction parameter is $\Lambda=0.01$ and the number of bosons is $N=10$. Shown are snapshots for the densities per particle at $t=10t_{Rabi}$ (first and fourth columns), $20t_{Rabi}$ (second and fifth columns), and $30t_{Rabi}$ (third and sixth columns) for the initial states  $\Psi_G$,  $\Psi_X$, $\Psi_Y$, and  $\Psi_V$.  The MCTDHB computation is performed with $M=6$ time-adaptive orbitals for $\Psi_{G}$ and $\Psi_{X}$, and with $M=10$ time-adaptive orbitals for $\Psi_{Y}$ and $\Psi_{V}$. The quantities shown are dimensionless.}
\label{Fig10}
\end{figure*}

\clearpage

\section{Concluding remarks}

In the present work, we have studied the tunneling dynamics of initially coherent bosonic clouds  in a two-dimensional double-well potential. The bosonic systems are prepared either as the ground, transversely or longitudinally excited, or vortex state in the left well of a symmetric 2D double-well potential. Although, the tunneling  dynamics of the ground and longitudinally excited states have one-dimensional manifestations, here we examine their two-dimensional analogs by solving the full many-body Schr\"{o}dinger equation. Moreover, we study the transversely excited and vortex states which do not have any one-dimensional analog, and require at least a two-dimensional geometry to be realized.  

Explicitly, we have performed the numerical simulations  to study the  dynamics of the $\Psi_G$, $\Psi_X$, $\Psi_Y$, and $\Psi_V$ states based on a well-known method, MCTDHB. We observe the dynamical behavior of a few physical quantities such as the survival probability in the left well, depletion and fragmentation, and the many-particle position, momentum, and angular-momentum expectation values and variances   of each of the bosonic clouds when they tunnel back and forth  in the double-well potential. To show the impact of  growing degree of fragmentation with time, we compare the  respective quantities  at the many-body level of theory with their respective mean-field results.

We have shown that apart from the vortex state, all other initial states can tunnel through the barrier without destroying their initial structures at the mean-field level as well as at the many-body level. But the vortex state distorts its structure while  it tunnels and  produces two vortex dipoles which rotate around the minima of the corresponding well both at the mean-field and many-body levels. However, the shape and orientation of the dipoles, in the long-time dynamics, are found to be different at the mean-field level in comparison with the many-body level when one observes the density oscillations of $\Psi_V$.  We find that the creation of the vortex dipoles  occurs due to   differences in the tunneling frequencies of  $\Psi_X$ and $\Psi_Y$.  Moreover, the effect of the many-body correlation  appears   even in the dynamics of the most basic quantity, i.e., the survival probability, in terms of the collapse  of the density oscillations which can not be seen  using the  Gross-Pitaevskii theory.  Also, the collapse rates are found to be different for the different initial states considered here. Therefore, the loss of coherence and development of fragmentation  demonstrate the clear signature  of  many-body correlations on the dynamics of the bosonic clouds in  the two-dimensional BJJ. We notice that the rate of loss of coherence or development of fragmentation is maximum for $\Psi_Y$  and minimum for $\Psi_X$. Examining the first few natural orbitals for all the states, we observe that the presence of transverse excitations  enhances the loss of coherence for $\Psi_Y$ and $\Psi_V$,  and impacts the dynamics of physical quantities discussed in this work.

Based on the time-evolution of the survival probability and fragmentation, we have further discussed how the many-body correlations  affect some basic quantum mechanical observables and their variances. Precisely, we present the interconnection of  many-particle  variances, $\dfrac{1}{N}\Delta_{\hat{X}}^2(t)$,  $\dfrac{1}{N}\Delta_{\hat{Y}}^2(t)$,  $\dfrac{1}{N}\Delta_{\hat{P}_X}^2(t)$, $\dfrac{1}{N}\Delta_{\hat{P}_Y}^2(t)$, and  $\dfrac{1}{N}\Delta_{\hat{L}_Z}^2(t)$ with the density oscillations and fragmentation.  The  many-body variances incorporate the  depletion and fragmentation,  generally leading to  different values with the respective mean-field variances. It is  observed that the time-evolution of  each variance vary due to the different  initial structures of the bosonic cloud. The distinctive  feature of the breathing-mode oscillations in addition to the density oscillations are found in the time evolution of $\dfrac{1}{N}\Delta_{\hat{X}}^2(t)$, $\dfrac{1}{N}\Delta_{\hat{P}_X}^2(t)$, and  $\dfrac{1}{N}\Delta_{\hat{L}_Z}^2(t)$.  The breathing-mode oscillations are the most prominent for the initial states $\Psi_X$ and $\Psi_V$. We show that the existence of  the transverse degrees-of-freedom, although   nearly frozen, can  have significant influence on  angular-momentum properties in the system. It is clear from the investigation that the information of the many-body features  can not be extracted  from only the shape of the density profile of the system, but it requires a close analysis of the  natural orbitals and microscopic mechanism of the fragmentation.   The present investigation shows that the tunneling dynamics of the ground, excited, and vortex states in two-dimension bosonic Josephson junction  is very rich and many-body theory is required to accurately represent their dynamics.   We believe that our work will motivate researchers to study the out-of-equilibrium tunneling dynamics of more complex and intricate objects.

\section*{Acknowledgments}
This research was supported by the Israel Science Foundation (Grants No. 600/15 and No. 1516/19).
Computation time on the High Performance Computing system Hive of the Faculty of Natural Sciences at the University of Haifa and computational resources at the High Performance Computing Center Stuttgart (HLRS) are gratefully acknowledged.
\clearpage




\clearpage
\section*{Supplemental material for Impact of the transverse direction on the many-body tunneling dynamics in a two-dimensional bosonic Josephson junction}

In this supplemental material, we augment the main text with further details. In Section V, we present a brief mathematical description on the many-particle variance discussed in the main text. In Section VI, we show the long-time dynamics of the quantum mechanical quantities along with their numerical convergence  with respect to the number of time-adaptive orbitals  and discrete-variable representation  grid points. In Section VII, we show the consistency of the preparation of the ground state.

\section{Many-particle variance}
The quantum variance of an observable $\hat{A}$ for a system in a state $|\Psi(t)\rangle$ determines the  quantum resolution with which the observable can be measured. The variance of $\hat{A}$ is measured by the combination of the expectation values of $\hat{A}$ and the square of $\hat{A}$. Here the expectation value of $\hat{A}$ $=\sum_{j=1}^N \hat{a}({r_j})$ is solely made of one-body operators but the expectation of the square of $\hat{A}$, ${\hat{A}}^2=\sum_{j=1}^N\hat{a}^2({r_j})+\sum_{j<k}2\hat{a}({r_j})\hat{a}({r_k})$, is a mixture of one- and two-body operators. The variance  can be expressed as \cite{Alon2019b}
\begin{eqnarray}\label{11}
\dfrac{1}{N}\Delta_{\hat{A}}^2&(t)&=\dfrac{1}{N}[\langle\Psi(t)|\hat{A}^2|\Psi(t)\rangle-\langle\Psi(t)|\hat{A}|\Psi(t)\rangle^2] \nonumber \\
&=& \dfrac{1}{N}\Bigg\{\sum_j n_j(t)\int d\textbf{r}\phi_j^*(\textbf{r}; t)\hat{a}^2({\textbf{r}})\phi_j(\textbf{r}; t)-\left[\sum_j n_j(t)\int d\textbf{r}\phi_j^*(\textbf{r}; t)\hat{a}({\textbf{r}})\phi_j(\textbf{r}; t)\right]^2 \nonumber \\ &+& \sum_{jpkq}\rho_{jpkq}(t)\left[\int d\textbf{r}\phi_j^*(\textbf{r}; t)\hat{a}({\textbf{r}})\phi_k(\textbf{r}; t)\right] \left[\int d\textbf{r}\phi_p^*(\textbf{r}; t)\hat{a}({\textbf{r}})\phi_q(\textbf{r}; t)\right]\Bigg\},
\end{eqnarray}

where $\{\phi_j(\textbf{r}; t)\}$ are the natural orbitals, $\{n_j(t)\}$  the natural occupations, and $\rho_{jpkq}(t)$ are the elements of the   reduced two-particle density matrix, $\rho(\textbf{r}_1, \textbf{r}_2, \textbf{r}_1^\prime, \textbf{r}_2^\prime; t)= \sum \limits_{jpkq}\rho_{jpkq}(t)\phi_j^*(\textbf{r}_1^\prime; t) \phi_p^*(\textbf{r}_2^\prime; t)$ 
$\phi_k(\textbf{r}_1; t) \phi_q(\textbf{r}_2; t).$  For one-body operators which are local in  position space, the variance described in Eq~\ref{11} becomes \cite{Lode2020}
 
\begin{eqnarray}\label{12}
\dfrac{1}{N}\Delta_{\hat{A}}^2&(t)&= \int d\textbf{r}\dfrac{\rho(\textbf{r};t)}{N}\hat{a}^2({\textbf{r}})-N\left[\int \dfrac{\rho(\textbf{r};t)}{N}\hat{a}({\textbf{r}}) \right]^2 \nonumber \\ &+& \int d\textbf{r}_1 d\textbf{r}_2 \dfrac{\rho^{(2)}(\textbf{r}_1, \textbf{r}_2, \textbf{r}_1, \textbf{r}_2; t)}{N}a(\textbf{r}_1)a(\textbf{r}_2).
\end{eqnarray} 

In our study,   the center-of-mass of the bosonic clouds are at the position $(a,b)=(-2,0)$ at  $t=0$. Eq~\ref{11} describes the variances when  the center-of-mass of the bosonic clouds are at (0, 0).  To calculate the  variances at $(a,b)$, we have used the general relation between the variances at $(a,b)$ and  at the origin. As mentioned in the main text, for the position and momentum operators, the variances do not change with the position of center-of-mass of the clouds \cite{Alon2019b}, i.e., $\dfrac{1}{N}\Delta_{{\hat{X}}}^2\Big|_{\Psi(a,b)}=\dfrac{1}{N}\Delta_{{\hat{X}}}^2\Big|_{\Psi(0,0)}$,  $\dfrac{1}{N}\Delta_{{\hat{Y}}}^2\Big|_{\Psi(a,b)}=\dfrac{1}{N}\Delta_{{\hat{Y}}}^2\Big|_{\Psi(0,0)}$, $\dfrac{1}{N}\Delta_{{\hat{P}_X}}^2\Big|_{\Psi(a,b)}=\dfrac{1}{N}\Delta_{{\hat{P}_X}}^2\Big|_{\Psi(0,0)}$, and  $\dfrac{1}{N}\Delta_{{\hat{P}_Y}}^2\Big|_{\Psi(a,b)}=\dfrac{1}{N}\Delta_{{\hat{P}_Y}}^2\Big|_{\Psi(0,0)}$. But the situation becomes a more involved for the variance of the angular-momentum operator, which takes the form as \cite{Alon2019b}
 
\begin{eqnarray}\label{13}
\dfrac{1}{N}\Delta_{{\hat{L}_Z}}^2\Big|_{\Psi(a,b)}&=&\dfrac{1}{N}\Delta_{{\hat{L}_Z}}^2\Big|_{\Psi(0,0)}+a^2\dfrac{1}{N}\Delta_{{\hat{P}_Y}}^2\Big|_{\Psi(0,0)}+b^2\dfrac{1}{N}\Delta_{{\hat{P}_X}}^2\Big|_{\Psi(0,0)} \nonumber \\
&+& a [\langle \Psi(0,0)|{\hat{L}_Z}{\hat{P}_Y}+{\hat{P}_Y}{\hat{L}_Z}|\Psi(0,0)\rangle - 2\langle\Psi(0,0)|{\hat{L}_Z}|\Psi(0,0)\rangle\langle|\Psi(0,0)|{\hat{P}_Y}|\Psi(0,0)\rangle]\nonumber \\
&-& b [\langle \Psi(0,0)|{\hat{L}_Z}{\hat{P}_X}+{\hat{P}_X}{\hat{L}_Z}|\Psi(0,0)\rangle - 2\langle\Psi(0,0)|{\hat{L}_Z}|\Psi(0,0)\rangle\langle|\Psi(0,0)|{\hat{P}_X}|\Psi(0,0)\rangle]\nonumber \\
&-& 2ab [\langle \Psi(0,0)|{\hat{P}_Y}{\hat{P}_X}|\Psi(0,0)\rangle - \langle\Psi(0,0)|{\hat{P}_Y}|\Psi(0,0)\rangle\langle|\Psi(0,0)|{\hat{P}_X}|\Psi(0,0)\rangle].
\end{eqnarray}

Because of symmetry,  Eq.~\ref{13} boils down to $\dfrac{1}{N}\Delta_{{\hat{L}_Z}}^2\Big|_{\Psi(a,b)}=\dfrac{1}{N}\Delta_{{\hat{L}_Z}}^2\Big|_{\Psi(0,0)}+a^2\dfrac{1}{N}\Delta_{{\hat{P}_Y}}^2\Big|_{\Psi(0,0)}$ for our considered systems in this work (see the subsection III C  in the main text). The final form of $\dfrac{1}{N}\Delta_{{\hat{L}_Z}}^2\Big|_{\Psi(a,b)}$ is used  directly to calculate the angular-momentum variance at $t=0$, see Table I of the main text.

\section{Long-time dynamics and convergence of quantities}
Here we check the numerical convergence for the  long-time dynamics of our results discussed in the main text  with respect to the number of time-adaptive orbitals and density of the grid points for the ground $(\Psi_G)$, longitudinally-excited $(\Psi_X)$,  transversely-excited $(\Psi_Y)$, and vortex $(\Psi_V)$ states.

\subsection{Convergence with the number of time-adaptive orbitals}
In our work, we have performed  the computations for the $\Psi_G$ and  $\Psi_X$ states with $M=6$ time-adaptive orbitals, while for   the $\Psi_Y$ and  $\Psi_V$ states using $M=10$ time-adaptive orbitals. To check the convergence with the orbital numbers, we  repeat our computations with $M=10$ and $M=12$ orbitals for $\Psi_G$,  $\Psi_X$ and $\Psi_Y$,  $\Psi_V$, respectively.  As discussed in the main text, we have prepared the initial states of the bosonic clouds in the left well of a symmetric double-well with $N=10$ bosons. The interaction parameter is $\Lambda=0.01$, also see Section VII below. The many-body Hamiltonian is represented by $64\times 64$ exponential discrete-variable-representation grid points in a box size $[-10,10)\times [-10,10)$. 

\renewcommand{\thefigure}{S\arabic{figure}}

\setcounter{figure}{0}
\begin{figure*}[!h]
{\includegraphics[trim = 0.1cm 0.5cm 0.1cm 0.2cm, scale=.60]{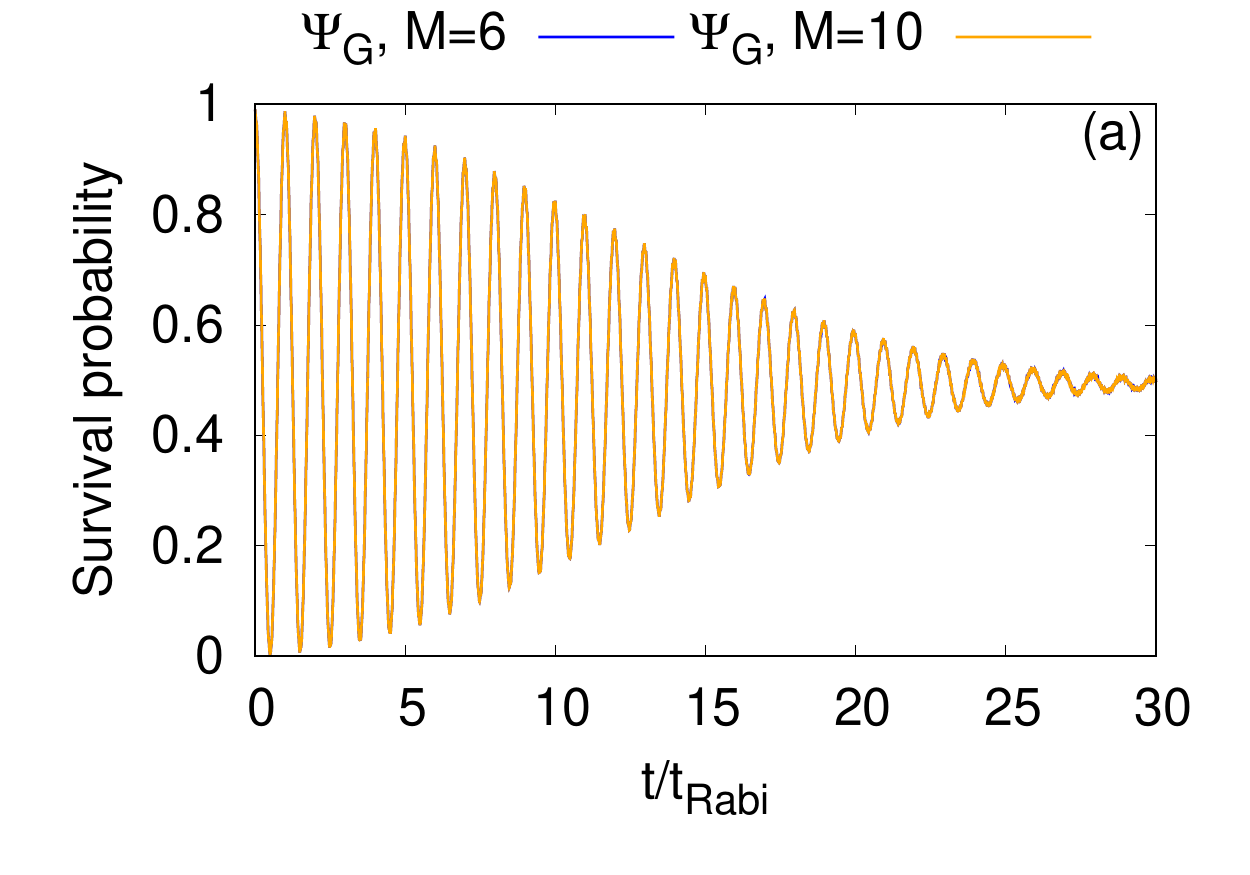}}
{\includegraphics[trim = 0.1cm 0.5cm 0.1cm 0.2cm, scale=.60]{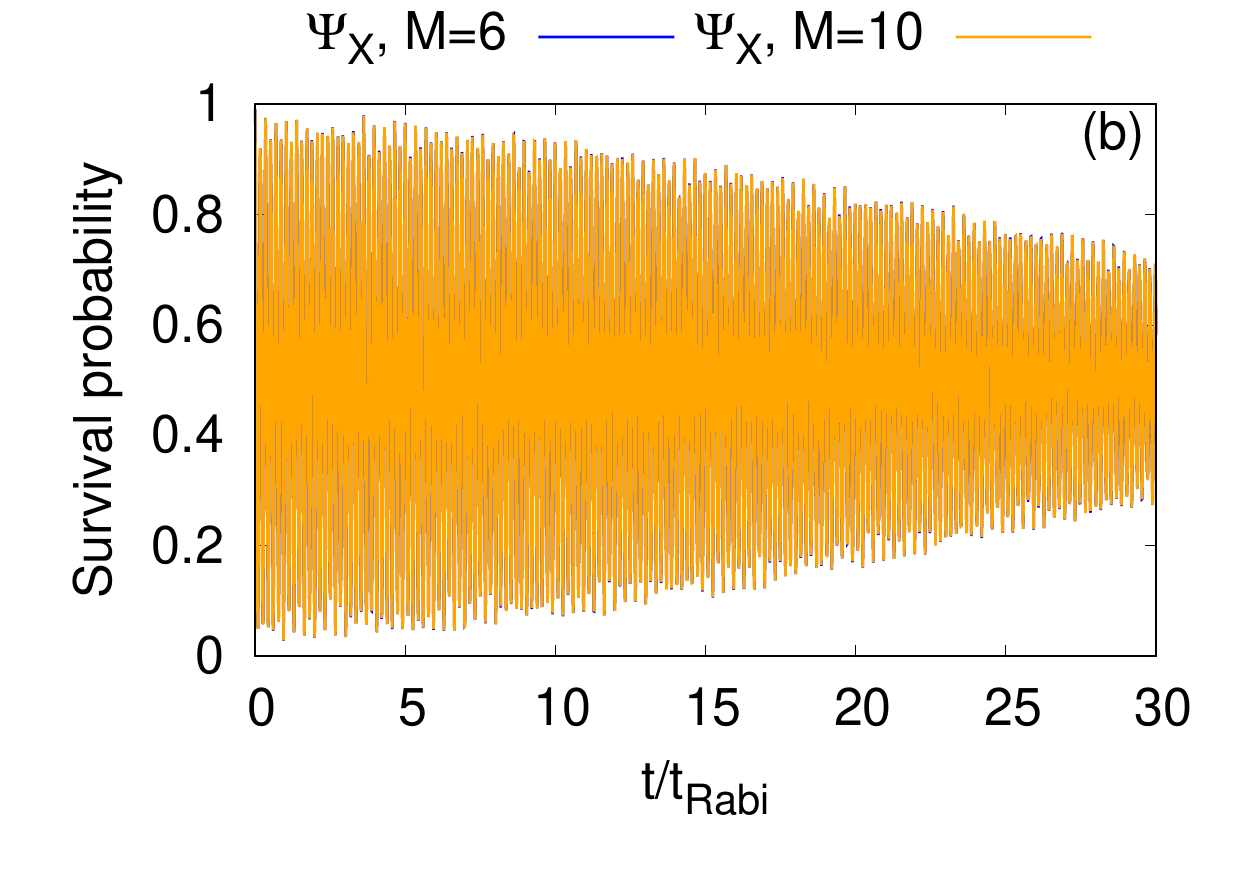}}\\
\vglue 0.25 truecm 
{\includegraphics[trim = 0.1cm 0.5cm 0.1cm 0.2cm, scale=.60]{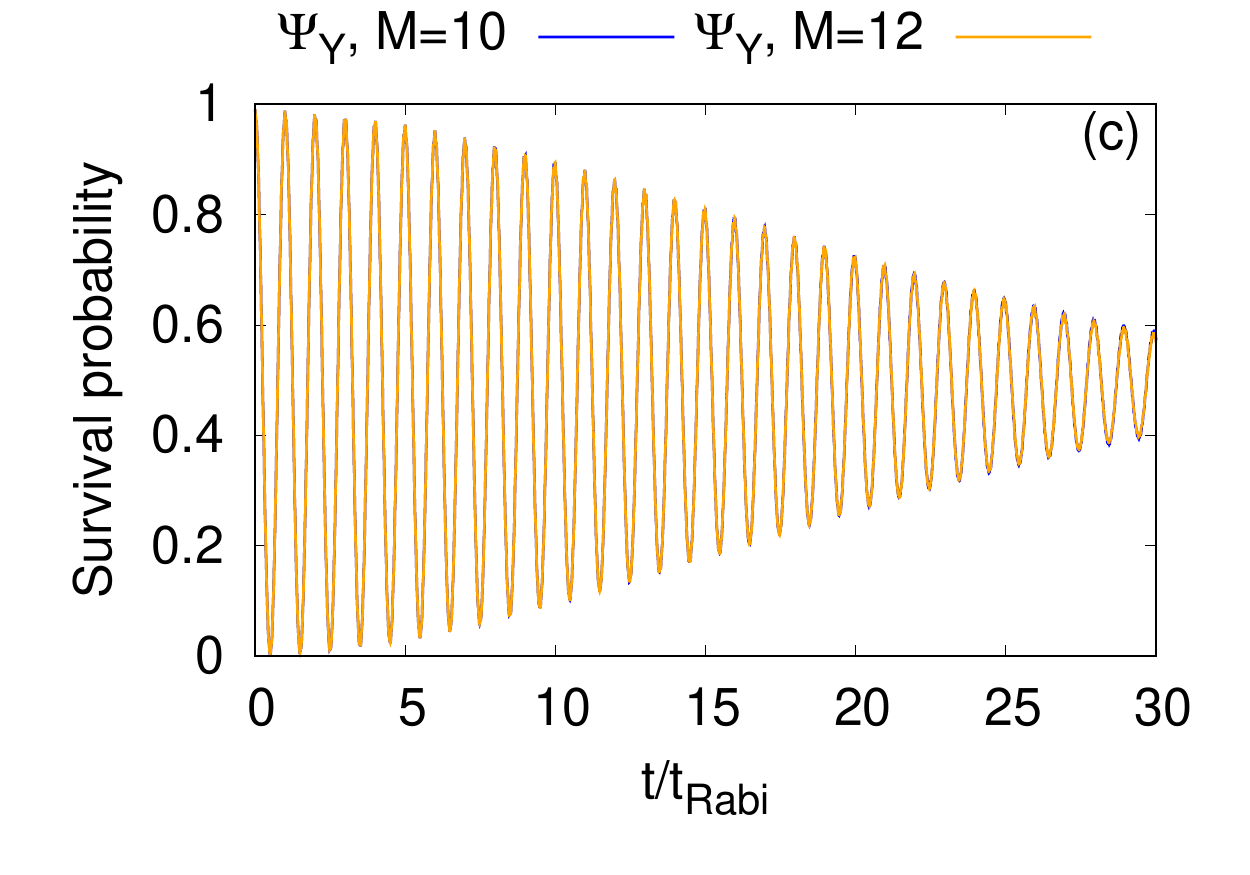}}
{\includegraphics[trim = 0.1cm 0.5cm 0.1cm 0.2cm, scale=.60]{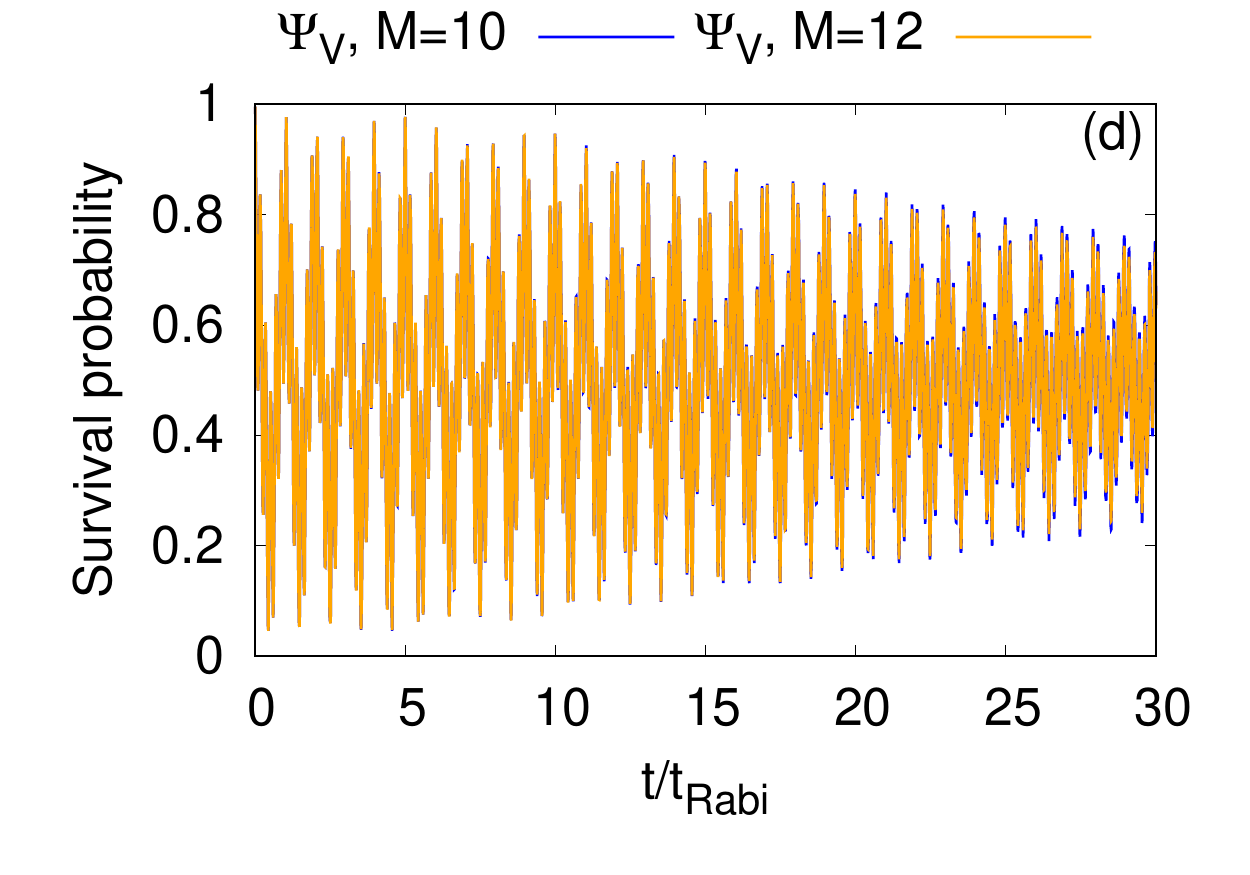}}\\
\caption{\label{fig11}Convergence  of the survival probability, $P_L(t)$, with respect to the number of  time-adaptive orbitals for the initial states (a) $\Psi_G$, (b) $\Psi_X$, (c) $\Psi_Y$, and (d) $\Psi_V$ of $N=10$ interacting bosons with $\Lambda=0.01$ in the symmetric double-well trap. The many-body results are computed using the MCTDHB method. The convergence are verified with $M=6$,  $10$ time-adaptive orbitals for  the states, $\Psi_{G}$ and $\Psi_{X}$. While we demonstrate the convergence of the results for $\Psi_{Y}$ and $\Psi_{V}$  using $M=10$,  $12$ time-adaptive orbitals. The quantities shown  are dimensionless.}
\end{figure*}

\begin{figure*}[!h]
{\includegraphics[trim = 0.1cm 0.5cm 0.1cm 0.2cm, scale=.60]{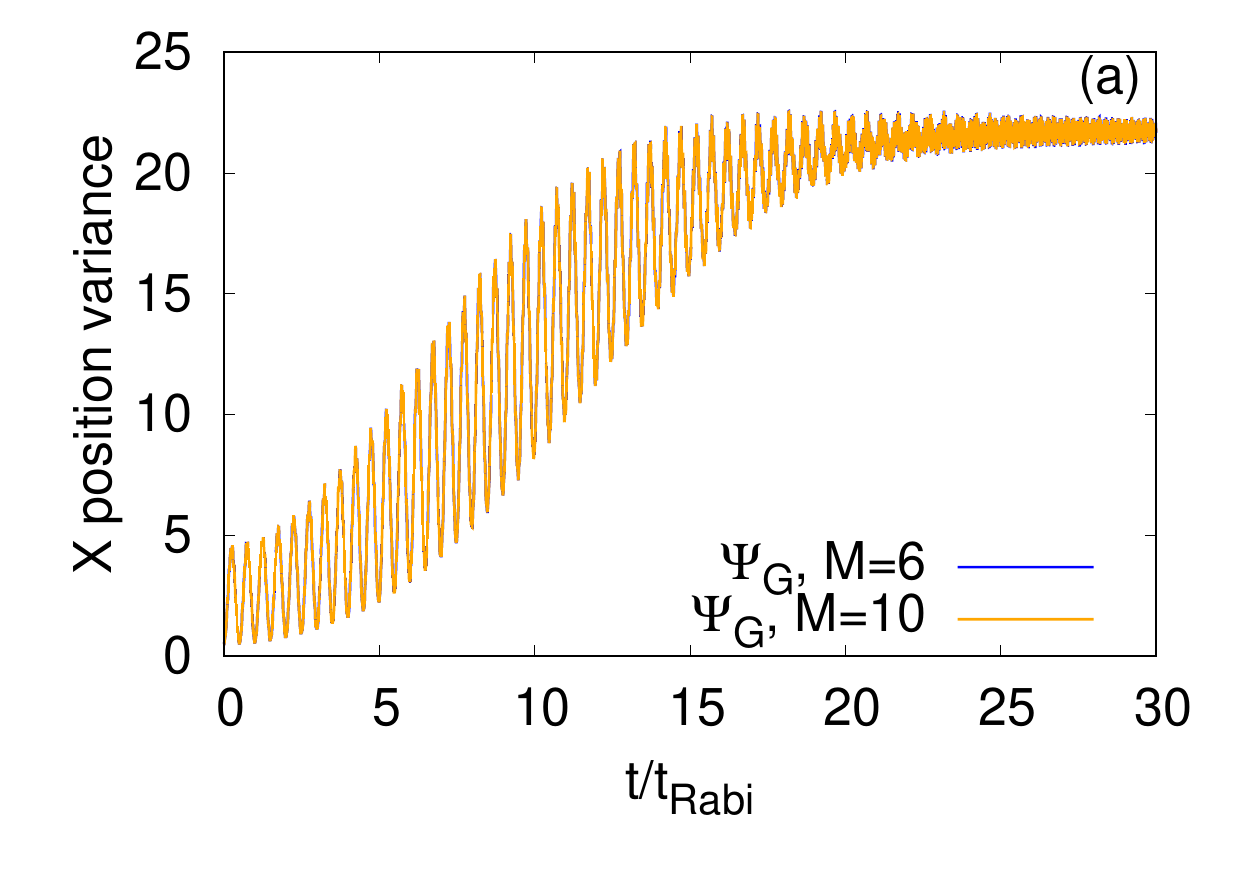}}
{\includegraphics[trim = 0.1cm 0.5cm 0.1cm 0.2cm, scale=.60]{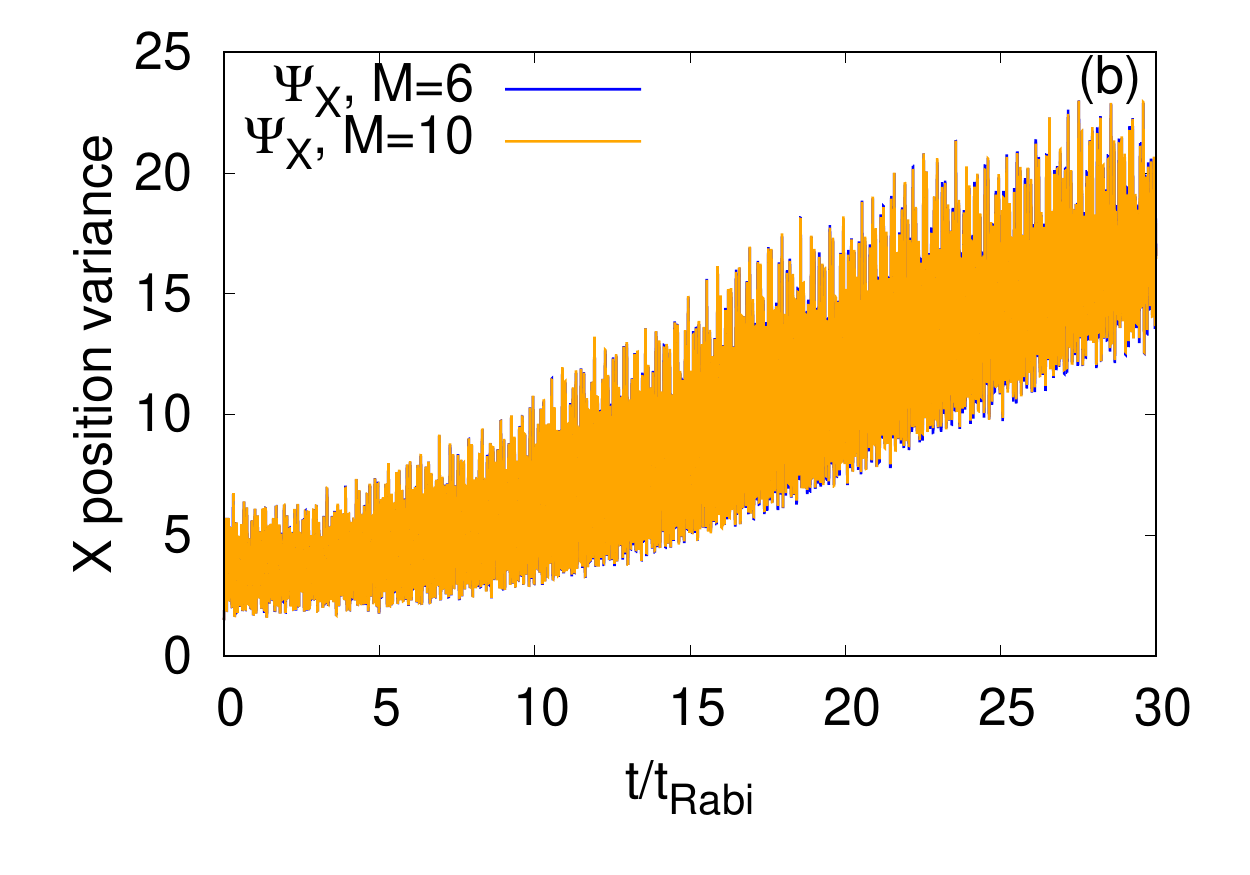}}\\
\vglue 0.25 truecm
{\includegraphics[trim = 0.1cm 0.5cm 0.1cm 0.2cm, scale=.60]{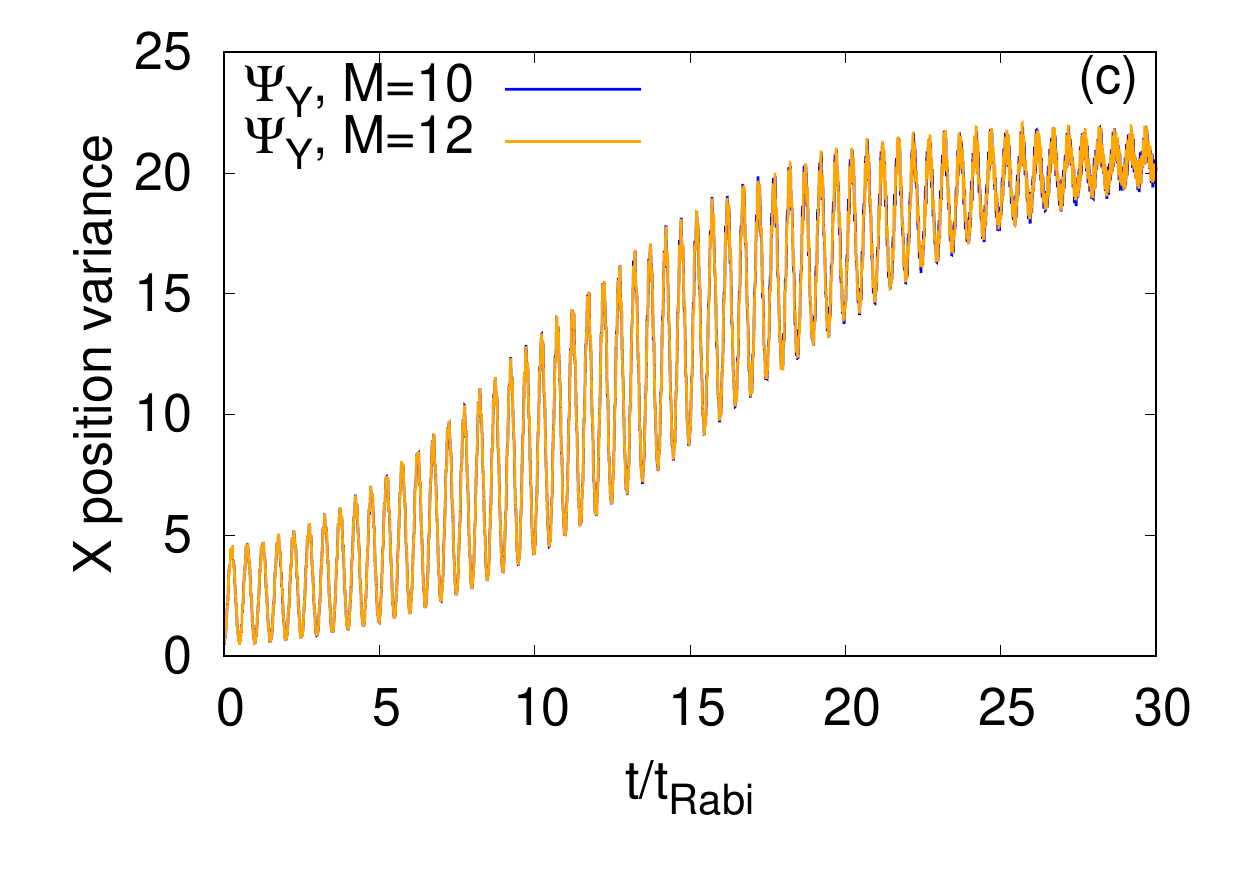}}
{\includegraphics[trim = 0.1cm 0.5cm 0.1cm 0.2cm, scale=.60]{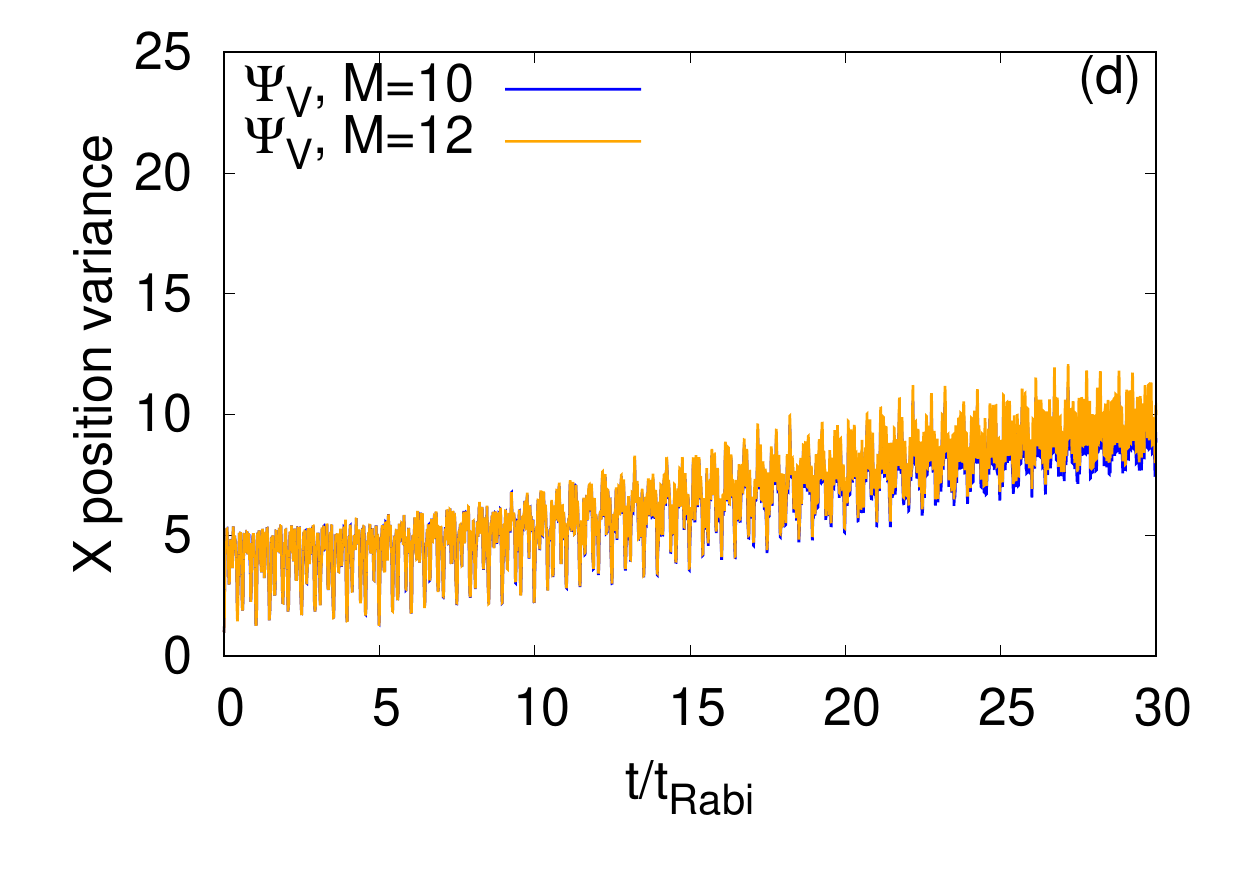}}\\
\caption{\label{fig12}Convergence of the time-dependent many-body position variance per particle along $x$-direction, $\dfrac{1}{N}\Delta_{\hat{X}}^2(t)$, in a symmetric double-well trap with the number of time-adaptive orbitals for the initial states (a) $\Psi_G$, (b) $\Psi_X$, (c) $\Psi_Y$, and (d) $\Psi_V$ for $N=10$ interacting bosons with $\Lambda=0.01$. The many-body $\dfrac{1}{N}\Delta_{\hat{X}}^2(t)$ are computed using the MCTDHB method. The convergence are verified with $M=6$,  $10$ time-adaptive orbitals for  the states, $\Psi_{G}$ and $\Psi_{X}$. While we demonstrate the convergence of the results for $\Psi_{Y}$ and $\Psi_{V}$  using $M=10$,  $12$ time-adaptive orbitals.   See the text for more details. The quantities shown  are dimensionless.}
\end{figure*}

\begin{figure*}[!h]
{\includegraphics[trim = 0.1cm 0.5cm 0.1cm 0.2cm, scale=.60]{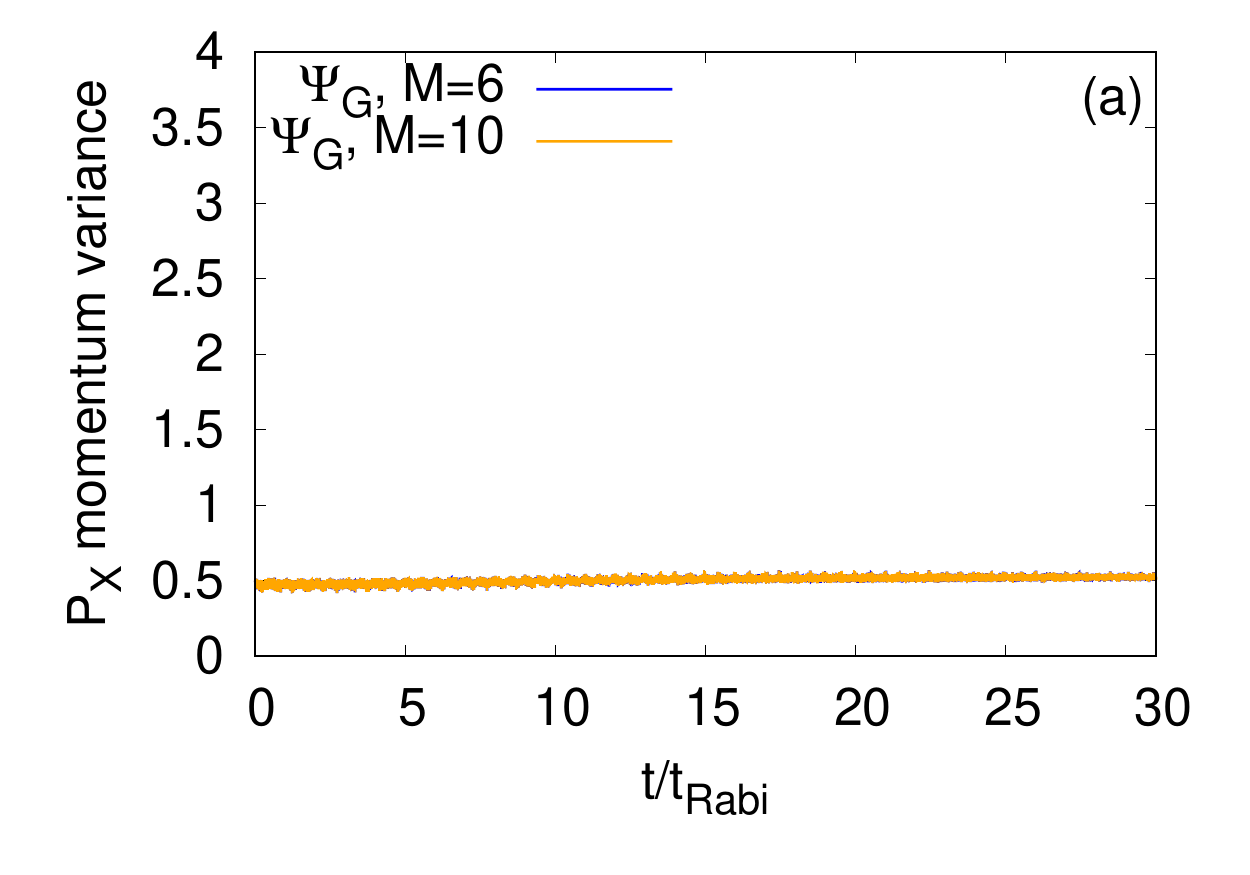}}
{\includegraphics[trim = 0.1cm 0.5cm 0.1cm 0.2cm, scale=.60]{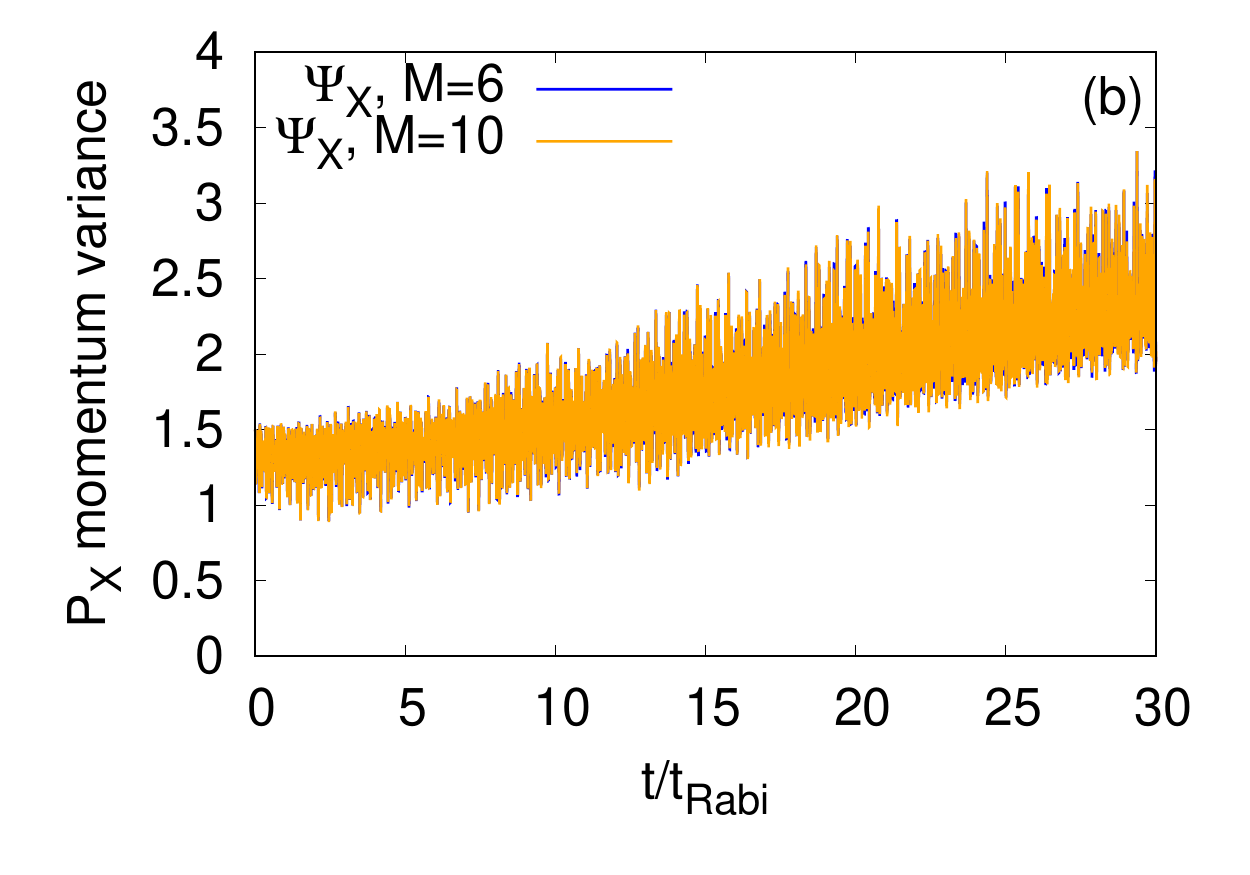}}\\
\vglue 0.25 truecm
{\includegraphics[trim = 0.1cm 0.5cm 0.1cm 0.2cm, scale=.60]{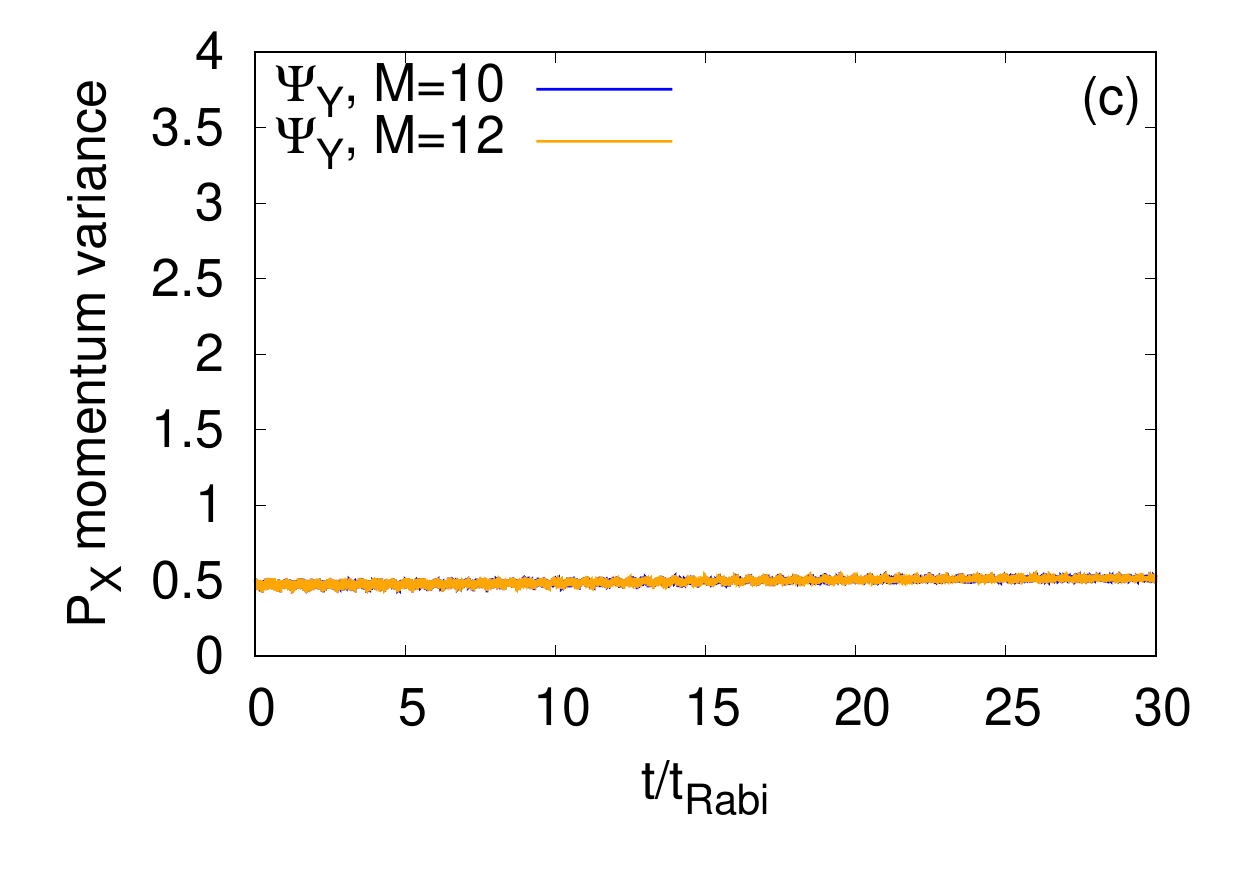}}
{\includegraphics[trim = 0.1cm 0.5cm 0.1cm 0.2cm, scale=.60]{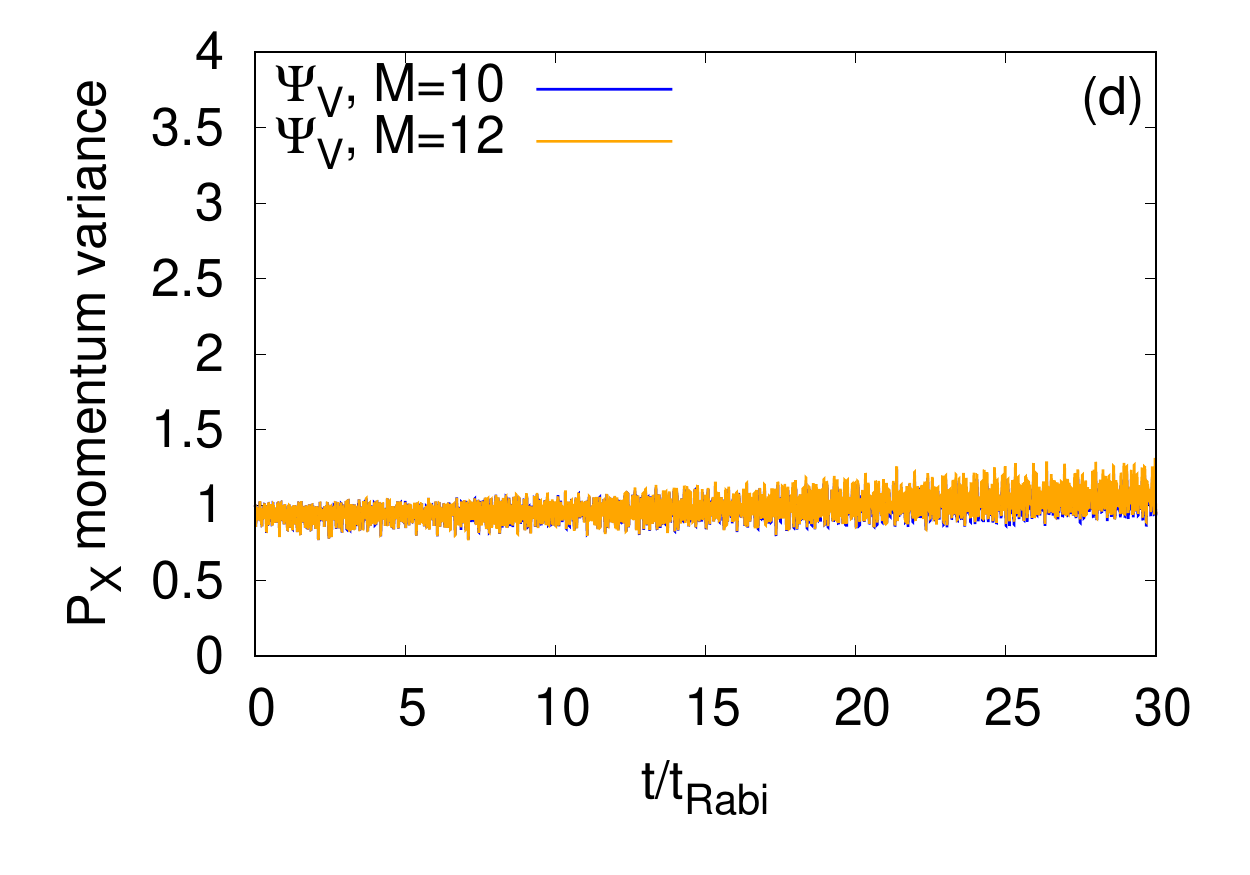}}\\
\caption{\label{fig13}Convergence of the time-dependent many-body momentum variance per particle along $x$-direction, $\dfrac{1}{N}\Delta_{{\hat{P}_X}}^2(t)$, in a symmetric double-well trap with the number of time-adaptive orbitals for the initial states (a) $\Psi_G$, (b) $\Psi_X$, (c) $\Psi_Y$, and (d) $\Psi_V$ for $N=10$ interacting bosons with $\Lambda=0.01$. The many-body $\dfrac{1}{N}\Delta_{{\hat{P}_X}}^2(t)$ are computed using the  MCTDHB method. The convergence are verified with $M=6$,  $10$ time-adaptive orbitals for  the states, $\Psi_{G}$ and $\Psi_{X}$. While we demonstrate the convergence of the results for $\Psi_{Y}$ and $\Psi_{V}$  using $M=10$,  $12$ time-adaptive orbitals.  See the text for more details. The quantities shown  are dimensionless.}
\end{figure*}

\begin{figure*}[!h]
{\includegraphics[scale=.80]{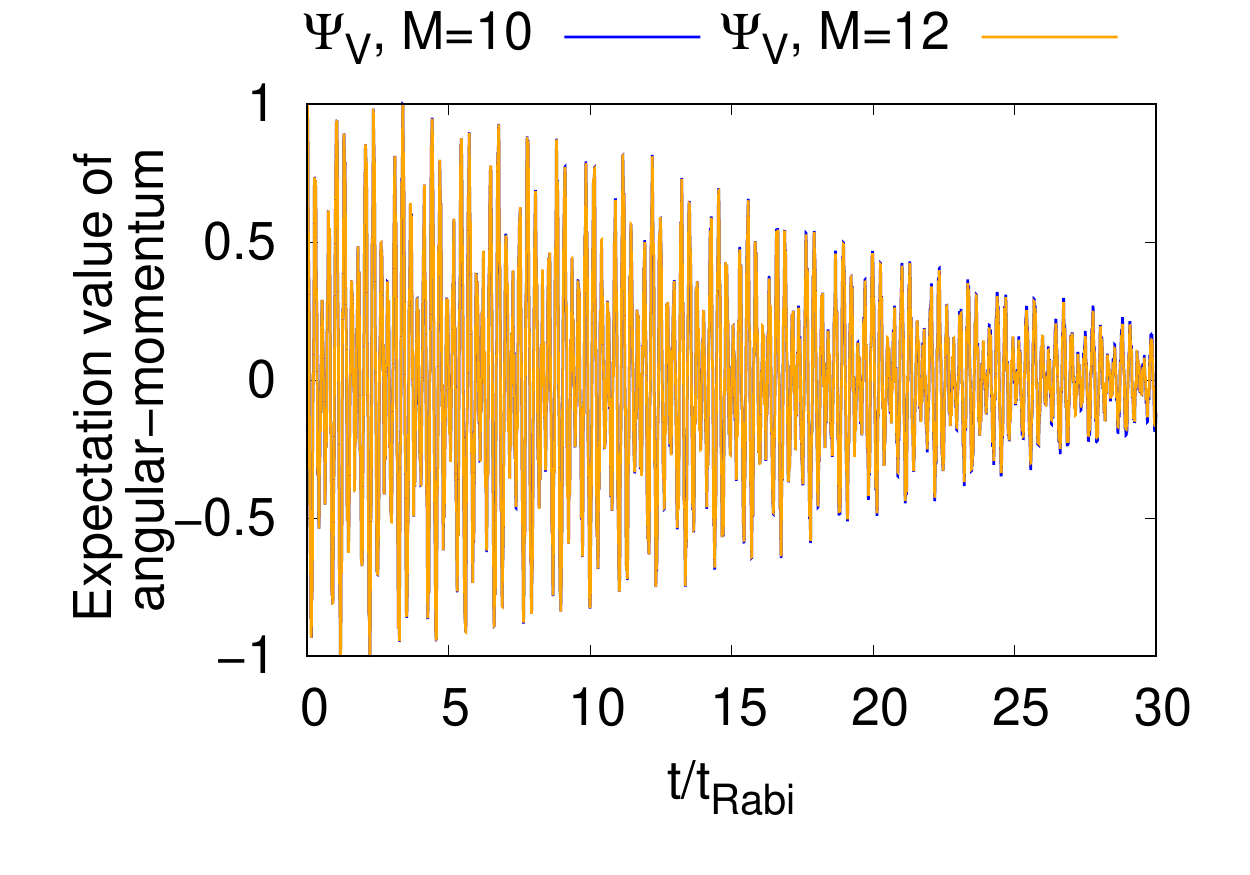}}
\caption{\label{fig14}Long-time dynamics of the angular-momentum expectation value   per particle, $\dfrac{1}{N}\langle\Psi_V|{{\hat{L}_Z}}|\Psi_V\rangle$,  in a symmetric double-well for the vortex state, $\Psi_V$. The number of bosons is  $N=10$. The interaction parameter is $\Lambda=0.01$.  The convergence is verified with  $M=10$ and $12$ time-adaptive orbitals. The mean-field $\dfrac{1}{N}\langle\Psi_V|{{\hat{L}_Z}}|\Psi_V\rangle$ would not produce any collapse (not shown) as presented  here (see Fig. 5 in the main text).  See the text for more details. The quantities shown  are dimensionless.}
\end{figure*}

We  demonstrate  the numerical convergence with the orbital numbers of the many-particle  $P_L(t)$,   $\dfrac{1}{N}\Delta_{\hat{X}}^2(t)$,  $\dfrac{1}{N}\Delta_{{\hat{P}_X}}^2(t)$,  $\dfrac{1}{N}\langle\Psi_V|{{\hat{L}_Z}}|\Psi_V\rangle$, and   $\dfrac{1}{N}\Delta_{{\hat{L}_Z}}^2(t)$   in Fig.~\ref{fig11}, \ref{fig12}, \ref{fig13}, \ref{fig14}, and \ref{fig15}, respectively.  The variances of the position and momentum operators along the $y$-direction for all  states have very small fluctuations (of the order of $10^{-3}$) with a function of time,  and they practically overlap with  the corresponding mean-field results (see subsection IIIC in the main text). Therefore, we have not shown  explicitly the convergences  of  variances of the position and momentum operators along the $y$-direction in our presentation which, of course, converge as well.

As discussed in the main text, the collapse of   $P_L(t)$ is prominent in the the long-time dynamics  for all  initial states. Here also we observe the collapse in the overall oscillation of $P_L(t)$ when computed with $M=10$ and $M=12$ time adaptive orbitals for $\Psi_G$, $\Psi_X$  and $\Psi_Y$,  $\Psi_V$, respectively.  For all  initial states, $P_L(t)$ show a complete overlap  when their respective time-adaptive orbital numbers are increased indicating that the dynamics of $P_L(t)$ is already converged for $M=6$ and $M=10$ orbitals for  $\Psi_G$, $\Psi_X$  and $\Psi_Y$,  $\Psi_V$, respectively.  We verify that the small amplitude and high frequency oscillations of $P_L(t)$ for $\Psi_{V}$ computed using $M=10$ and $12$ time-adaptive orbitals fall on top of each other.

The long-time dynamics  of $\dfrac{1}{N}\Delta_{\hat{X}}^2(t)$ for all  initial states and their  convergence with the number of time-adaptive orbitals  are presented in Fig.~\ref{fig12}. The  dynamics of $\dfrac{1}{N}\Delta_{\hat{X}}^2(t)$ obtained from larger number of time-adaptive orbitals falls on top of the respective variances with smaller number of time-adaptive orbitals exhibiting the convergence with the orbital numbers. The effect of increased degree of the fragmentation can be seen for each of the initial states as discussed in the main text.        The consequences of the density oscillations and breathing mode oscillations (occurred due to the effect of coupling to the lowest
energy band and the higher excited states) are also being observed in the long-time dynamics in terms of two kind of oscillations, i.e., small frequency with large amplitude and high frequency with small amplitude oscillations.  A noticeable difference in the long-time dynamics is that $\dfrac{1}{N}\Delta_{\hat{X}}^2(t)$ reaches   its equilibrium which is more evident for $\Psi_G$ and $\Psi_Y$. This equilibration-like effect comes to the picture if the density oscillations collapse and the fragmentation reaches its plateau. Therefore, here we find a consistent behavior of  $\dfrac{1}{N}\Delta_{\hat{X}}^2(t)$ with the survival probability and fragmentation also for an excited state like $\Psi_Y$.

Similar to  $\dfrac{1}{N}\Delta_{\hat{X}}^2(t)$, we find that the converged results of $\dfrac{1}{N}\Delta_{{\hat{P}_X}}^2(t)$ with the orbital numbers, see Fig.~\ref{fig13}.  $\dfrac{1}{N}\Delta_{{\hat{P}_X}}^2(t)$ of $\Psi_{G}$ and $\Psi_{Y}$ keep on fluctuating with a smaller amplitude with respect to $\Psi_{X}$ and $\Psi_{V}$ even in the long-time dynamics. The presence of the breathing mode oscillations in the dynamics of $\dfrac{1}{N}\Delta_{{\hat{P}_X}}^2(t)$ is prominent for $\Psi_{X}$ and $\Psi_{V}$. A comparative study of $\dfrac{1}{N}\Delta_{{\hat{P}_X}}^2(t)$ for all  initial states shows that $\dfrac{1}{N}\Delta_{{\hat{P}_X}}^2(t)$ for $\Psi_X$ is the most influenced by the many-body effect, see Fig. 7 of the main text. 

In Fig.~\ref{fig14}, we demonstrate the convergence with the orbital number of $\dfrac{1}{N}\langle\Psi|{{\hat{L}_Z}}|\Psi\rangle$ for the vortex state. The main text shows the beginning of the collapse of $\dfrac{1}{N}\langle\Psi|{{\hat{L}_Z}}|\Psi\rangle$ due to the many-body correlations, which continues in the long-time dynamics presented here. It shows that the average angular-momentum tends to zero in the many-body long-time dynamics  for a  vortex state. By comparing Fig~\ref{fig11} (d) and Fig~\ref{fig14}, it is found that the decay of  $\dfrac{1}{N}\langle\Psi|{{\hat{L}_Z}}|\Psi\rangle$ is faster than that of  $P_L(t)$ for the vortex state.

\begin{figure*}[!h]
{\includegraphics[trim = 0.1cm 0.5cm 0.1cm 0.2cm, scale=.60]{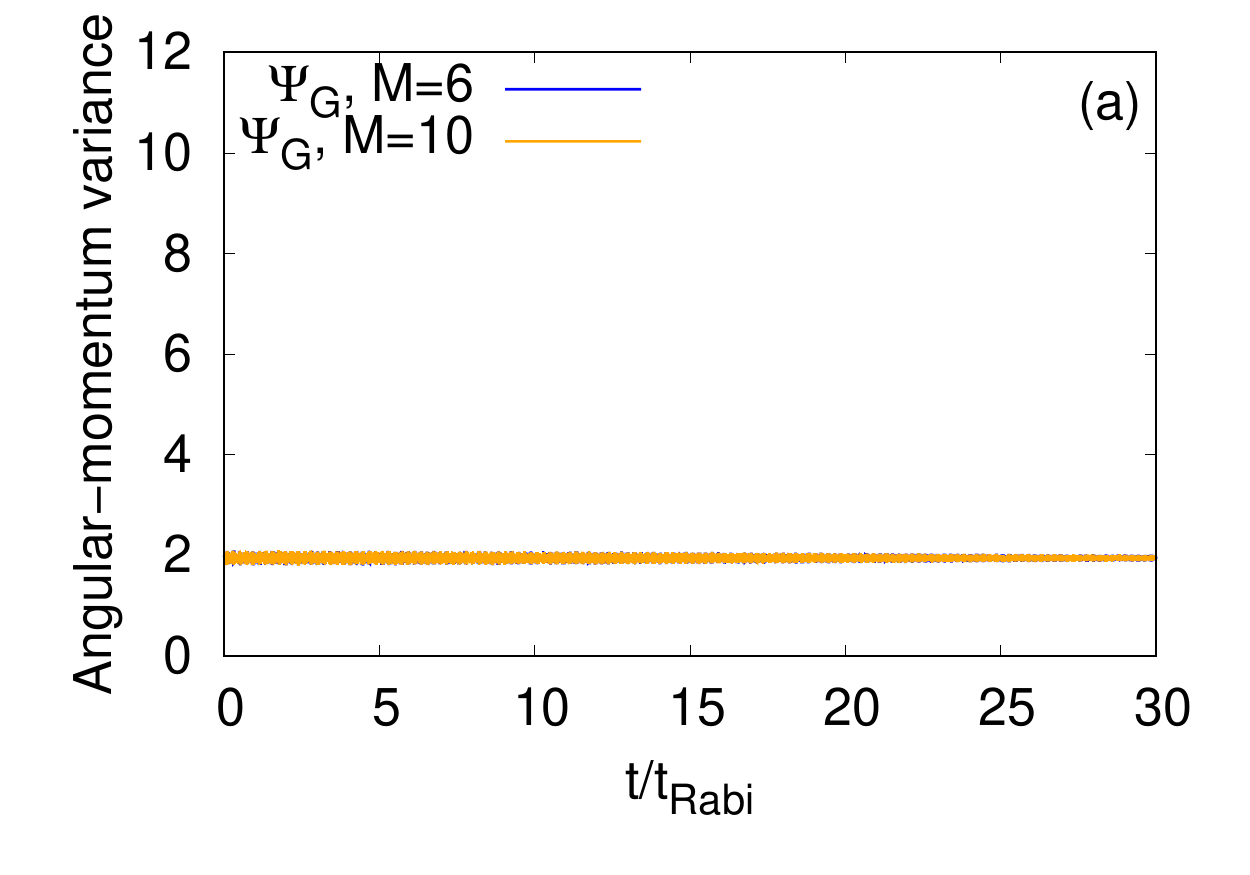}}
{\includegraphics[trim = 0.1cm 0.5cm 0.1cm 0.2cm, scale=.60]{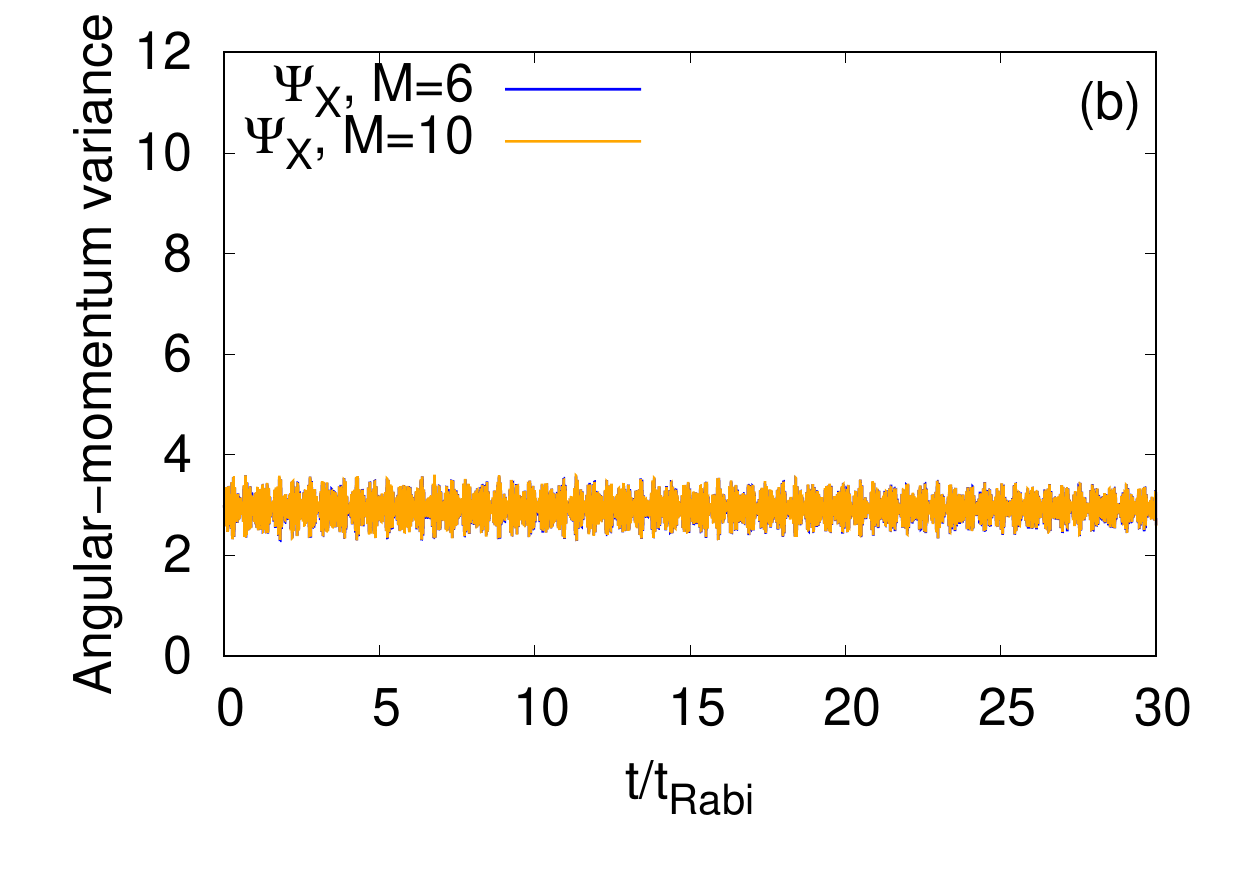}}\\
\vglue 0.25 truecm
{\includegraphics[trim = 0.1cm 0.5cm 0.1cm 0.2cm, scale=.60]{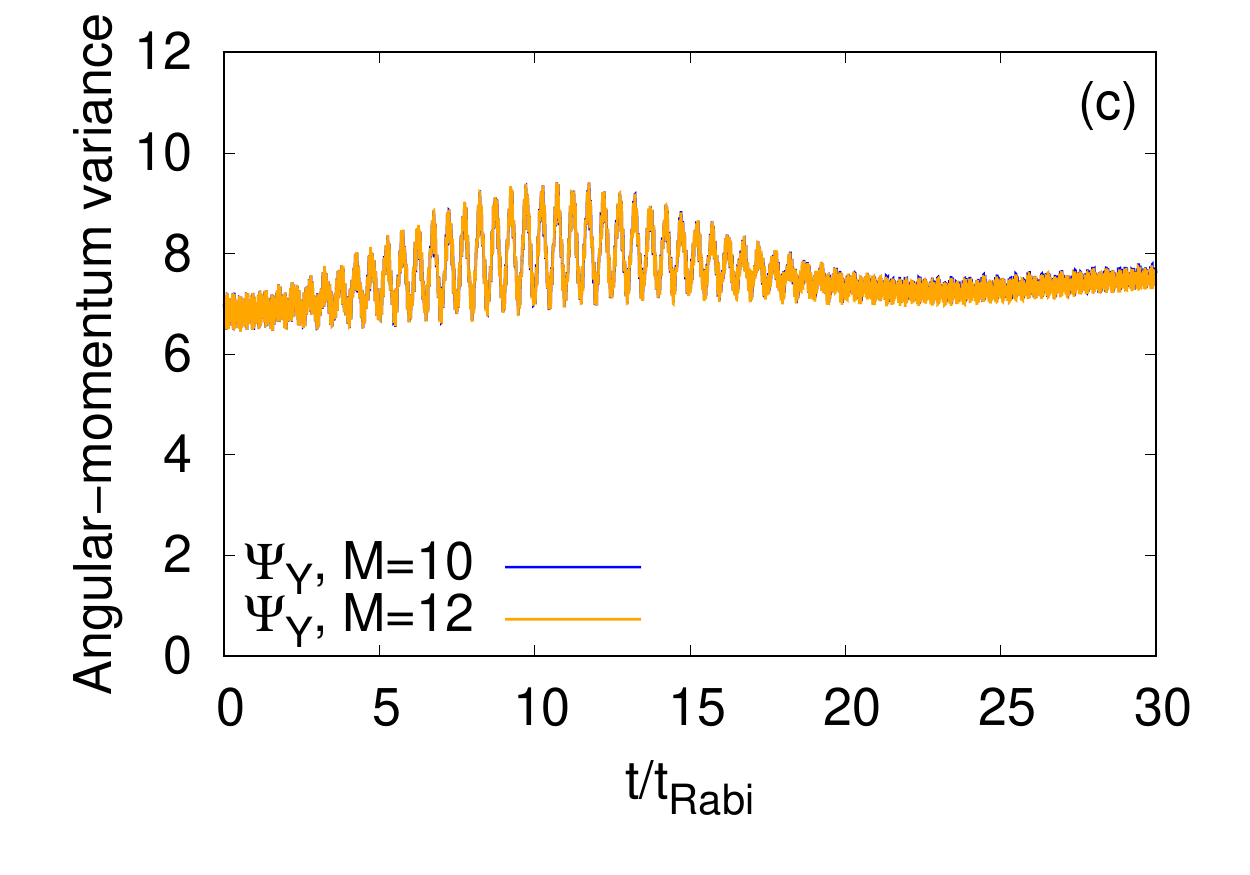}}
{\includegraphics[trim = 0.1cm 0.5cm 0.1cm 0.2cm, scale=.60]{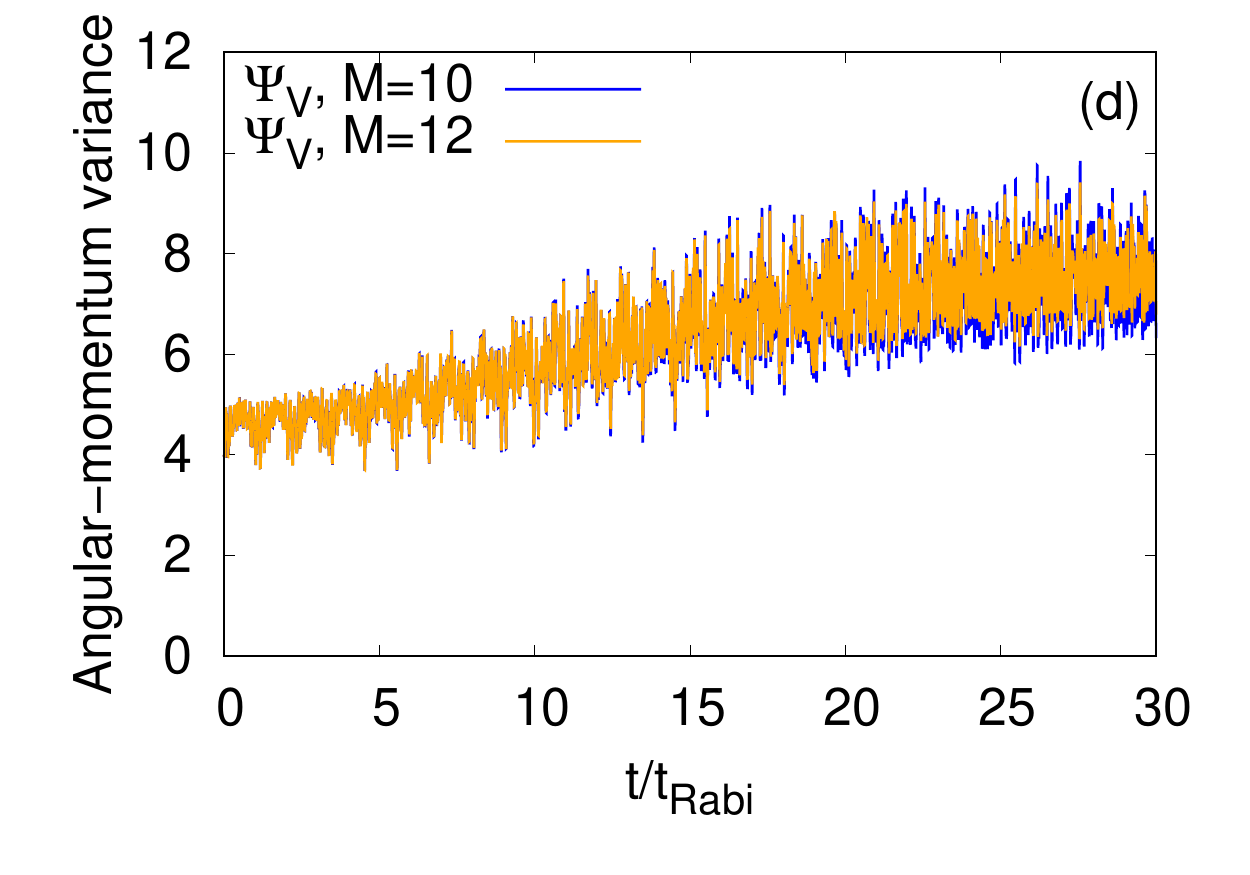}}\\
\caption{\label{fig15}Convergence of the time-dependent many-body angular-momentum variance per particle, $\dfrac{1}{N}\Delta_{{\hat{L}_Z}}^2(t))$, in a symmetric double-well trap with the number of time-adaptive  orbitals  for the initial states (a) $\Psi_G$, (b) $\Psi_X$, (c) $\Psi_Y$, and (d) $\Psi_V$ for $N=10$ interacting bosons with $\Lambda=0.01$.  The many-body $\dfrac{1}{N}\Delta_{{\hat{P}_X}}^2(t)$ are computed using the MCTDHB method. The convergence are verified with $M=6$, $10$ time-adaptive orbitals for  the states, $\Psi_{G}$ and $\Psi_{X}$. While we demonstrate the convergence of the results for $\Psi_{Y}$ and $\Psi_{V}$  using $M=10$,  $12$ time-adaptive orbitals. See the text for more details. The quantities shown  are dimensionless.}
\end{figure*}

Fig.~\ref{fig15} presents the long-time dynamics  of the many-body variance of the most sensitive quantum mechanical observable presented in this work, i.e.   $\dfrac{1}{N}\Delta_{{\hat{L}_Z}}^2(t)$. Simultaneously, the figure ensures the convergence with the orbital numbers for each  of the initial state. As mentioned in the dynamics of $\dfrac{1}{N}\Delta_{{\hat{P}_X}}^2(t)$, here also we observe a prominent breathing oscillations accompanied by  density oscillations. For $\Psi_G$ and $\Psi_X$, fluctuations of $\dfrac{1}{N}\Delta_{{\hat{L}_Z}}^2(t)$ continue their trend in the long-time dynamics as observed in the short-time dynamics. But for $\Psi_Y$ and $\Psi_V$, the many-body $\dfrac{1}{N}\Delta_{{\hat{L}_Z}}^2(t)$ almost reach an equilibration  in the long-time dynamics after the short-time growth dynamics.

In Fig.~\ref{fig16}, we plot the natural  occupancy of the orbitals per particle, $\dfrac{n_j(t)}{N}$, for  the four initial states, $\Psi_G$, $\Psi_X$, $\Psi_Y$, and $\Psi_V$.   Results are obtained with  $M=6$, $10$ time-adaptive orbitals for $\Psi_G$ and $\Psi_X$, and $M=10$, $12$ time-adaptive orbitals for  $\Psi_Y$ and $\Psi_V$. We find that the results for $\Psi_G$ and $\Psi_X$ with $M=6$ and $M=10$ completely fall on top of  each other for  the two largest occupation numbers, $\dfrac{n_1(t)}{N}$ and $\dfrac{n_2(t)}{N}$.  The natural occupations  $\dfrac{n_{3}(t)}{N}$ to $\dfrac{n_{6}(t)}{N}$ of $\Psi_G$ and $\Psi_X$  are very small in magnitude (less than $10^{-3}$) and   almost  completely overlap in comparison  when computed using $M=6$ and 10 time-adaptive orbitals.  In case of  $\Psi_{Y}$, the first eight orbitals are showing fully converged results with the number of orbitals. Small deviations can be observed for  $\dfrac{n_9(t)}{N}$ and $\dfrac{n_{10}(t)}{N}$ of $\Psi_Y$ when one compares the results, computed from  $M=10$ and 12 time-adaptive orbitals.  The latter exhibit in comparison very small occupations (less than $10^{-3}$).   Finally, as shown in the figure, the occupations of all the natural orbitals of $\Psi_{V}$ are well converged with the  number of orbitals.  The results signify that the fragmentation dynamics of all objects studied in the present work are well converged and whenever the transverse excitations exist in the system,   more  natural orbitals are required to represent the  dynamics accurately.

We discussed in the main text that the largest  natural  orbitals of $\Psi_{G}$ and $\Psi_{X}$ show only excitations in the $x$-direction, the direction along which the barrier is formed,  with no-node in the $y$-direction.  Therefore, here we discuss the first four highest natural orbitals of $\Psi_{Y}$ and $\Psi_{V}$, as their occupancies are greater than $10^{-1}$ in the long-time dynamics. The results  at $t=10t_{Rabi}$, $20t_{Rabi}$ and $30t_{Rabi}$ for  $\Psi_{Y}$ and $\Psi_{V}$ are presented in  Fig.~\ref{fig17} and ~\ref{fig18}, respectively. For $\Psi_{Y}$, the 1st and 4th natural orbitals look like the 1st and 2nd excited states in $y$, respectively. Interestingly, the 2nd and 3rd natural orbitals of $\Psi_{Y}$ show a change in order in time. The natural orbitals of $\Psi_{V}$ show comparatively complex structures as they have the combined effect of $\Psi_{X}$ and $\Psi_{Y}$. The 1st and 2nd natural orbitals of $\Psi_{V}$ have one and zero nodes in the $x$-$y$ plane, respectively.  Similar to $\Psi_{Y}$, we find that there is a change in order  of natural orbitals for $\Psi_{V}$ in time but this happen between the 3rd and 4th natural orbitals.

\begin{figure*}[!h]
{\includegraphics[trim = 1cm 0.0cm 0.1cm 0.5cm, scale=.60]{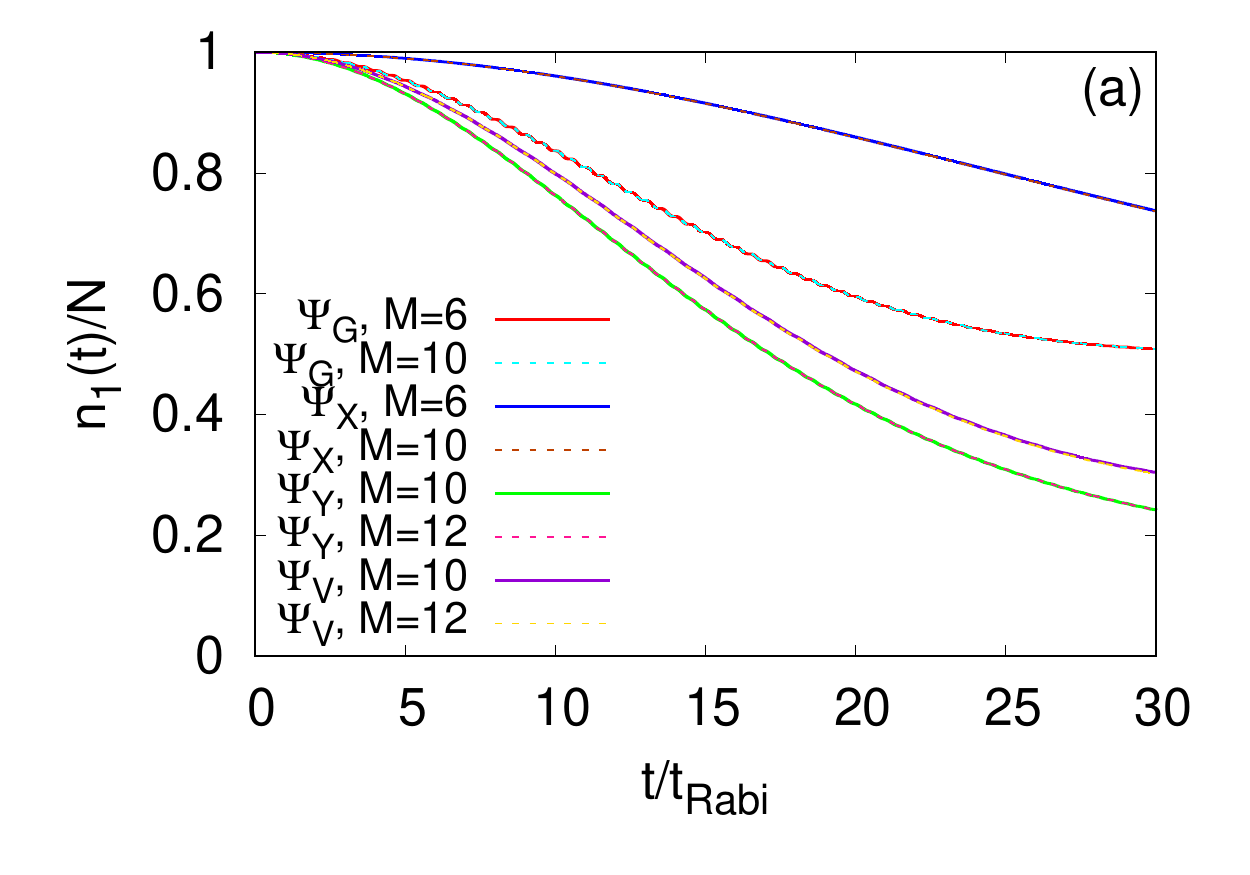}}
{\includegraphics[trim = 0.5cm 0.0cm 0.1cm 0.5cm, scale=.60]{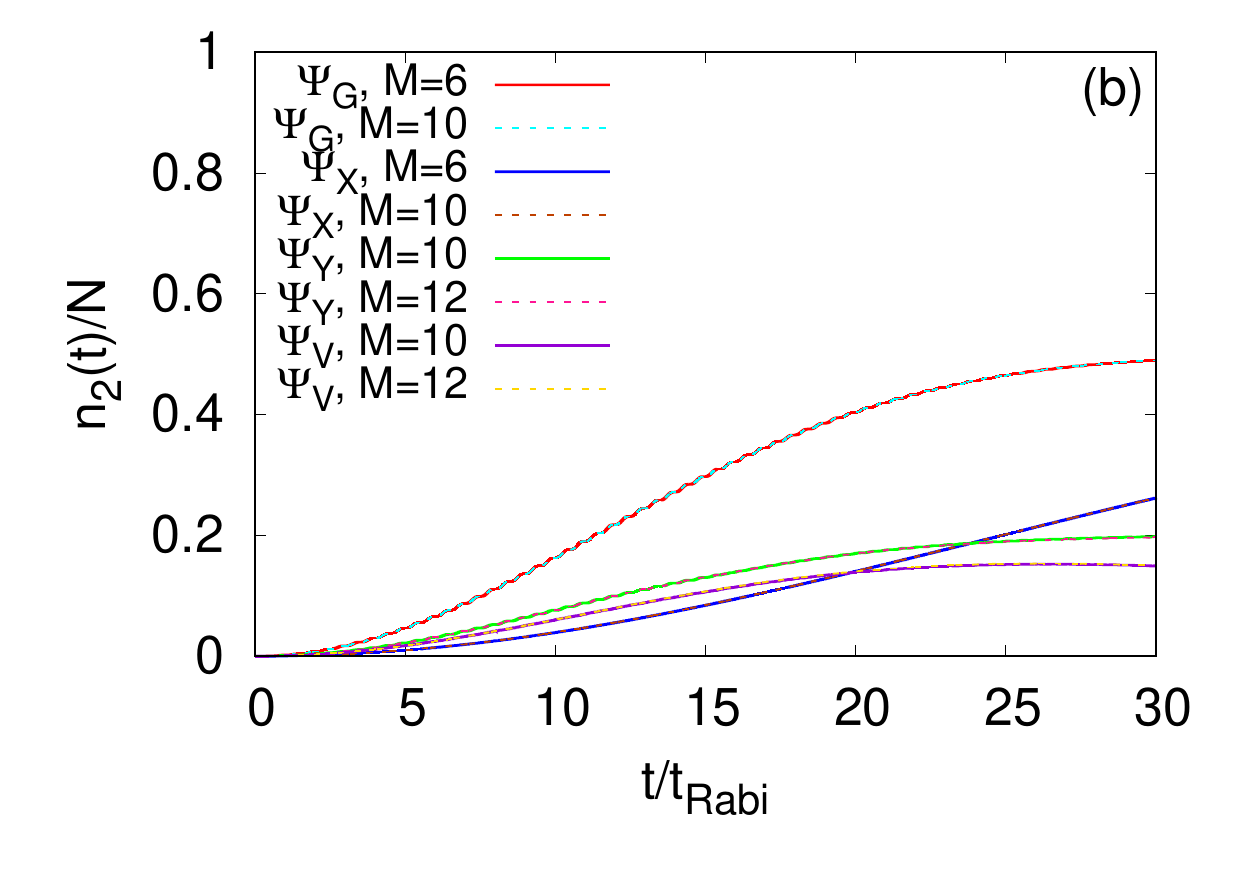}}\\
\vglue 0.25 truecm
{\includegraphics[trim =  1cm 0.0cm 0.1cm 1.1cm,scale=.60]{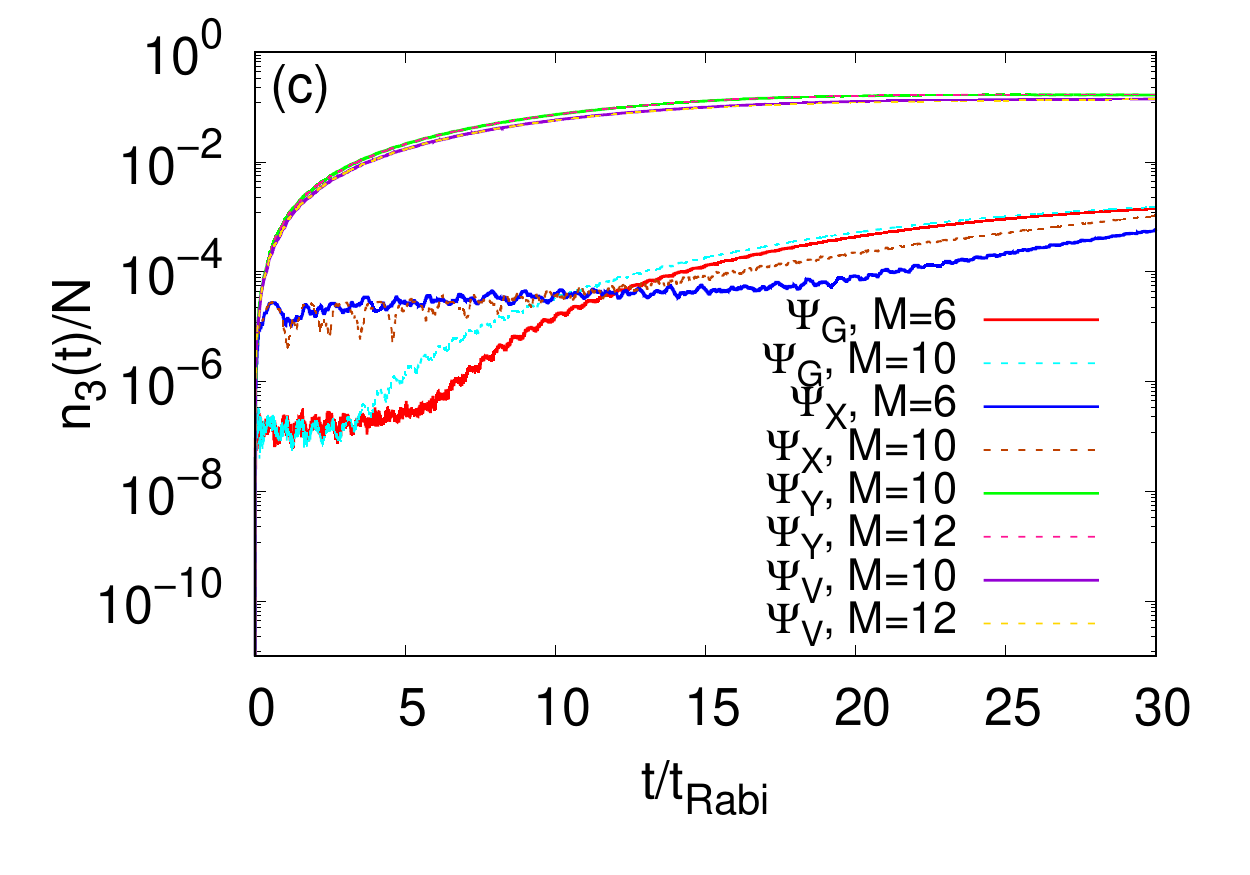}}
{\includegraphics[trim =  0.5cm 0.0cm 0.1cm 1.1cm, scale=.60]{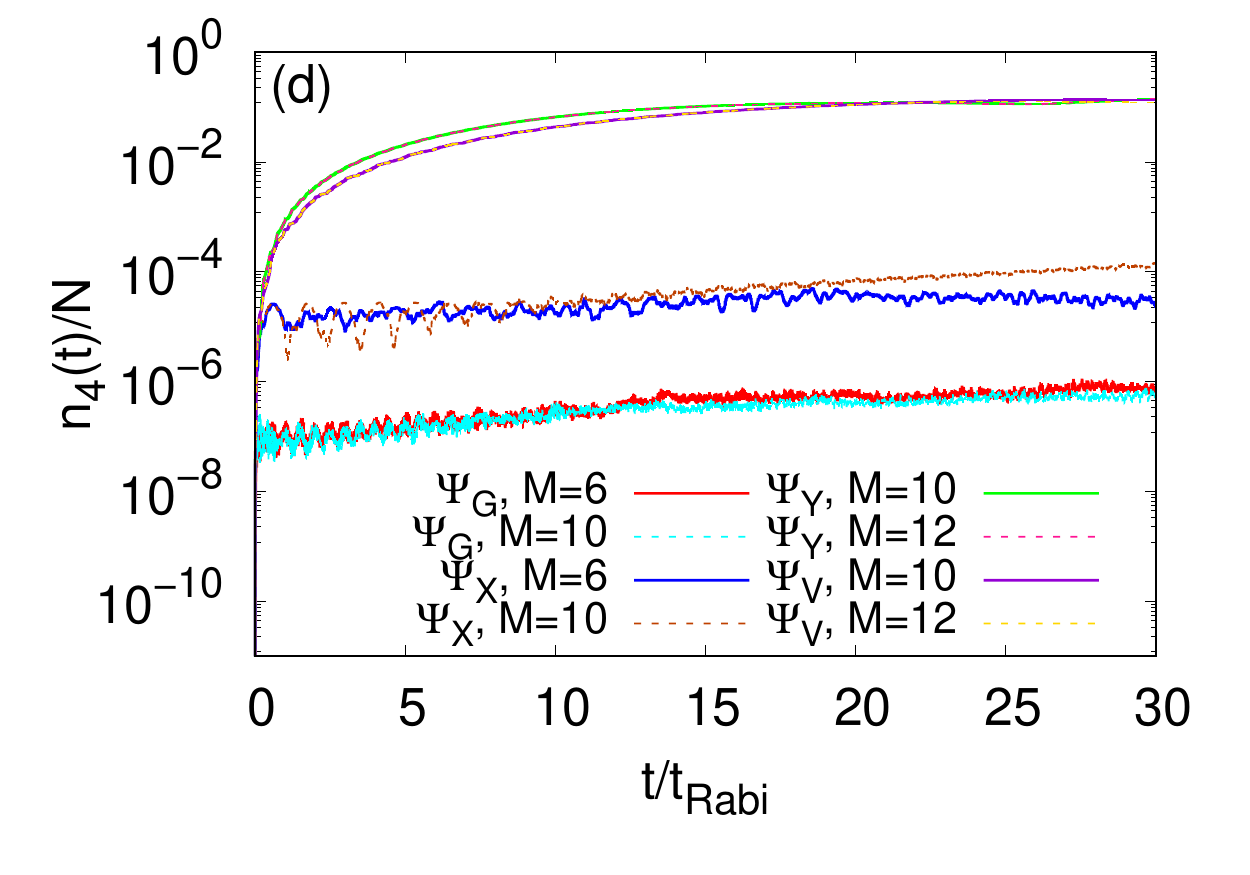}}\\
\vglue 0.25 truecm
{\includegraphics[trim =  1cm 0.0cm 0.1cm 1.1cm,scale=.60]{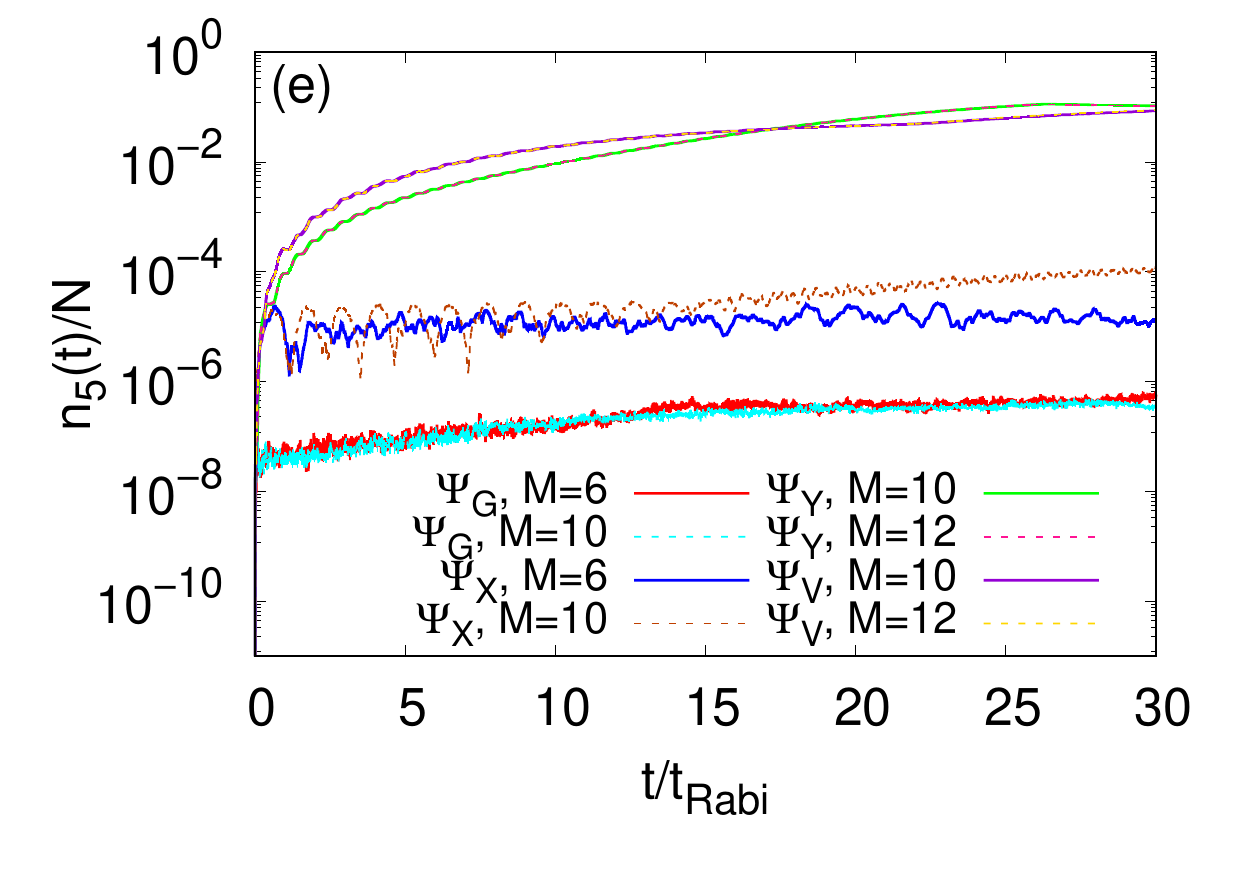}}
{\includegraphics[trim =  0.5cm 0.0cm 0.1cm 1.1cm, scale=.60]{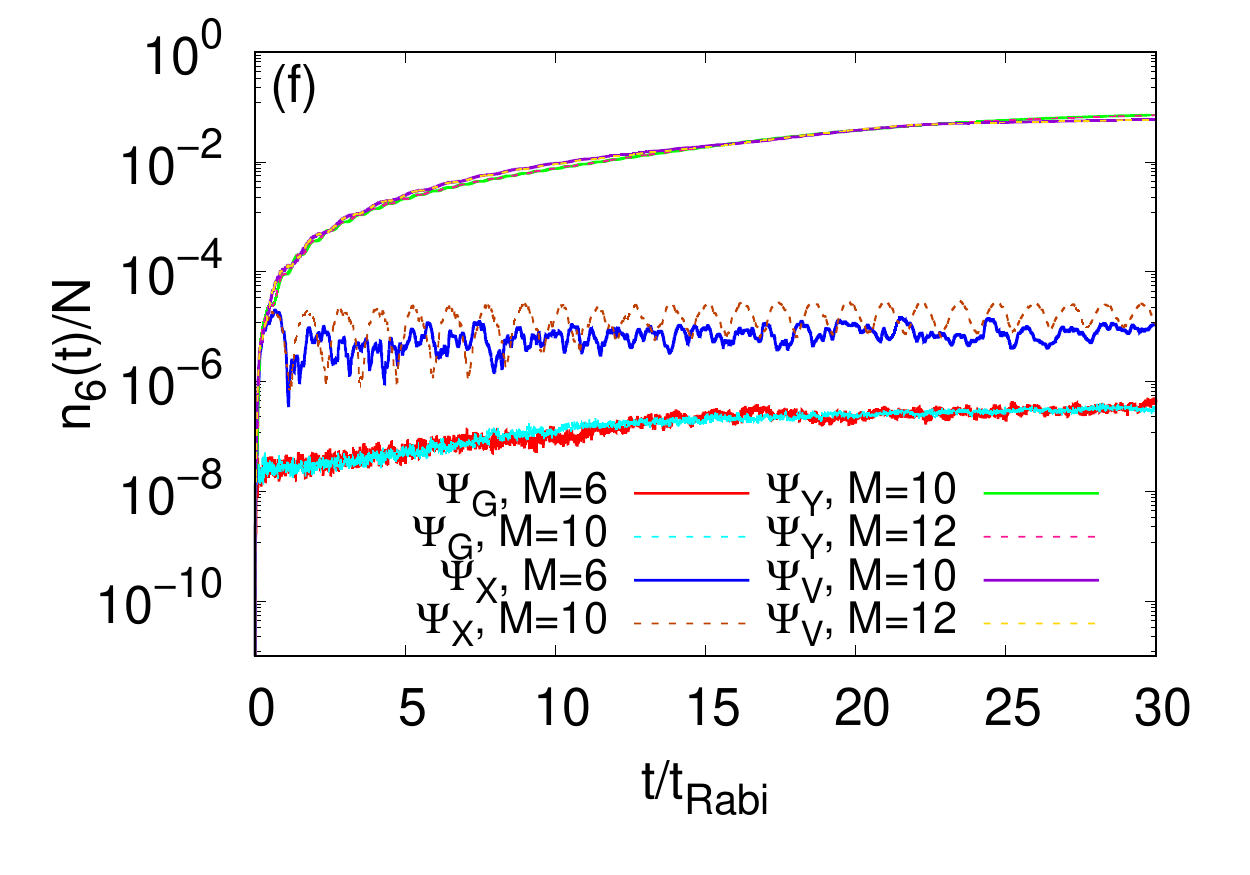}}\\
\vglue 0.25 truecm
{\includegraphics[trim =  1cm 0.0cm 0.1cm 1.1cm,scale=.60]{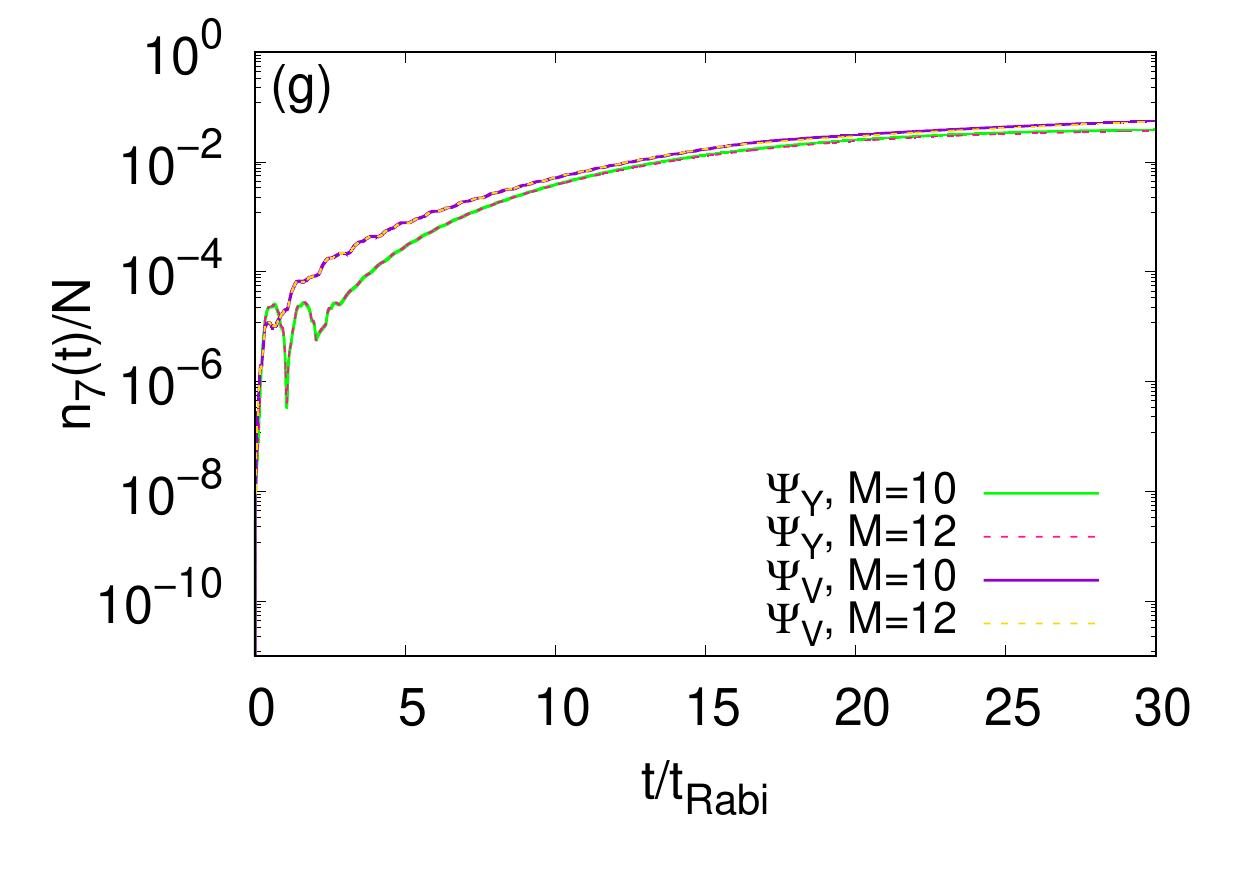}}
{\includegraphics[trim =  0.5cm 0.0cm 0.1cm 1.1cm, scale=.60]{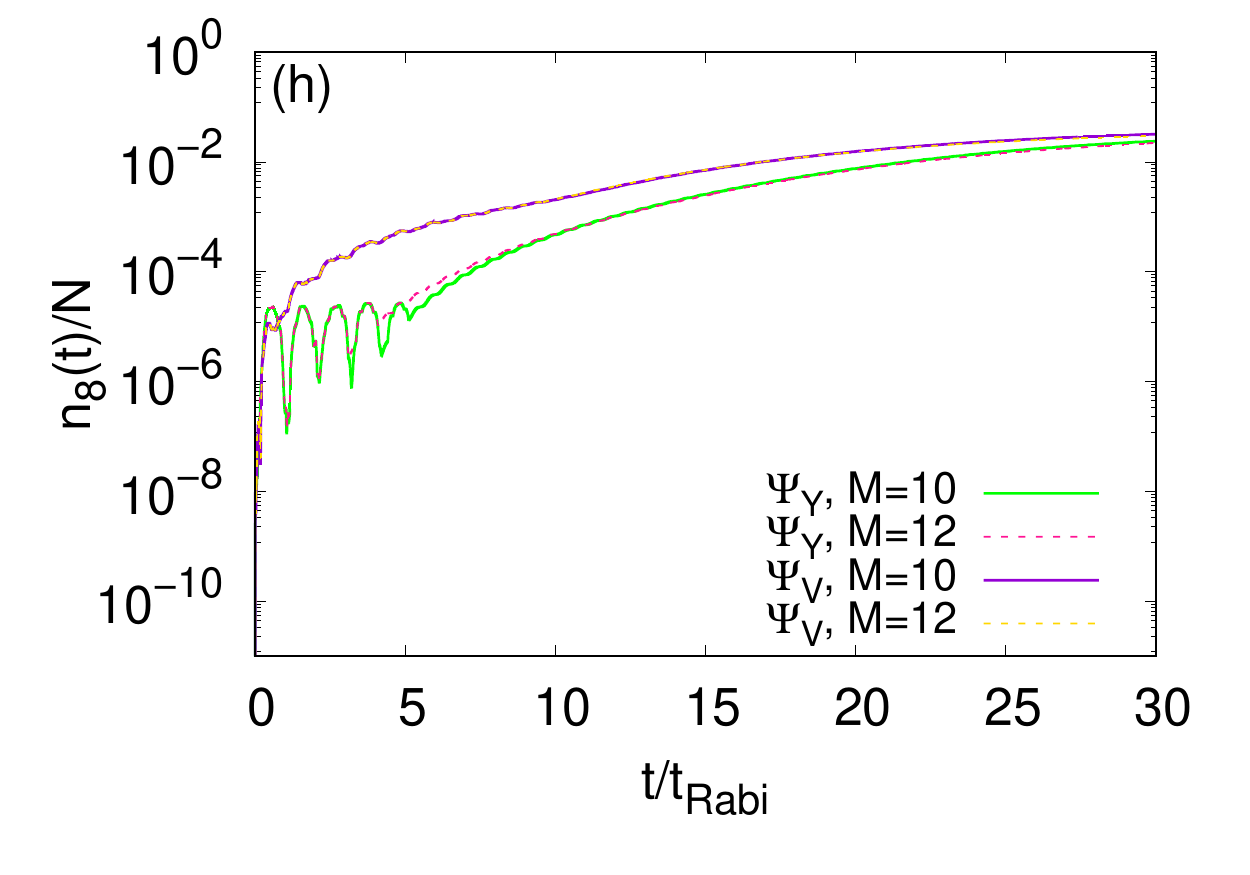}}\\
\caption{Continued}
\end{figure*}

 \begin{figure*}[!h]
 \ContinuedFloat
{\includegraphics[trim = 1cm 0.0cm 0.1cm 0.5cm, scale=.60]{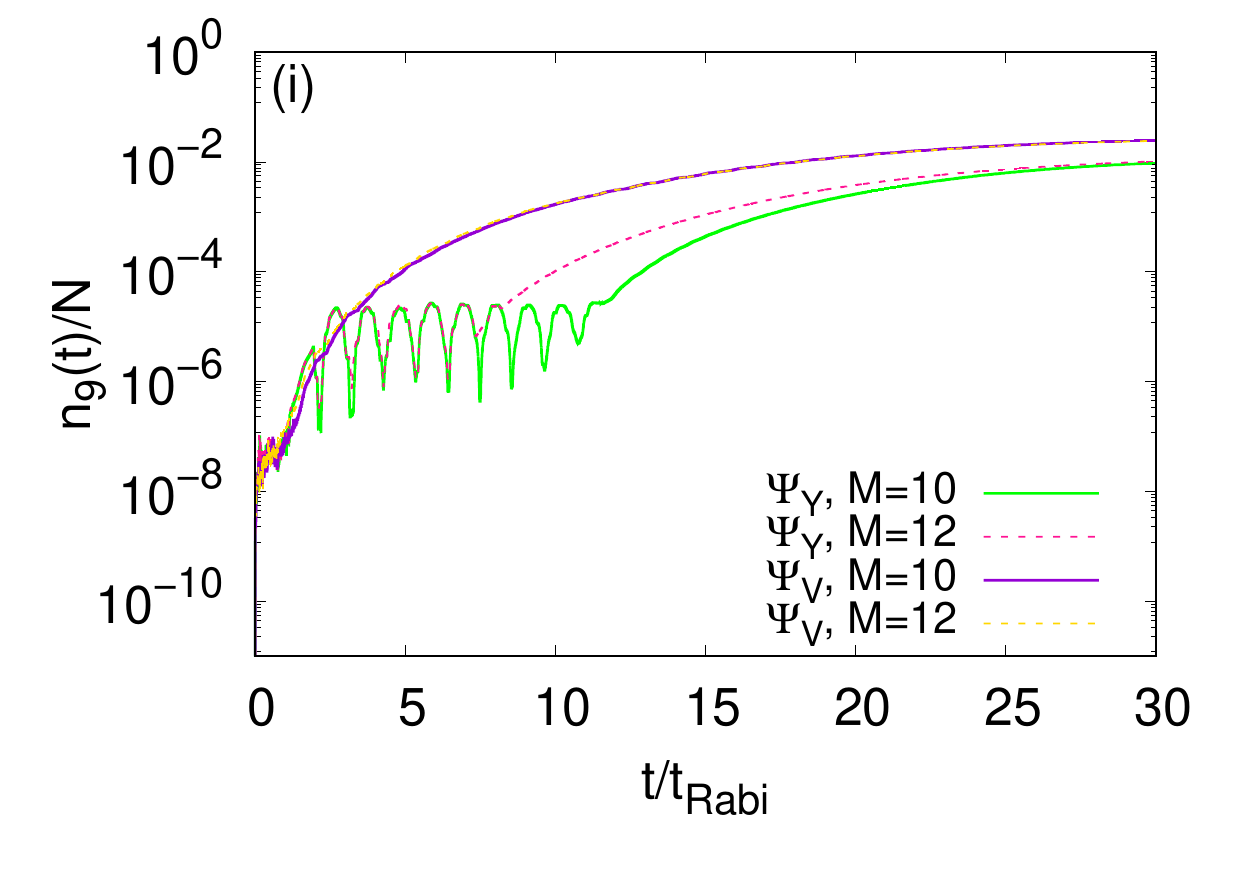}}
{\includegraphics[trim = 0.5cm 0.0cm 0.1cm 0.5cm, scale=.60]{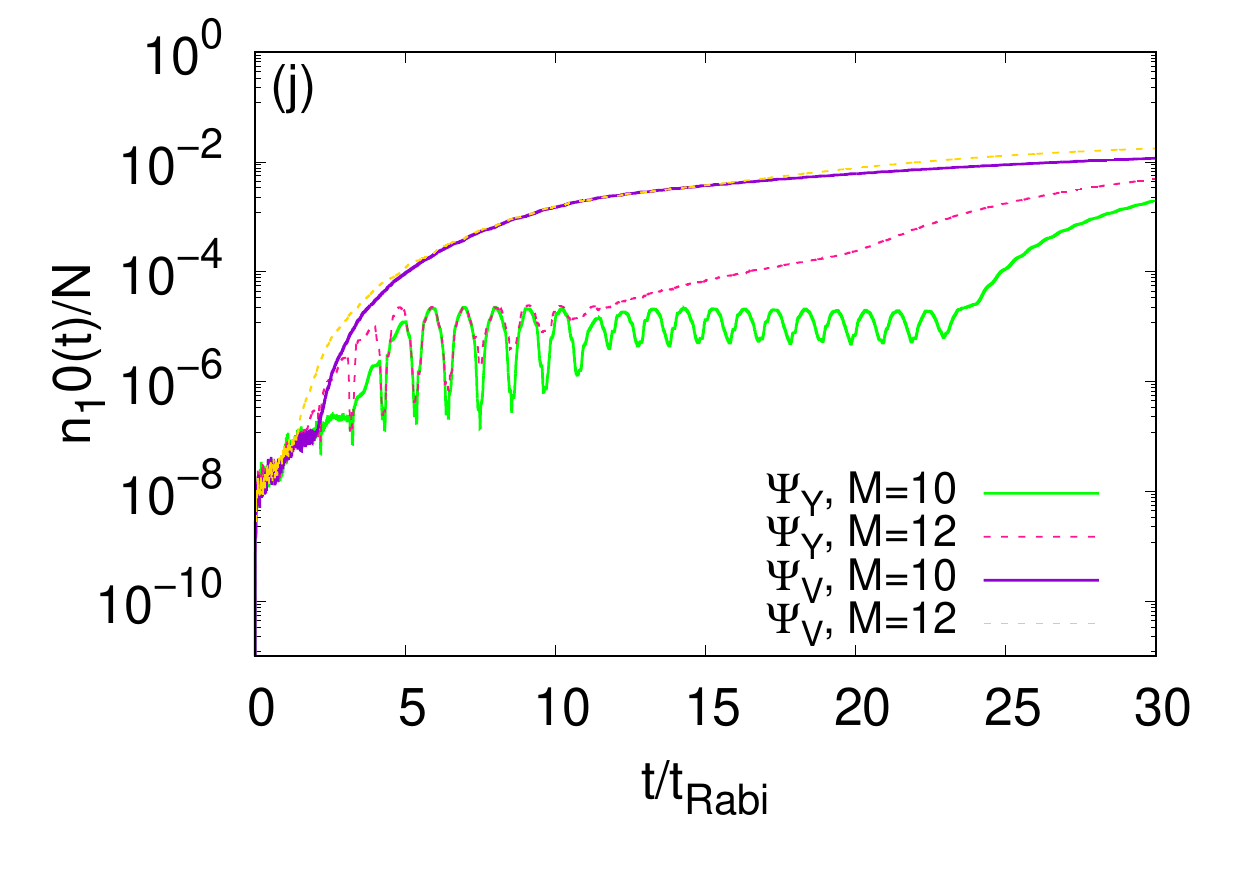}}\\
\caption{\label{fig16}Convergence of the natural occupation numbers per particle, $\dfrac{n_j(t)}{N}$,   as a function of time with the number of time-adaptive  orbitals  for the initial states $\Psi_G$, $\Psi_X$, $\Psi_Y$, and $\Psi_V$ in the symmetric 2D double-well trap. The number  of bosons is $N=10$. The interaction parameter is $\Lambda=0.01$. The many-body results are computed using the MCTDHB method. The convergence are verified with $M=6$, $10$ time-adaptive orbitals for  the states, $\Psi_{G}$ and $\Psi_{X}$. While we demonstrate the convergence of the results for $\Psi_{Y}$ and $\Psi_{V}$  using $M=10$,  $12$ time-adaptive orbitals.   Convergence of the time-dependent occupation number from top to bottom (largest to smallest) is demonstrated for all bosonic clouds. The plots reveal that at long propagation times a large number of self-consistent orbitals are needed to accurately represent the tunneling dynamics  of the considered states. Color codes are explained in each panel. See the text for more details. The quantities shown  are dimensionless.}
\end{figure*}
\begin{figure*}[!h]
{\includegraphics[trim = 4.9cm 0.5cm 3.1cm 0.2cm,scale=.75]{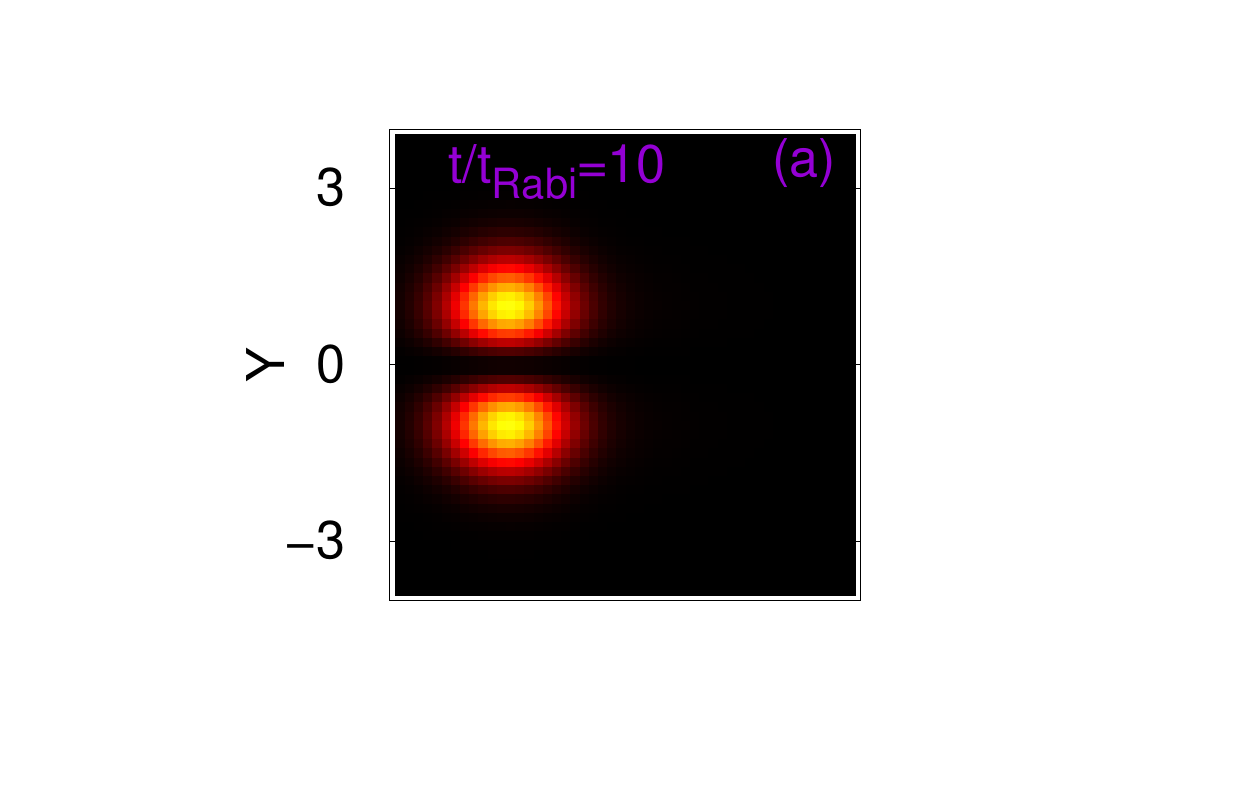}}
{\includegraphics[trim =  4.9cm 0.5cm 3.1cm 0.2cm, scale=.75]{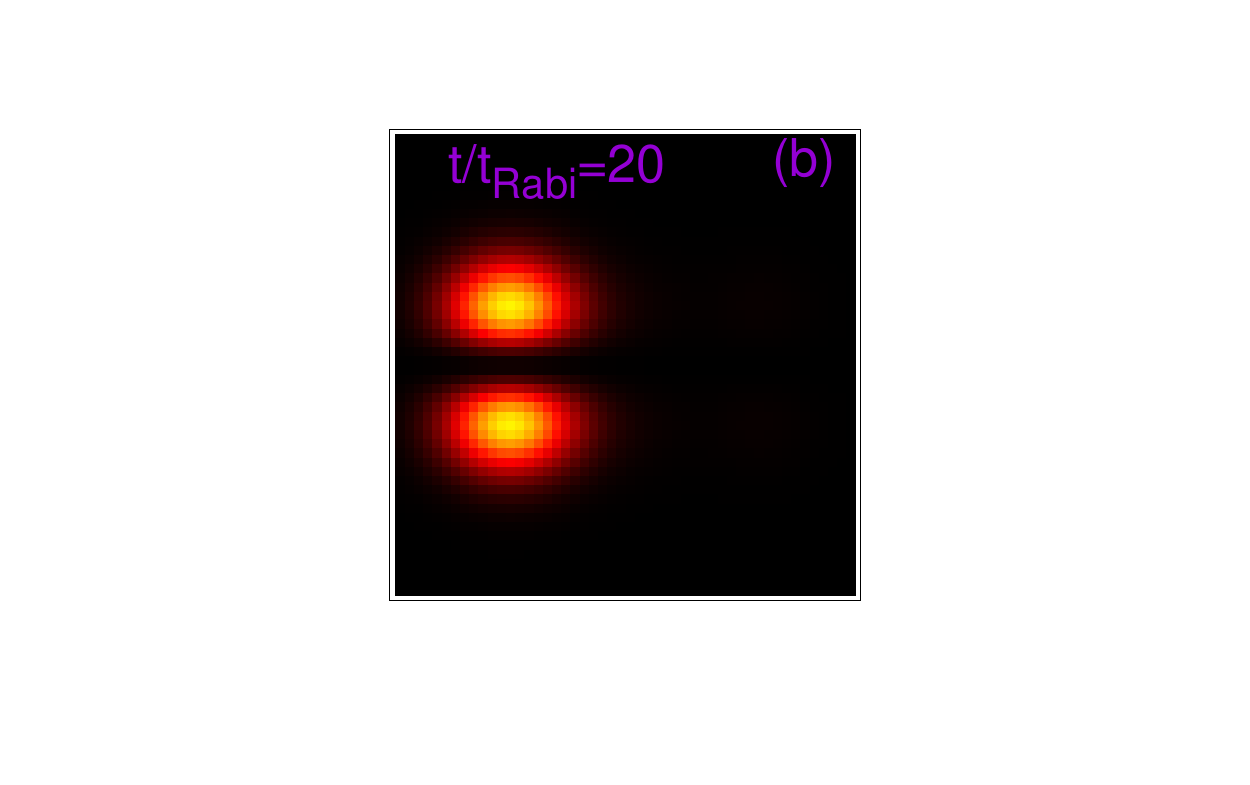}}
{\includegraphics[trim =  4.9cm 0.5cm 3.1cm 0.2cm, scale=.75]{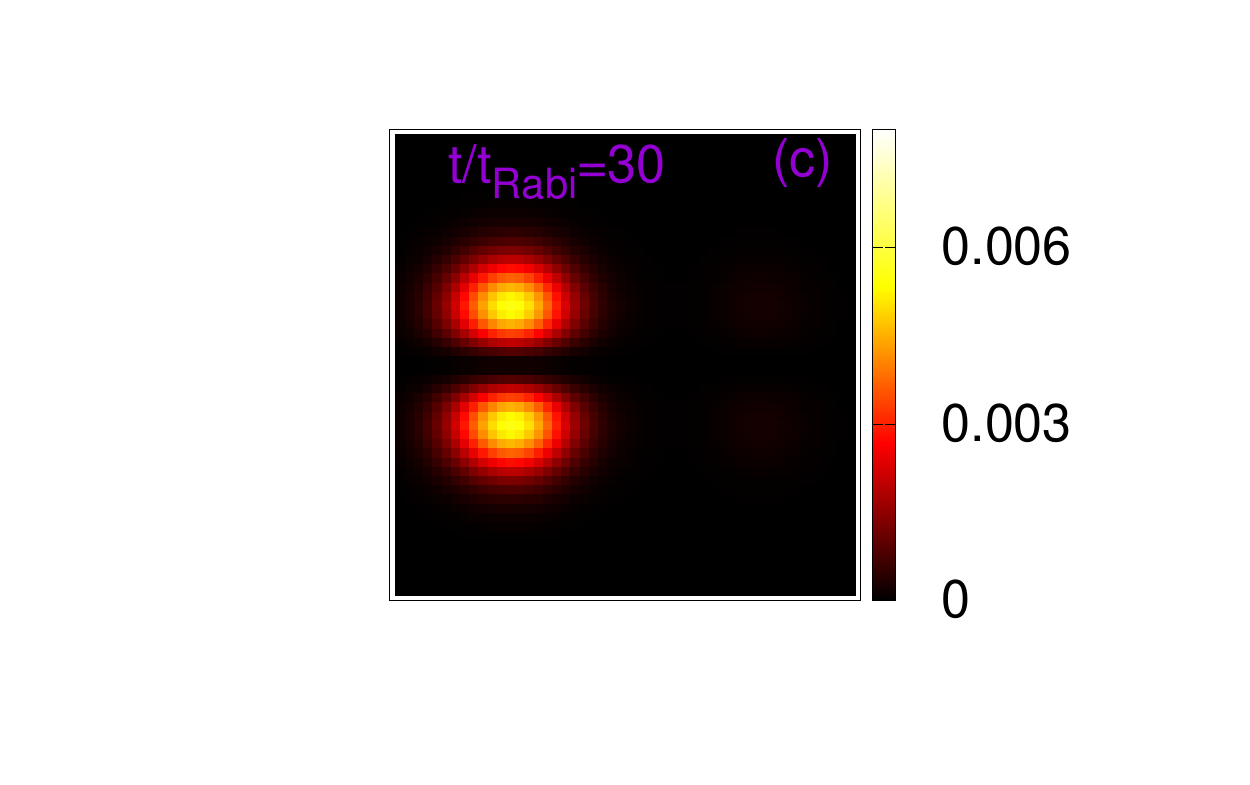}}\\
{\includegraphics[trim = 4.9cm 0.5cm 3.1cm 2.5cm,scale=.75]{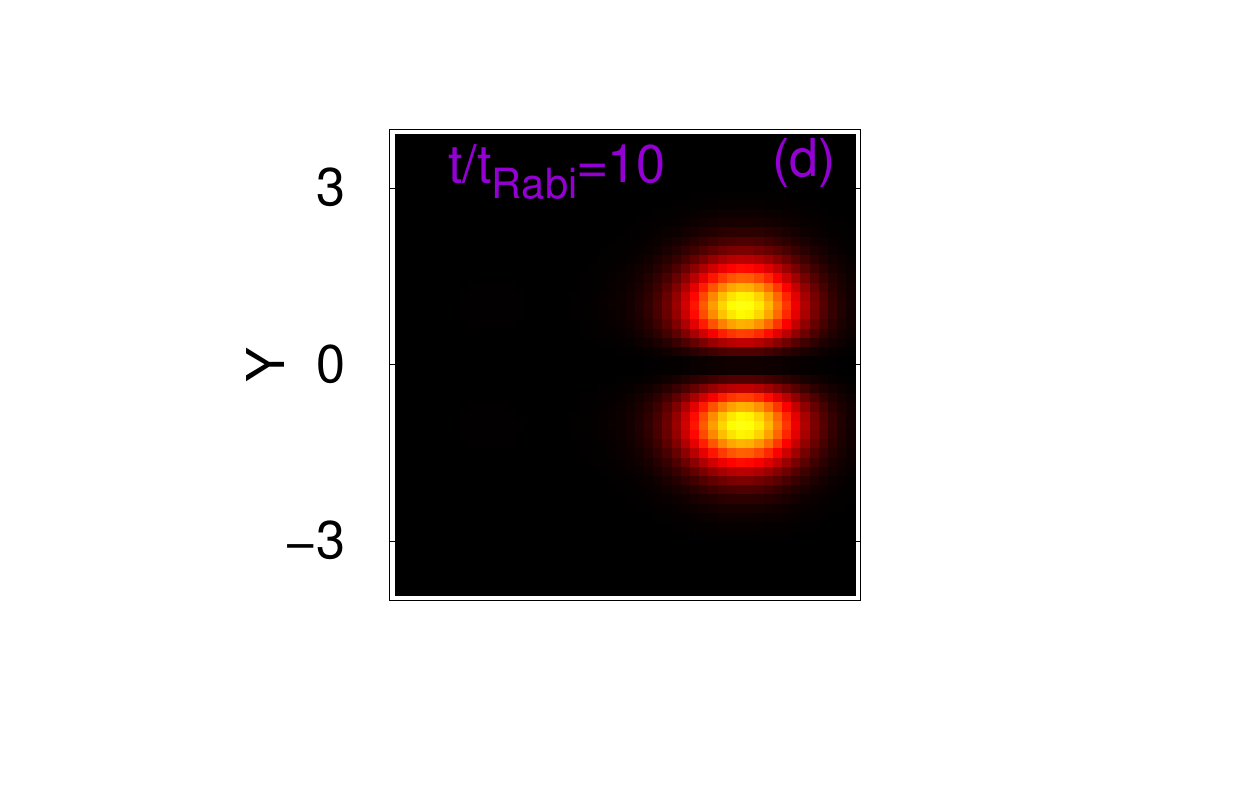}}
{\includegraphics[trim =  4.9cm 0.5cm 3.1cm 2.5cm, scale=.75]{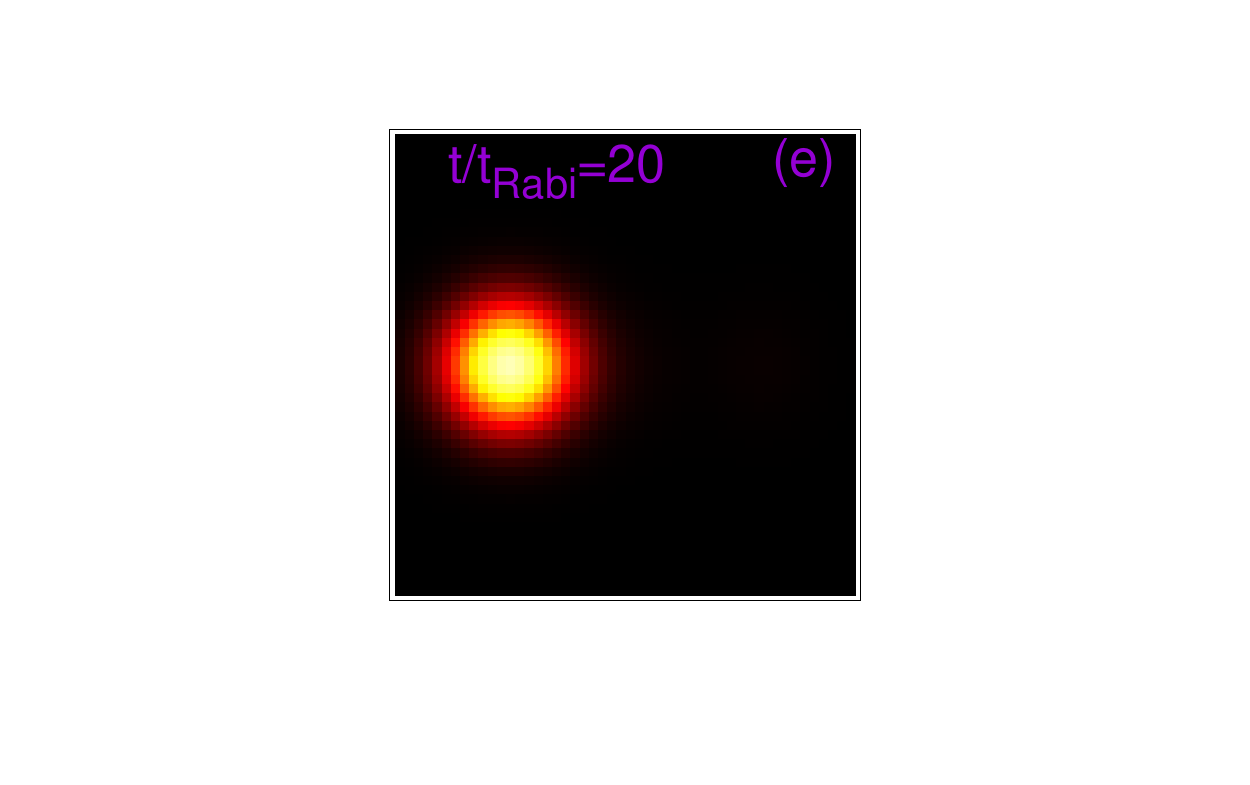}}
{\includegraphics[trim =  4.9cm 0.5cm 3.1cm 2.5cm, scale=.75]{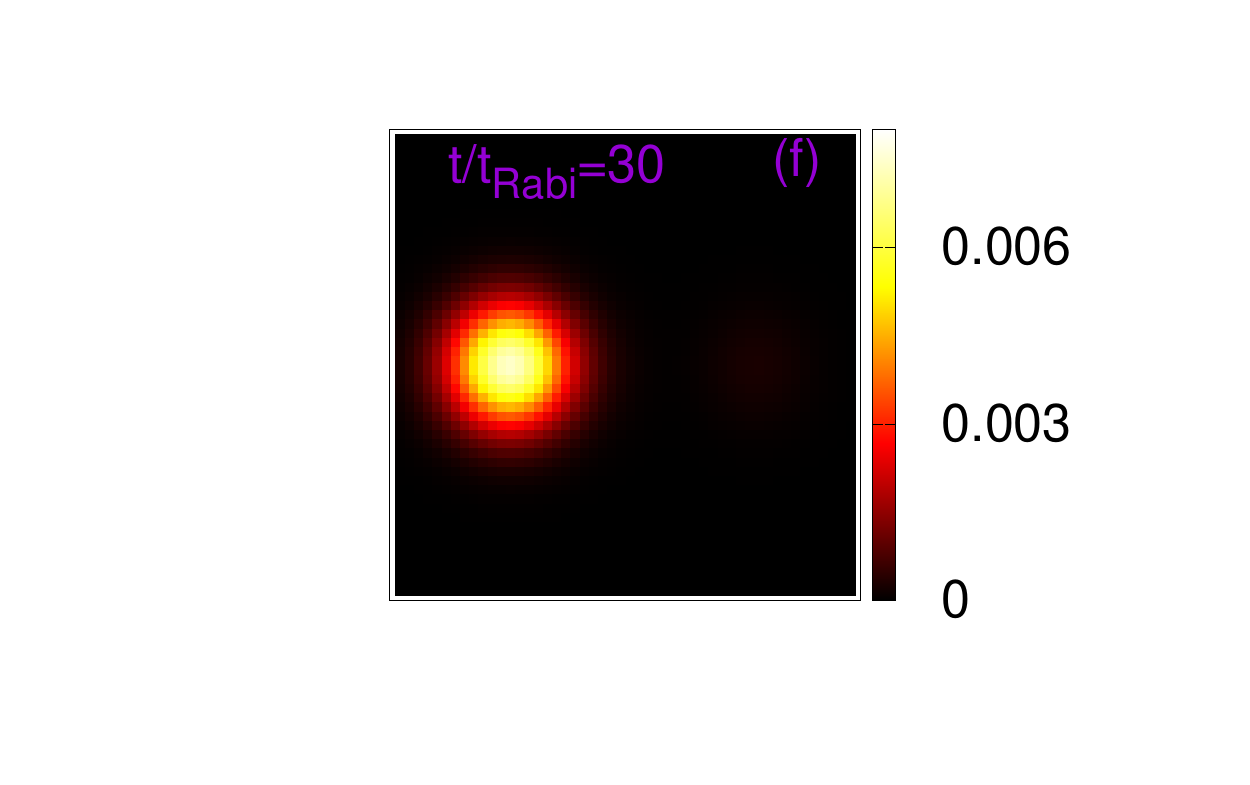}}\\
{\includegraphics[trim = 4.9cm 0.5cm 3.1cm 2.5cm,scale=.75]{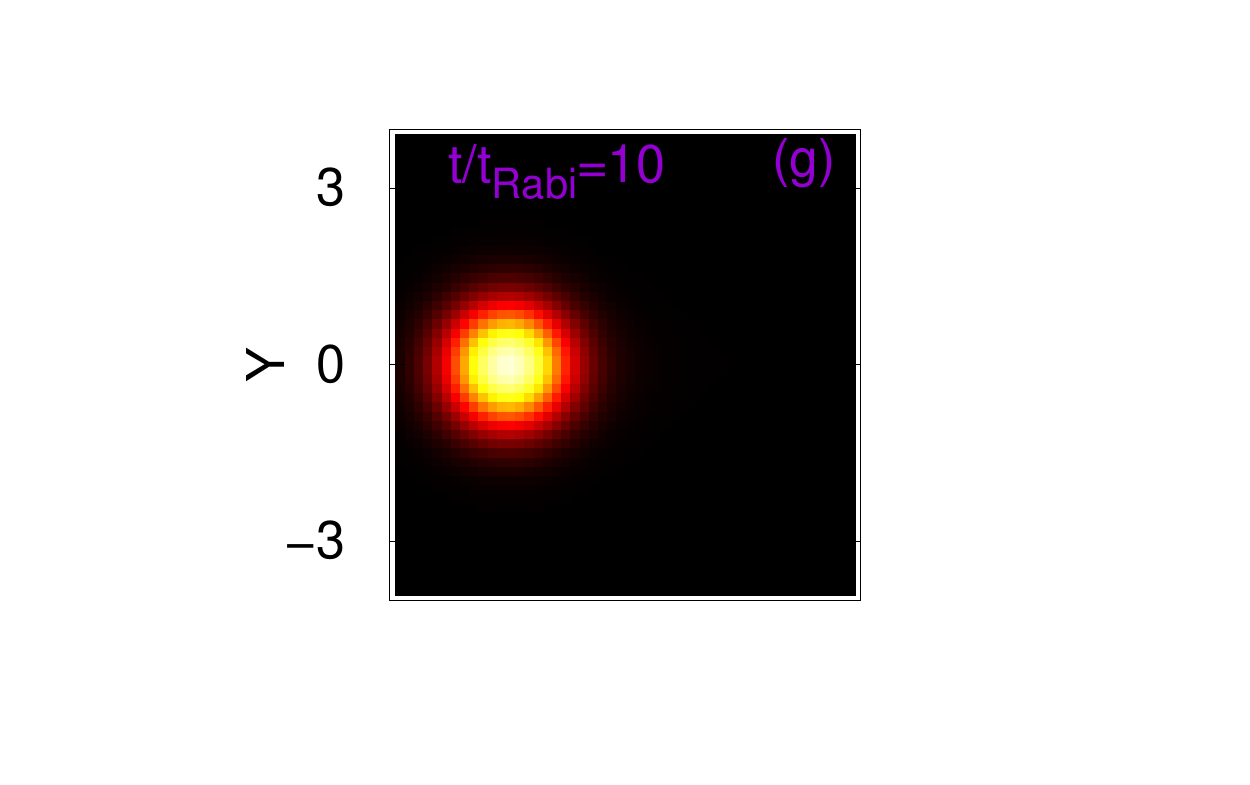}}
{\includegraphics[trim =  4.9cm 0.5cm 3.1cm 2.5cm, scale=.75]{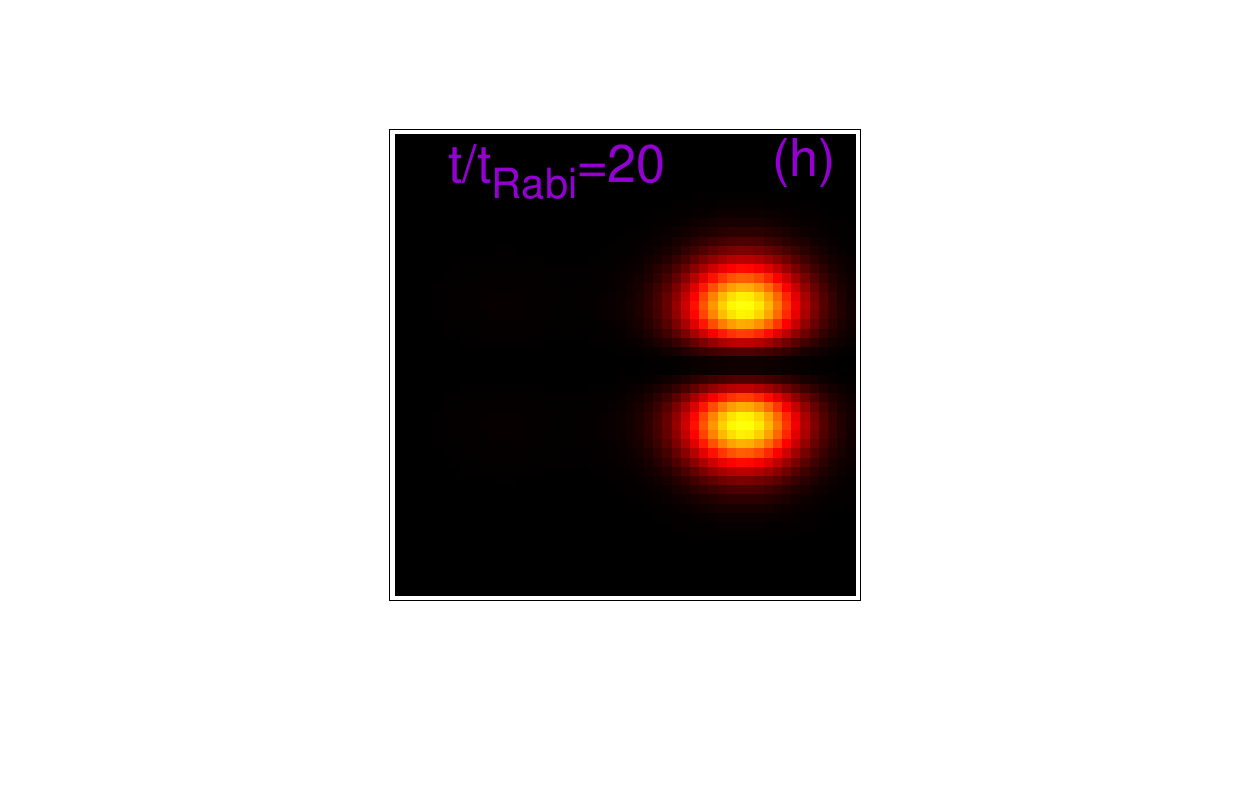}}
{\includegraphics[trim =  4.9cm 0.5cm 3.1cm 2.75cm, scale=.75]{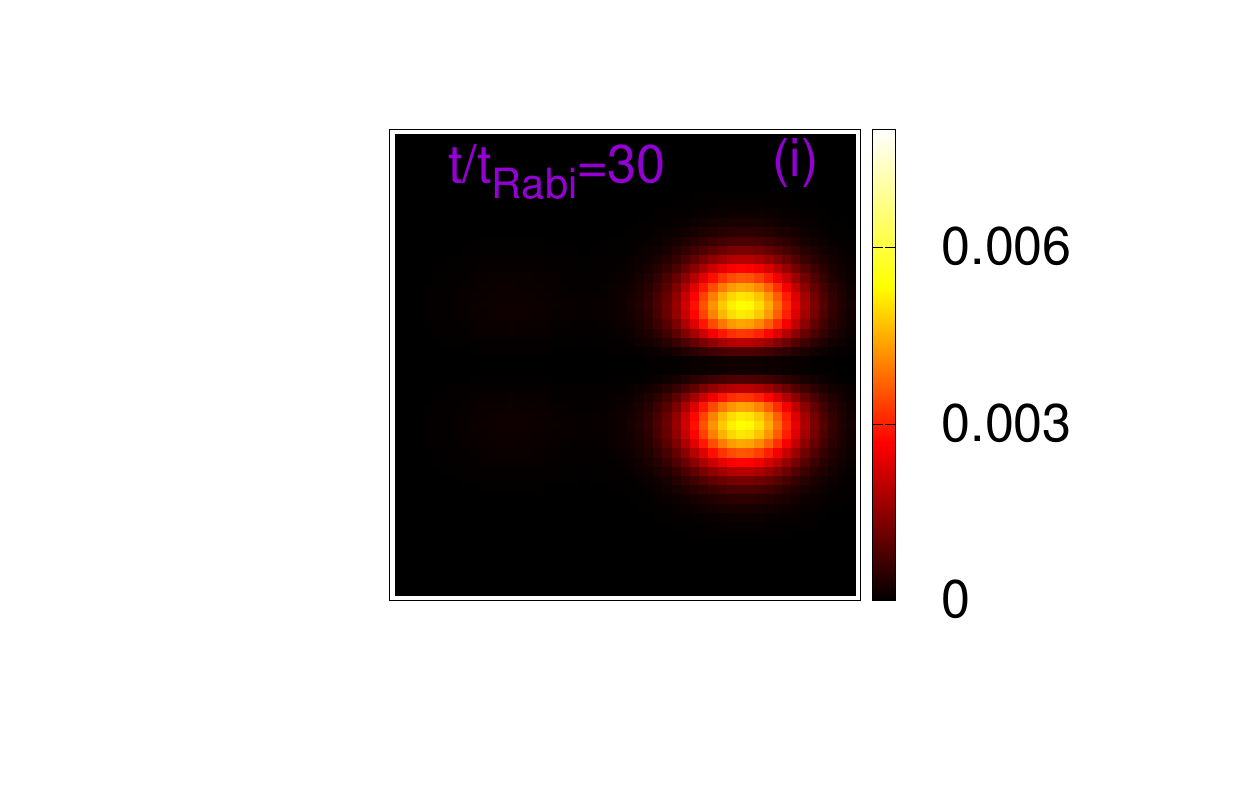}}\\
{\includegraphics[trim = 4.9cm 0.5cm 3.1cm 2.5cm,scale=.75]{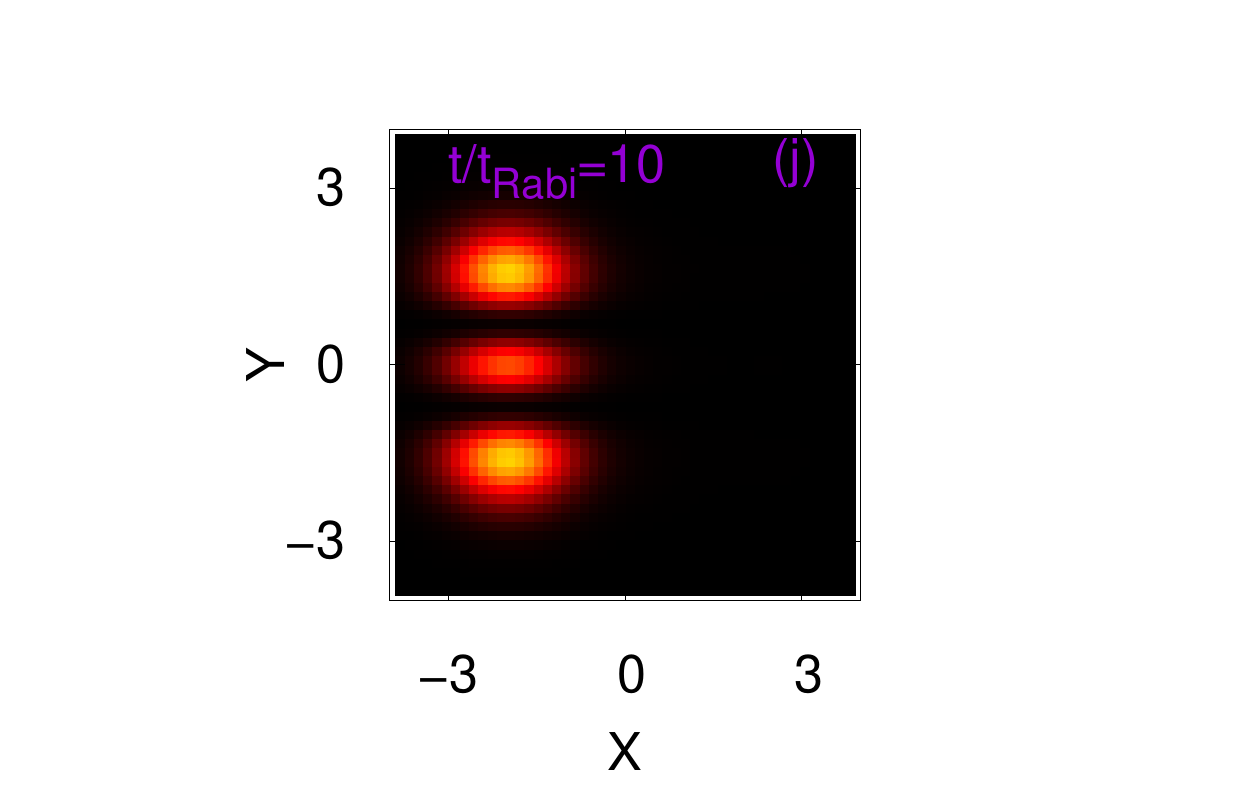}}
{\includegraphics[trim =  4.9cm 0.5cm 3.1cm 2.5cm, scale=.75]{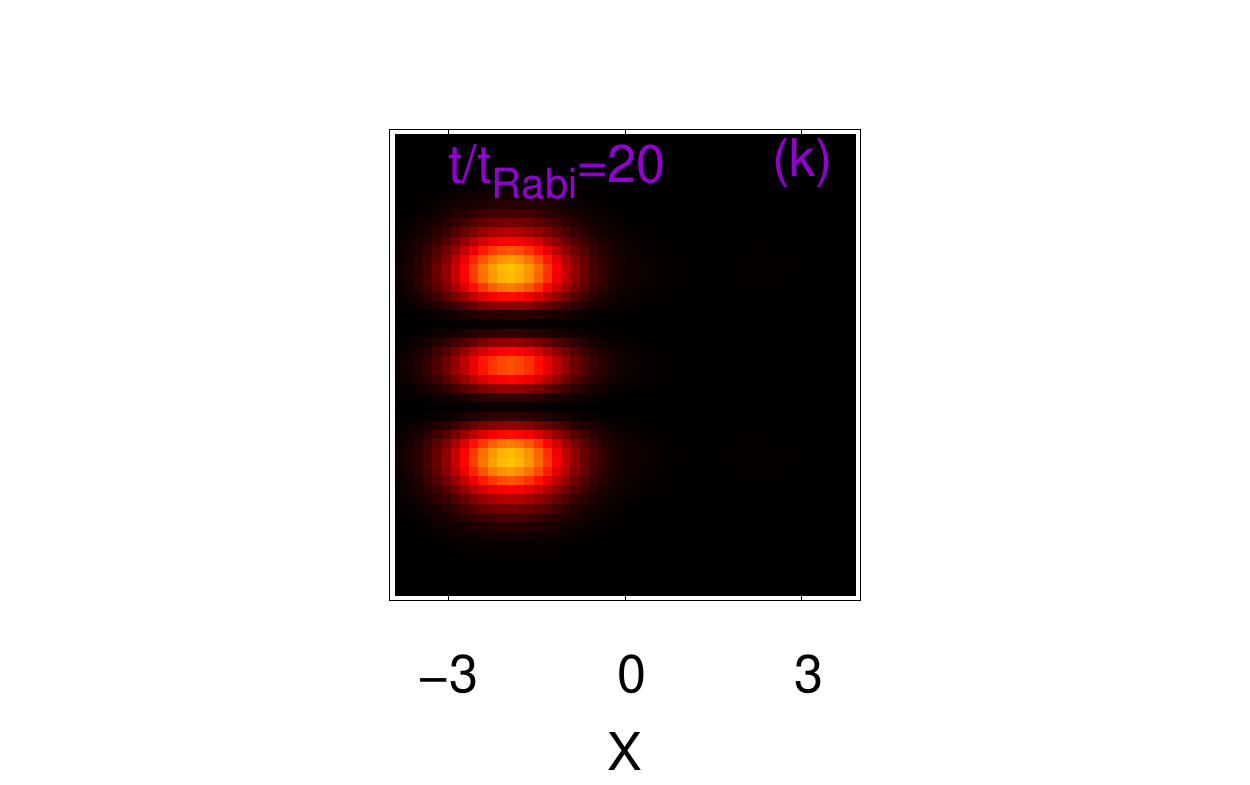}}
{\includegraphics[trim =  4.9cm 0.5cm 3.1cm 2.5cm, scale=.75]{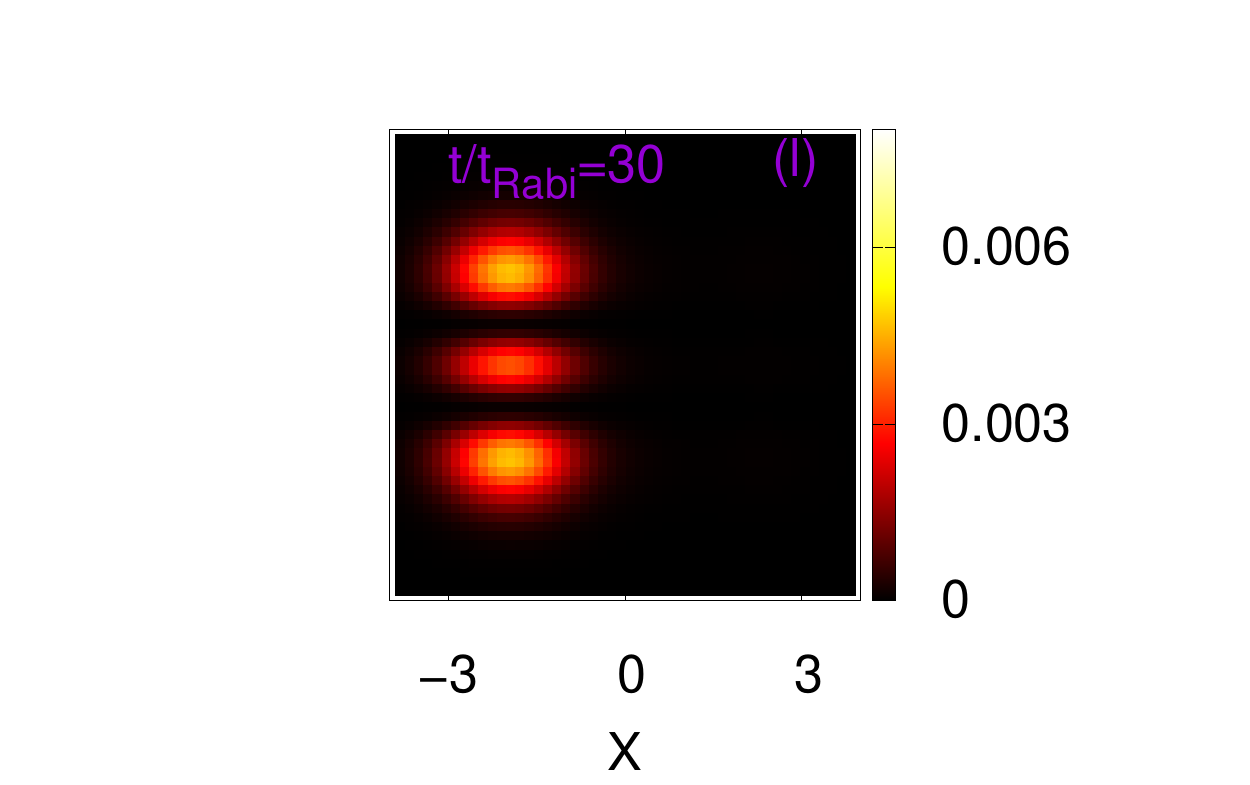}}\\
\caption{Time evolution of  the natural orbital densities, $|\phi_j(\textbf{r})|^2$ where $j=1$, 2, 3, and 4 (row wise),  in a symmetric 2D double-well for $\Psi_{Y}$. The interaction parameter is $\Lambda=0.01$ and the number of bosons is $N=10$.   The MCTDHB computation is performed with  with $M=10$ time-adaptive orbitals. Shown are snapshots  at $t = 10t_{Rabi}$ (first column), $20t_{Rabi}$ (second   column), and $30t_{Rabi}$ (third  column). See the text for more details. The quantities shown are dimensionless.}
\label{fig17}
\end{figure*}

\begin{figure*}[!h]
{\includegraphics[trim = 4.9cm 0.5cm 3.1cm 0.2cm,scale=.75]{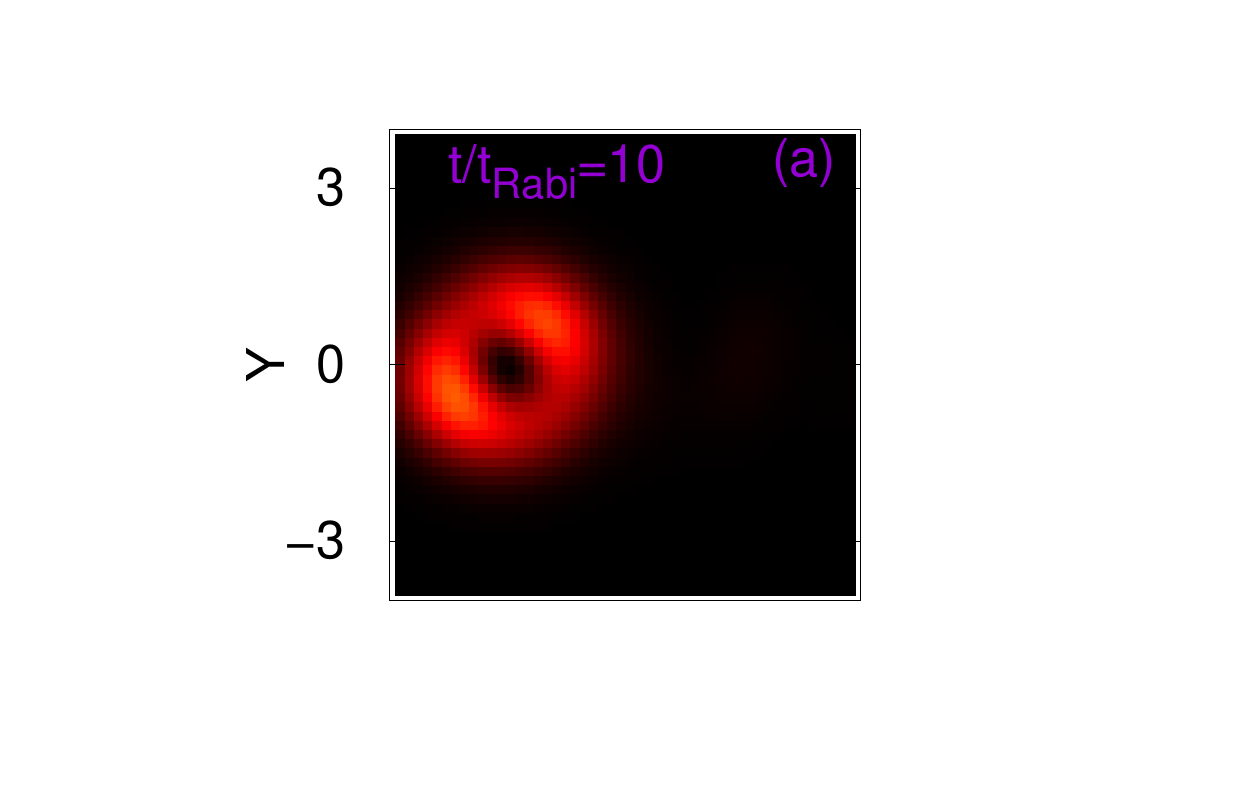}}
{\includegraphics[trim =  4.9cm 0.5cm 3.1cm 0.2cm, scale=.75]{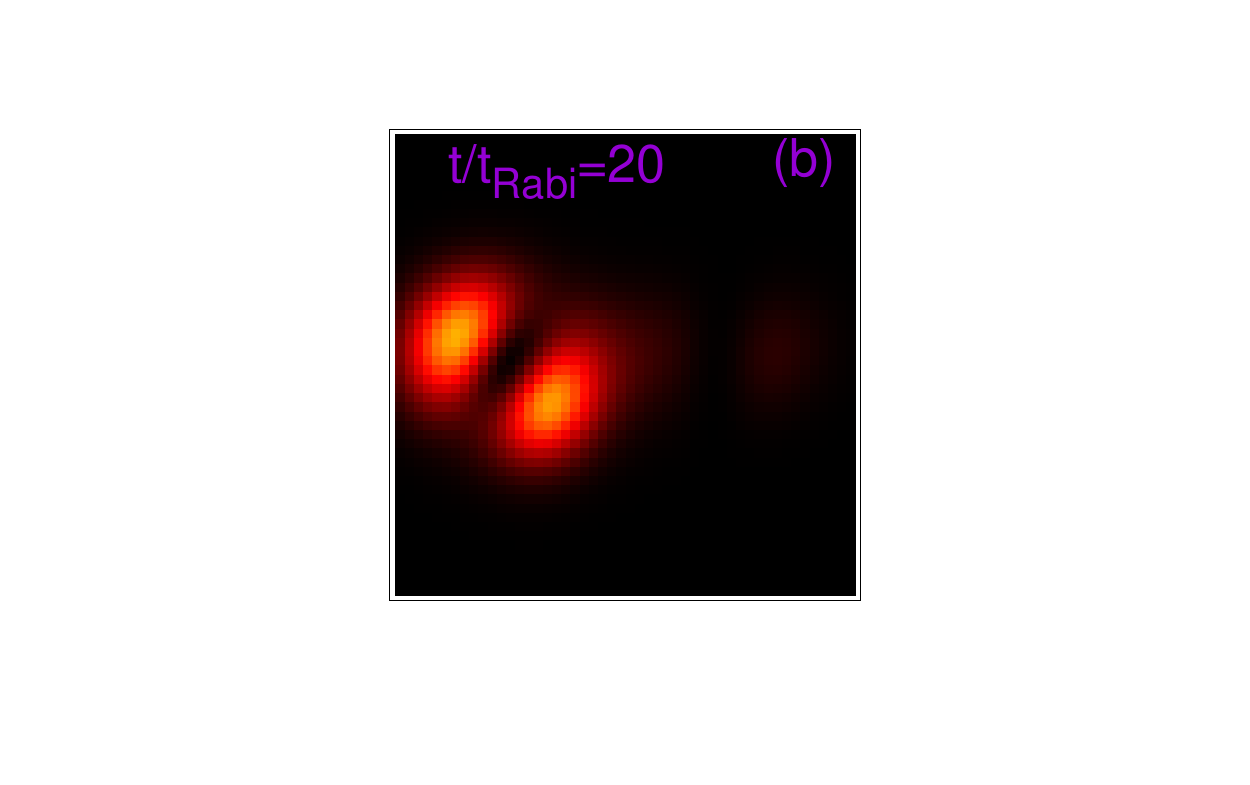}}
{\includegraphics[trim =  4.9cm 0.5cm 3.1cm 0.2cm, scale=.75]{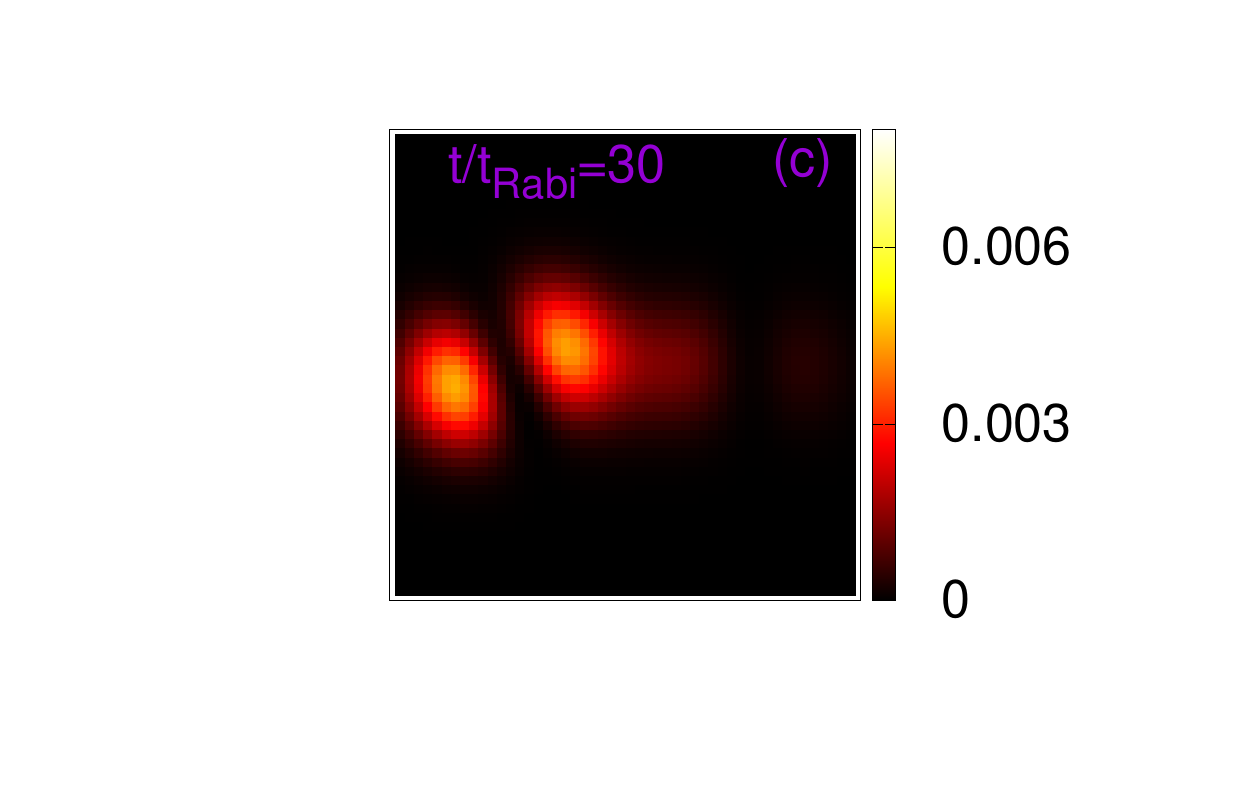}}\\
{\includegraphics[trim = 4.9cm 0.5cm 3.1cm 2.5cm,scale=.75]{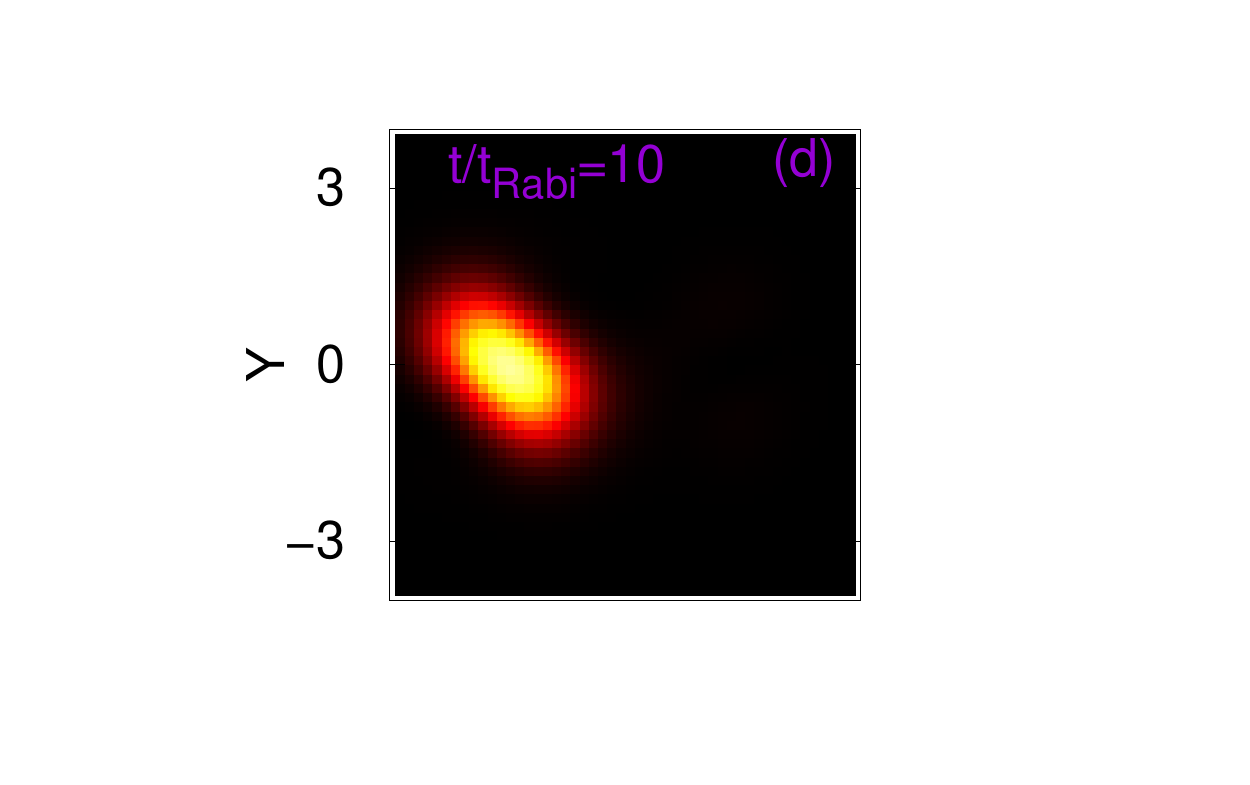}}
{\includegraphics[trim =  4.9cm 0.5cm 3.1cm 2.5cm, scale=.75]{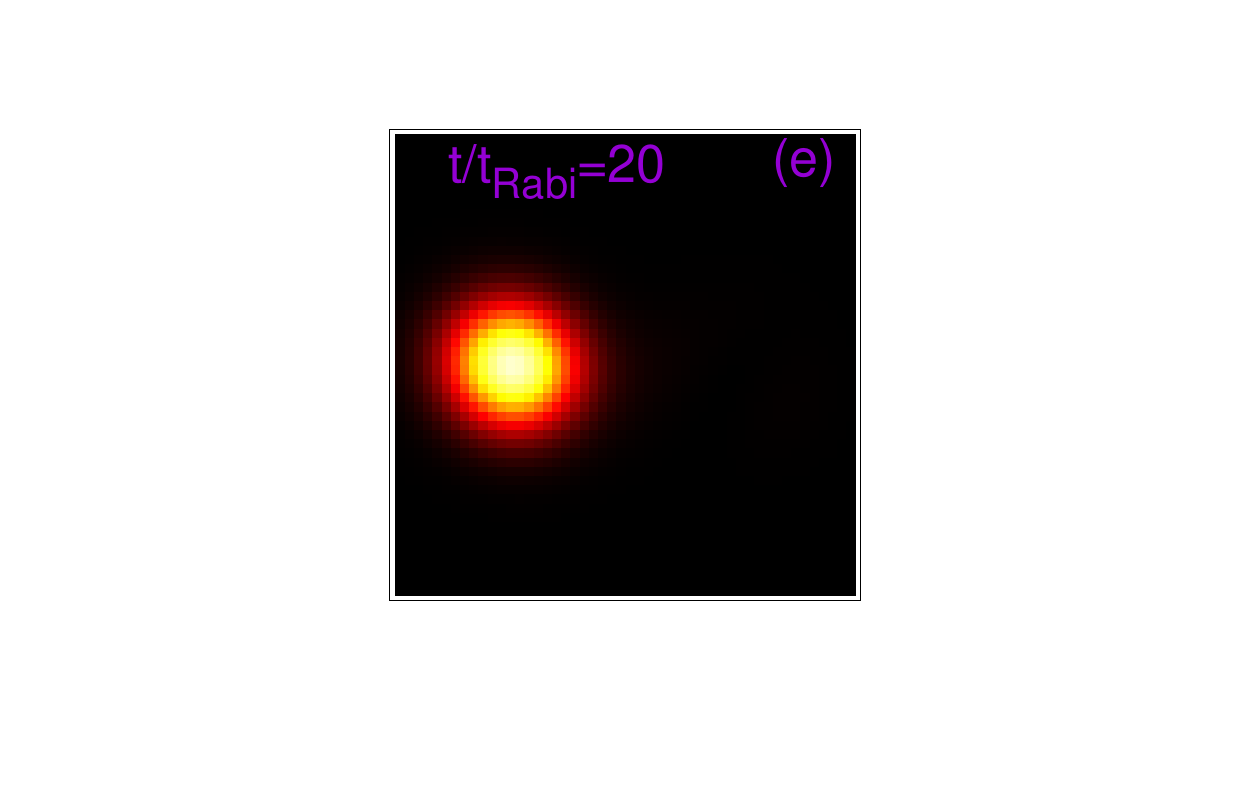}}
{\includegraphics[trim =  4.9cm 0.5cm 3.1cm 2.5cm, scale=.75]{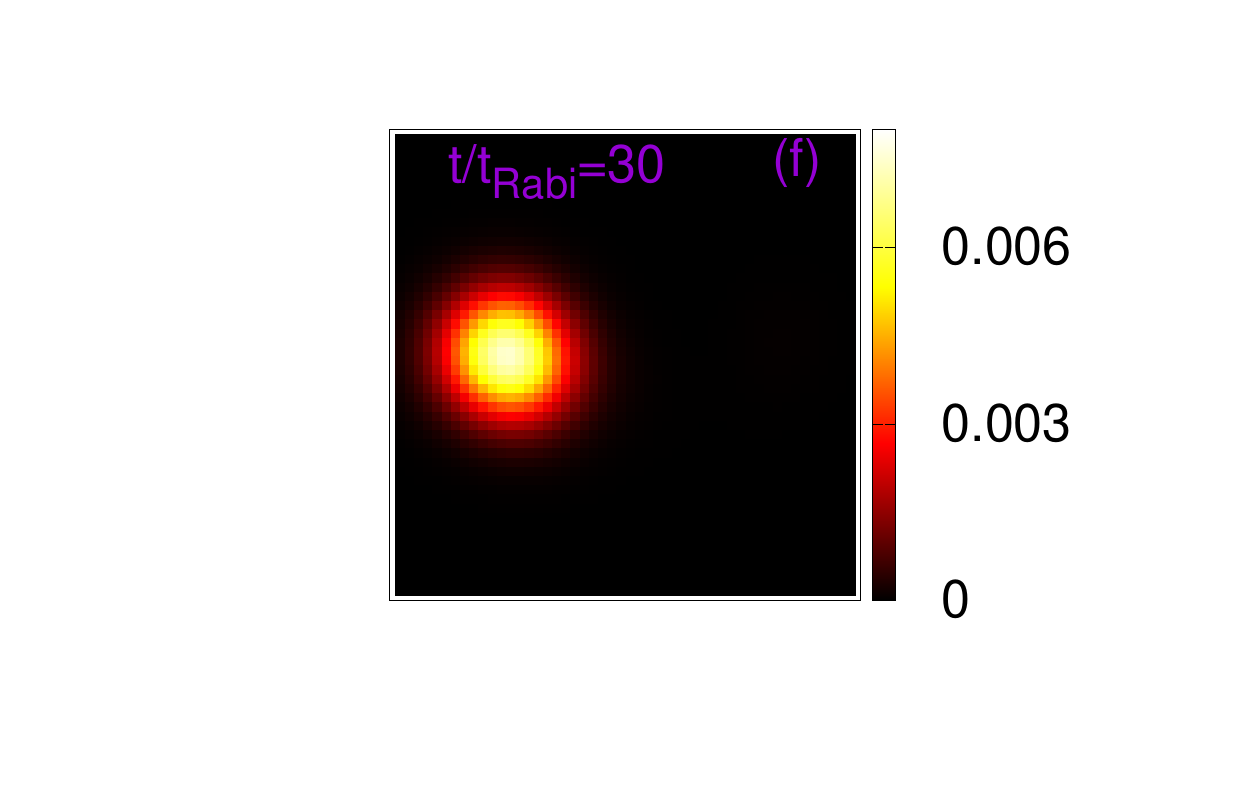}}\\
{\includegraphics[trim = 4.9cm 0.5cm 3.1cm 2.5cm,scale=.75]{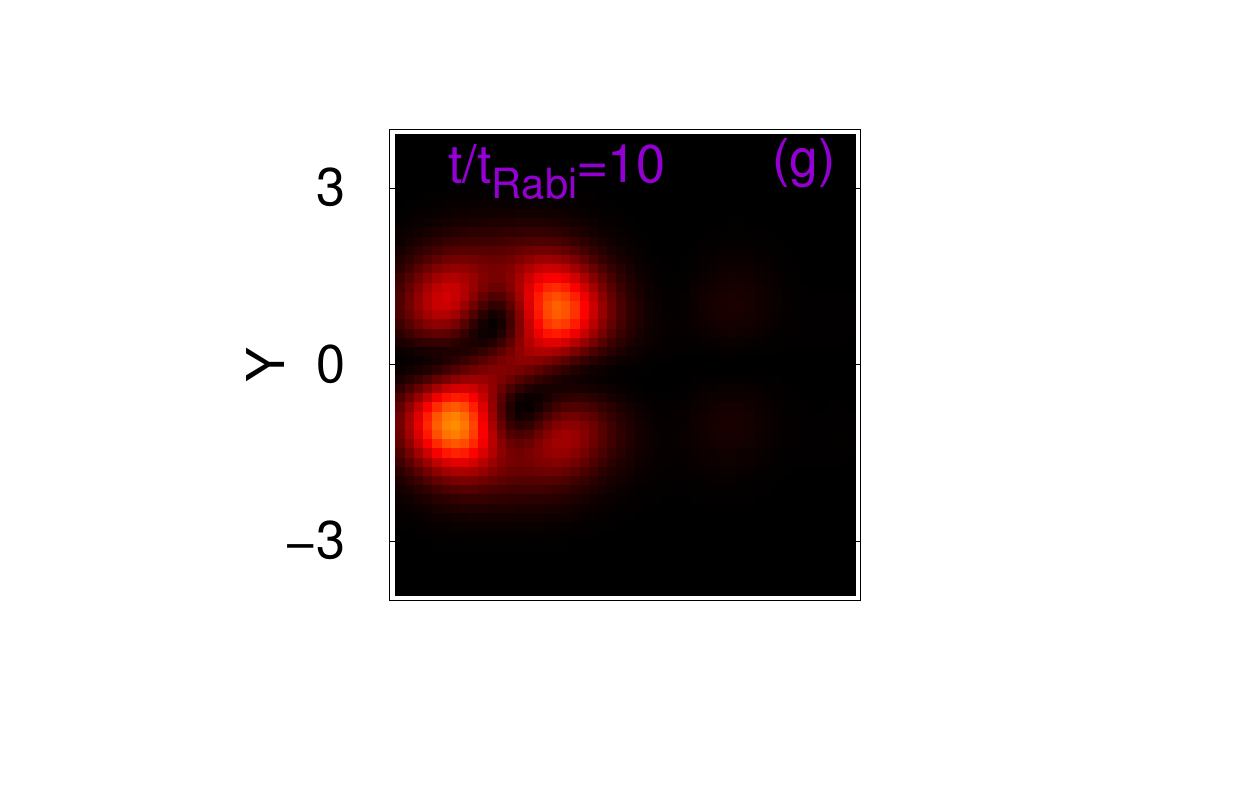}}
{\includegraphics[trim =  4.9cm 0.5cm 3.1cm 2.5cm, scale=.75]{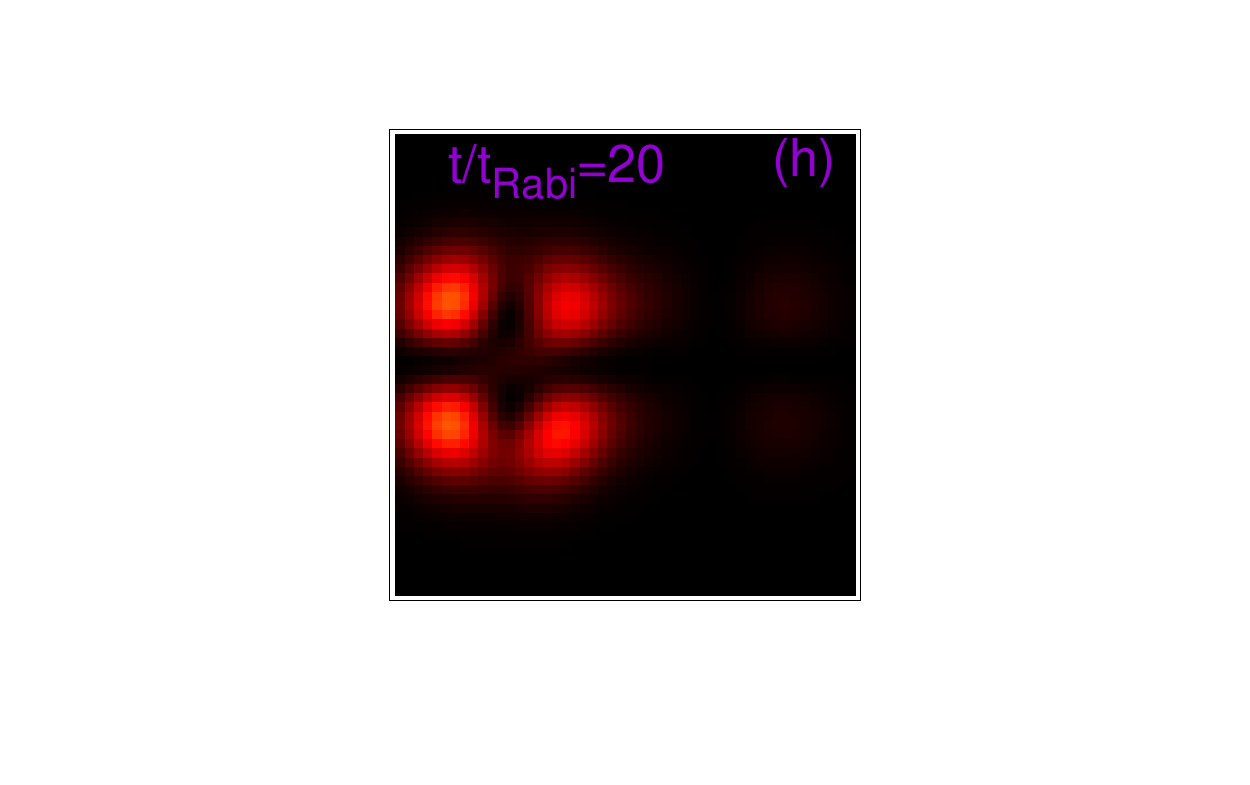}}
{\includegraphics[trim =  4.9cm 0.5cm 3.1cm 2.5cm, scale=.75]{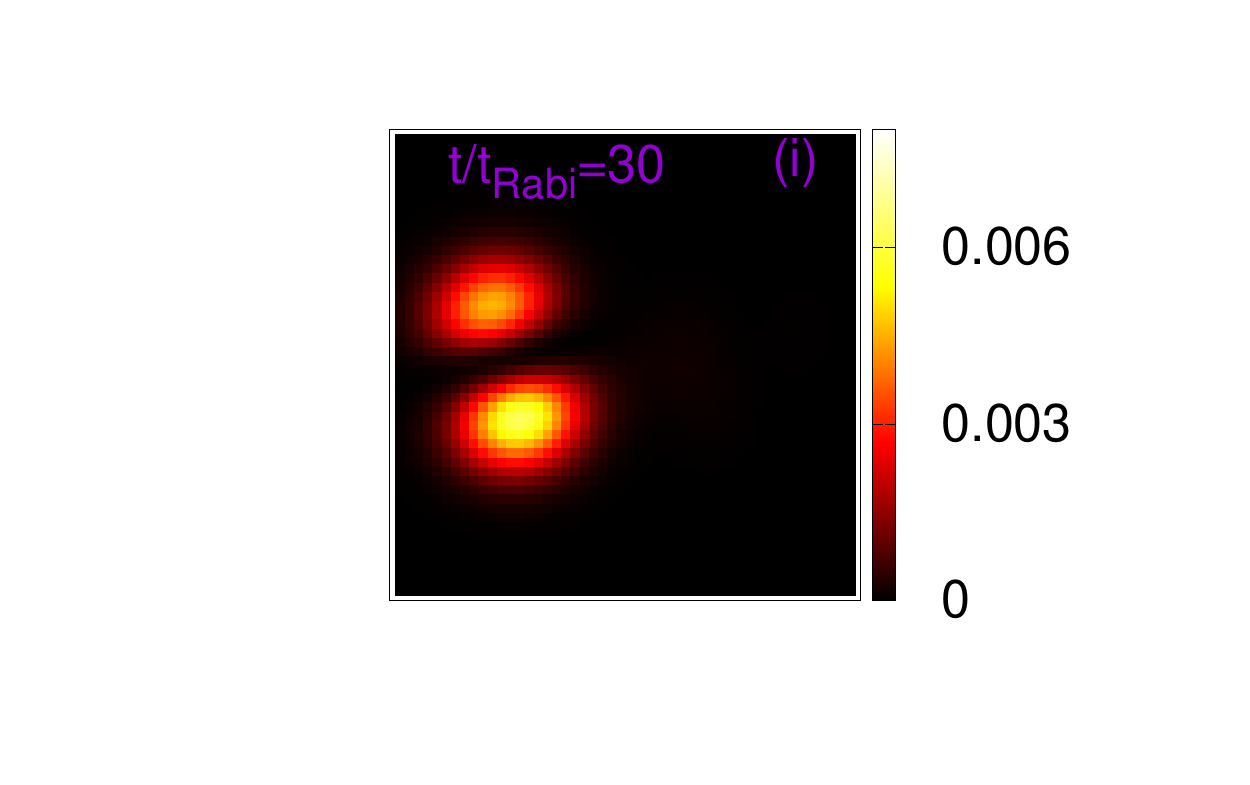}}\\
{\includegraphics[trim = 4.9cm 0.5cm 3.1cm 2.5cm,scale=.75]{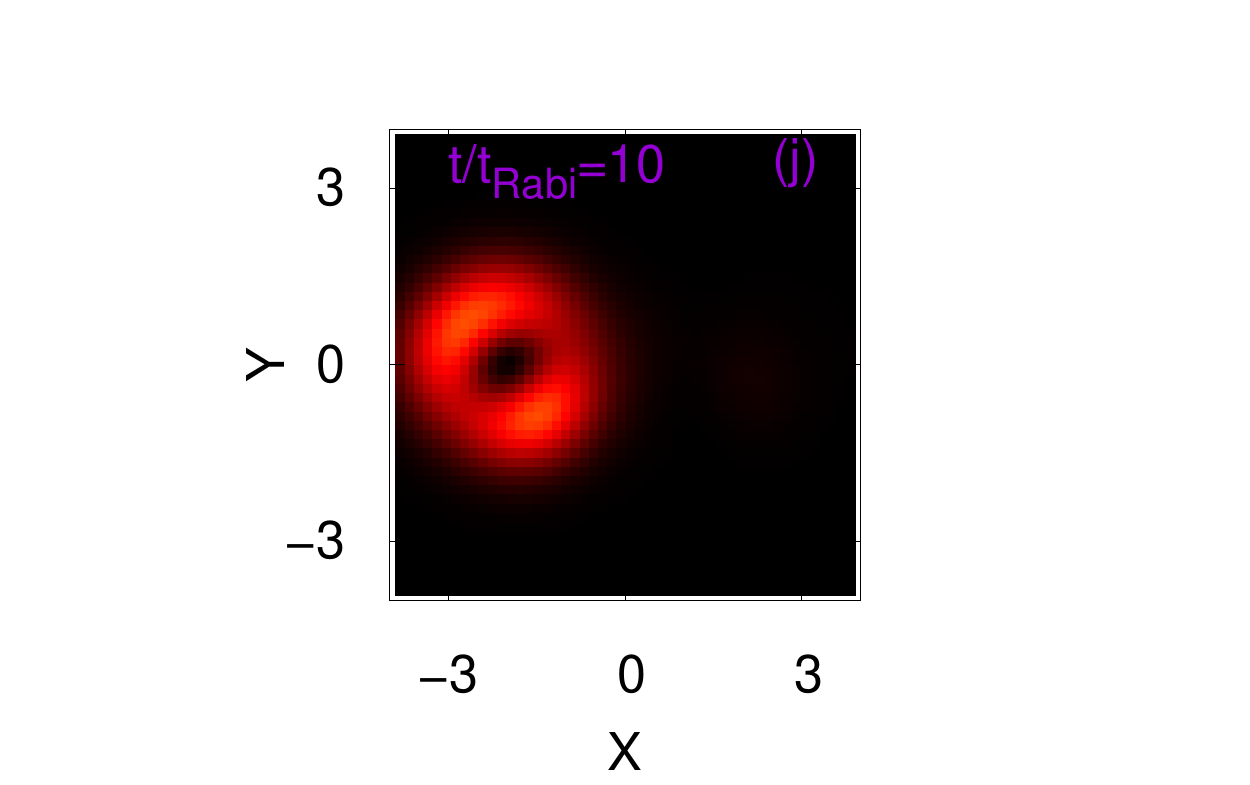}}
{\includegraphics[trim =  4.9cm 0.5cm 3.1cm 2.5cm, scale=.75]{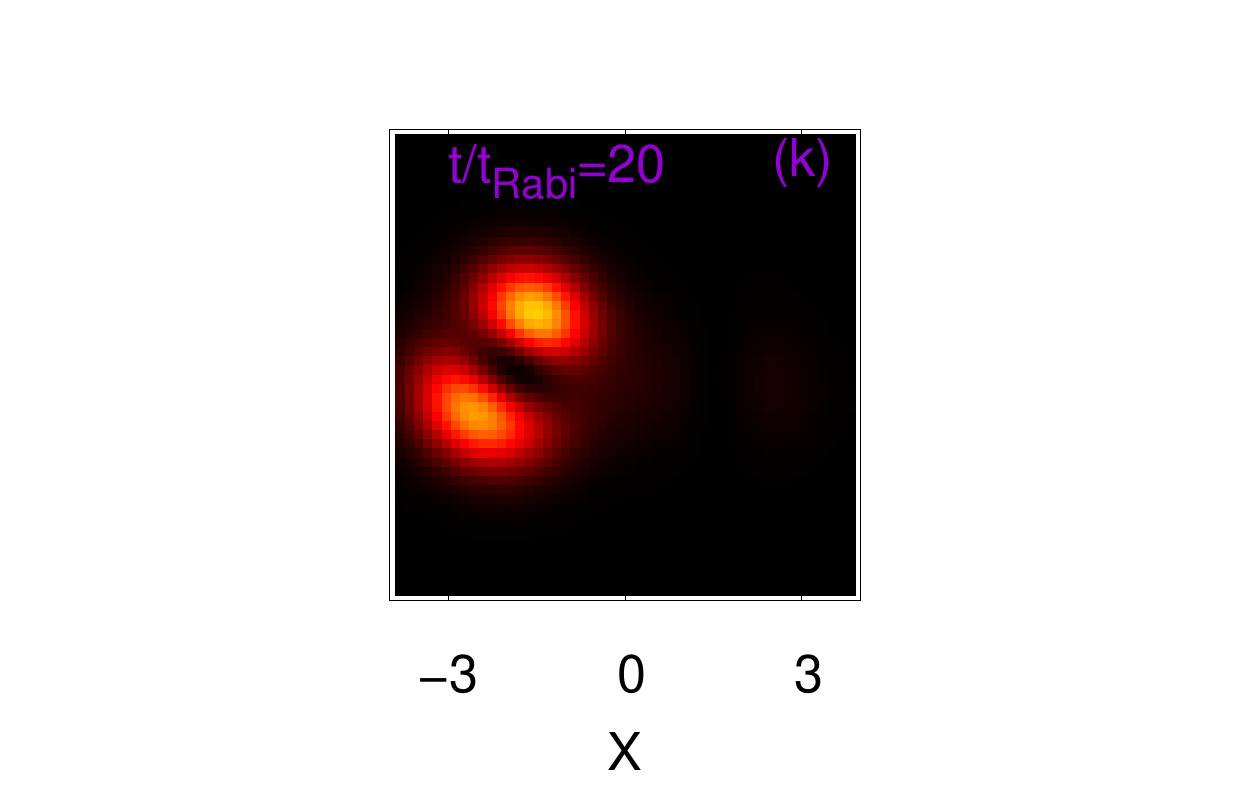}}
{\includegraphics[trim =  4.9cm 0.5cm 3.1cm 2.5cm, scale=.75]{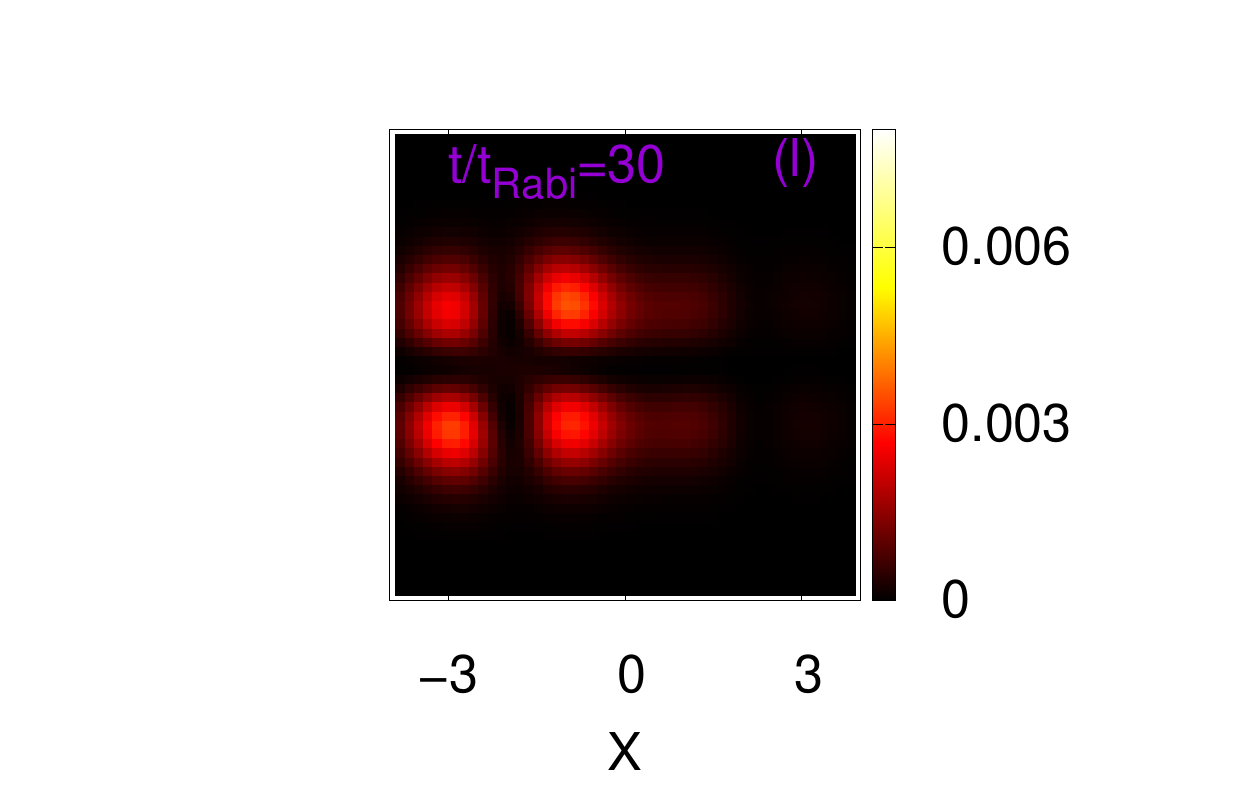}}\\
\caption{Time evolution of  the natural orbital densities, $|\phi_j(\textbf{r})|^2$ where $j=1$, 2, 3, and 4 (row wise),  in a symmetric 2D double-well for $\Psi_{V}$. The interaction parameter is $\Lambda=0.01$ and the number of bosons is $N=10$.   The MCTDHB computation is performed with  with $M=10$ time-adaptive orbitals. Shown are snapshots  at $t = 10t_{Rabi}$ (first column), $20t_{Rabi}$ (second   column), and $30t_{Rabi}$ (third  column). See the text for more details. The quantities shown are dimensionless.}
\label{fig18}
\end{figure*}
\clearpage
\subsection{Convergence with  the number of grid points}
\begin{figure*}[!h]
{\includegraphics[trim = 0.1cm 0.5cm 0.1cm 0.2cm, scale=.60]{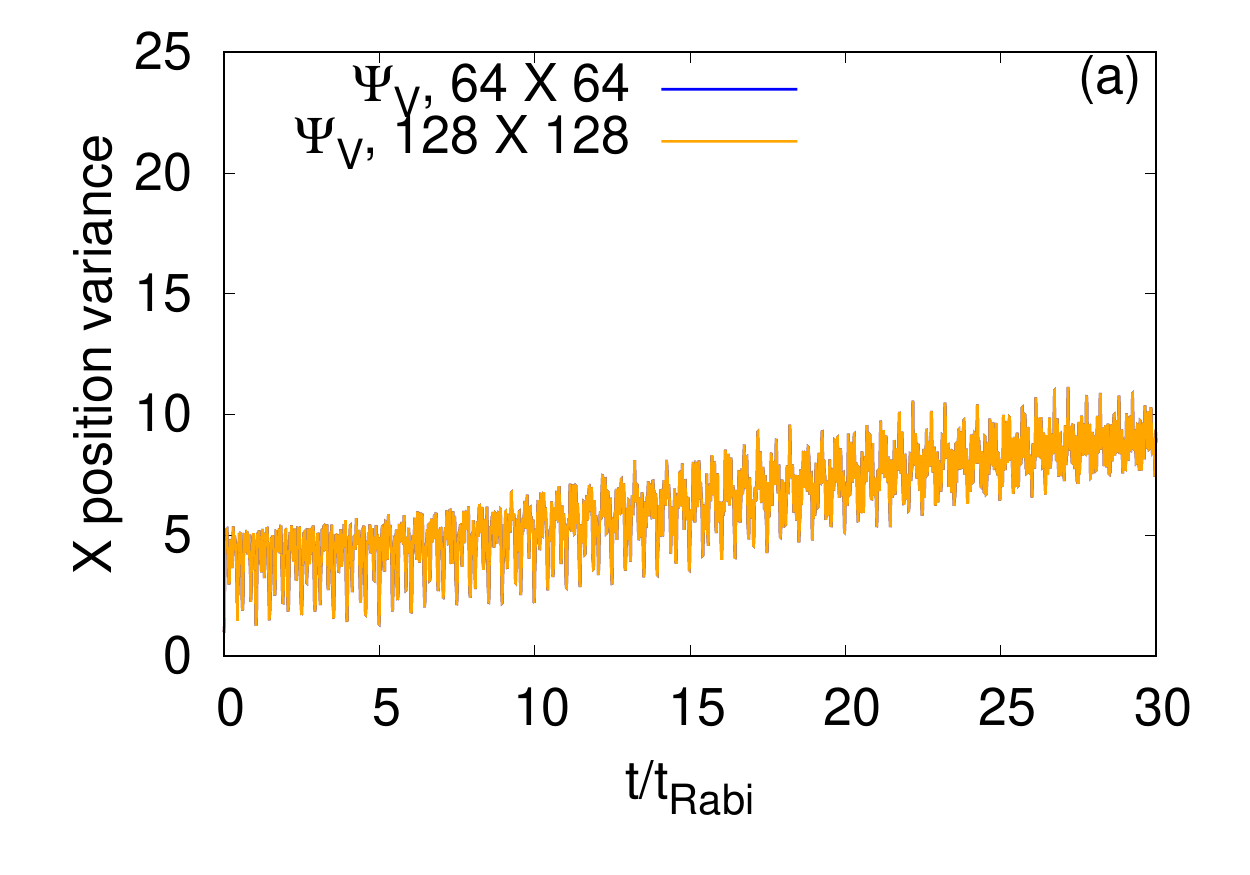}}
{\includegraphics[trim = 0.1cm 0.5cm 0.1cm 0.2cm, scale=.60]{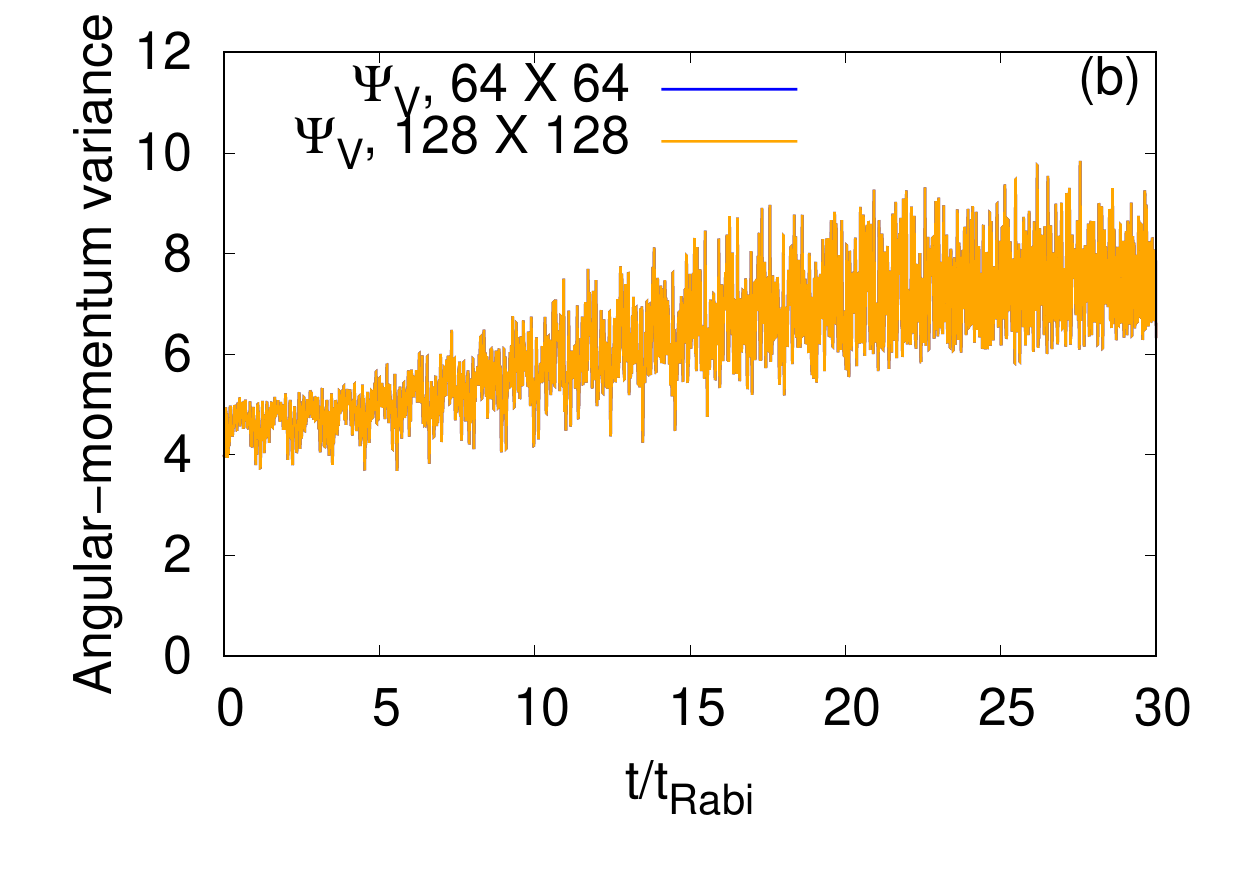}}\\
\caption{\label{fig19}Convergence of the (a)  time-dependent many-body position variance per particle along the $x$-direction, $\dfrac{1}{N}\Delta_{\hat{X}}^2(t)$ and  (b) many-body angular-momentum variance per particle, $\dfrac{1}{N}\Delta_{\hat{L}_Z}^2(t)$, in a symmetric 2D double-well trap with the number of grid points for the vortex state. The number of interacting bosons is $N=10$. The interaction parameter is  $\Lambda=0.01$. The many-body results are computed using the MCTDHB method. The convergence are demonstrated with $64\times 64$ and $128\times 128$ grid points. See the text for more details. The quantities shown  are dimensionless.}
\end{figure*}

In the main text, we have computed all  quantities, survival probability, fragmentation, expectation value, and variance, with $64 \times 64$ grid points. In order to verify the convergence with the grid points, we repeat our computation with $128 \times 128$ grid points for all objects. To demonstrate the convergence with the grid points, we choose two many-body variances which have high sensitivity, $\dfrac{1}{N}\Delta_{\hat{X}}^2(t)$ and $\dfrac{1}{N}\Delta_{\hat{L}_Z}^2(t)$ of the vortex state and plot the results.   Fig.~\ref{fig19} exhibits that increasing the  density of the grid points does not have any visible effect on the results presented in this work which  signifies the convergence with the number of grid points.

\section{Consistency of the initial-state preparation for the dynamics}
\begin{figure*}[!h]
{\includegraphics[trim = 0.1cm 0.5cm 0.1cm 0.2cm, scale=.60]{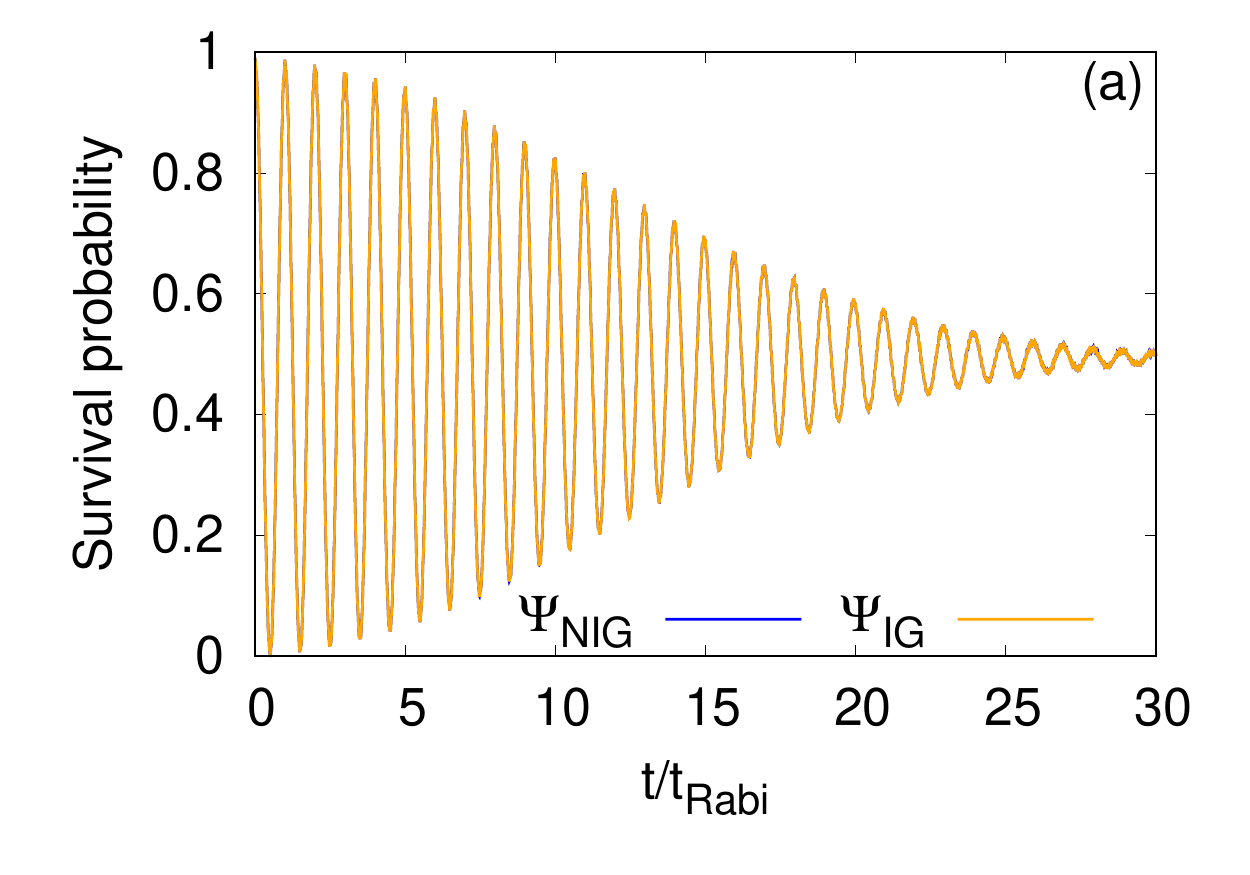}}
{\includegraphics[trim = 0.1cm 0.5cm 0.1cm 0.2cm, scale=.60]{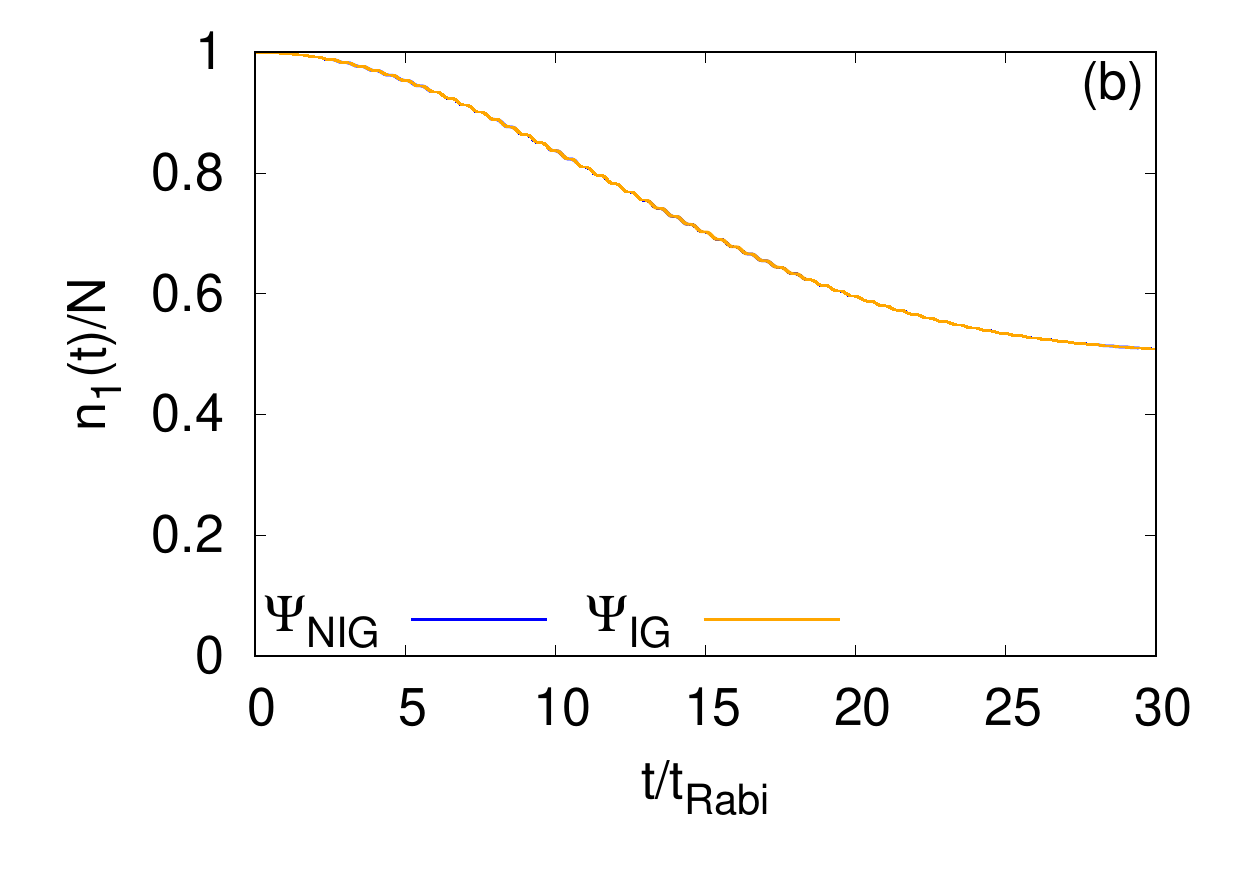}}\\
{\includegraphics[trim = 0.1cm 0.5cm 0.1cm 0.2cm, scale=.60]{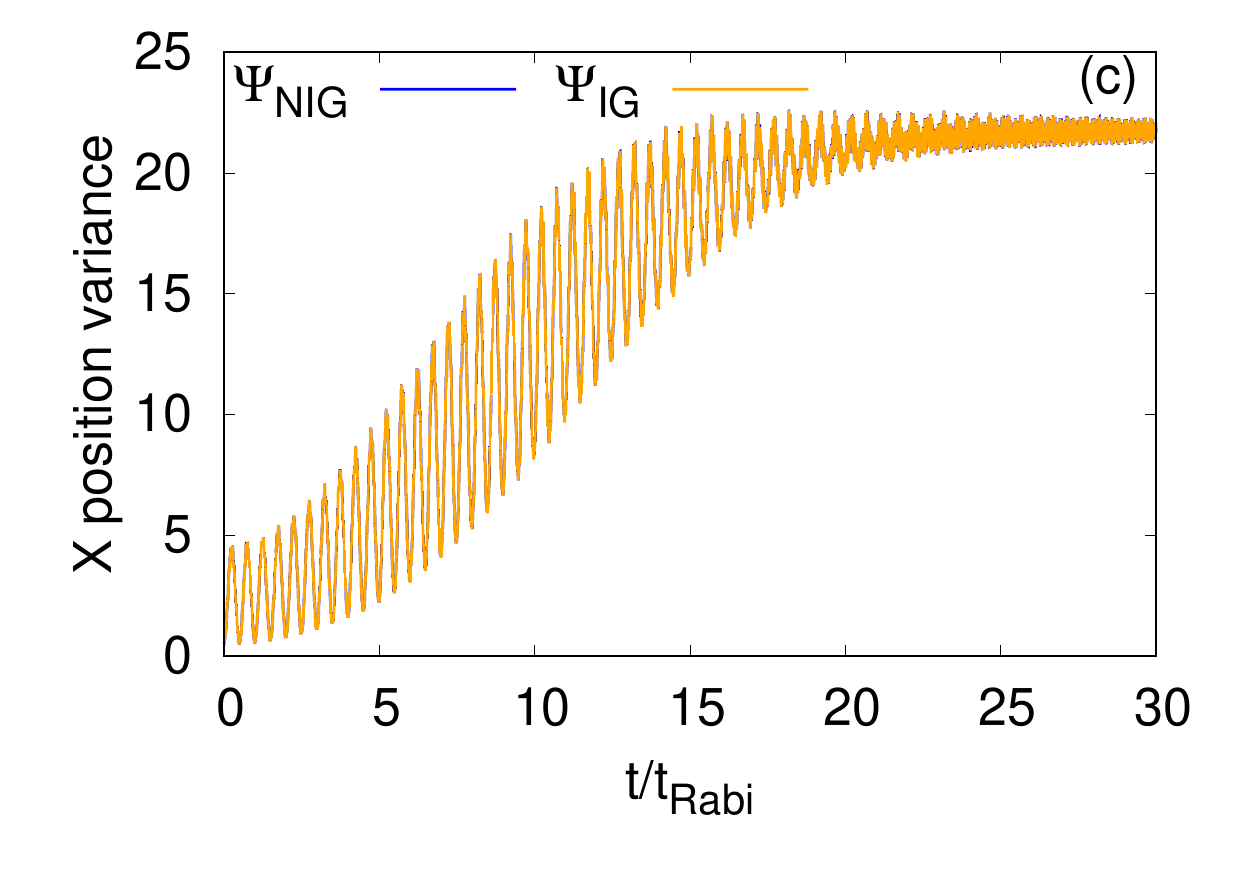}}
{\includegraphics[trim = 0.1cm 0.5cm 0.1cm 0.2cm, scale=.60]{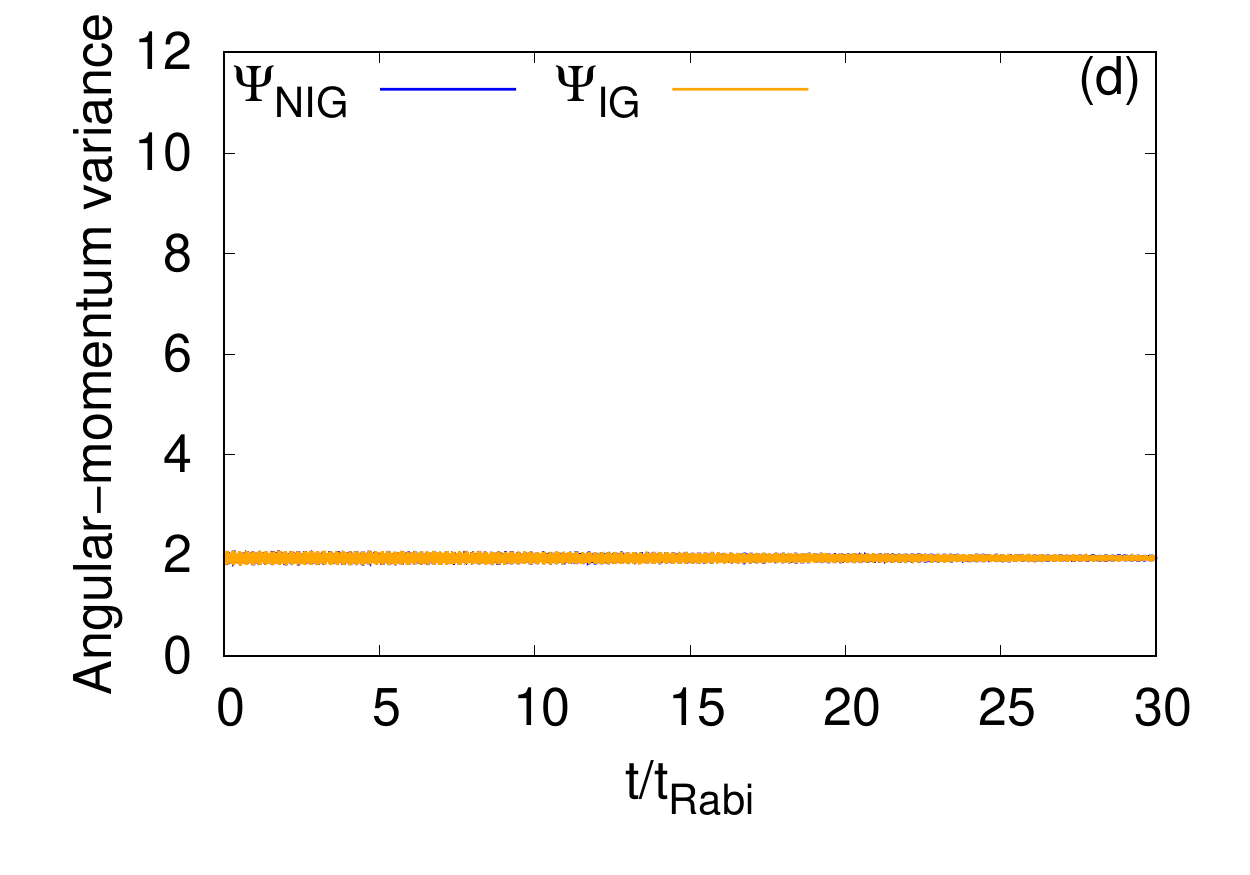}}\\
\caption{\label{fig20}Equivalence of the (a)  survival probability, (b) occupation of the first natural orbital, and variances of the (c) position operator and (d) angular-momentum operator for $M=6$ time-adaptive orbitals of the ground state, prepared in two different ways. Here $\Psi_{\text{NIG}}$ and $\Psi_{\text{IG}}$ refer to the non-interacting and interacting initial ground state, respectively. See the text for more details. The quantities shown  are dimensionless.}
\end{figure*}

Here we  check the consistency of the ground state by preparing it in two different ways. One of the way to obtain the initial ground state in the left well, $V_{L}=\dfrac{1}{2}(x+ 2)^2+\dfrac{1}{2}y^2$, of a symmetric double-well by propagating the MCTDHB equations of motion in imaginary time \cite{Streltsov2007, Lode2020, Alon2008, Grond2009, Grond2011, Streltsov2013, Streltsova2014, Klaiman2014, Fischer2015, Tsatsos2015, Schurer2015, Lode2016, Klaiman2016, Weiner2017, Lode2017, Lode2018, Klaiman2018, Alon2018, Chatterjee2019, Alon2019a, Alon2019b, Bera2019, Lin2020, Package_1, Package_2} with  the mean-field interaction parameter  $\Lambda=0.01$. In order to investigate the real-time tunneling phenomenon of  the ground state, we  suddenly change the  trapping potential to a symmetric double-well $V_T(x,y)$ (see the main text). The wavefuntion of the ground state prepared in this way by quenching only the trapping potential is referred to as $\Psi_{\text{IG}}$ in Fig.~\ref{fig20}. Another way to produce the ground state is considering a non-interacting Gaussian wavefunction in the left well of a symmetric double-well, and simultaneously quench the mean-field interaction from $\Lambda=0$ to $\Lambda=0.01$ and trap potential from $V_L(x,y)$ to $V_T(x,y)$. The wavefunction of the ground state obtained by the later process is termed as $\Psi_{\text{NIG}}$ in Fig.~\ref{fig20}. The second process is applied to produce the ground,  excited and vortex states in the main text. We have computed  the different physical quantities at the many-body level of the ground state  using two different procedures with $M=6$ time-adaptive orbitals. The many-body Hamiltonian is represented by $64\times 64$ exponential discrete-variable-representation grid points in a box size $[-10,10)\times [-10,10)$.

In Fig.~\ref{fig20}, we  demonstrate the numerical equivalence of the long-time survival probability, $P_L(t)$, the loss of coherence  in terms of occupation number of the first natural orbital, $\dfrac{n_1(t)}{N}$, position variance along the $x$-direction,  $\dfrac{1}{N}\Delta_{\hat{X}}^2(t)$, and angular-momentum variance, $\dfrac{1}{N}\Delta_{{\hat{L}_Z}}^2(t)$, of the ground state computed using two different procedures presented above. We have already discussed in the main text that the variance of any quantum operator is much more sensitive to the many-body effects compared to  the survival probabilities and the loss of coherence of the state. Moreover, as the angular-momentum is the combination of the position and momentum,  $\dfrac{1}{N}\Delta_{{\hat{L}_Z}}^2(t)$ is the most sensitive and it requires more numerical resources compared to  the other quantities discussed in this work. Therefore, the convergences of $\dfrac{1}{N}\Delta_{\hat{X}}^2(t)$ and  $\dfrac{1}{N}\Delta_{{\hat{L}_Z}}^2(t)$  automatically imply the convergences of the $P_L(t)$ and $n_1(t)$. 

 Fig.~\ref{fig20} shows the complete overlap of  $P_L(t)$, $\dfrac{n_1(t)}{N}$, $\dfrac{1}{N}\Delta_{\hat{X}}^2(t)$, and $\dfrac{1}{N}\Delta_{{\hat{L}_Z}}^2(t)$  of the ground state computed using two different procedures. As there are very small oscillations  occurring with time for the variances of the position and momentum operators along the $y$-direction for the ground state,  we have not shown them here explicitly. We have verified that the respective curves sit atop each other as well. Based on the consistency of the dynamics when the ground state is prepared in the two different ways, we have further investigated the dynamics of the excited and vortex states in the main text. The long-time dynamics  in Fig.~\ref{fig20} shows that the collapse in $P_L(t)$ is consistent with the equilibration of $\dfrac{n_1(t)}{N}$, $\dfrac{1}{N}\Delta_{\hat{X}}^2(t)$ and $\dfrac{1}{N}\Delta_{{\hat{L}_Z}}^2(t)$.




\begin{thebibliography}{}
\bibitem{Streltsov1995}
Anderson, M. H., Ensher, J. R., Matthews, M. R., Wieman, C. E. \& Cornell, E. A.  Observation of Bose-Einstein condensation in a dilute atomic vapor.  Science \textbf{269}, 198 (1995).
\bibitem{Bradley1995}
Bradley, C. C., Sackett, C. A., Tollett, J. J. \& Hulet, R. G.  Evidence of Bose-Einstein condensation in an atomic gas with attractive interactions. Phys. Rev. Lett. \textbf{75}, 1687 (1995).
\bibitem{Davis1995}
Davis, K. B., Mewes, M. -O., Andrews, M. R., van Druten, N. J., Durfee, D. S., Kurn, D. M. \& Ketterle, W.  Bose-Einstein condensation in a gas of sodium atoms. Phys. Rev. Lett. \textbf{75}, 3969 (1995).

\bibitem{Lewenstein2007}
Lewenstein, M., Sanpera, A., Ahufinger, V., Damski,  B., Sen(De), A. \& Sen, U.  Ultracold atomic gases in optical lattices: mimicking condensed matter physics and beyond. Adv. Phys. \textbf{56}, 243  (2007).

\bibitem{Lewenstein2012}
Lewenstein, M., Sanpera, A. \& Ahufinger, V.  Ultracold Atoms in Optical Lattices: Simulating Quantum Many-Body Systems (Oxford: Oxford University Press) (2012).
\bibitem{Lee2006}
Lee, C.  Adiabatic Mach-Zehnder interferometry on a quantized Bose-Josephson junction. Phys. Rev. Lett. \textbf{97}, 150402 (2006).

\bibitem{Hild2014}
Hild, S.,  Fukuhara, T.,   Schau\ss{}, P.,  Zeiher, J.,   Knap, M.,  Demler, E.,  Bloch, I. \&  Gross, C. Far-from-Equilibrium Spin Transport in Heisenberg Quantum Magnets.  Phys. Rev. Lett. \textbf{113}, 147205 (2014).  

\bibitem{Fukuhara2013}
 Fukuhara, T., Schau\ss{}, P.,  Endres, M.,  Hild, S.,  Cheneau, M., Bloch, I. \&  Gross, C. Microscopic observation of magnon bound states and their dynamics.  Nature (London) \textbf{502}, 76 (2013).  
 
 \bibitem{Buchler2005}
   B\"{u}chler, H. P., Hermele, M., Huber, S. D., Fisher, M. P. A. \& Zoller, P. Atomic Quantum Simulator for Lattice Gauge Theories and Ring Exchange Models. Phys. Rev. Lett. \textbf{95}, 040402 (2005).  
   
\bibitem{Jaksch2002}
  Jaksch, D., Venturi, V., Cirac, J. I., Williams, C. J. \& Zoller, P. Creation of a Molecular Condensate by Dynamically Melting a Mott Insulator.  Phys. Rev. Lett. \textbf{89}, 040402 (2002). 
\bibitem{Ferlaino2011}   
 Ferlaino,  F.,  Zenesini, A.,  Berninger, M.,  Huang, B.,  N\"{a}gerl, H. -C. \&  Grimm, R.  Efimov Resonances in Ultracold Quantum Gases. Few-Body Syst. \textbf{51}, 113 (2011).

\bibitem{Josephson1962} 
 Josephson, B. D. Possible new effects in superconductive tunnelling. Phys. Lett. \textbf{1}, 251 (1962).
\bibitem{Davis2002}
  Davis, J. C.  \&    Packard, R. E. Superfluid $^3$He Josephson weak links. Rev.  Mod.  Phys. \textbf{74},  741 (2002).
\bibitem{Smerzi1997}
Smerzi, A., Fantoni, S., Giovanazzi, S. \& Shenoy, S. R.  Quantum coherent atomic tunneling between two trapped Bose-Einstein condensates. Phys. Rev. Lett. \textbf{79}, 4950  1997
  
 \bibitem{Albiezet2005}
 Albiezet, M.,  Gati, R.,  F\"{o}lling, J.,  Hunsmann, S.,   Cristiani, M. \&  Oberthaler, M. K. Direct Observation of Tunneling and Nonlinear Self-Trapping in a Single Bosonic Josephson Junction. Phys. Rev. Lett. \textbf{95}, 010402 (2005).

\bibitem{Gati2007}
Gati, R. \& Oberthaler, M. K.  A bosonic Josephson junction. J. Phys. B: At. Mol. Opt. Phys. \textbf{40}, R61 (2007).



\bibitem{Dobrzyniecki2016}
Dobrzyniecki, J. \& Sowi\'nski, T.  Exact dynamics of two ultra-cold bosons confined in a one-dimensional double-well potential. Eur. Phys. J. D  \textbf{70}, 83 (2016).
\bibitem{Menotti2001}
Menotti, C., Anglin, J. R., Cirac, J. I. \& Zoller,  P.  Dynamic splitting of a Bose-Einstein condensate Phys. Rev. A  \textbf{63}, 023601 (2001).

\bibitem{Salgueiro2007}
Salgueiro, A. N., de Toledo Piza, A. F. R., Lemos, G. B., Drumond, R., Nemes, M. C. \&  Weidem\"{u}ller, M.  Quantum dynamics of bosons in a double-well potential: Josephson oscillations, self-trapping and ultralong tunneling times. Eur. Phys. J. D \textbf{44}, 537 (2007).


\bibitem{Zollner2008}
Z\"{o}llner, S., Meyer, H. -D. \& Schmelcher, P.  Few-boson dynamics in double wells: from single-atom to correlated pair tunneling.  Phys. Rev. Lett.  \textbf{100}, 040401 (2008).
\bibitem{Carr2010}
Carr, L. D., Dounas-Frazer, D. R. \& Garcia-March, M.  A.  Dynamical realization of macroscopic superposition states of cold bosons in a tilted double well Eurphys. Lett. \textbf{90}, 10005 (2010).
\bibitem{He2012}
He, Q. -Y., Reid, M. D., Opanchuk, B., Polkinghorne, R., Rosales-Z\'arate, L. E. C. \& Drummond, P. D.  Quantum dynamics in ultracold atomic physics. Front. Phys. \textbf{7}, 16 (2012).
\bibitem{Liu2015}
Liu, Y. \& Zhang, Y.  Two atoms in a double well: exact solution with a Bethe ansatz. Phys. Rev. A  \textbf{91}, 053610 (2015).
\bibitem{Dobrzyniecki2018}
Dobrzyniecki, J., Li, X., Nielsen, A. E. B. \& Sowi\'nski, T.  Effective three-body interactions for bosons in a double-well confinement. Phys. Rev. A  \textbf{97}, 013609 (2018).

\bibitem{Dobrzyniecki2018a}
Dobrzyniecki, J. \& Sowi\'nski, T.  Effective two-mode description of a few ultra-cold bosons in a double-well potential. Phys. Lett. A \textbf{382},  394 (2018).
\bibitem{Ferrini2008}
Ferrini, G., Minguzzi, A. \& Hekking, F. W. J.  Number squeezing, quantum fluctuations, and oscillations in mesoscopic Bose Josephson junctions. Phys. Rev. A  \textbf{78}, 023606 (2008).
\bibitem{Jia2008}
Jia, X. Y., Li, W. D. \& Liang, J. Q.  Nonlinear correction to the boson Josephson-junction model. Phys. Rev. A \textbf{78}, 023613 (2008).
\bibitem{Burchinati2017}
Burchinati, A., Fort, C. \& Modugno, M.  Josephson plasma oscillations and the Gross-Pitaevskii equation: bogoliubov approach versus two-mode model. Phys. Rev. A \textbf{95}, 023627 (2017).

\bibitem{Pawlowski2011}
 Paw\l{}owski, K.,  Zi\ifmmode \acute{n}\else \'{n}\fi{}, P., Rza\ifmmode \dot{z}\else \.{z}\fi{}ewski, K. \&  Trippenbach, M. Revivals in an attractive Bose-Einstein condensate in a double-well potential and their decoherence. Phys. Rev. A \textbf{83}, 033606 (2011).
 \bibitem{Griffin2020}
Griffin, A., Nazarenko, S. \& Proment, D. Breaking of Josephson junction oscillations and onset of quantum turbulence in Bose–Einstein condensates. J. Phys. A: Math. Theor. \textbf{53}, 175701 (2020).

\bibitem{Gillet2014}
Gillet, J., Garcia-March, M. A., Busch, T. \& Sols,  F.  Tunneling, self-trapping, and manipulation of higher modes of a Bose-Einstein condensate in a double well. Phys. Rev. A \textbf{89}, 023614 (2014).
\bibitem{Levy2007}
Levy, S., Lahoud, E., Shomroni, I. \& Steinhauer, J.  The a.c. and d.c. Josephson effects in a Bose-Einstein condensate. Nature (London) \textbf{449}, 579 (2007).
\bibitem{Sakmann2009}
Sakmann, K., Streltsov, A. I., Alon, O. E. \& Cederbaum, L. S.  Exact quantum dynamics of a bosonic Josephson junction. Phys. Rev. Lett.  \textbf{103}, 220601 (2009).
\bibitem{Sakmann2010}
Sakmann, K., Streltsov, A. I., Alon, O. E. \& Cederbaum, L. S.  Quantum dynamics of attractive versus repulsive bosonic Josephson junctions: Bose-Hubbard and full-Hamiltonian results. Phys. Rev. A \textbf{82}, 013620 (2010).

\bibitem{Sakmann2014}
Sakmann, K., Streltsov, A. I., Alon, O. E. \& Cederbaum, L. S.  Universality of fragmentation in the Schr\"{o}dinger dynamics of bosonic Josephson junctions. Phys. Rev. A \textbf{89}, 023602 (2014).

\bibitem{Halder2018}
 Haldar, S. K. \&  Alon, O. E.  Impact of the range of the interaction on the quantum dynamics of a bosonic Josephson junction, Chemical Physics  \textbf{509}, 72 (2018).

\bibitem{Halder2019}
Haldar, S. K. \&  Alon, O. E. Many-body quantum dynamics of an asymmetric bosonic Josephson junction.  New J. Phys. \textbf{21}, 103037 (2019).

\bibitem{Klaiman2016}
Klaiman, S., Streltsov, A. I. \&  Alon, O. E.  Uncertainty product of an out-of-equilibrium many-particle system. Phys. Rev. A \textbf{93}, 023605 (2016).

\bibitem{Ananikian2006}
Ananikian, D. \& Bergeman, T.  Gross-Pitaevskii equation for Bose particles in a double-well potential: two-mode models and beyond, Phys. Rev. A \textbf{73}, 013604 (2006).
\bibitem{Spagnolli2017}
Spagnolli, G.,  Semeghini, G.,  Masi, L.,   Ferioli, G.,   Trenkwalder, A.,  Coop, S.,  Landini, M.,  Pezz\`e, L.,  Modugno, G.,  Inguscio, M.,   Smerzi, A. \&  Fattori, M.  Crossing over from attractive to repulsive interactions in a tunneling bosonic Josephson junction. Phys. Rev.  Lett. \textbf{118}, 230403 (2017).

\bibitem{Arovas2008}
 Arovas P. \&  Auerbach, A. Quantum tunneling of vortices in two-dimensional superfluids. Phys. Rev.  B \textbf{78}, 094508 (2008).
 \bibitem{Martin2007}
 Martin, M.,  Scott, R. G. \&  Fromhold, T. M.   Transmission and reflection of Bose-Einstein condensates incident on a Gaussian tunnel barrier.   Phys. Rev. A \textbf{75}, 065602 (2007).
\bibitem{Fialko2012}
 Fialko, O.,  Bradley, A. S. \&  Brand, J. Quantum Tunneling of a Vortex Between Two Pinning Potentials. Phys. Rev. Lett. \textbf{108}, 015301 (2012).

\bibitem{Salgueiro2009}
 Salgueiro, J. R.,  Zacar\'es, M.,  Michinel, H. \&  Ferrando, A. Vortex replication in Bose-Einstein condensates trapped in double-well potentials. Phys. Rev. A \textbf{79}, 033625 (2009).
 \bibitem{March2015}
Garcia-March, M.  A. \& Carr, L. D. Vortex macroscopic superposition in ultracold bosons in a double-well potential. Phys. Rev. A \textbf{91}, 033626 (2015).

\bibitem{Beinke2015}
Beinke, R., Klaiman, S., Cederbaum, L. S., Streltsov, A. I. \&  Alon, O. E.  Many-body tunneling dynamics of Bose-Einstein condensates and vortex states in two spatial dimensions.  Phys. Rev. A \textbf{92}, 043627 (2015).

\bibitem{Wen2010}
 Wen, L., Xiong, H., \& Wu, B. Hidden vortices in a Bose-Einstein condensate in a rotating double-well potential. Phys. Rev. A \textbf{82}, 053627 (2010). 
 \bibitem{Montgomery2010}
  Montgomery, T. W. A.,   Scott, R. G.,   Lesanovsky, I., \&  Fromhold T. M. Spontaneous creation of nonzero-angular-momentum modes in tunnel-coupled two-dimensional degenerate Bose gases. Phys. Rev. A \textbf{81}, 063611 (2010).
  
  \bibitem{Schmiegelow2016}
 Schmiegelow, C. T., Schulz, J., Kaufmann, H., Ruster, T., Poschinger,  U. G.,  \& Schmidt-Kaler,  F.  Transfer  of  optical orbital  angular  momentum  to  a  bound  electron.  Nature Communications \textbf{7}, 12998 (2016). 
 \bibitem{Bhowmik2016}
 Bhowmik, A., Mondal, P. K., Majumder, S., \& Deb, B. Interaction of atom with nonparaxial laguerre-gaussian beam:Forming  superposition  of  vortex  states  in  Bose-Einstein condensates. Phys. Rev. A \textbf{93}, 063852 (2016).
  \bibitem{Bhowmik2018}
 Bhowmik, A. \&  Majumder, S. Tuning of non-paraxial effects of the Laguerre-Gaussian beam interacting with the two-component Bose–Einstein condensates. J. Phys. Commun. \textbf{2}, 125001 (2018).
 
 
 \bibitem{Streltsov2007}
Streltsov, A. I., Alon, O. E. \& Cederbaum, L. S.  Role of excited states in the splitting of a trapped interacting Bose-Einstein condensate by a time-dependent barrier. Phys. Rev. Lett. \textbf{99}, 030402 (2007).
\bibitem{Alon2008}
Alon, O. E., Streltsov, A. I. \& Cederbaum, L. S.  Multiconfigurational time-dependent Hartree method for bosons: many-body dynamics of bosonic systems.  Phys. Rev. A \textbf{77}, 033613 (2008).
 
 \bibitem{Lode2020}
 Lode,  A. U. J., L\'ev\^eque, C.,  Madsen, L. B.,   Streltsov, A. I. \&  Alon, O. E. Colloquium: Multiconfigurational time-dependent Hartree approaches for indistinguishable particles. Rev.  Mod. Phys. \textbf{92}, 011001 (2020). 

\bibitem{Klaiman2015}
 Klaiman, S. \&  Alon, O. E. Variance as a sensitive probe of correlations. Phys. Rev. A \textbf{91}, 063613 (2015).
\bibitem{Klaiman2014}
 Klaiman, S.,  Lode, A. U. J.,  Streltsov, A. I.,  Cederbaum, L. S. \&  Alon, O. E. Breaking the resilience of a two-dimensional Bose-Einstein condensate to fragmentation. Phys. Rev. A  \textbf{90}, 043620 (2014).
\bibitem{Doganov2013}
 Doganov, R. A.,   Klaiman, S.,   Alon, O. E.,   Streltsov, A. I. \&  Cederbaum, L. S. Two trapped particles interacting by a finite-range two-body potential in two spatial dimensions. Phys. Rev. A \textbf{87}, 033631 (2013).
 \bibitem{Christensson2009}
 Christensson, J.,  Forss\'en, C., \AA{}berg, S., \& Reimann, S. M. Effective-interaction approach to the many-boson problem. Phys. Rev. A \textbf{79}, 012707 (2009).

 \bibitem{Beinke2018}
 Beinke, R., Cederbaum, L. S. \&   Alon, O. E. Enhanced many-body effects in the excitation spectrum of a weakly interacting rotating Bose-Einstein condensate. Phys. Rev. A \textbf{98}, 053634 (2018).




\bibitem{Grond2009}
Grond, J., Schmiedmayer, J. \& Hohenester, U.  Optimizing number squeezing when splitting a mesoscopic condensate. Phys. Rev. A \textbf{79}, 021603(R) (2009).
\bibitem{Grond2011}
Grond, J., Betz, T., Hohenester, U., Mauser, N. J., Schmiedmayer, J. \& Schumm, T.  The Shapiro effect in atom chip-based bosonic Josephson junctions. New J. Phys. \textbf{13}, 065026 (2011).
\bibitem{Streltsov2013}
Streltsov, A. I. Quantum systems of ultracold bosons with customized interparticle interactions. Phys. Rev. A \textbf{88}, 041602(R) (2013).
\bibitem{Streltsova2014}
 Streltsova, O. I.,  Alon, O. E., Cederbaum,  L. S. \&  Streltsov, A. I. Generic regimes of quantum many-body dynamics of trapped bosonic systems with strong repulsive interactions. Phys. Rev. A \textbf{89}, 061602(R) (2014).


\bibitem{Fischer2015}
 Fischer, U. R.,  Lode, A. U. J. \&  Chatterjee, B. Condensate fragmentation as a sensitive measure of the quantum many-body behavior of bosons with long-range interactions. Phys. Rev. A \textbf{91}, 063621 (2015).



\bibitem{Tsatsos2015}
 Tsatsos M. C. \&  Lode, A. U. J.  Resonances and dynamical fragmentation in a stirred Bose-Einstein condensate. J. Low Temp. Phys. \textbf{181}, 171 (2015).

\bibitem{Schurer2015}
Schurer, J. M., Negretti, A. \& Schmelcher, P.  Capture dynamics of ultracold atoms in the presence of an impurity ion.  New J. Phys. \textbf{17}, 083024 (2015).
\bibitem{Lode2016}
Lode, A. U. J. \& Bruder, C.  Dynamics of Hubbard Hamiltonians with the multiconfigurational time-dependent Hartree method for indistinguishable particles. Phys. Rev. A \textbf{94} 013616 (2016).

\bibitem{Weiner2017}
 Weiner, S. E.,  Tsatsos, M. C.,  Cederbaum, L. S. \&  Lode, A. U. J. Phantom vortices: hidden angular momentum in ultracold dilute Bose-Einstein condensates. Sci Rep \textbf{7}, 40122 (2017).
\bibitem{Lode2017}
Lode, A. U. J. \& Bruder, C.  Fragmented superradiance of a Bose-Einstein condensate in an optical cavity. Phys. Rev. Lett. \textbf{118}, 013603 (2017).
 
\bibitem{Lode2018}
Lode, A. U. J., Diorico. F. S., Wu, R., Molignini, P.,  Papariello, L., Lin, R., L\'ev\^eque, C., Exl,  L., Tsatsos, M. C., Chitra, R. \& Mauser, N. J. Many-body physics in two-component Bose-Einstein condensates in a cavity: fragmented superradiance and polarization. New J. Phys. \textbf{20},  055006 (2018).

\bibitem{Klaiman2018}
 Klaiman, S.,  Beinke, R.,  Cederbaum, L. S.,  Streltsov, A. I. \&  Alon, O. E.  Variance of an anisotropic Bose-Einstein condensate. Chemical Physics \textbf{509}, 45  (2018).

\bibitem{Alon2018}
Alon, O. E. \& Cederbaum, L. S.   Attractive Bose-Einstein condensates in anharmonic traps: Accurate numerical treatment and the intriguing physics of the variance.  Chemical Physics \textbf{515}, 287 (2018).
\bibitem{Chatterjee2019}
Chatterjee, B., Tsatsos, M. C. \& Lode, A. U. J.  Correlations of strongly interacting one-dimensional ultracold dipolar few-boson systems in optical lattices. New J. Phys. \textbf{21}, 033030 (2019).


\bibitem{Alon2019a}
Alon, O. E.  Condensates in annuli: dimensionality of the variance. Molecular Physics \textbf{117},  2108 (2019).

\bibitem{Alon2019b}
Alon, O. E.  Analysis of a Trapped Bose-Einstein Condensate in Terms of Position, Momentum, and Angular-Momentum Variance. Symmetry \textbf{11}, 1344 (2019).

\bibitem{Bera2019}
Bera, S., Chakrabarti, B., Gammal, A.,  Tsatsos, M. C.,   Lekala, M. L.,   Chatterjee, B.,  L\'ev\^eque, C. \&  Lode, A. U. J.  Sorting Fermionization from Crystallization in Many-Boson Wavefunctions. Sci. Rep. \textbf{9}, 17873 (2019).
\bibitem{Lin2020}
 Lin, R.,  Molignini, P.,  Papariello, L.,  Tsatsos, M. C.,   L\'ev\^{e}que, C.,  Weiner, S. E.,   Fasshauer, E.,   Chitra, R. \&  Lode A. U. J. MCTDH-X: The multiconfigurational time-dependent Hartree method for indistinguishable particles software.  Quantum Sci. Technol. \textbf{5}, 024004  (2020).
 \bibitem{Package_1}
Streltsov, A. I. \& Streltsova, O. I. 2015 MCTDHB-Lab,  version 1.5, 2015 (\href{http://mctdhb-lab.com}{http://mctdhb-lab.com}).
\bibitem{Package_2}
Streltsov, A. I., Cederbaum, L. S.,  Alon, O. E.,  Sakmann, K., Lode, A. U. J., Grond, J.,  Streltsova, O. I.,  Klaiman, S. \& Beinke, R. The Multiconfigurational Time-Dependent Hartree for Bosons Package, Version 3.x,  \href{http://mctdhb.org}{http://mctdhb.org}.
\bibitem{Lode2012}
Lode,  A. U. J.,  Sakmann, K., Alon, O. E.,  Cederbaum, L. S. \& Streltsov, A. I. Numerically exact quantum dynamics of bosons with time-dependent interactions of harmonic type. Phys. Rev. A \textbf{86}, 063606 (2012).

 \bibitem{Erdos2007} 
 Erd\ifmmode \mbox{\H{o}}\else \H{o}\fi{}s,  L.,   Schlein, B. \& Yau, H. -T. Rigorous Derivation of the Gross-Pitaevskii Equation.  Phys. Rev. Lett. \textbf{98}, 040404 (2007).
 \bibitem{Coleman2000}
 Coleman, A. J. \& Yukalov, V. I. Reduced Density Matrices: Coulson's Challenge; Lectures Notes in Chemistry; Springer: Berlin, Germany, 2000; Volume 72.
 \bibitem{Sakmann2008}
 Sakmann, K.,   Streltsov, A. I.,   Alon, O. E. \&  Cederbaum L. S. Reduced density matrices and coherence of trapped interacting bosons. Phys. Rev. A \textbf{78}, 023615 (2008).



\bibitem{Pigneur2018}
Pigneur, M., Berrada, T., Bonneau, M., Schumm, T., Demler, E. \& Schmiedmayer, J.  Relaxation to a phase-locked equilibrium state in a one-dimensional bosonic Josephson junction. Phys. Rev. Lett. \textbf{120}, 173601 (2018).



 


 
  
 
 
\end{thebibliography}

\begin{thebibliography}{}
\bibitem{Alon2019b}
Alon, O. E.  Analysis of a Trapped Bose-Einstein Condensate in Terms of Position, Momentum, and Angular-Momentum Variance. Symmetry \textbf{11}, 1344 (2019).
\bibitem{Lode2020}
Lode, A. U. J.,  L\'ev\^eque, C.,  Madsen, L. B., Streltsov, A. I. \&  Alon, O. E. Colloquium: Multiconfigurational time-dependent Hartree approaches for indistinguishable particles. Rev. Mod. Phys. \textbf{92}, 011001 (2020).

\bibitem{Klaiman2016}
Klaiman, S., Streltsov, A. I. \&  Alon, O. E.  Uncertainty product of an out-of-equilibrium many-particle system. Phys. Rev. A \textbf{93}, 023605 (2016).

\bibitem{Streltsov2007}
Streltsov, A. I., Alon, O. E. \& Cederbaum, L. S.  Role of excited states in the splitting of a trapped interacting Bose-Einstein condensate by a time-dependent barrier. Phys. Rev. Lett. \textbf{99}, 030402 (2007).
\bibitem{Alon2008}
Alon, O. E., Streltsov, A. I. \& Cederbaum, L. S.  Multiconfigurational time-dependent Hartree method for bosons: many-body dynamics of bosonic systems.  Phys. Rev. A \textbf{77}, 033613 (2008).


\bibitem{Grond2009}
Grond, J., Schmiedmayer, J. \& Hohenester, U.  Optimizing number squeezing when splitting a mesoscopic condensate. Phys. Rev. A \textbf{79}, 021603(R) (2009).
\bibitem{Grond2011}
Grond, J., Betz, T., Hohenester, U., Mauser, N. J., Schmiedmayer, J. \& Schumm, T.  The Shapiro effect in atom chip-based bosonic Josephson junctions. New J. Phys. \textbf{13}, 065026 (2011).
\bibitem{Streltsov2013}
Streltsov, A. I. Quantum systems of ultracold bosons with customized interparticle interactions. Phys. Rev. A \textbf{88}, 041602(R) (2013).
\bibitem{Streltsova2014}
 Streltsova, O. I.,  Alon, O. E., Cederbaum,  L. S. \&  Streltsov, A. I. Generic regimes of quantum many-body dynamics of trapped bosonic systems with strong repulsive interactions. Phys. Rev. A \textbf{89}, 061602(R) (2014).
\bibitem{Klaiman2014}
 Klaiman, S.,  Lode, A. U. J.,  Streltsov, A. I.,  Cederbaum, L. S. \&  Alon, O. E. Breaking the resilience of a two-dimensional Bose-Einstein condensate to fragmentation. Phys. Rev. A  \textbf{90}, 043620 (2014).

\bibitem{Fischer2015}
 Fischer, U. R.,  Lode, A. U. J. \&  Chatterjee, B. Condensate fragmentation as a sensitive measure of the quantum many-body behavior of bosons with long-range interactions. Phys. Rev. A \textbf{91}, 063621 (2015).



\bibitem{Tsatsos2015}
 Tsatsos M. C. \&  Lode, A. U. J.  Resonances and dynamical fragmentation in a stirred Bose-Einstein condensate. J. Low Temp. Phys. \textbf{181}, 171 (2015).

\bibitem{Schurer2015}
Schurer, J. M., Negretti, A. \& Schmelcher, P.  Capture dynamics of ultracold atoms in the presence of an impurity ion.  New J. Phys. \textbf{17}, 083024 (2015).
\bibitem{Lode2016}
Lode, A. U. J. \& Bruder, C.  Dynamics of Hubbard Hamiltonians with the multiconfigurational time-dependent Hartree method for indistinguishable particles. Phys. Rev. A \textbf{94} 013616 (2016).

\bibitem{Weiner2017}
 Weiner, S. E.,  Tsatsos, M. C.,  Cederbaum, L. S. \&  Lode, A. U. J. Phantom vortices: hidden angular momentum in ultracold dilute Bose-Einstein condensates. Sci Rep \textbf{7}, 40122 (2017).
\bibitem{Lode2017}
Lode, A. U. J. \& Bruder, C.  Fragmented superradiance of a Bose-Einstein condensate in an optical cavity. Phys. Rev. Lett. \textbf{118}, 013603 (2017).
 
\bibitem{Lode2018}
Lode, A. U. J., Diorico. F. S., Wu, R., Molignini, P.,  Papariello, L., Lin, R., L\'ev\^eque, C., Exl,  L., Tsatsos, M. C., Chitra, R. \& Mauser, N. J. Many-body physics in two-component Bose-Einstein condensates in a cavity: fragmented superradiance and polarization. New J. Phys. \textbf{20},  055006 (2018).

\bibitem{Klaiman2018}
 Klaiman, S.,  Beinke, R.,  Cederbaum, L. S.,  Streltsov, A. I. \&  Alon, O. E.  Variance of an anisotropic Bose-Einstein condensate. Chemical Physics \textbf{509}, 45  (2018).

\bibitem{Alon2018}
Alon, O. E. \& Cederbaum, L. S.   Attractive Bose-Einstein condensates in anharmonic traps: Accurate numerical treatment and the intriguing physics of the variance.  Chemical Physics \textbf{515}, 287 (2018).
\bibitem{Chatterjee2019}
Chatterjee, B., Tsatsos, M. C. \& Lode, A. U. J.  Correlations of strongly interacting one-dimensional ultracold dipolar few-boson systems in optical lattices. New J. Phys. \textbf{21}, 033030 (2019).


\bibitem{Alon2019a}
Alon, O. E.  Condensates in annuli: dimensionality of the variance. Molecular Physics \textbf{117},  2108 (2019).


\bibitem{Bera2019}
Bera, S., Chakrabarti, B., Gammal, A.,  Tsatsos, M. C.,   Lekala, M. L.,   Chatterjee, B.,  L\'ev\^eque, C. \&  Lode, A. U. J.  Sorting Fermionization from Crystallization in Many-Boson Wavefunctions. Sci. Rep. \textbf{9}, 17873 (2019).
\bibitem{Lin2020}
 Lin, R.,  Molignini, P.,  Papariello, L.,  Tsatsos, M. C.,   L\'ev\^{e}que, C.,  Weiner, S. E.,   Fasshauer, E.,   Chitra, R. \&  Lode A. U. J. MCTDH-X: The multiconfigurational time-dependent Hartree method for indistinguishable particles software.  Quantum Sci. Technol. \textbf{5}, 024004  (2020).
 
  \bibitem{Package_1}
Streltsov, A. I. \& Streltsova, O. I. 2015 MCTDHB-Lab,  version 1.5, 2015 (\href{http://mctdhb-lab.com}{http://mctdhb-lab.com}).
\bibitem{Package_2}
Streltsov, A. I., Cederbaum, L. S.,  Alon, O. E.,  Sakmann, K., Lode, A. U. J., Grond, J.,  Streltsova, O. I.,  Klaiman, S. \& Beinke, R. The Multiconfigurational Time-Dependent Hartree for Bosons Package, Version 3.x,  \href{http://mctdhb.org}{http://mctdhb.org}.
\end{thebibliography}

\end{document}